\documentclass[journal]{IEEEtran}
\usepackage{verbatim}
\usepackage{tikz}
\usetikzlibrary{spy}
\usetikzlibrary{arrows,shapes}
\usepackage{cite}
\usepackage{amsmath,amssymb,latexsym,bm}
\usepackage{paralist}
\usepackage{graphicx}
\usepackage{float}
\usepackage{algorithmic}
\usepackage{algorithm}
\usepackage{cases}
\usepackage{subfigure}
\usepackage{caption2}
\usepackage{epstopdf}
\usepackage{makecell,multirow,diagbox}
\usepackage{mathrsfs}



\usepackage{xcolor}
\usepackage{tikz}
\usetikzlibrary{spy}
\usepackage{wrapfig}
\usetikzlibrary{arrows}
\tikzstyle{block}=[draw opacity=0.7,line width=1.4cm]
\tikzstyle{process} = [rectangle, minimum width=3cm, minimum height=1cm, text centered, draw=black, fill = yellow!50]

\usepackage[colorlinks=true]{hyperref}
\hypersetup{urlcolor=blue, citecolor=red,linkcolor=blue,}

\begin{document}
	\title{Deep Learning with Adaptive Hyper-parameters for Low-Dose CT Image Reconstruction}
	\author{Qiaoqiao Ding, Yuesong Nan, Hao Gao, and Hui Ji%
		\IEEEcompsocitemizethanks{
			\IEEEcompsocthanksitem Q. Ding\IEEEauthorrefmark{1} (e-mail: matding@nus.edu.sg), Y. Nan\IEEEauthorrefmark{1} (e-mail: matny@nus.edu.sg), and H. Ji\IEEEauthorrefmark{1} (e-mail: matjh@nus.edu.sg) are with Department  of Mathematics, National University of Singapore. 
			119076, SINGAPORE
			\IEEEcompsocthanksitem  H. Gao\IEEEauthorrefmark{2} (e-mail: hao.gao.2012@gmail.com)  is with  Department of Radiation Oncology, Winship Cancer Institute of Emory University,
			30322, USA		 
		}
	}
	\maketitle
	\thispagestyle{empty}
	
	\begin{abstract}
		Low-dose CT (LDCT) imaging is preferred in many applications to reduce the object's exposure to X-ray radiation. In recent years, one promising approach to image reconstruction in LDCT is the so-called optimization-unrolling-based deep learning approach, which replaces pre-defined image prior by learnable  adaptive prior in some model-based iterative image reconstruction  scheme (MBIR).  
		While it is known that setting appropriate hyper-parameters in MBIR is challenging yet important to the reconstruction quality, it does not receive enough attention in the development of deep learning methods. 	
		This paper proposed a deep learning method for LDCT reconstruction that unrolls a half-quadratic splitting scheme. The proposed method not only introduces learnable image prior  built on framelet filter bank, but also
		learns a network that automatically adjusts the  hyper-parameters to fit noise  level and the data for processing. As a result, only one universal model needs to be trained in our method to process the data taken under different dose levels.
		Experimental evaluation on clinical patient dataset showed that the proposed method outperformed both conventional and deep-learning-based solutions by a large margin.
	\end{abstract}

	\section{Introduction}
	\IEEEPARstart{X}{-ray} Computed Tomography (CT) is to provide high-resolution three-dimensional (3D) images of  internal anatomical structures using X-ray scanning and computational tomographic imaging techniques. As  excessive exposure to radiation from X-ray CT scanner may increase the risk of damage to issue,  many techniques have been developed to use less radiation than conventional CT scan, including decreasing the number of projection views~\cite{sidky2006accurate} and  lowering the X-ray tube current~\cite{whiting06pop}. The latter is the so-called  low-dose CT (LDCT). As a result, the signal-to-noise ratio (SNR)  of the measurements collected   is much lower than that of conventional CT, which has a negative impact on the quality of the reconstructed image.
	
	Given measurements from LDCT with low SNR, the images reconstructed by conventional filtered back-projection (FBP) method  are often  of poor quality with noticeable artifacts. In such a setting, the  so-called  model-based iterative reconstruction (MBIR) methods \cite{thibault2007three,nuyts13mtp} are more widely used for image reconstruction. In MBIR, image reconstruction is usually formulated as  an optimization problem regularized by certain functional for suppressing noise magnification during reconstruction. Such a regularization term is derived from some  pre-defined prior on the image. In the past, many different priors have been proposed  for MBIR-based  reconstruction, \emph{e.g.} total variation (TV)~\cite{zhang2005total,sidky2008image,chen2008prior}, wavelet-based sparsity prior~\cite{jia2011gpu,gao20124d}, nonlocal sparsity prior \cite{jia20104d}, and low-rank-based patch prior~\cite{gao2011robust,gao2011multi,cai2014cine,chen2015synchronized}. The resulting optimization problems from these MBIR-based methods usually require some iterative scheme to find an approximate solution. Instead of using some man-crafted prior for regularization, there are also some works that learn image prior for regularization, \emph{e.g.,} dictionary learning~\cite{xu2012low} and sparsifying transform learning~\cite{bai2017z}. In recent years,  there has been a rapid progress on the development of deep learning methods for image reconstruction in LDCT. Most existing methods, \emph{e.g.}~\cite{gupta2018cnn,he2018optimizing,adler2018learned}, are based on the so-called optimization unrolling, which replaces the step involving the image prior by a learnable function built on a deep neural network (NN).   The performance of these optimization-unrolling-based deep learning methods is very promising  in  LDCT image reconstruction.
	
	For MBIR, there are usually several hyper-parameters involved in the iterative scheme. It is known that in addition to the choice of image prior, the setting of  these hyper-parameters also plays an important role in the reconstruction; see~\cite{ramani2012regularization,hansen1992analysis,xu2012low}. In other words, how to automate the selection of appropriate hyper-parameters is
	an important yet challenging task in MBIR. The optimal values of these hyper-parameters  depend on many factors~\cite{conn1991globally,jin2012iterative,hansen1992analysis,golub1979generalized,ramani2012regularization,wang2019global}, including:
	\vspace{0.1cm}
	\begin{itemize}
		\item SNR of measurement data;
		\item Content of target image for reconstruction;
		\item Global convergence property and fast convergence rate.
	\end{itemize}
   \vspace{0.1cm}
	Based on these factors, there has been an enduring effort on developing automatic selection strategies of hyper-parameters for the classic non-learning-based regularization methods, \emph{e.g.} \cite{conn1991globally,jin2012iterative,golub1979generalized,ramani2012regularization,bai2017z,wang2019global}.

	 Similar to classic MBIR-based approaches, there are also several hyper-parameters involved in existing optimization-unrolling-based deep learning methods, which are either
	manually tuned-up~\cite{gupta2018cnn,mardani2017recurrent} or treated as one part of network weights to be learned during the training~\cite{he2018optimizing,sun2016deep,adler2018learned}. As a result, the setting of these hyper-parameter is optimized only for one specific noise level of measurement data. When processing the data with different noise levels, they need to train different networks w.r.t. different noise levels. Such a practice can be inconvenient and sometimes difficult in practical usage when noise level of data is unknown. 
	Furthermore,  these hyper-parameters are fixed as constants for different measurements which correspond to the images with different contents.  Such a constant treatment on hyper-parameters in existing deep learning methods~\cite{wu2017iterative,gupta2018cnn} certainly is not optimal. Motivated by the practical value of processing data with unknown noise level and the need for better image quality in LDCT imaging, this paper aims at developing an adaptive and automated hyper-parameter selection mechanism in deep-learning-based image reconstruction methods, as well as other refinements for further performance improvement.
	
	In this paper, we proposed an image reconstruction method in LDCT whose image prior and hyper-parameter selection are collaboratively learned by two NNs. The proposed method is based on the unrolling of the so-called half-quadratic splitting method \cite{geman1995nonlinear}. Specifically, the proposed NN is composed by  $K$ stages, which corresponds to ${K}$ outer iterations in the optimization method. Each stage contains two blocks:
	\begin{itemize}
		\item {\bf Inversion block} reconstructs an image using both the  measurement and the  estimate from the previous stage, whose hyper-parameters are predicted by a multi-layer perception neural network (MLP). 
		\item {\bf De-noising block} removes the artifacts of the estimate passed from the inverse block by a convolutional neural network (CNN).
	\end{itemize}
	The de-noising block, same as many existing approaches, is implemented using a CNN. However, the design of inversion block is less attentive yet crucial. In a nutshell, we are focusing on the construction of a powerful inversion block which enables the adaption to both noise level and image contents. The main difference is highlighted as follows:

		\begin{itemize}
		\item Hyper-parameters involved inner-loop is predicted by a learnable NN for the adaptivity of both dose levels and image contents.
		\item Fine-grained high-pass filter bank is adopted instead of often-seen gradient operator or learnable filter bank for better sub-band reconstruction and image content adaption.
	\end{itemize}
	Extensive experimental studies showed these designs not only bring performance advantage in LDCT image reconstruction, in compared with existing methods by a large margin, but also allows one to train a single NN model to process the data with unknown dose level.


		\subsection{Related work}
	Due to space limitation, we only discuss  most related methods for LDCT image reconstruction, \emph{e.g.}, deep-learning-based methods. One type of such methods
	uses deep NN to post-process the images reconstructed from some existing methods. By treating the artifacts as noise in the reconstructed images, these methods
	train an NN-based image denoiser to remove the artifacts from images for better quality.
	The denoising NN is trained by using the pairs of the image reconstructed from LDCT and its counterpart from conventional CT. Different NN architectures have been proposed for  such a denoiser, \emph{e.g.}, CNN \cite{chen2017lowS}, residual encoder CNN \cite{chen2017low}, residual network \cite{han2016deep,li2017low}, U-net \cite{han2016deep,jin2017deep}, GAN \cite{wolterink2017generative,yang2018low}, multi-resolution deep convolutional framelets neural network~\cite{ye2018deep}. Instead of directly denoising images,
	Kang \emph{et al.}~\cite{kang2017deep,kang2017wavelet} proposed to denoise wavelet transform coefficients of the images using an NN-based denoiser. Overall, in such a post-processing approach, deep learning does not get involved in the reconstruction and data consistence is omitted. 
	
	In recent years, a more popular deep-learning-based  approach  is built on optimization unrolling, which introduces NNs in  the iterations of the MBIR by replacing image-prior-based operations with the function modeled using learnable NN~\cite{chen2018learn}. In other words, the pre-defined image prior is replaced by a learnable prior using NN.
	There are two ways to train the NN. One is plugging a pre-trained NN into the iterations of some MBIR; see \emph{e.g.}~\cite{wu2017iterative,gupta2018cnn}. 
	The other is training the end-to-end NNs together with the MBIR; see \emph{e.g.}~\cite{sun2016deep,adler2018learned,he2018optimizing,pelt2013fast,hauptmann2018model,bazrafkan2019deep,aggarwal2018modl,gilton2019neumann}.

	In addition to the training scheme, another main difference among different optimization-unrolling-based deep learning methods lies in what numerical solver is chosen for unrolling. Based on the alternating direction method of multipliers (ADMM), the ADMM-net is proposed in \cite{sun2016deep} for image reconstruction in compressed-sensing-based Magnetic Resonance Imaging (MRI).
	For consistent CT image reconstruction, Gupta \emph{et al.}~\cite{gupta2018cnn} proposed to unroll the proximal gradient methods with CNN-based learnable prior.
	For LDCT image reconstruction, the ADMM is used in \cite{he2018optimizing} and the  primal-dual hybrid gradient (PDHG) method is used in \cite{adler2018learned} for unrolling with learnable image prior.
	
	Similar to most iterative schemes of MBIR, how to set appropriate  hyper-parameters is very crucial to the quality of reconstructed images in these optimization-unrolling-based deep learning methods; see \emph{e.g.}~ \cite{gupta2018cnn,wu2017iterative}.  Different from conventional methods, there are few systematic studies that address the problem of hyper-parameter setting in deep learning methods for image reconstruction. 
	
	In existing deep learning methods, the hyper-parameters are either manually tuned-up~\cite{gupta2018cnn,wu2017iterative} or treated as a part of learnable NN parameters~\cite{he2018optimizing,adler2018learned}. Both treatments are sub-optimal. The former  can only take a few trials on the hyper-parameters, since training the NN is a very time-consuming process. The later usually cannot find optimal values for hyper-parameters either, as these hyper-parameters are treated the same as millions of other NN weights in the optimization. In addition, for measurement with different dose levels, these deep learning methods need to train different models to fit a specific noise level for optimal performance. There will be a performance hit if only one model is trained for processing the data with different dose levels.

	\section{Measurement Model and problem formulation}
	\label{Measurement Model}
	In CT imaging with a mono-energetic source, the projection measurements from CT scan follow the Poisson distribution~\cite{ding2018statistical}, which can be expressed as: 
	\begin{eqnarray}
	\bar{\bm{y}}_i\sim {\rm{Poisson}}\{I_i\exp(-[\bm{A}\bm{\bm{x}}]_i)\}+ \mathcal{N}(0,\sigma_e^2),
	\label{elemodel}
	\end{eqnarray}
	where  $\bar{\bm{y}}=[y_i]_{i=1}^{N_d}$ represents the vector of measured projections.
	$I_i$ represents the incident X-ray intensity incorporating X-ray source illumination and the detector efficiency.
	$\bm{A}=[a_{i,j}]_{i,j}$ is the $N_d \times N_p$ system matrix, and $\bm{x}=[x_j]_{j=1}^{N_p}$ denotes the attenuation map.
	 The quantity $[\bm{A}\bm{x}]_i=\sum_{j=1}^{N_p}a_{ij}x_{j}$ refers to the line integral of the attenuation map $\bm{x}$ along the $i$-th X-ray.
	 $\mathcal{N}$ refers to normal distribution, and the quantity $\sigma_e^2$ denotes the variance of the background electronic
	 noise which is considered to be  stable for a
	 commercial CT scanner and  has been converted to photon units \cite{LinComparison}.

%

	It is noted that the noise level varies for different target images. Given a target image $\bm{x}$, its noise level is controlled by $I_i$,  \emph{i.e.}, the measure data is corrupted with noise which becomes larger when   dose level $I_i$ decreases.
	However, even with a fixed dose level $I_i$, for different patients or different parts of human-body, their corresponding noise levels are  different too.
	
	To reconstruct the attenuation map $\bm x$, one can first run the correction and the logarithm  transform on the noisy measurements $\bar{\bm{y}}$ to generate the so-called 
	sinogram $\bm{y}=[y_i]_{i=1}^{N_d}$, whose relation to $\bm{A}\bm{x}$ is often expressed as:
	\begin{equation}\label{eqn:prob}
	\bm{y}=\bm{A}\bm{x}+\epsilon,
	\end{equation}
	where $\epsilon= \mathcal{N}(0,\sigma^2)$ denotes the noise. 
	The linear system  (\ref{eqn:prob}) is ill-posed, and certain regularization needs to be imposed to resolve the solution ambiguity and suppress noise magnification. Let $p(\bm{x};\lambda)$ denotes the prior distribution function of $\bm{x}$ with distribution parameter $\lambda$. The maximum a posterior (MAP) estimation of $\bm{x}$ is then the minimum of the  cost function given by
	\begin{equation}\label{eqn:model}
	\min_{\bm{x}}\frac 1 2\|\bm{A}\bm{x}-\bm{y}\|^2_{\sigma^2}- \log p(\bm{x};\lambda), 
	\end{equation}
	where the first is data fidelity term and the second is  regularization term on $\bm{x}$ derived from its prior distribution. 
	The model~(\ref{eqn:model}) is often called  penalized weighted least-squares (PWLS)	image reconstruction model\cite{thibault2007three,sauer1993a,fessler2000statistical}, which is a  widely used one in CT image reconstruction.
	
	Many regularization terms $-\log p(\bm x;\lambda)$ have been proposed for LDCT image reconstruction, \emph{e.g.}   TV \cite{sidky2008image},   nonlocal TV \cite{jia20104d} and framelet~\cite{jia2011gpu}. These regularization functionals are usually not directly defined in image domain, but in the domain of image gradients or their generalizations. For example,
	both the TV and wavelet-transform-based regularizations
	take the form of $\lambda\|\Gamma \bm x\|_1$, where $\Gamma$ is the gradient operator $\nabla$ (TV) or wavelet transform $W$ (wavelet).
	The prior distribution parameter $\lambda$ is also called the regularization parameter, which needs to be set in advance.
	
	There are many numerical scheme for solving (\ref{eqn:model}) with  $-\log(p(\bm x;\lambda))=\lambda \rho(\Gamma \bm x)$, \emph{e.g.}  ADMM and PDHG. In this paper, our work is based on the half-quadratic splitting method \cite{geman1995nonlinear}, which solve the problem (\ref{eqn:model}) by introducing an auxiliary variable $\bm  z$ with following two steps: 
\begin{subequations}
	\label{HQ}
	\begin{align}
		\label{HQ1}
		& \textbf{Inversion:}~ \bm{x}^k=\arg\min_{\bm{x}}\frac{1}{\sigma^2}\|\bm{A}\bm{x}-\bm{y}\|_{2}^2+\frac 1 {\sigma_1^k}\|
		\Gamma \bm x-\bm{z}^k\|_2^2, \\
		\label{HQ2}
		& \textbf{Denoising:}~\bm{z}^{k+1}=\arg\min_{\bm{z}}  \lambda \rho(\bm z) +\frac 1 {\sigma_1^k}\|\Gamma \bm{x}^k-\bm{z}\|_2^2,
	\end{align}
\end{subequations}
\noindent where $\{\sigma_1^k\}_k$ is the parameter sequences of the algorithm. 
The noise variance $\sigma$, the hyper-parameters $\{\sigma_1^k\}_k$, and the regularization parameter $\lambda$, will make a noticeable impact on  the convergence behavior and the quality of the result.

There are two blocks in the iterative scheme (\ref{HQ}). The first is an inversion block (\ref{HQ1}), which reconstructs an image from the measurement $\bm y$ and the estimate  from the previous iteration. The second is a denoising block (\ref{HQ2}), which refines the estimate from the inverse block using the the prior-based regularization. The performance of the scheme $(\ref{HQ})$ depends on the answers to the following two questions.
\begin{enumerate}
	\item  What prior $p(\bm x;\lambda)$ fits the target image $\bm x$ well?
	\item  What values of the hyper-parameters are optimal for estimating  $\bm x$?
\end{enumerate}
Most existing optimization-unrolling-based methods  focus on the replacement of the denoising block (\ref{HQ2}) using a NN-based denoiser to address the first problem.
However, the second problem has not attracted much attention in spite of the importance of the hyper-parameters\cite{gupta2018cnn}. 
The current treatment on the setting of hyper-parameters is done by  either manual selection or learning them as one part of  NN weights.

This paper aims at designing a deep learning method for LDCT reconstruction that not only uses a CNN-based denoiser to replace (\ref{HQ2}), but also introduces a MAP-based predictor in the inversion block (\ref{HQ1}) to automate the adaptive setting of hyper-parameters to the SNR of measurement data and the target image $\bm x$. 
The proposed deep learning method not only provides very competitive performance for LDCT image reconstruction, but also is universal such that  a single trained NN model can be used for the reconstruction of CT measurements with varying dose levels.

\section{Method}
\label{method}
In this section, we give a detailed discussion on the proposed deep learning method for LDCT reconstruction.
\vspace{-10pt} 
\subsection{Detailed  algorithm of the iterative scheme (\ref{HQ})}
\label{method,model}
Recall that each iteration of the scheme (\ref{HQ}) has two blocks: inversion block and denoising block.
For the inversion block, the problem has an analytical solution given by
$$
\bm{x}^k=  (\bm A^\top\bm A+\frac{\sigma^2} {\sigma_1^k} \Gamma^\top \Gamma)^{-1}(\bm A^\top \bm y+ \frac{\sigma^2} {\sigma_1^k}\Gamma^\top \bm  z^k). 
$$
In conventional methods such as TV method, the operator $\Gamma$ is image gradient operator $\nabla$ that convolves the image $\bm x$ by two filters: $[1,-1]$ and 
$[1,-1]^\top$.
Motivated by the performance improvement of framelet transform over $\nabla$  in many image recovery tasks, we propose to use the filter bank of  spline framelet transform \cite{dong2010mra}. In our implementation, the filter bank of linear B-spline framelet transform is adopted, which contains totally 8 2D high-pass filters 
\begin{eqnarray}
\{\bm{f}_i\}_{i=1}^8:=\{h_{k_1}h_{k_2}^\top\}_{0\leq k_1,k_2\leq2}\setminus\{h_0h_0^\top\},
\end{eqnarray}
composed by the tensor product of  three 1D filters:
\begin{eqnarray}
h_0=[\frac{1}{4},\frac{1}{2},\frac{1}{4}]^\top,h_1=[-\frac{1}{4},\frac{1}{2},\frac{1}{4}]^\top,h_2=[\frac{\sqrt{2}}{4},0,-\frac{\sqrt{2}}{4}]^\top.
\end{eqnarray}
It can be seen that such a filter bank is composed of 
more  fine-grained 2D filters on image gradients with various difference orders and along different directions. Then, the iterative scheme (\ref{HQ}) can be re-formulated  as 
\begin{align}
&\bm{x}^{k}=\arg\min_{\bm{x}}\|\bm{A}\bm{x}-\bm{y}\|_2^2+\sum_{i=1}^L\beta^k_i\|\bm{f}_i\otimes\bm{x}-\bm{z}^k_i\|_2^2, \label{sub1}\\
&\bm{z}_i^{k+1}=\arg\min_{\bm{z}_i}\|\bm{f}_i\otimes\bm{x}^k-\bm{z}_i\|_2^2+\alpha^k\rho(\bm{z}_i),\,\,1\leq i\leq L,\label{sub2}
\end{align}
where $\beta^k_i=\frac{\sigma^2}{\sigma^k_i}$, $\alpha^k_i=\frac{1}{\lambda_i\sigma^k_i}$.
Again,	 the inversion block (\ref{sub1}) has an analytical solution expressed by
\begin{equation}
\bm{x}^{k}=\left(\bm{A}^\top\bm{A}+\sum_{i=1}^L\beta_i^k\bm{F}_i^\top\bm{F}_i\right)^{-1}\left(\bm{A}^\top\bm{y}+\sum_{i=1}^{L}\beta_i^k\bm{F}_i^\top\bm{z}_i^k\right),
\label{solution1}
\end{equation}
where $\bm F_i$ denote the matrix form of the convolution operator with the filter $\bm f_i$.

In short, the inversion block (\ref{sub1}) reconstructs $\bm x$ from measurement data $\bm y$ and the estimate of $\{\bm z_i\}_{i=1}^L$ in high-pass channels using the least squares estimator, and the denoising block (\ref{sub2}) refines the estimate of $\{\bm z_i\}_{i=1}^L$ in high-frequency channels by treating the artifacts as noise and suppressing it via imposed regularization.

\subsection{Inversion block with an MLP-based adaptive predictor for hyper-parameters}
\label{Inversion}
Recall that in the analytical solution of the inversion block~(\ref{solution1}), there is a sequence of hyper-parameters 
$\{\beta_i^k\}_{i=1}^L$. The setting of such a sequence makes significant impact on the intermediate output, which in turn makes noticeable impact on the quality of the final result. In this section, we present a NN-based solution to  predict $\{\beta_i^k\}_{i=1}^L$ for optimal performance.

Indeed, the step (\ref{sub1}) can be interpreted as an MAP estimator under the assumption  that both $\bm{y}$ and $\{\bm z_i\}_{i=1}^L$ are the measurements of $\bm A\bm x$ and $\bm F_i\bm x$ corrupted by 
additive  Gaussian white noise:
\begin{subequations}\label{eqn:noise}
	\begin{numcases}
	{} \bm{y}-\bm{A}\bm{x}=\bm{n}\sim \mathcal{N}(0,\sigma^2), \\
	\bm{z}_i^k-\bm{F}_i\bm{x}=\bm{\epsilon}_i^k\sim \mathcal{N}(0,(\sigma_i^k)^2), \,\,1\leq i\leq L.
	\end{numcases}
\end{subequations}
\noindent In such a simplified case, one can have an explicit  solution to the hyper-parameters: $\beta^k_i:=(\sigma_i^k)^{-2}\sigma^2.$
Such an observation is also utilized in \cite{zheng2017sparse} for automating the setting of hyper-parameters. 

In above formula, the ground truth $\bm{x}$ is unknown, which hinders the estimation of hyper-parameters sequence $\{\beta_i^k\}$. Our observation is that for
unrolling-based framework, the estimated intermediate reconstruction results $\bm{x}^{k-1}$
can serve as surrogates for unknown ground truth $\bm{x}$ in stage $k$. Mathematically, such approximation
can be done via
\begin{subequations}
	\begin{numcases}
	{} \bm{n}=\bm{y}-\bm{A}\bm{x}^{k-1}+\bm{A}\bm{\delta}^{k-1}, \\
	\bm{\epsilon}_i^k=\bm{z}_i^k-\bm{F}_i\bm{x}^{k-1}+\bm{F}_i\bm{\delta}^{k-1}, \,\,1\leq i\leq L.
	\end{numcases}
\end{subequations}
with error $\bm{\delta}^{k-1}=\bm{x}^{k-1} - \bm{x}$. By taking the $\ell_2$ norm, the parameters $\{\sigma^2, (\sigma_i^k)^2\}$ can be estimated from the following relationship,
\begin{subequations}
	\begin{numcases}
	{} \sigma^2 = N_d \bm{h}_0(\|\bm{y}-\bm{A}\bm{x}^{k-1}\|_2^2) \\
	(\sigma_i^k)^2 =N_p \bm{h}_i(\|\bm{z}_i^k-\bm{F}_i\bm{x}^{k-1}\|_2^2), \, 1\leq i\leq L.
	\end{numcases}
\end{subequations}
where $\{\bm{h}_i\}_{i=0}^L$ denotes the correction functions that rectify the input $\ell_2$ norm of the residues due to the presence of unknown error $\bm{\delta}^{k-1}$. 
Thus, the parameters can be estimated via
\begin{equation}
\beta_i^k=\frac{\sigma^2}{(\sigma_i^k)^2} = \frac{N_d\bm{h}_0(\|\bm{y}-\bm{A}\bm{x}^{k-1}\|_2^2)}{N_p \bm{h}_i(\|\bm{z}_i^k-\bm{F}_i\bm{x}^{k-1}\|_2^2)}, \quad \,1\leq i\leq L
\end{equation}
Instead of direct modeling $\{\bm{h}_i\}_{i=0}^L$, we choose a non-linear function, MLP, to model such relationship for simplicity, \emph{i.e} we  automate the estimation of the hyper-parameter sequence $\{\beta_i^k\}$
using a learnable NN. 
By setting $\bm{r}_0^k=\bm{y}-\bm{A}\bm{x}^{k-1}$ and $\bm{r}_i^k=\bm{z}_i^k-\bm{F}_i\bm{x}^{k-1}$, the NN, denoted by $\mathcal{P}_{\textrm{mlp}}^k(\cdot; \theta_{\mathcal P}^k)$ with parameters $\theta_{\mathcal P}^k$,  takes
the known value $\{\|\bm{r}_i^k\|\}_{i=0}^L$ as the input and outputs the estimation of $\{\beta_i^k\}$:
\begin{equation}
\mathcal{P}_{\textrm{mlp}}^k(\cdot,\theta_{\mathcal{P}}^k): [\|\bm{r}_i^k\|^2_2]_{i=0}^L\in\mathbb{R}^{L+1} \rightarrow  [\beta_i^k]_{i=1}^L\in\mathbb{R}^L.
\end{equation}
An MLP is implemented for modeling the mapping $\mathcal{P}^k$, which contains 3 layers. Each layer is composed by  one fully connected layer followed by one ReLU function. See Fig.~\ref{fig:MLPp} for the diagram of the proposed hyper-parameter predictor $\mathcal{P}^k$. 
\tikzstyle{arrow} = [->,>=stealth,blue,line width=0.05cm]
\tikzstyle{Block} = [rectangle, minimum width=2cm, minimum height=.5cm, text centered, draw=blue, fill = blue!20]
\tikzstyle{BBlock} = [rectangle, minimum width=1cm, minimum height=.5cm, text centered, draw=blue, fill = blue!20]
\tikzstyle{BBBlock} = [rectangle, minimum width=3.5cm, minimum height=.5cm, text centered, draw=blue, fill = blue!20]
\tikzstyle{cir} = [circle, text centered, draw=blue, fill = blue!20]
\tikzstyle{Format} = [draw, thin, draw =white,fill=white]
\tikzstyle{circ} = [circle,scale=0.5, draw=black, fill = black]
\begin{figure}[!htp]
	\centering
	\scalebox{0.65}{
		\begin{tikzpicture}[node distance=0.5cm]
		\path[->] node[Format] (Input) {$\bm{x}^{k-1},\bm{z}^k$};
		\path[->] node[Block,right of=Input,xshift=2.0cm,yshift=1.8cm] (Res1) {$\bm{r}^k_0=\bm{y}-\bm{A}\bm{x}^{k-1}$};
		\path[->] node[Block,below of=Res1,yshift=-0.6cm] (Res2) {$\bm{r}^k_1=\bm{z}_1^k-\bm{F}_1\bm{x}^{k-1} $};
		\path[->] node[Block,below of=Res1,yshift=-2.5cm] (Res3) {$\bm{r}^k_L=\bm{z}_L^k-\bm{F}_L\bm{x}^{k-1}$};
		\draw [arrow] (Input) -- (Res1.west);
		\draw [arrow] (Input) -- (Res2.west);
		\draw [arrow] (Input) -- (Res3.west);
		
		\path[->] node[circ,below of=Res2,yshift=-0.6cm] (dot11){}; 
		\path[->] node[circ,below of=dot11,yshift=-0.2cm] (dot12){}; 
		\path[->] node[circ,below of=dot12,yshift=-0.2cm] (dot13){}; 
		
		\path[->] node[BBlock,right of=Res1,xshift=3.0cm] (ResNorm1) {$\|\bm{r}^k_0\|_2^2$};
		\path[->] node[BBlock,right of=Res2,xshift=3.0cm] (ResNorm2) {$\|\bm{r}^k_1\|_2^2$};
		\path[->] node[BBlock,right of=Res3,xshift=3.0cm] (ResNorm3) {$\|\bm{r}^k_L\|_2^2$};
		\draw [arrow] (Res1) -- (ResNorm1);
		\draw [arrow] (Res2) -- (ResNorm2);
		\draw [arrow] (Res3) -- (ResNorm3);
		
		\path[->] node[circ,below of=ResNorm2,yshift=-0.6cm] (dot21){}; 
		\path[->] node[circ,below of=dot21,yshift=-0.2cm] (dot22){}; 
		\path[->] node[circ,below of=dot22,yshift=-0.2cm] (dot23){}; 
		
		\path[->] node[cir,right of=ResNorm1,xshift=1.2cm] (w1){};
		\path[->] node[cir,right of=ResNorm2,xshift=1.2cm] (w2){};
		\path[->] node[cir,right of=ResNorm3,xshift=1.2cm] (w3){};
		
		\foreach \x in{1,2,3}
		{\foreach \y in{1,2,3}
			{\draw[-{stealth[sep=2pt]},blue,line width=0.05cm](ResNorm\x.east)--(w\y);
			}
		}
		\path[->] node[circ,below of=w2,yshift=-0.6cm] (dot31){}; 
		\path[->] node[circ,below of=dot31,yshift=-0.2cm] (dot32){}; 
		\path[->] node[circ,below of=dot32,yshift=-0.2cm] (dot33){}; 	
		
		\path[->] node[BBBlock,right of=ResNorm2,xshift=2.0cm,yshift=-0.5cm,rotate=-90] (Relu1) {ReLU};
		\draw [arrow] (w1) -- ([yshift=-0.15cm, xshift=-0.2cm]Relu1.west);
		\draw [arrow] (w2) -- ([yshift=-1.25cm, xshift=-0.2cm]Relu1.west);
		\draw [arrow] (w3) -- ([yshift=-3.15cm, xshift=-0.2cm]Relu1.west);
		
		\path[->] node[circ,right of=Relu1,xshift=0.5cm] (cdot11){}; 
		\path[->] node[circ,right of=cdot11,xshift=0.2cm] (cdot12){}; 
		\path[->] node[circ,right of=cdot12,xshift=0.2cm] (cdot13){}; 
		
		\path[->] node[cir,right of=w1,xshift=3.0cm] (ww1){};
		\path[->] node[cir,right of=w2,xshift=3.0cm] (ww2){};
		\path[->] node[cir,right of=w3,xshift=3.0cm] (ww3){};
		{\foreach \z in{1,2,3}
			{\draw[-{stealth[sep=2pt]},blue,line width=0.05cm]([yshift=3.35cm, xshift=1.5cm]Relu1.east)--(ww\z);}
		} 
		{\foreach \z in{1,2,3}
			{\draw[-{stealth[sep=2pt]},blue,line width=0.05cm]([yshift=2.25cm, xshift=1.5cm]Relu1.east)--(ww\z);}
		}
		{\foreach \z in{1,2,3}
			{\draw[-{stealth[sep=2pt]},blue,line width=0.05cm]([yshift=0.35cm, xshift=1.5cm]Relu1.east)--(ww\z);}
		}
		\path[->] node[circ,below of=ww2,yshift=-0.6cm] (dot41){}; 
		\path[->] node[circ,below of=dot41,yshift=-0.2cm] (dot42){}; 
		\path[->] node[circ,below of=dot42,yshift=-0.2cm] (dot43){}; 
		
		\path[->] node[BBBlock,right of=ww2,xshift=0.5cm,yshift=-0.5cm,rotate=-90] (Relu2) {ReLU}; 
		\draw [arrow] (ww1) -- ([yshift=-0.15cm, xshift=-0.2cm]Relu2.west);
		\draw [arrow] (ww2) -- ([yshift=-1.25cm, xshift=-0.2cm]Relu2.west);
		\draw [arrow] (ww3) -- ([yshift=-3.15cm, xshift=-0.2cm]Relu2.west);
		\path[->] node[Format,right of=Relu2,xshift=0.5cm] (Output) {$\bm{\beta}^k$};
		\draw [arrow] (Relu2) -- (Output);
		\end{tikzpicture}
	}
	\caption{Diagram of the MLP for predicting hyper-parameter sequence $\bm{\beta}^k=\{\beta_i^k\}_{i=1}^L$.}
	\label{fig:MLPp}
\end{figure}
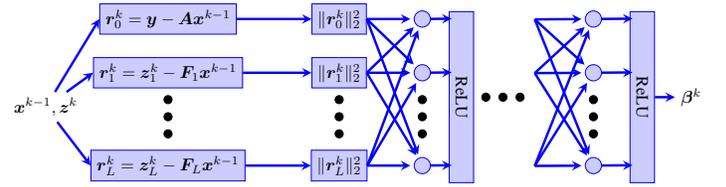

Once  the  sequence  $\{\beta_i^k\}_{i=1}^L$ is predicted, we can update the estimate using the analytical solution
(\ref{solution1}).
We denote such a solution by 
\begin{equation}
\mathcal{I}^k_{\textrm{Inv}}: [\bm{y}, \{\beta_i^k\}_{i=1}^L,\{\bm z_i^k\}_{i=1}^L] \rightarrow \bm{x}^{k}.
\end{equation}
In summary, the inversion block can be expressed as  
\begin{equation}\label{eqn:inv}
\mathscr{L}^k(\cdot\,; \theta_{\mathcal P}^k):\,
[\bm y, \{\bm z_i\}_{i=1}^L]\rightarrow \bm x^k,
\end{equation}
where $\theta_{\mathcal P}^k$ is the  learnable  weights.

Note that 
one can numerically solve (\ref{solution1}) by conjugate gradient. 
To give the backward gradient for the backpropagation algorithm, let $\ell:\mathbb{R}^{N_P}\rightarrow \mathbb{R}$,  then the derivatives of  $\ell(\bm x^k)$ with respect to $\{\beta_i^k\}_{i=1}^L,\{\bm z_i^k\}_{i=1}^L $
can be correspondingly computed by:
\begin{subequations}
	\label{backpro}
	\begin{align}
		& \frac{\partial \ell(\bm x^k)}{\partial \bm z_i^k}= \beta_i^k \bm{F}_i(\bm{A}^\top\bm{A}+\sum_{i=1}^L\beta_i^k\bm{F}_i^\top\bm{F}_i)^{-1}\frac{\partial \ell(\bm x^k)}{\partial \bm x^k}, \\
		&\frac{\partial \ell(\bm x^k)}{\partial \beta_i^k}=(\bm{F}^\top_i\bm z_i^k - \bm{F}^\top_i\bm{F}_i\bm x^k )^T(\bm{A}^\top\bm{A}+\sum_{i=1}^L\beta_i^k\bm{F}_i^\top\bm{F}_i)^{-1}\frac{\partial \ell(\bm x^k)}{\partial \bm x^k}.
	\end{align}
\end{subequations}

\subsection{Denoising block with CNN-based adaptive prior}
\label{Denoising}
For the denoising block, following many existing methods, we also adopt a CNN to learn the function (\ref{sub2}) such that the estimate passed from the inversion block can be refined using learnable prior that is adaptive to the target image.
It is noted that the measurements of an image in different high-pass channels are highly correlated. Independently running  denoising in these high-pass channels is not a good practice as the inherent correlation is lost in such a process. Thus, the proposed CNN take the estimate  ${\bm x}^k$ as the input and output a denoised version $\bm{\tilde{x}}^k$, which is then fed to $L$ high-pass channels to have an estimate $\bm {z}_i^{k+1}$:
\begin{equation}
\bm{z}_i^{k+1}=\{\bm{f}_i\otimes\bm{\tilde{x}}^k\}_i, \,\,1\leq i\leq L.
\end{equation}
Such a procedure reserves the correlations of an image among different high-pass channels.

Furthermore, instead of only taking the previous estimate ${\bm x}^k$ as the input,
The CNN for denoising block takes all previous estimates $\{\bm{x}^0,\bm{x}^1\cdots\bm{x}^k\}$ as the input, which can alleviate the so-called vanishing gradient in training~\cite{huang2017densely}. Another benefit of doing so is that  the fusion of these previous estimates provide more information for the refinement, as these are different estimates of the truth with different types of artifacts. The final version of the mapping of the CNN-based denoising process can be expressed as 
\begin{equation}
\mathcal{D}^{k}_{\textrm{cnn}}(\cdot,\theta_\mathcal{D}^{k}): [\bm{x}^0,\bm{x}^1,\cdots,\bm{x}^k] \rightarrow  \bm{\tilde{x}}^{k},
\end{equation}
where $\theta_\mathcal{D}^k$ denotes the parameters of denoising NN, $\mathcal{D}_{\textrm{cnn}}^k$. The output of the whole denoising block is then 
\begin{equation}
\bm{z}_i^{k+1} =\bm{f}_i\otimes\mathcal{D}_{cnn}^k([\bm{x}^0,\bm{x}^1,\cdots, \bm{x}^k],\theta_\mathcal{D}^k).
\end{equation}
For each stage of the iteration, we use 17-block standard CNN with the structure Conv$\rightarrow$BN$\rightarrow$ReLU, except the first block and the last block.
The BN layer is omitted for the first and  last block.  For all the Conv layers in the CNN, the kernel size is set as  $3\times3$.
The channel size is set to 64. See Fig.~\ref{fig:2} for the diagram of the CNN-based denoising block.
In summary, the denoising block can be expressed as  
\begin{equation}\label{eqn:dn}
{\mathscr{D}}^k(\cdot\,; \theta^k_{\mathcal D}):\,
[\bm x^0,\bm x^1,\cdots, \bm x^{k}]\rightarrow \bm z^{k+1},	
\end{equation}
where $\theta_{\mathcal D}^k$ is the set of learnable  weights w.r.t. the CNN-based denoiser.
\tikzstyle{arrow} = [->,>=stealth,blue,line width=0.05cm]
\tikzstyle{arrowS} = [-,>=stealth,blue,line width=0.05cm]
\tikzstyle{Block} = [rectangle, minimum width=2cm, minimum height=.5cm, text centered, draw=blue, fill = blue!20]
\tikzstyle{BBlock} = [rectangle, minimum width=1cm, minimum height=.3cm, text centered, draw=blue, fill = blue!20]
\tikzstyle{FormatS} = [draw, thin, draw =white,fill=white,text width = 0.1cm]
\tikzstyle{Format} = [draw, thin, draw =white,fill=white]
\tikzstyle{circ} = [circle, scale=0.3, draw=black, fill = black]
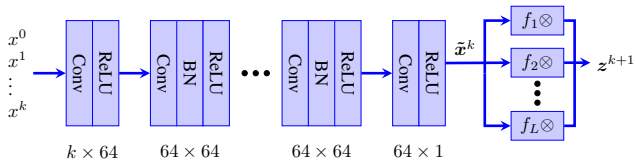
\begin{figure}[!htp]
	\centering
	\scalebox{0.7}{
		\begin{tikzpicture}[node distance=0.5cm]
		\path[->] node[FormatS] (Input){$x^0$ $x^1$ $\vdots$ $x^k$};		
		\path[->] node[Block,right of=Input,rotate=-90,yshift=0.8cm] (layer1) {Conv};
		\path[->] node[Block,right of=layer1,rotate=-90] (layer2) {ReLU};
		\path[->] node[Block,right of=layer2,rotate=-90,yshift=0.6cm] (layer3) {Conv};
		\path[->] node[Block,right of=layer3,rotate=-90] (layer4) {BN};
		\path[->] node[Block,right of=layer4,rotate=-90] (layer5) {ReLU};
		\path[->] node[Block,right of=layer5,rotate=-90,yshift=1cm] (layer6) {Conv};
		\path[->] node[Block,right of=layer6,rotate=-90] (layer7) {BN};
		\path[->] node[Block,right of=layer7,rotate=-90] (layer8) {ReLU};
		\path[->] node[Block,right of=layer8,rotate=-90,yshift=0.6cm] (layer9) {Conv};
		\path[->] node[Block,right of=layer9,rotate=-90] (layer10) {ReLU};		
		\path[->] node[BBlock,right of=layer10,xshift=1.5cm,yshift=1.0cm] (ww1){$f_1\otimes$};
		\path[->] node[BBlock,right of=layer10,xshift=1.5cm,yshift=0.2cm] (ww2){$f_2\otimes$};
		\path[->] node[BBlock,right of=layer10,xshift=1.5cm,yshift=-1.0cm] (ww3){$f_L\otimes$};
		\path[->] node[circ,right of=layer5,xshift=1.3cm] (dot1){}; 
		\path[->] node[circ,right of=dot1,xshift=0.2cm] (dot2){}; 
		\path[->] node[circ,right of=dot2,xshift=0.2cm] (dot3){}; 
		\path[->] node[Format,right of=ww2,xshift=1.0cm] (OutPut) {$\bm{z}^{k+1}$};
		\path[->] node[Format,below of=layer1,yshift=-1.0cm] () {$~~~~k\times 64$};
		\path[->] node[Format,below of=layer4,yshift=-1.0cm] () {$64\times 64$};
		\path[->] node[Format,below of=layer7,yshift=-1.0cm] () {$64\times 64$};
		\path[->] node[Format,below of=layer9,yshift=-1.0cm] () {$~~~~64\times1$};
		\draw [arrow] ([xshift=0.25cm]Input.east) -- (layer1);
		\draw [arrow] (layer2) -- (layer3);
		\path[->] node[circ,below of=ww1,yshift=-3.5cm] (vdot1){}; 
		\path[->] node[circ,below of=vdot1,yshift=-0.1cm] (vdot2){}; 
		\path[->] node[circ,below of=vdot2,yshift=-0.1cm] (vdot3){}; 
		\draw [arrow] (layer8) -- (layer9);
		\draw [arrow] ([yshift=1.2cm, xshift=0.25cm]layer10.east) --node [above,black] {$\bm{\tilde{x}}^{k}$~~~~ } (ww2.west);
		\draw [arrow] ([yshift=1.0cm, xshift=1.0cm]layer10.east) |-([yshift=-0.0cm, xshift=-0.0cm]ww1.west);
		\draw [arrow] ([yshift=1.0cm, xshift=1.0cm]layer10.east) |-([yshift=-0.0cm, xshift=-0.0cm]ww3.west);
		\draw [arrow] (ww2.east) --([yshift=-0.0cm, xshift=-0.0cm]OutPut.west);
		\draw [arrowS] ([yshift=0.0cm, xshift=0.0cm]ww1.east) -|([yshift=-0.0cm, xshift=-0.3cm]OutPut.west);
		\draw [arrowS] ([yshift=0.0cm, xshift=0.0cm]ww3.east) -|([yshift=-0.0cm, xshift=-0.3cm]OutPut.west);
		\end{tikzpicture}
	}
	\caption{Diagram of the CNN for denoising.}
	\label{fig:2}
\end{figure}
\subsection{The overall architecture of the proposed method}
\label{Overall model}
The proposed deep learning method for LDCT contains totally ${K}+1$ stages, denoted by $\{\mathcal{S}^k\}_{k=0}^{{K}}$. Let $\mathscr{L}^k$ and $\mathscr{D}^k$ denote the inversion block and denoising block defined by (\ref{eqn:inv}) and (\ref{eqn:dn}).
Then,  each stage corresponds to one iteration in MBIR as follows.
\begin{eqnarray}
\mathcal{S}^0(\cdot;\Theta^0):&\quad\bm{y}  \stackrel{\mathscr{L}^0}{\longrightarrow} \bm{x}^0,\\
\mathcal{S}^k(\cdot;\Theta^k):&\quad [\bm{y},\{\bm{x}^\ell\}_{\ell=0}^{k-1}]\stackrel{\mathscr{D}^{k-1}}{\longrightarrow} [\bm y, \bm z^{k-1}]\stackrel{\mathscr{L}^k}{\longrightarrow} \bm{x}^k,
\end{eqnarray}
for $k=1,2,\ldots, K$, where $\Theta^k:=\{\theta_\mathcal{D}^{k-1}, \theta_\mathcal{P}^k\}$ denotes the weights at the $k$-th stage, including the weights of $\mathcal{D}^{k-1}_{\textrm{cnn}}$, and $\mathcal{P}^k_{\textrm{mlp}}$. 
In stage $\mathcal{S}^0$, we initialize $\bm{z}^0=\bm{0}$ and $\beta_i^0=0.005$ for $1\leq i\leq L$. 
It can be seen that when the measurement $\bm$ is fed to the proposed NN, it generates a sequence through the $K+1$ stages:
$$\{\bm x^0,\bm x^1,\cdots, \bm x^K\},
$$
The final output of the whole NN is defined as $\bm x^\ast:=\bm x^K$.
See Fig. \ref{fig:1} for the outline of the proposed NN, termed as \textit{AHP-Net}, for LDCT reconstruction.
\tikzstyle{startstop} =[draw, thin,draw =blue, fill=blue!20]
\tikzstyle{format} = [draw, thin, draw =blue,fill=blue!20]
\tikzstyle{process} = [rectangle, minimum width=1.5cm, minimum height=.6cm, text centered, draw=blue, fill = blue!20]
\tikzstyle{arrow} = [->,>=stealth,blue,line width=0.05cm]
\tikzstyle{Out} = [draw, thin,draw =blue, fill=blue!20]
\tikzstyle{circ} = [circle,scale=0.5, draw=black, fill = black]
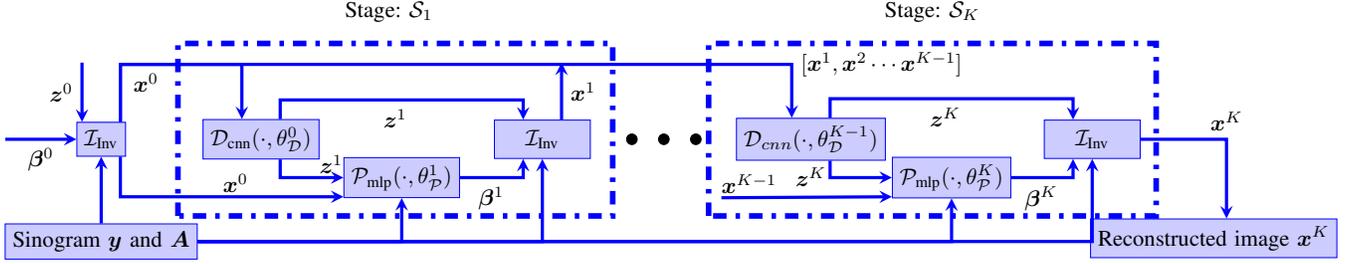
\begin{figure*}[!htp]
	\scalebox{0.85}{
		\centering
		\begin{tikzpicture}[node distance=2cm]
		\path[->] node[format] (Inv0){$\mathcal{I}_{\textrm{Inv}}$};
		\path[->] node[startstop, below of=Inv0,yshift=-0.6cm] (Input) at (0,1) {Sinogram $\bm{y}$ and $\bm{A}$};
		\path[->] node[format, right of=Inv0,xshift=0.5cm] (Cnn1) {$\mathcal{D}_{\textrm{cnn}}(\cdot,\theta_\mathcal{D}^0)$};
		\path[->] node[process, right of=Cnn1,xshift=0.2cm,yshift=-0.6cm] (mlp1) {$\mathcal{P}_{\textrm{mlp}}(\cdot,\theta_\mathcal{P}^1)$};
		\path[->] node[process, right of=mlp1,xshift=0.2cm,yshift=0.6cm] (Inv1) {$\mathcal{I}_{\textrm{Inv}}$};
		\path[->] node[format, right of=Inv1,xshift=2.2cm] (Cnn2) {$\mathcal{D}_{cnn}(\cdot,\theta_\mathcal{D}^{{{K}-1}})$};
		\path[->] node[process, right of=Cnn2,xshift=0.2cm,yshift=-0.6cm] (mlp2) {$\mathcal{P}_{\textrm{mlp}}(\cdot,\theta_\mathcal{P}^{{{K}}})$};
		\path[->] node[process, right of=mlp2,xshift=0.2cm,yshift=0.6cm] (Inv2) {$\mathcal{I}_{\textrm{Inv}}$};
		\path[->] node[Out, below of=Inv2,yshift=0.45cm,xshift=1.9cm] (OutPut) {Reconstructed image  $\bm{x}^{{{K}}}$ };
		\draw [arrow] (Input) -- (Inv0);
		\draw [arrow] ([xshift=0.3cm]Inv0.south) |-node [below,black,xshift=1.8cm,yshift=0.5cm] {$\bm{x}^{0}$} ([yshift=-0.3cm]mlp1.west);
		\draw [arrow] ([xshift=0.3cm]Cnn1.south)  |-node [below,black,xshift=0.8cm,yshift=0.5cm] {$\bm{z}^{1}$} (mlp1);
		\draw [arrow] (mlp1)  -|node [below,black,xshift=-0.5cm] {$\bm{\beta}^{1}$} ([xshift=-0.3cm]Inv1.south);
		\path[->] node[circ,right of=Inv1,xshift=0.8cm] (dot1){}; 
		\path[->] node[circ,right of=dot1,xshift=-1cm] (dot2){}; 
		\path[->] node[circ,right of=dot2,xshift=-1cm] (dot3){}; 
		\draw [arrow] ([xshift=0.3cm]Cnn2.south)  |-node [below,black,xshift=-0.3cm,yshift=0.3cm] {$\bm{z}^{{{K}}}$} (mlp2);
		\draw [arrow] (mlp2.east)  -|node [below,black,xshift=-0.5cm] {$\bm{\beta}^{{{K}}}$} ([xshift=-0.3cm]Inv2.south);
		\draw [arrow] (Inv2) -|node [above,black] {$\bm{x}^{{{K}}}$ } ([yshift=0.0cm, xshift=0.2cm]OutPut.north);
						\draw [arrow] ([xshift=-1.4cm,yshift=-0.58cm]Cnn2.south) --node [below,black,xshift=-0.9cm,yshift=0.5cm] {$\bm{x}^{K-1}$} ([yshift=-0.3cm]mlp2.west);		
		\draw [arrow] (Input) -|node [right] { } (Inv1);
		\draw [arrow] (Input) -|node [right] { } (mlp1);
		\draw [arrow] (Input) -|node [right] { } ([yshift=0.0cm, xshift=-0.0cm]Inv2.south);
		\draw [arrow] (Input) -|node [right] { } (mlp2);
		\draw[arrow]([yshift=0.0cm, xshift=0.3cm]Cnn1.north)--++(0, 0.3cm)-|node [below,black,xshift=-2.0cm,yshift=0.0cm] {$\bm{z}^1$}([yshift=0.0cm, xshift=-0.3cm]Inv1.north);
		\draw[line width=2.5pt,blue,dash pattern=on 6pt off 4pt on 2pt off 4pt,-] (1.2,-1.2)--(1.2,1.5)--(8.0,1.5)--(8.0,-1.2)--(1.2,-1.2);
		\draw[line width=2.5pt,blue,dash pattern=on 6pt off 4pt on 2pt off 4pt,-] (9.5,-1.2)--(9.5,1.5)--(16.5,1.5)--(16.5,-1.2)--(9.5,-1.2);
		\draw[arrow]([yshift=0.0cm, xshift=0.3cm]Cnn2.north)--++(0, 0.3cm)-|node [below,,black,xshift=-2.0cm,yshift=0.0cm] {$\bm{z}^{{{K}}}$}([yshift=0.0cm, xshift=-0.3cm]Inv2.north);
		\draw[arrow](-0.3,1.2)--node [left,black] {$\bm{z}^{0}$}([yshift=0.0cm, xshift=-0.3cm]Inv0.north);
		\draw[arrow](-1.5,0)--node [below,black] {$\bm{\beta}^{0}$}(Inv0.west);
		\draw[arrow] ([yshift=0.0cm, xshift=0.3cm]Inv0.north)--++(0, 0.9cm)-|node [below,black,xshift=-1.5cm,yshift=0.0cm] {$\bm{x}^{0}$}([yshift=0.0cm, xshift=-0.3cm]Cnn1.north);
		\draw[arrow] ([yshift=0.0cm, xshift=0.3cm]Inv1.north)--node [right,black] {$\bm{x}^{1}$ }(7.2, 1.2cm);
		\draw[arrow]([yshift=0.0cm, xshift=0.3cm]Inv0.north)--++(0, 0.9cm)-|node [right,black] {$[\bm{x}^{1},\bm{x}^{2}\cdots \bm{x}^{{{K}-1}}]$ }([yshift=0.0cm, xshift=-0.3cm]Cnn2.north);
		\node at	(4.5,2.0)	{Stage: $\mathcal{S}_1$};
		\node at	(13,2.0){Stage: $\mathcal{S}_{{{K}}}$};
		\end{tikzpicture}
	}
	\caption{Diagram of the proposed AHP-Net for LDCT image reconstruction.}
	\label{fig:1}
\end{figure*}

For the training of the proposed AHP-Net, we consider the training dataset $\{\bm{x}_{j},\bm{y}_{j}\}_{j=1}^J$ with $J$ training samples, where each $(\bm{x}_{j},\bm{y}_{j})$ denotes one pair of normal dose image and low dose sinogram data. 
The loss function  is defined as
\begin{equation}\label{eqn:loss}
\mathcal{L}(\Theta)=\frac{1}{J}\sum_{j=1}^{J}\left(\|\bm{x}_{j}^{{{K}}}-\bm{x}_{j} \|_2^2+\sum_{k=1}^{K-1}\mu_k\|\bm{x}^k_{j}-\bm{x}_{j} \|_2^2\right),
\end{equation}
where $\Theta:=\{\Theta^k\}_{k=0}^K$ is the whole set of NN parameters, and 
$$\bm x_j^k=\left(\mathcal{S}^k(\cdot\,;\Theta^k)\circ
\mathcal{S}^{k-1}(\cdot\,;\Theta^{k-1})\circ\cdots\circ
\mathcal{S}^0(\cdot\,;\Theta^0)
\right)(\bm y_j)$$
denotes the output of the AHP-Net at the $k$-th stage w.r.t. the input $\bm y_j$. The first term in  (\ref{eqn:loss}) is for encouraging the output of the NN close to the truth.
The second term is for ensuring the intermediate results not too far from the truth. The balancing parameters $\{\mu_k\}_{k=1}^{K-1}$ are set to $\frac{4}{5}$ throughout all experiments. Our optimization was done via  minimizing the loss function (\ref{eqn:loss}) using the Adam method. Once we finished the training of the model,  an estimation of NN parameters $\overline{\Theta}$ is obtained.
For a low dose  input data $\bm{y}$,  the image can be reconstructed by fed it into the NN with parameters $\overline{\Theta}$:
$$
\bm x^\ast:=\left(\mathcal{S}^k(\cdot\,;\overline{\Theta}^k)\circ
\mathcal{S}^{k-1}(\cdot\,;\overline{\Theta}^{k-1})\circ\cdots\circ
\mathcal{S}^0(\cdot\,;\overline{\Theta}^0)
\right)(\bm y).
$$

\subsection{Implementation}
The proposed AHP-Net is implemented with $K=3$ stages. 
For the proposed NN method, training is performed with PyTorch \cite{paszke2017automatic} framework on a NVIDIA Titan GPU. As for parameters in the Adam optimizer, we set momentum parameter $\beta=0.9$, mini-batch size to be $4$, and the learning rate to be $10^{-4}$. 
The model was trained for $50$ epochs.
The convolution weight of $\mathcal{D}_{\textrm{cnn}}$ for denoising were initialized with orthogonal matrices and the biases were initialized with zeros.
All weights in $\mathcal{P}_{\textrm{mlp}}$   were initialized to 1 and the biases to 0.
\section{Experiments}
\label{Experiments}
In this section, we conduct a comprehensive experimental evaluation of the proposed AHP-Net, and compare it to  other representative methods in LDCT image reconstruction.
\subsection{Dataset for evaluation}
To evaluate the performance of the proposed  method under different dose levels, we simulated low dose projection data from their normal-dose counterparts. 
The normal dose dataset included 6400 normal-dose  prostate  CT  images of $256\times 256$ pixels per image from 100 anonymized scans, where 80\%, and 20\% of the data is set for training and testing respectively.
The  LDCT projection data was simulated  by adding Poisson noise onto the normal-dose projection data as Section \ref{Measurement Model}.
The simulated geometry for projection data includes {fan-beam CT scanner}, flat-panel  detector of $0.388~\rm{mm}\times 0.388~\rm{mm}$ pixel size, 
$600$ projection views  evenly  spanning a $360^{\circ}$ circular orbit,  $512$ detector bins for each projection with $1~\rm{mm}$ pixel size, 
$100.0~\rm{cm}$ source to detector distance, and $50.0~\rm{cm}$  source to isocenter distance.

For the training dataset, 80 patients' simulated low-dose projection  and corresponding  normal dose images were selected as the pairs of  measurement and truth image.
For every patient's normal dose data, we simulated the low-dose measurement with uniform dose level.
Then, the  sinograms of different patients were obtained by taking logarithm on projection data $\bar{\bm{y}}$.
The remaining 20 scans were  selected  as testing dataset.  For  testing dataset, low dose measurement was simulated by the same way as training dataset. Throughout all experiments,  5120 pairs were included in  training dataset, and 1280 pairs in testing dataset.
\subsection{Training scheme}	
Two types of training schemes were conducted in this paper. 
\begin{itemize}
	\item \emph{Different models for different dose levels.} \,\,  In this training scheme, for each dose level, all deep learning  methods trained one specific model using the measurements with the same dose level, and also tested it  on the measurements with the same dose level. Totally 4 dose levels were evaluated: $I_i=1\times10^5, 5\times10^4,1\times10^4,5\times10^3$.
	\item 	 \emph{One universal  model for different dose levels.}\,\, In this training scheme, all deep learning methods trained a single model using the measurements with varying dose levels, and also tested on the measurements with varying dose levels. In this case, each measurement in training dataset was generated with its dose level randomly selected from  $\{1\times10^5,7.5\times10^4, 5\times10^4, 2.5\times10^4, 1\times10^4,7.5\times10^3,5\times10^3\}$, and each measurement in validation/testing dataset was generated by with its dose level
	$I_i=1\times10^5, 5\times10^4,1\times10^4,5\times10^3$ respectively.
\end{itemize}
\subsection{Comparison Method}
The performance of the proposed methods was evaluated in
comparison with classic FBP method, TV method, KSVD~\cite{zhang2010discriminative}, BM3D~\cite{dabov2007image}, FBPConvNet~\cite{jin2017deep}, MoDL~\cite{aggarwal2018modl},
Neumann Network~\cite{gilton2019neumann}, Projected Gradient Descent (PGD)~\cite{gupta2018cnn} and Learned Primal-Dual (Learned-PD)~\cite{adler2018learned}. We only brief these methods and more details can be found in Appendix \ref{APP:com}.
\begin{itemize}
	\item The TV method is a well-established classic method, which is often solved via the ADMM solver:
		\begin{align}
		&\bm{x}^{k+1}=\arg\min_{\bm{x}}\frac{1}{2}\|\bm{Ax}-\bm{y}\|_2^2+\frac{\mu}{2}\|\nabla\bm{x}-\bm{z}^{k}+\frac{\bm{p}^k}{\mu}\|_2^2,\nonumber\\
		&	\bm{z}^{k+1}=\arg\min_{\bm{z}}\lambda\|\bm{z}\|_1+\frac{\mu}{2}\|\bm{z}-(\nabla\bm{x}^{k+1}+\frac{\bm{p}^k}{\mu})\|_2^2,\nonumber\\
		&\bm{p}^{k+1}=\bm{p}^{k}+\mu(\nabla\bm{x}^{k+1}-\bm{z}^{k+1} ).\nonumber
		\end{align}
%
	The parameters $\lambda,\mu$ in the TV method are manually tuned-up for optimal performance under different dose levels.
	\item 	
	FBPConvNet~\cite{jin2017deep}  is one representative deep-learning-based method for CT reconstruction that use the deep NN as a post-process technique, where an NN with U-net-based CNN is trained to directly denoise the image reconstructed by the FBP method.
	
	\item  MoDL~\cite{aggarwal2018modl} is a deep learning method proposed for MRI reconstruction, which also can be used for LDCT reconstruction.  MoDL unrolls the following iterative scheme
	\begin{eqnarray}
	\bm{x}^{k}&=&(\bm{A}^T\bm{A}+\lambda\bm{I})^{-1}(\bm{A}^T\bm{y}+\lambda \tilde{\bm{x}}^{k-1}), \nonumber\\
	\tilde{\bm{x}}^{k}&=&\mathcal{D}_{\textrm{cnn}}(\bm{x}^k;\theta^k), \nonumber
	\end{eqnarray}
	where $\mathcal{D}_{\textrm{cnn}}$ is a CNN-based denoiser and $\lambda$ is learned as an NN weight.
	\item {Neumann-Net}~\cite{gilton2019neumann} is proposed for linear inverse problem in generic imaging, whose iterative scheme is as follows:
	$$ \bm{x}^{K}= \sum_{k=0}^{K-1}\bm{x}^{k}+\eta(\bm{I}-\eta {\bm A}^\top {\bm A})\bm{x}^{K-1} -\eta\mathcal{D}_{\textrm{cnn}}( \bm{x}^{K-1};\theta^{K-1}),$$
	where $\mathcal{D}_{\textrm{cnn}}$ is a CNN-based mapping and $\bm{x}^{0}=\bm{A}^T\bm{y}$ .
	
	\item {PGD}~\cite{gupta2018cnn} is proposed for CT reconstruction with 
	the following iterative scheme:
	$$\bm{x}^{k+1}=(1-\alpha_k)\bm{x}^k+\alpha_k\mathcal{D}_{\textrm{cnn}}(\bm{x}^k-\gamma ({\bm A}^\top {\bm A}\bm{x}^k-{\bm A}^\top\bm{y})),$$
	where $\mathcal{D}_{\textrm{cnn}}$ is a CNN-based mapping and $\alpha_k$ and $\gamma$ are parameters.
	\item {Learned-PD}~\cite{adler2018learned} is the unrolling-based method proposed for CT reconstruction. 
	By setting  the initial value as $\bm{x}^0$ and introducing the dual variable $\bm{h}^0$, the iterative scheme based on  PDHG algorithm is derived:
	\begin{eqnarray}
	\bm{h}^k&=&\mathcal{D}^{\theta_d}_{\textrm{cnn}}(\bm{h}^{k-1}+\sigma\bm{A}\bar{\bm{x}}^{k-1},\bm{y} ), \nonumber\\
	\bm{x}^k&=&\mathcal{D}^{\theta_p}_{\textrm{cnn}}(\bm{x}^{k-1}-\tau\bm{A}^T\bm{h}^{k-1}),\nonumber\\
	\bar{\bm{x}}^{k}&=&\bm{x}^{k-1}+\theta(\bm{x}^k-\bm{x}^{k-1} ),\nonumber
	\end{eqnarray}
	where $\theta_d$ and $\theta_p$ are the parameters of the dual proximal and the primal proximal respectively, $\sigma$ and $\tau$ are  the step lengths, and $\theta$
	is the 	overrelaxation parameter.

	
\end{itemize}

\section{ Results}
\label{results}

In this section the proposed AHP-Net is evaluated on the simulated prostate CT data.
\subsection{Quantitative comparison on dataset}
Three metrics, peak signal to noise ratio (PSNR), root mean square error (RMSE)  and structural similarity index measure (SSIM) \cite{wang2004image}, are used for quantitative evaluation of image quality. Recall that
PSNR is defined as 
\begin{equation}
{\rm{PSNR}}(\bm{x},\bm{x}^\ast)=-10\log_{10}\left(\frac{\|\bm{x}-\bm{x}^\ast\|_2^2}{\max_i|x_i|^2}\right),\nonumber
\end{equation}
where $\bm{x}^\ast$ denotes reconstructed image and $\bm{x}$ denotes ground truth, \emph{i.e.} normal dose image.

See Table \ref{ComparisionU} for the quantitative comparison of the results from different methods, where different models were trained for processing the data w.r.t. different dose levels.  It can be seen that the performance of Neumann-Net is noticeably worse than the other. One possible reason might be that Neumann-Net need more iterations rather that only 3 stages.
Three deep learning methods, FBPConvNet, MoDL and PGD, improved the reconstruction results.
The proposed AHP-Net and Learned-PD are two best performers among all methods. In the case of relatively high dose level, $I_i=1 \times 10^5, 5\times 10^4$, the performances of these two methods are very close. In the case of relatively low dose level, $I_i=1 \times 10^4, 5\times 10^3$, the proposed AHP-Net outperformed Learned-PD by a noticeably margin, \underline{$1.7-2.5$} dB advantage in PSNR.

\begin{table*}
	\centering
	\caption{Quantitative comparison (Mean$\pm$STD) of the results reconstructed by different methods on testing dataset, where different models are trained for different dose levels in deep learning methods. }
		
      \begin{tabular}{ccccccc}
		\hline
		\hline
		
		Dose level  & Index   & FBP & TV& KSVD & BM3D   & FBPConvNet \\

		\hline
		\multirow{3}{*}{$1 \times 10^5$}&PSNR    &$37.15\pm2.04$      &$39.75\pm2.18$ &$35.61\pm3.46$      &$37.27\pm2.15$   &$38.14 \pm2.02$   \\   	                                    
		                              &RMSE    &$19.06\pm1.10$      &$14.18\pm1.38$ &$24.31\pm2.34$      &$20.07\pm1.35$   &$17.05 \pm1.48$   \\
		                              &SSIM    &$0.9336\pm0.01$     &$0.9685\pm0.01$&$0.9149\pm0.02$     &$0.9297\pm0.02$  &$0.9788\pm0.01$  \\
		\hline
		\multirow{3}{*}{$5\times10^4$}&PSNR    &$35.78\pm2.05$      &$38.85\pm2.18$ &$34.84\pm2.09$      &$34.31\pm1.97$   &$38.11\pm2.01$     \\ 
		                              &RMSE    &$22.34\pm1.55$      &$15.72\pm1.54$ &$25.07\pm3.45$      &$21.16\pm2.08$   &$17.13 \pm1.82 $  \\ 
		                              &SSIM    &$0.8963\pm0.02$     &$0.9594\pm0.01$&$0.8947\pm0.02$     &$0.9165\pm0.02$  &$0.9751\pm0.01$   \\
		\hline
		\multirow{3}{*}{$1 \times 10^4$} &PSNR    &$30.23\pm2.22$      &$34.21\pm2.31$ &$30.37\pm2.16$      &$31.46\pm2.32$   &$35.42\pm2.12$    \\
		                              &RMSE    &$42.60\pm5.74$      &$26.97\pm4.01$ &$39.58\pm5.03$      &$34.91\pm4.65$   &$23.35 \pm2.42$   \\
		                              &SSIM    &$0.6883\pm0.06$     &$0.8645\pm0.05$&$0.7351\pm0.07$     &$0.8046\pm0.06$  &$0.9616\pm0.01$  \\      
		\hline
		\multirow{3}{*}{$5\times10^3$}&PSNR    &$26.77\pm2.32$      &$30.42\pm2.43$ &$27.53\pm2.47$       &$29.42\pm2.07$  &$33.39\pm2.01 $   \\
		                              &RMSE    &$63.63\pm10.38$     &$41.94\pm7.83$ &$54.90\pm9.67$       &$44.17\pm8.74$  &$29.51\pm3.19$    \\
		                              &SSIM    &$0.5401\pm0.07$     &$0.7178\pm0.08$&$0.6124\pm0.09$      &$0.7267\pm0.08$ &$0.9453\pm0.01$   \\\hline
		Dose level  & Index   & NeumannNet & MoDL & PGD& Learned-PD   &AHP-Net                \\
		\hline
		\multirow{3}{*}{$1 \times 10^5$} &PSNR & $35.17\pm2.14$   &$38.94\pm3.33$ &$38.27\pm2.03$ &$\bm{41.16\pm 2.12}$  &$41.12\pm2.69$\\   	                                    
		                              &RMSE    & $24.20\pm3.87$   &$16.88\pm4.58$&$16.76\pm 1.12$  &$\bm{12.08\pm1.39}$     &$12.41\pm3.31$\\
		                              &SSIM    & $0.9442\pm0.02$  &$0.9842\pm0.02$&$0.9590\pm 0.01$ &$0.9863\pm 0.00$       &$\bm{0.9875\pm0.01}$\\
		\hline
		\multirow{3}{*}{$5\times10^4$}&PSNR    & $34.81\pm2.17$   &$38.09\pm2.80$ &$38.14\pm2.07$  &$39.25\pm 2.09$       &$\bm{39.55\pm3.08}$\\ 
		                              &RMSE    & $25.19\pm3.65$   &$17.80\pm6.79$&$17.03\pm1.37$ &$15.02\pm 1.57$      &$\bm{14.98\pm5.06}$\\ 
		                              &SSIM    &$0.9354\pm0.02$   &$0.9796\pm0.01$&$0.9531\pm0.01$ &$0.9808\pm0.01$      &$\bm{0.9819\pm0.01}$\\
		\hline
		\multirow{3}{*}{$1 \times 10^4$} &PSNR & $33.27\pm2.08$   &$36.20\pm2.55$ &$35.53\pm 2.24$  &$35.69\pm2.10$       &$\bm{38.20\pm2.50}$\\
		                              &RMSE    & $29.98\pm 3.77$  &$21.65\pm5.25$&$23.10\pm 2.84$ &$22.64\pm2.34$        &$\bm{17.26\pm4.60}$\\
		                              &SSIM    & $0.8736\pm 0.03$ &$0.9656\pm0.01$&$ 0.9602\pm0.01$ &$0.9637\pm0.01$       &$\bm{0.9738\pm 0.01}$\\      
		\hline
		\multirow{3}{*}{$5\times10^3$}&PSNR     & $32.16\pm2.07$   &$35.07\pm2.81$ &$34.91\pm2.15$ &$35.16\pm2.14$        &$\bm{36.93\pm2.42}$\\
		                              &RMSE    & $34.07\pm4.05$   &$25.23\pm13.45$&$24.73\pm 2.26$ &$24.10\pm2.74$       &$\bm{19.84\pm3.95}$\\
		                              &SSIM     &$0.8017\pm0.04$   &$0.9605\pm0.02$ &$ 0.9527\pm0.01$&$0.9594\pm0.01$      &$\bm{0.9673\pm 0.01}$\\ 
		 \hline
		\hline
	\end{tabular}

	\label{ComparisionU}
\end{table*}

See Table \ref{Comparision} for the quantitative comparison of   the results from different methods, where one single model was trained for processing the data with varying dose levels. Again,   Neumann-Net don't perform as well as the others.   The proposed AHP-Net is the best performer among all. The performance advantage of the proposed AHP-Net over the second best performer (Learned-PD) is around \underline{$1.5$} dB for higher dose levels and \underline{$2.2$} dB for lower dose levels.

Indeed, while the performance of the universal model trained for varying dose levels is compared to that of the individual models trained for each fixed dose level, the proposed AHP-Net see the smallest performance hit, which ranges from $0.3-0.6$dB in PSNR, in the case of low dose levels. Even so, by comparing Table~\ref{ComparisionU} and Table~\ref{Comparision},  the universal model of the AHP-Net still outperformed multiple models of other compared methods, in all tested dose levels. 

In short, the proposed AHP-Net showed its advantage  over existing representative non-learning  methods and deep learning methods, specially when the dose level is low. Also, the proposed AHP has its advantage in terms of practical usage, as it allows to train a single model to processing the data w.r.t. varying dose levels with very competitive performance. The outcome of  quantitative comparison clearly indicates the importance of the setting of hyper-parameters to optimization-unrolling-based deep learning methods and the effectiveness of our solution to it.

\begin{table*}[htbp!]
	\centering
	\caption{Quantitative comparison (Mean$\pm$STD) of  the results reconstructed by different methods on testing dataset, where one single model is trained for different dose levels in deep learning methods.}
	\scalebox{1.0}{
    \begin{tabular}{cccccccc}
		\hline
		\hline
		{Dose level}                  &Index              & FBPConvNet      &NeumannNet            &MoDL   &PGD &Learned-PD    &AHP-Net            \\
		\hline
		\multirow{3}{*}{$1 \times 10^5$}       &PSNR       &$37.44\pm1.94$   & $35.06\pm2.12$   &$34.30\pm6.21$  &$38.29\pm2.06$&$39.07\pm1.99$ &$\bm{40.54\pm2.42}$\\	                                    
		                              &RMSE       &$18.49\pm1.75$   & $24.50\pm3.80$   &$35.89\pm36.30$ &$16.75\pm1.43$&$15.34\pm1.54$ &$\bm{13.15\pm2.97}$\\
		                              &SSIM       &$0.9718\pm0.01$   & $0.9434\pm0.02$ &$0.9481\pm0.07$&$0.9776\pm0.01$&$0.9820\pm0.00$ &$\bm{0.9855\pm0.01}$\\
		\hline 
		\multirow{3}{*}{$5\times10^4$}&PSNR      &$37.06\pm1.96 $  & $34.94\pm2.12$   &$34.21\pm5.83$  &$38.04\pm2.07$&$38.46\pm2.02$&$\bm{40.07\pm2.40}$\\ 
		                              &RMSE      &$19.32\pm1.77$   & $24.82\pm3.77$   &$34.73\pm31.78$ &$17.25\pm1.48$&$16.45\pm1.64$ &$\bm{13.85\pm2.96}$\\ 
		                              &SSIM      &$0.9695\pm0.01$  &$0.9366\pm0.02$   &$0.9489\pm0.06$ &$0.9761\pm 0.02$ &$0.9792\pm0.01$&$\bm{0.9832\pm0.01}$\\
		\hline
		\multirow{3}{*}{$1 \times 10^4$}       &PSNR      &$34.76\pm2.11 $   & $33.82\pm2.07$   &$33.21\pm4.86$ &$36.15\pm2.17$&$35.88\pm2.14$ &$\bm{37.92\pm2.37}$\\
		                              &RMSE      &$25.19\pm2.63$    & $28.17\pm 3.60$  &$35.67\pm26.71$&$21.48\pm2.27$&$22.16\pm2.31$ &$\bm{17.65\pm3.14}$\\
		                              &SSIM      &$0.9508\pm 0.01$  & $0.8759\pm 0.03$ &$0.9447\pm0.05$&$0.9642\pm0.01$&$0.9641\pm 0.01$ &$\bm{0.9713\pm 0.01}$\\      
		\hline
		\multirow{3}{*}{$5\times10^3$}&PSNR      &$32.71\pm2.23$   & $32.22\pm2.10$   &$32.06\pm4.61$  &$33.94\pm2.33$&$34.12\pm 2.21$ &$\bm{36.39\pm2.38}$\\
		                              &RMSE      &$32.02\pm4.31$   & $33.78\pm3.96$   &$39.82\pm28.93$ &$27.87\pm4.38$&$27.15\pm3.22$  &$\bm{21.03\pm3.59}$\\
		                              &SSIM      &$0.9230\pm0.03$  &$0.7947\pm0.04$   &$0.9305\pm0.06$ &$0.9425\pm0.02$&$0.9498\pm0.01$ &$\bm{0.9603\pm 0.02}$\\
		\hline                                                                                                                       
		\hline
	\end{tabular}
		}
\label{Comparision}
\end{table*}
	\begin{figure}[htbp!]
	\begin{center}
		\begin{tabular}{c@{\hspace{10pt}}c@{\hspace{10pt}}c@{\hspace{0pt}}c@{\hspace{0pt}}c@{\hspace{0pt}}c@{\hspace{0pt}}c}
			\includegraphics[width=.45\linewidth,height=.3\linewidth]{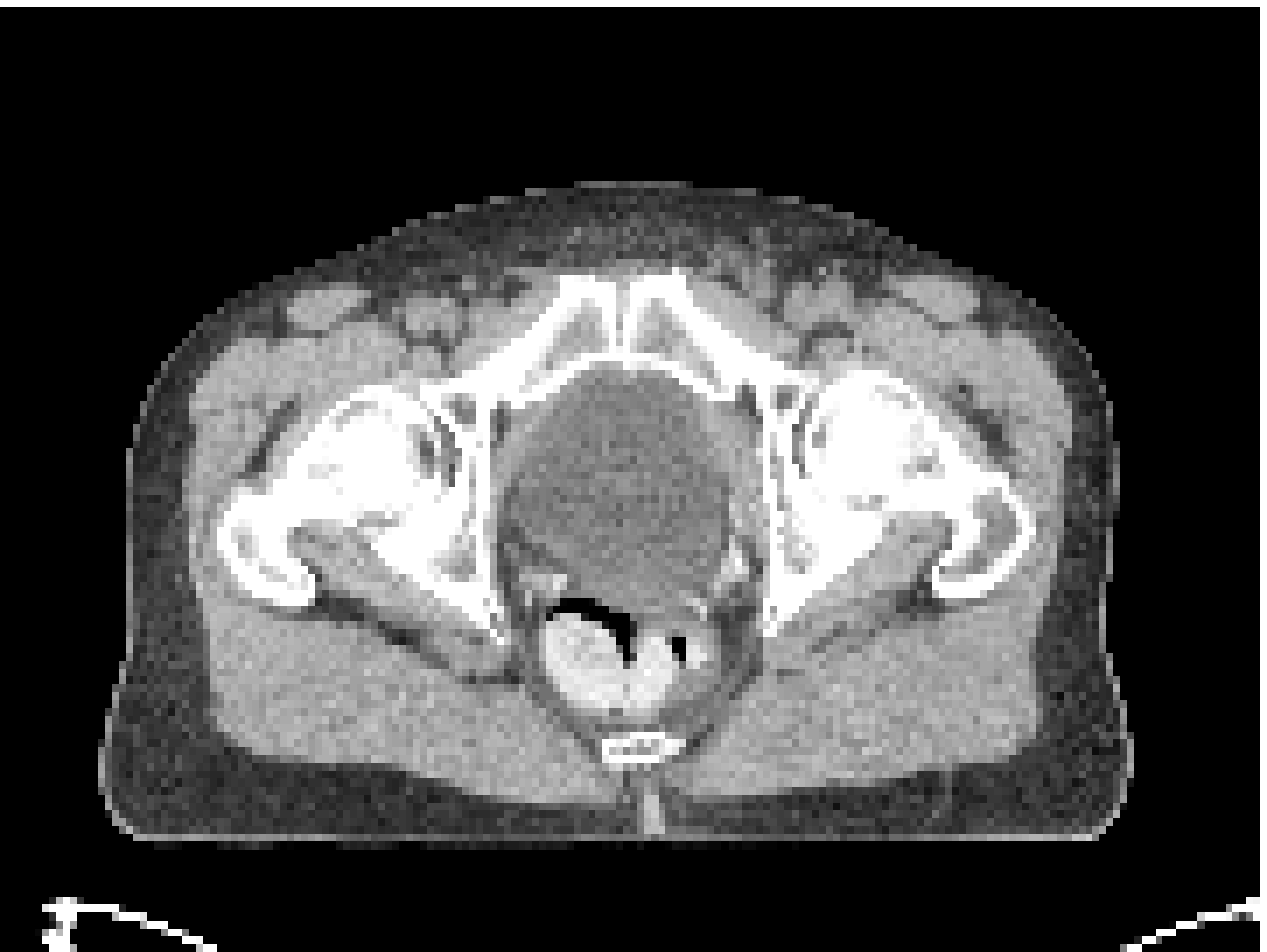}&
			\begin{tikzpicture}
			\node[anchor=south west,inner sep=0] (image) at (0,0) {\includegraphics[width=.3\linewidth,height=.3\linewidth]{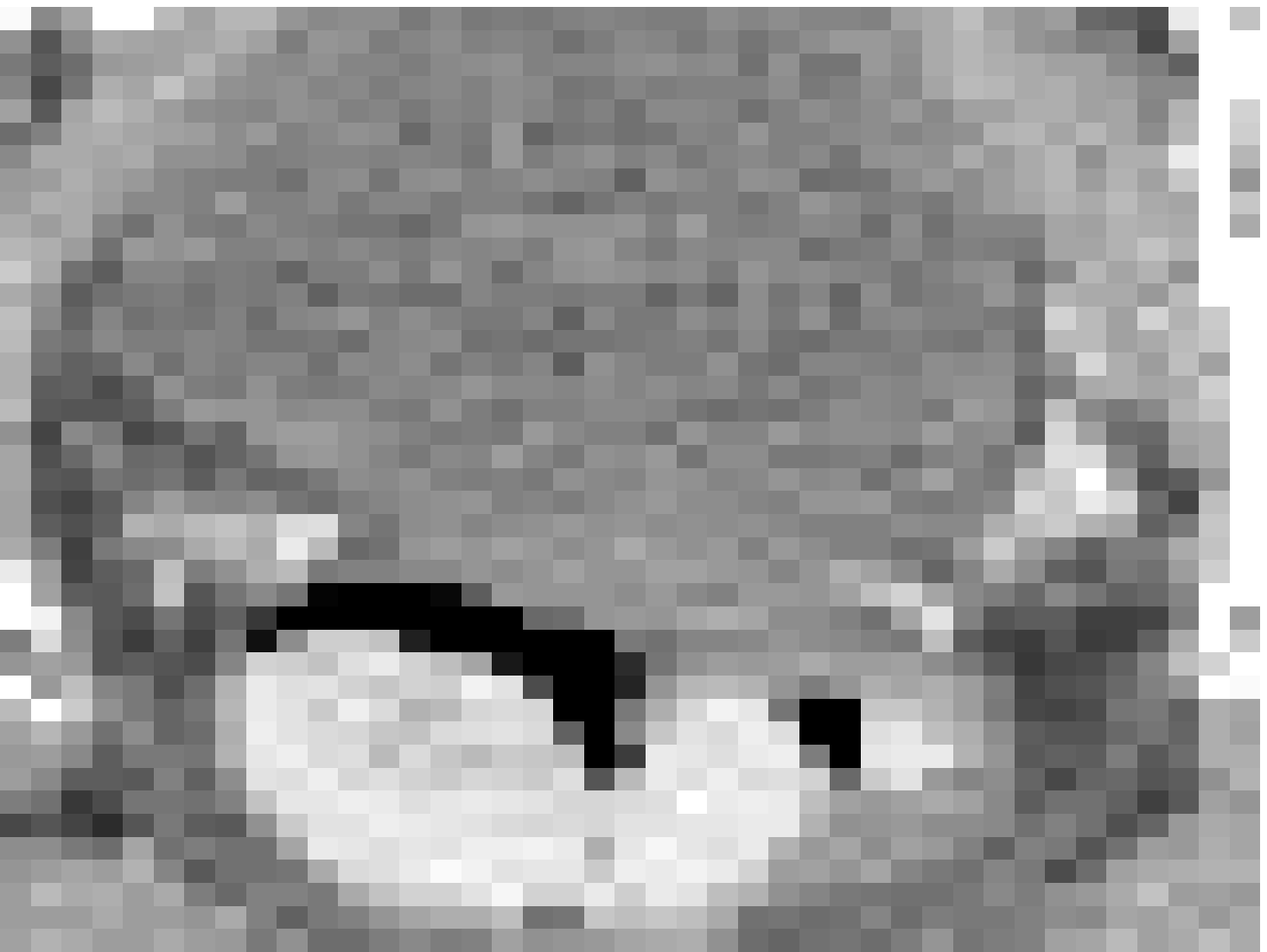}};
			\draw [-stealth, line width=2pt, cyan] (1.05,1.95) -- ++(-0.45,-0.45);
			\draw [-stealth, line width=2pt, cyan] (2.5,0.675) -- ++(-0.675,-0.0);
			\end{tikzpicture}\\
			NDCT&
			ROI
		\end{tabular}
		\caption{Normal dose CT image and zoomed region (ROI).
		}
		\label{Truth}
	\end{center}
\end{figure}
\subsection{Visual comparison of some examples}
Due to space limitation, only the images reconstructed by different methods from one sample measurement are shown for visual comparison. The dose level of the measurement is $I_i=1 \times 10^4$. The displayed window is set
to $[-150, 150]$~HU for all figures. More results w.r.t. other dose levels can be found in Appendix \ref{APP:VC} for visual comparison.

\begin{figure*}
	\begin{center}
		\begin{tabular}{c@{\hspace{0pt}}c@{\hspace{0pt}}c@{\hspace{0pt}}c@{\hspace{0pt}}c@{\hspace{0pt}}c@{\hspace{0pt}}c}
			\includegraphics[width=.198\linewidth,height=.132\linewidth]{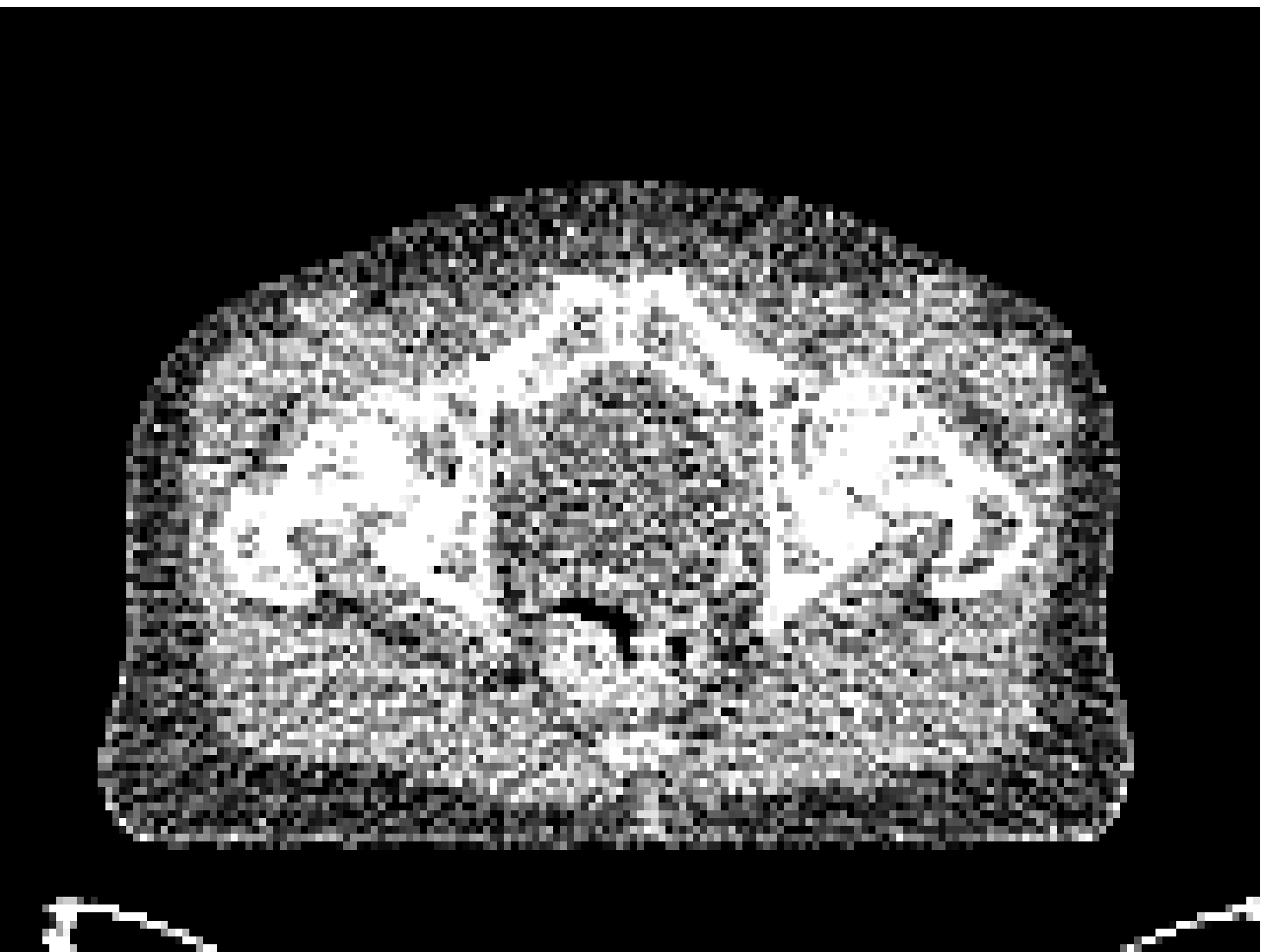}&
			\includegraphics[width=.198\linewidth,height=.132\linewidth]{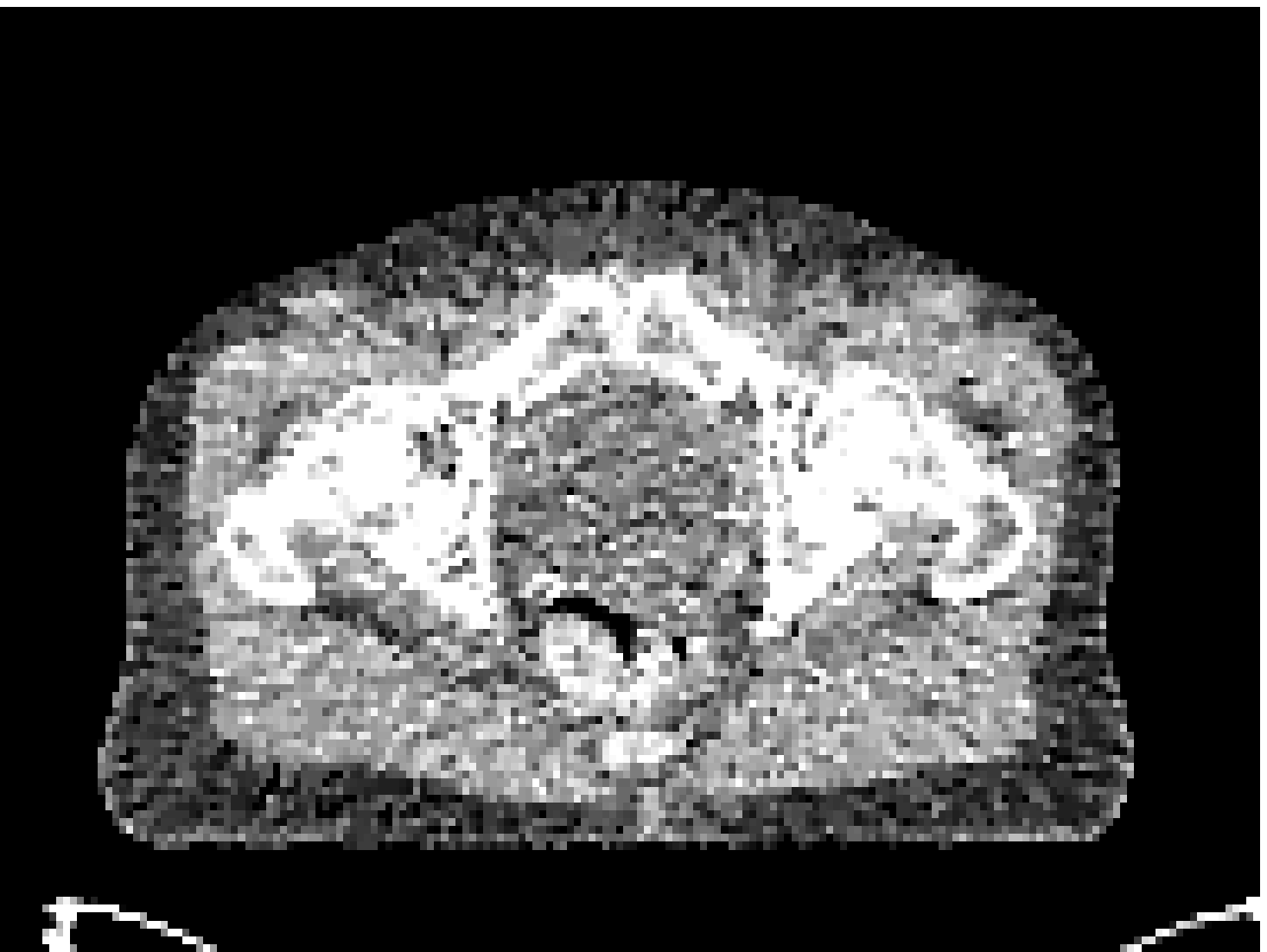}&			
			\includegraphics[width=.198\linewidth,height=.132\linewidth]{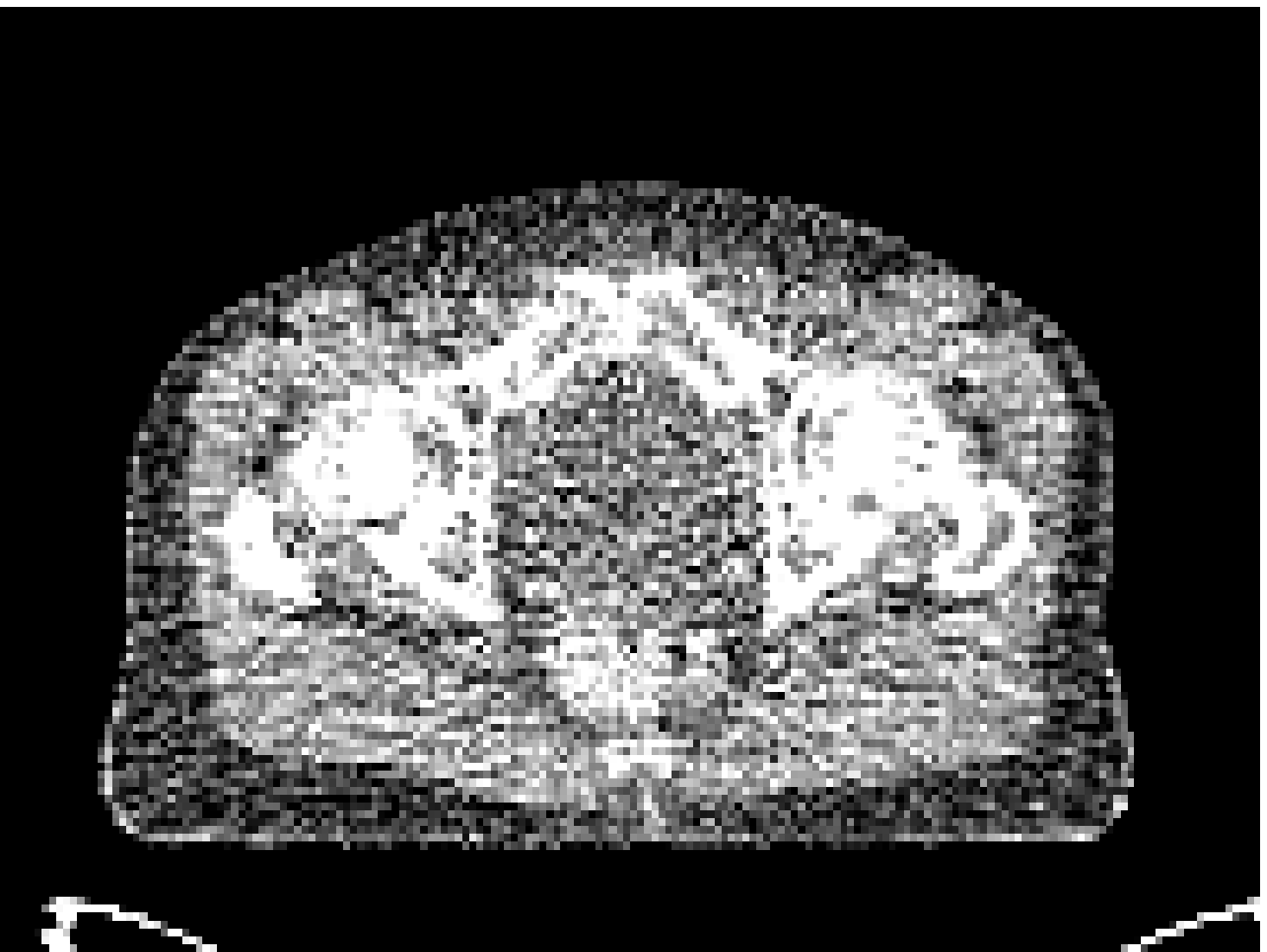}&
			\includegraphics[width=.198\linewidth,height=.132\linewidth]{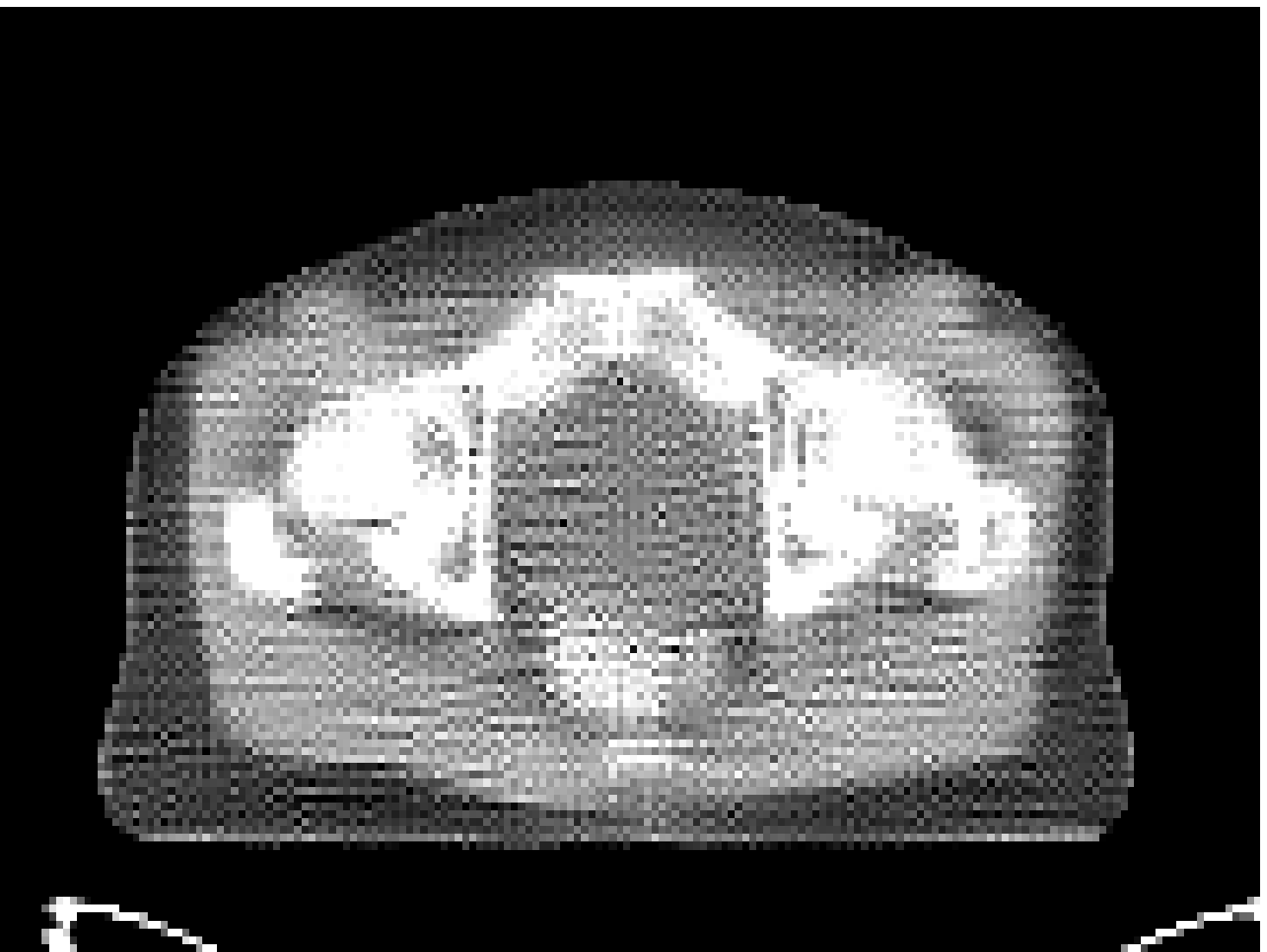}&									
			\includegraphics[width=.198\linewidth,height=.132\linewidth]{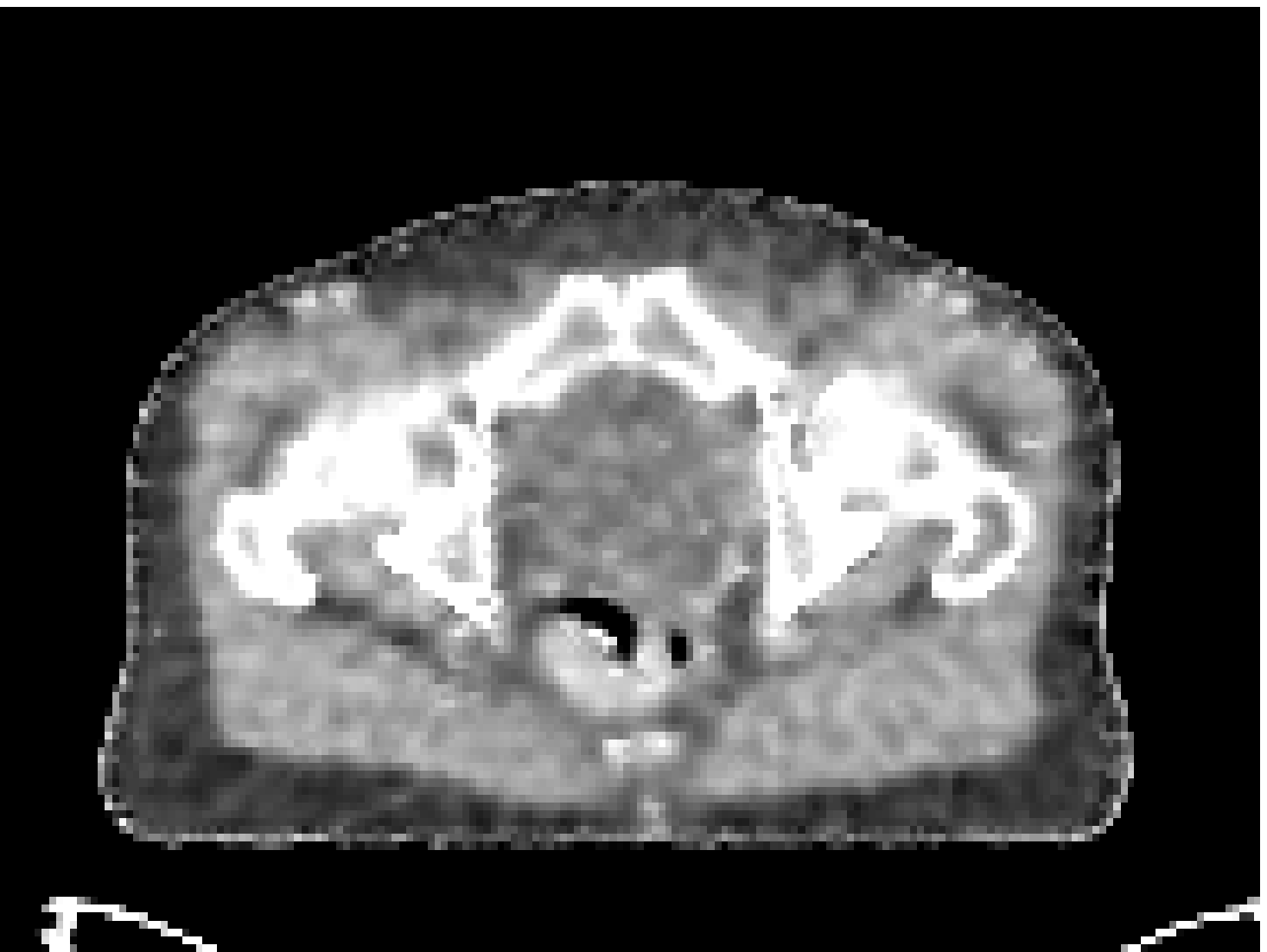}\\
			FBP&
			TV&
			KSVD&
			BM3D&
			FBPConvNet\\								
			\includegraphics[width=.198\linewidth,height=.132\linewidth]{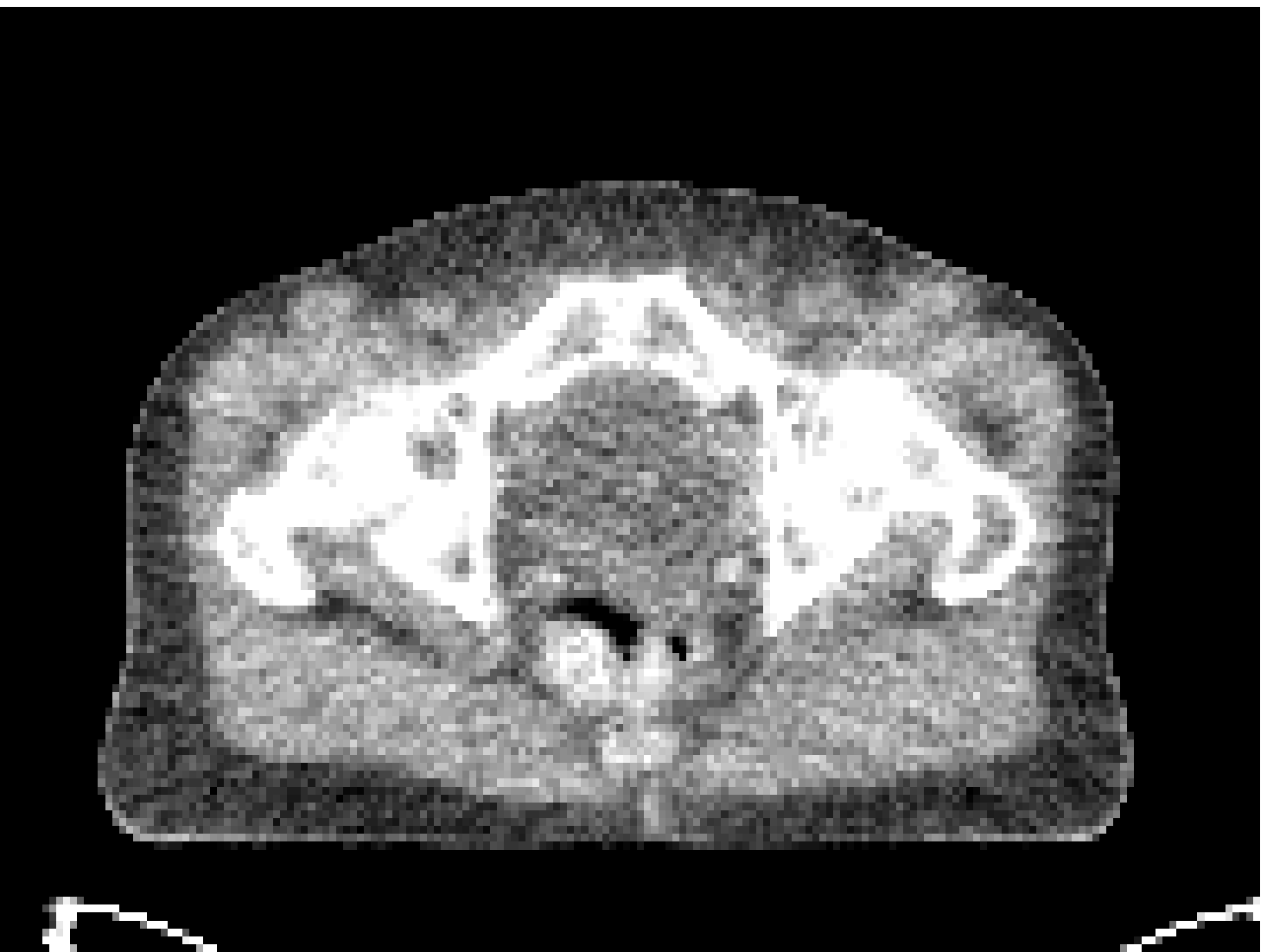}&		
			\includegraphics[width=.198\linewidth,height=.132\linewidth]{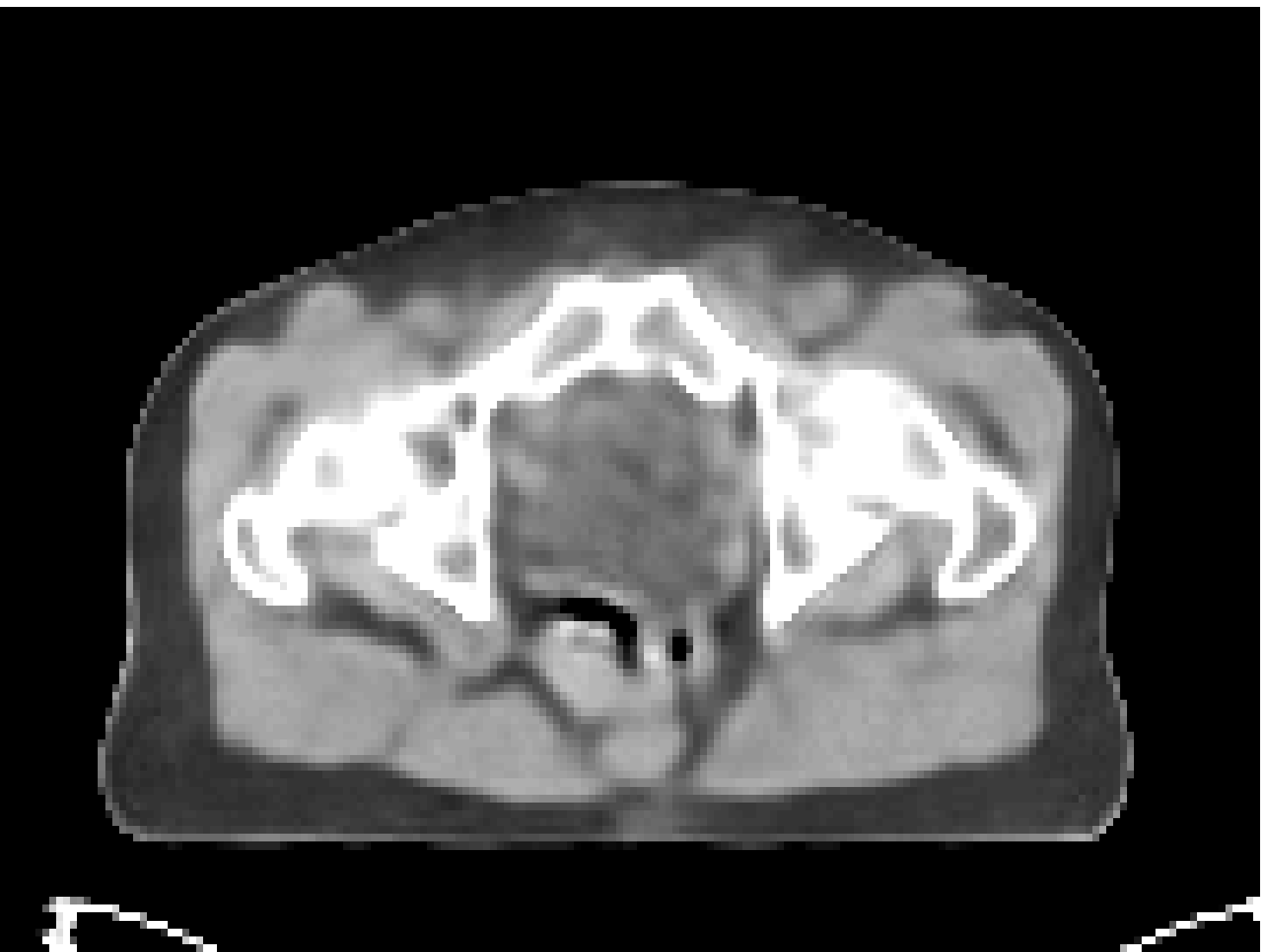}&	
			\includegraphics[width=.198\linewidth,height=.132\linewidth]{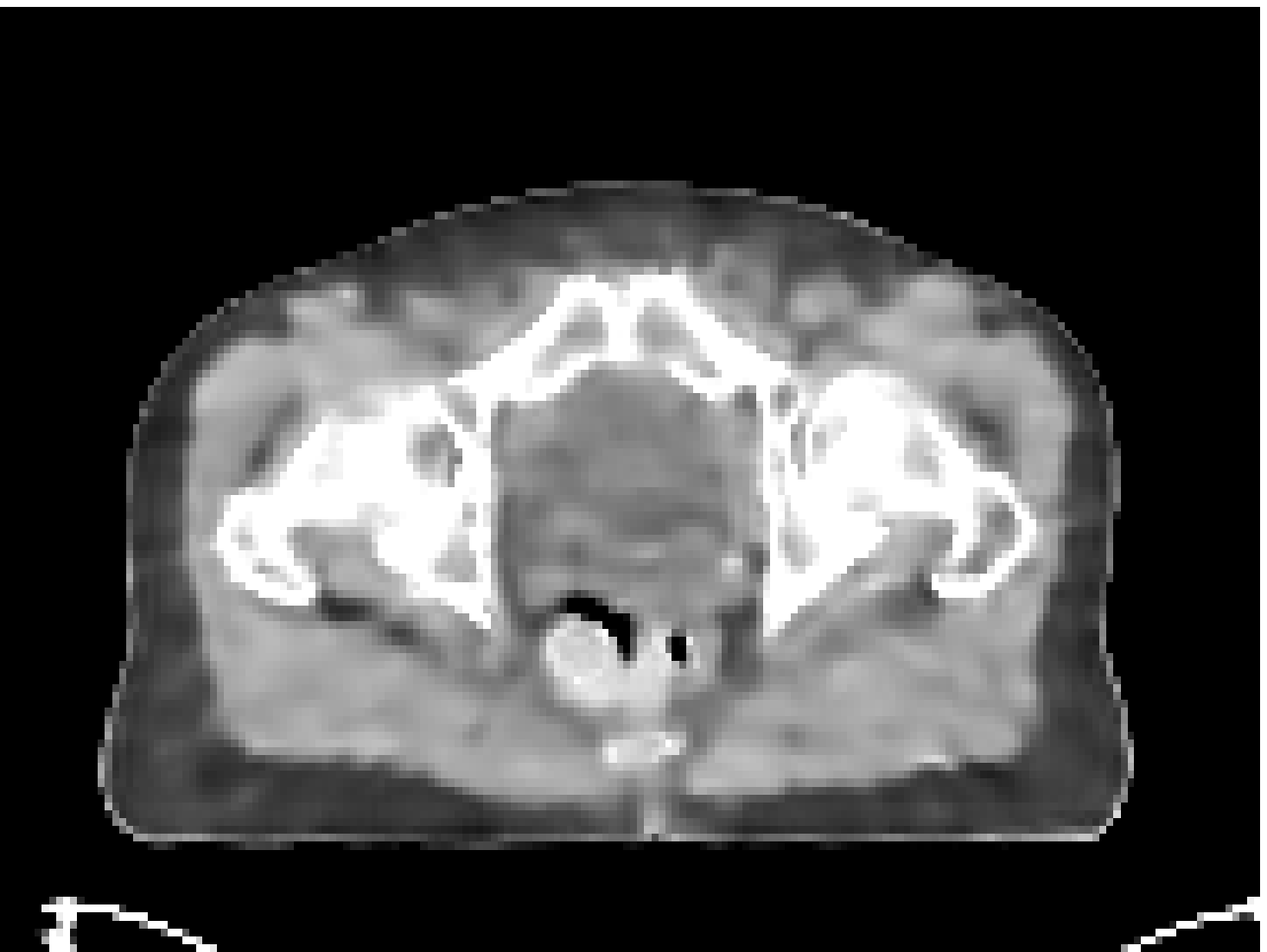}&		
			\includegraphics[width=.198\linewidth,height=.132\linewidth]{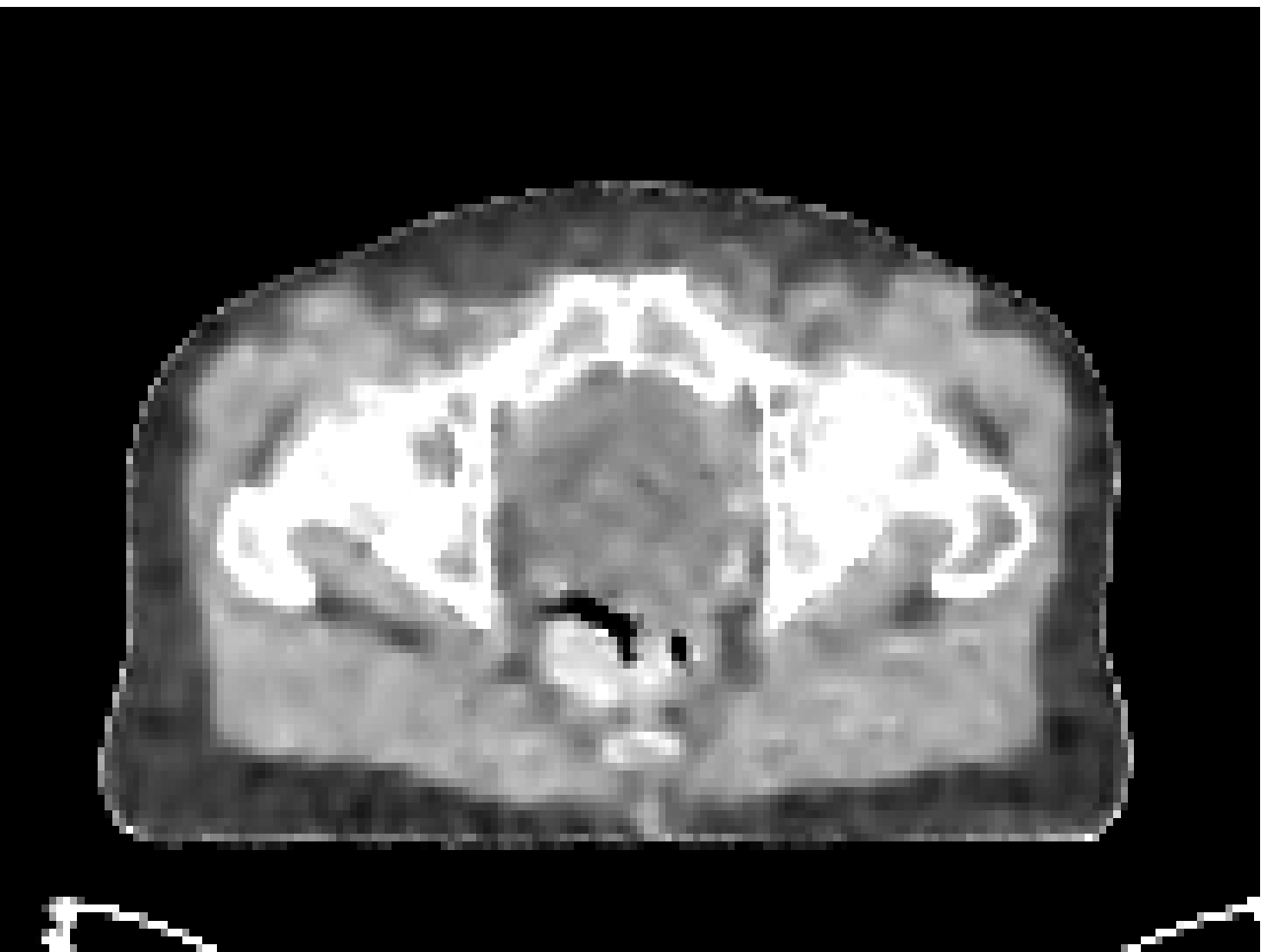}&								
			\includegraphics[width=.198\linewidth,height=.132\linewidth]{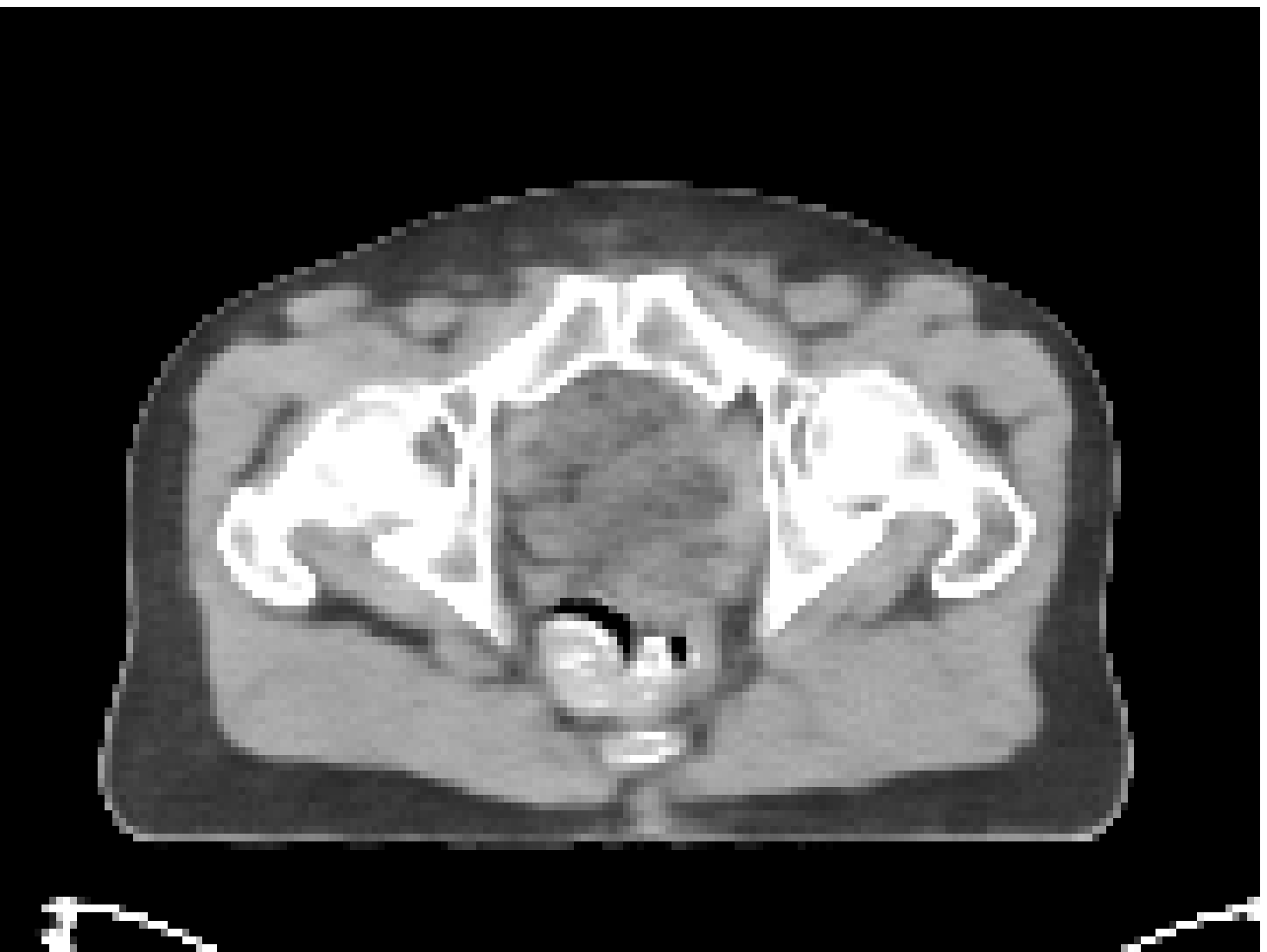}\\
			MoDL&
			Neumann-Net&
			PGD&
			Learn-PD&
			AHP-Net
		\end{tabular}
		\caption{Reconstruction results at dose level $I_i=5\times10^3$ by the models trained under same dose level.}
		\label{sliceU_5000}
	\end{center}
\end{figure*}

\begin{figure}
		\begin{tabular}{c@{\hspace{-2pt}}c@{\hspace{-2pt}}c@{\hspace{-2pt}}c@{\hspace{-2pt}}c@{\hspace{-2pt}}c@{\hspace{-2pt}}c}
			\begin{tikzpicture}
			\node[anchor=south west,inner sep=0] (image) at (0,0) {\includegraphics[width=.2\linewidth,height=.2\linewidth]{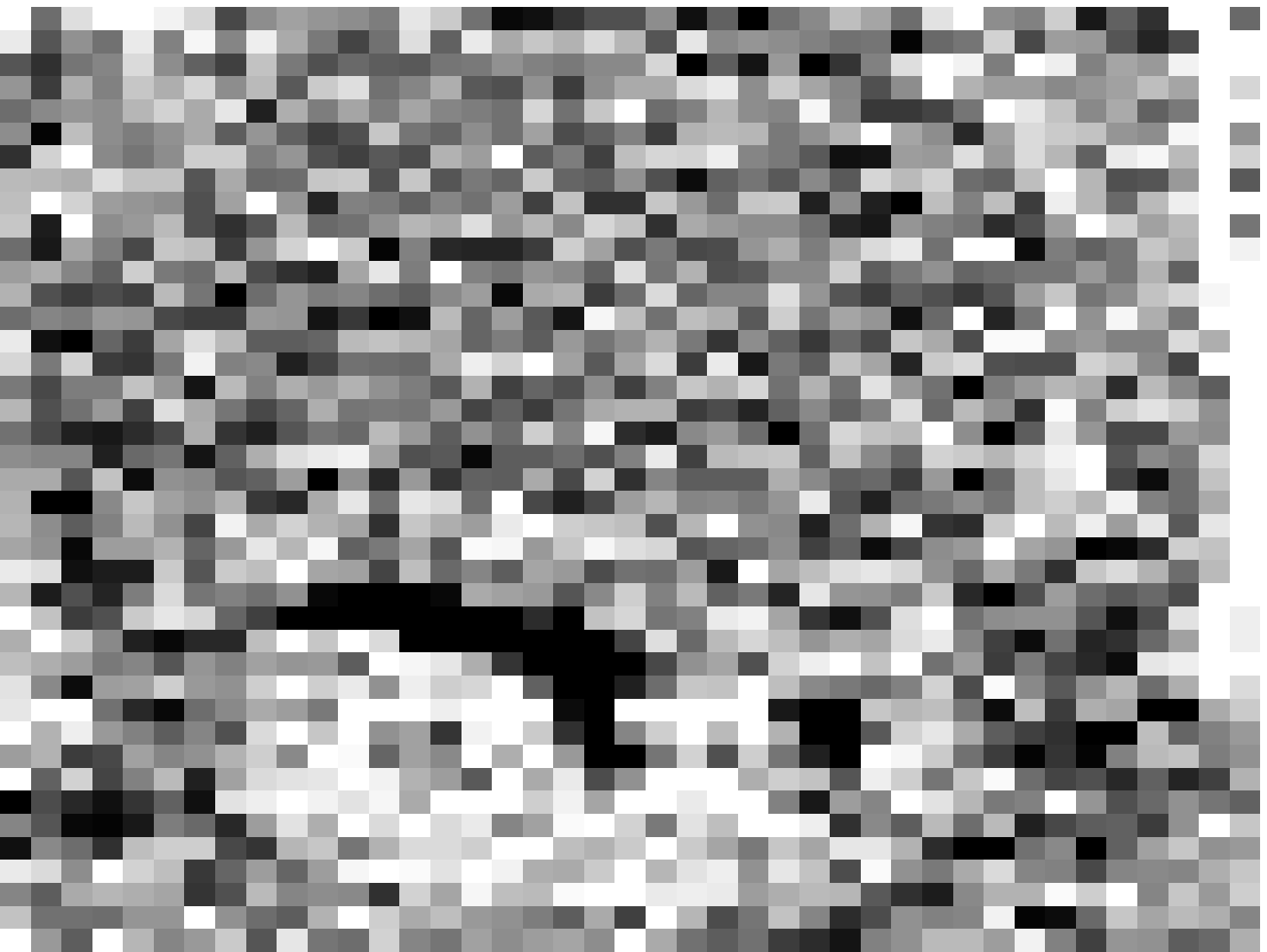}};
			\draw [-stealth, line width=2pt, cyan] (0.7,1.3) -- ++(-0.3,-0.3);
			\draw [-stealth, line width=2pt, cyan] (1.7,0.45) -- ++(-0.45,-0.0);
			\end{tikzpicture}&
			\begin{tikzpicture}
			\node[anchor=south west,inner sep=0] (image) at (0,0) {\includegraphics[width=.2\linewidth,height=.2\linewidth]{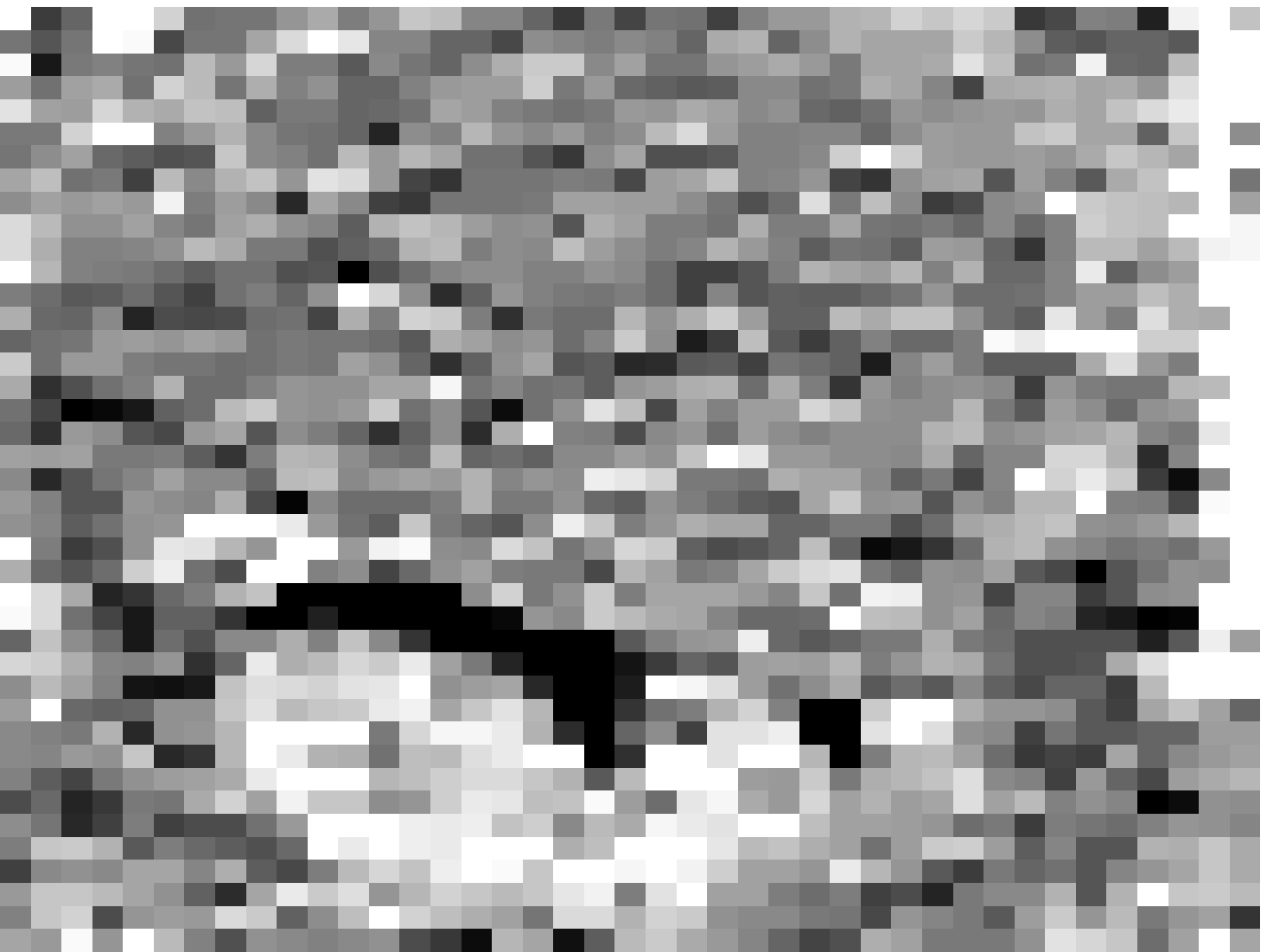}};
			\draw [-stealth, line width=2pt, cyan] (0.7,1.3) -- ++(-0.3,-0.3);
			\draw [-stealth, line width=2pt, cyan] (1.7,0.45) -- ++(-0.45,-0.0);
			\end{tikzpicture}&
			\begin{tikzpicture}
			\node[anchor=south west,inner sep=0] (image) at (0,0) {\includegraphics[width=.2\linewidth,height=.2\linewidth]{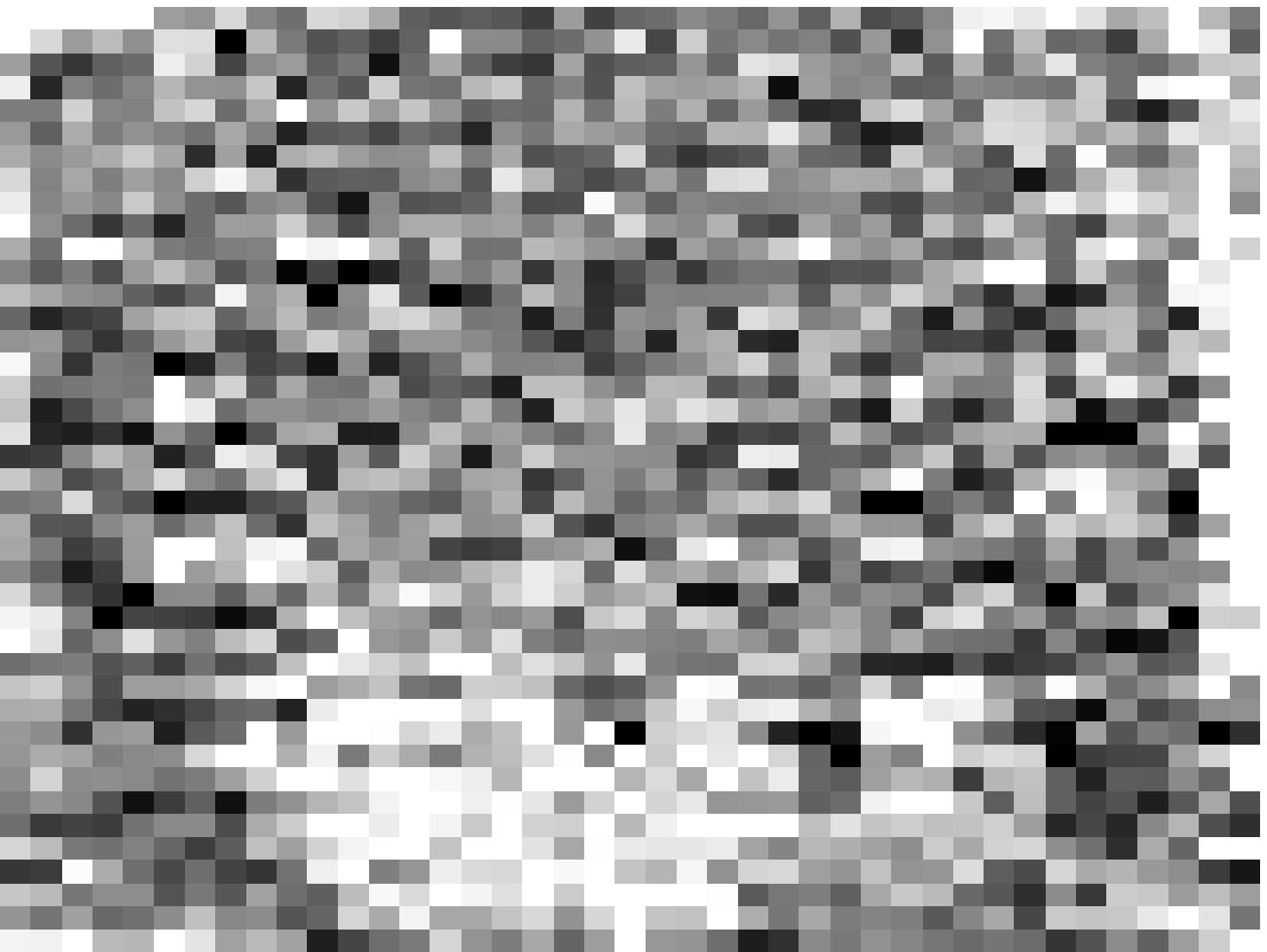}};
			\draw [-stealth, line width=2pt, cyan] (0.7,1.3) -- ++(-0.3,-0.3);
			\draw [-stealth, line width=2pt, cyan] (1.7,0.45) -- ++(-0.45,-0.0);
			\end{tikzpicture}&
			\begin{tikzpicture}
			\node[anchor=south west,inner sep=0] (image) at (0,0) {\includegraphics[width=.2\linewidth,height=.2\linewidth]{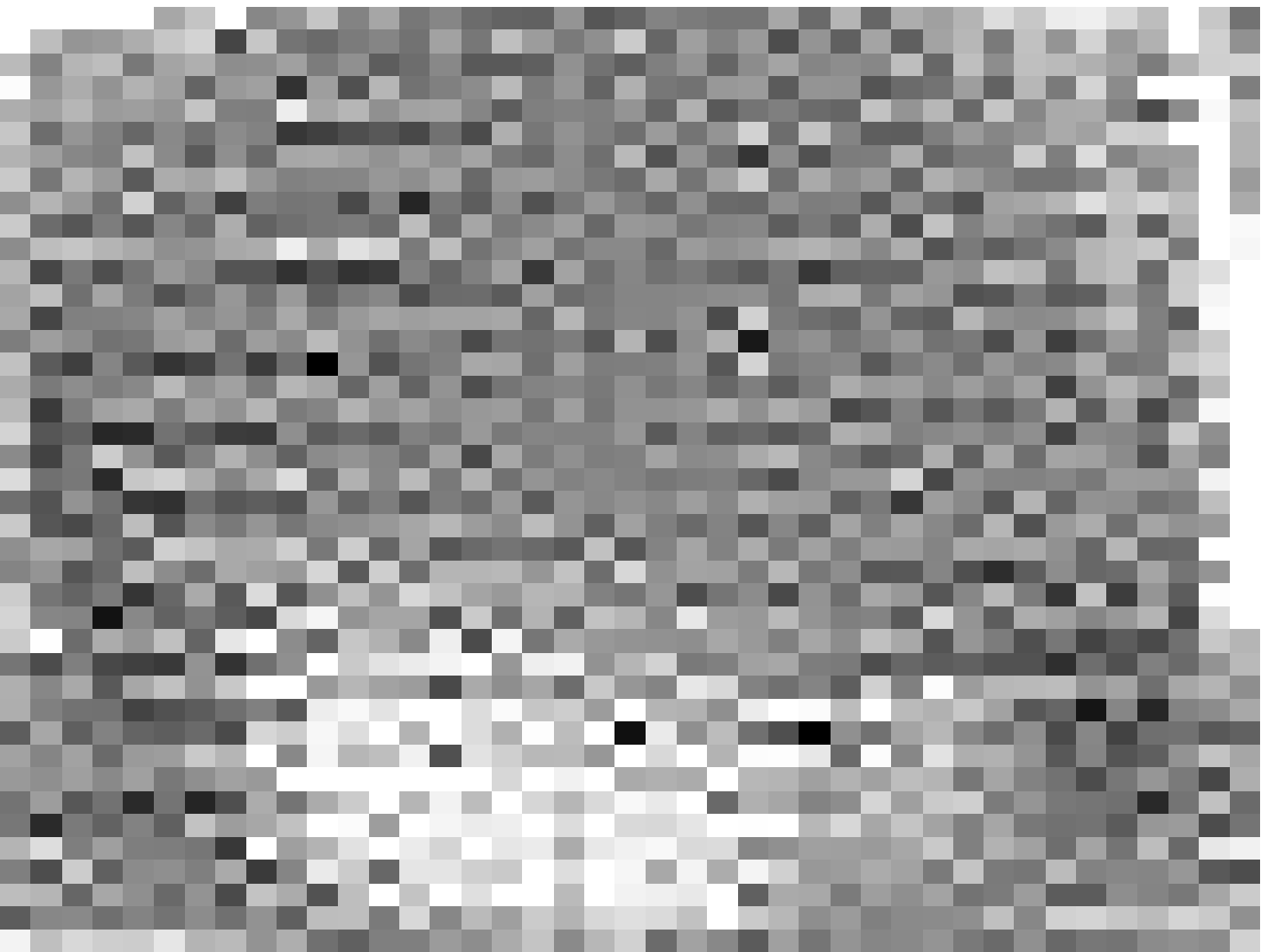}};
			\draw [-stealth, line width=2pt, cyan] (0.7,1.3) -- ++(-0.3,-0.3);
			\draw [-stealth, line width=2pt, cyan] (1.7,0.45) -- ++(-0.45,-0.0);
			\end{tikzpicture}&
			\begin{tikzpicture}
			\node[anchor=south west,inner sep=0] (image) at (0,0) {\includegraphics[width=.2\linewidth,height=.2\linewidth]{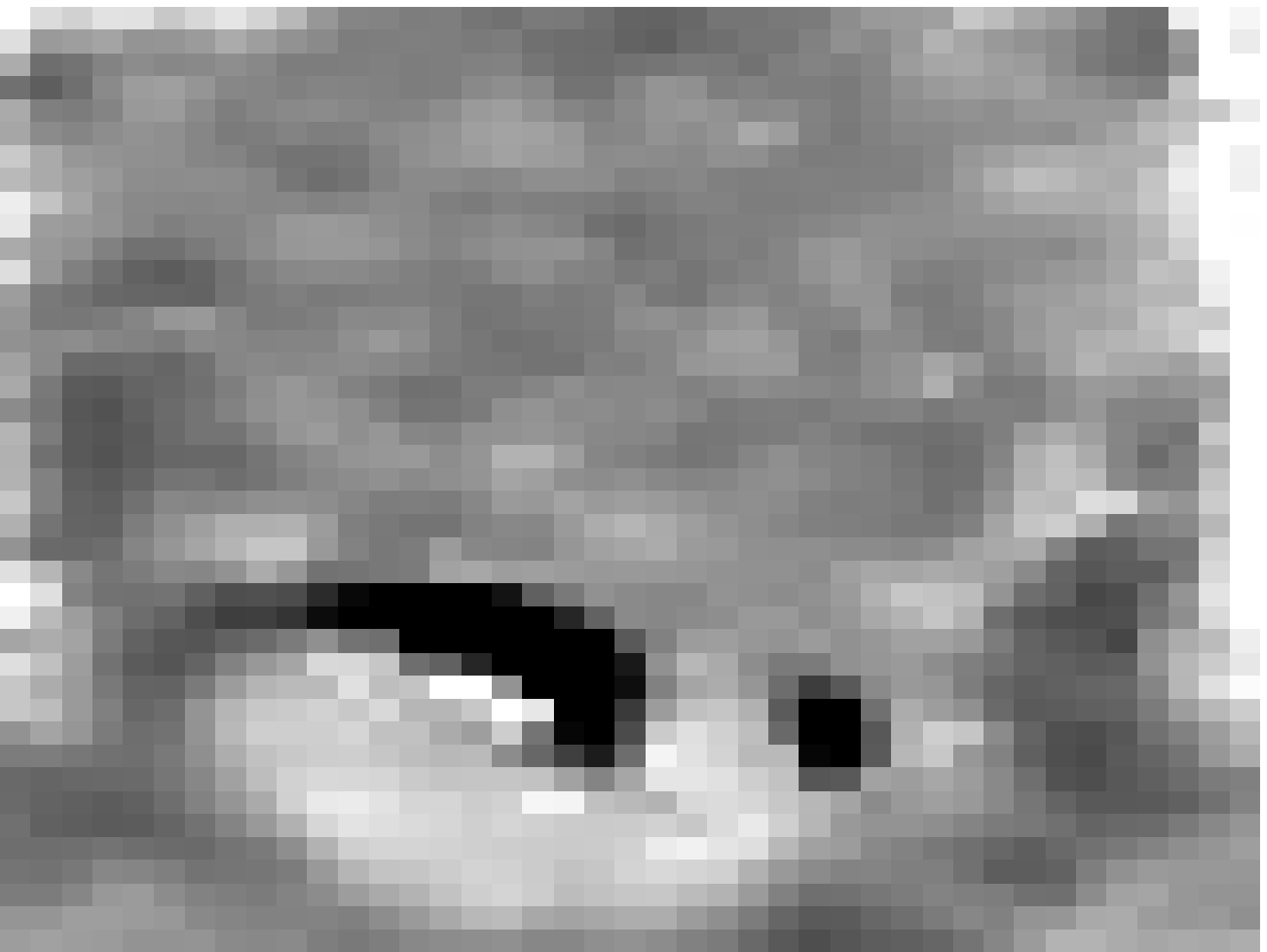}};
			\draw [-stealth, line width=2pt, cyan] (0.7,1.3) -- ++(-0.3,-0.3);
			\draw [-stealth, line width=2pt, cyan] (1.7,0.45) -- ++(-0.45,-0.0);
			\end{tikzpicture}\\
			FBP&		
			TV&
			KSVD&
			BM3D&
			FBPConvNet\\
			\begin{tikzpicture}
			\node[anchor=south west,inner sep=0] (image) at (0,0) {\includegraphics[width=.2\linewidth,height=.2\linewidth]{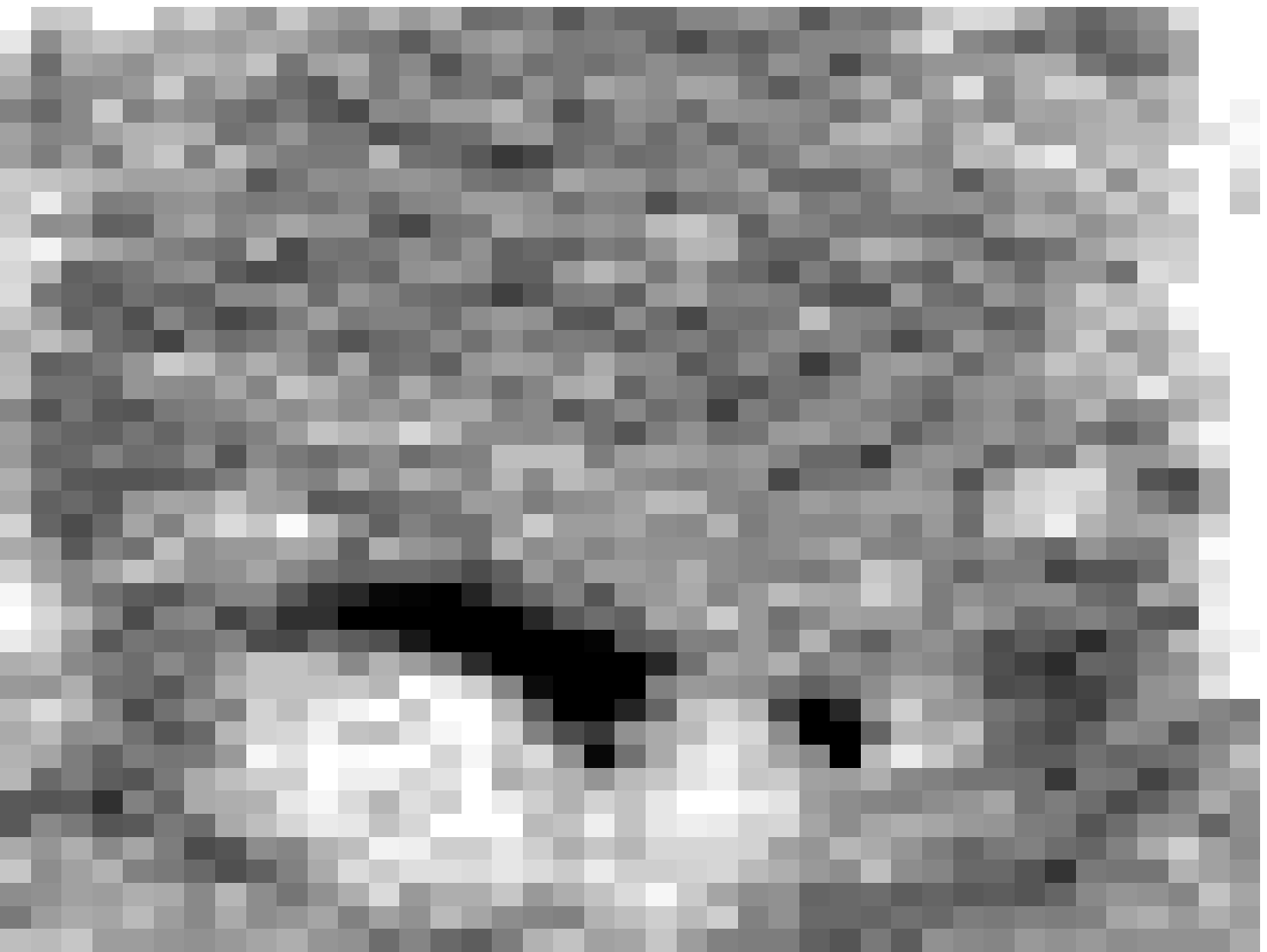}};
			\draw [-stealth, line width=2pt, cyan] (0.7,1.3) -- ++(-0.3,-0.3);
			\draw [-stealth, line width=2pt, cyan] (1.7,0.45) -- ++(-0.45,-0.0);
			\end{tikzpicture}&		
			\begin{tikzpicture}
			\node[anchor=south west,inner sep=0] (image) at (0,0) {\includegraphics[width=.2\linewidth,height=.2\linewidth]{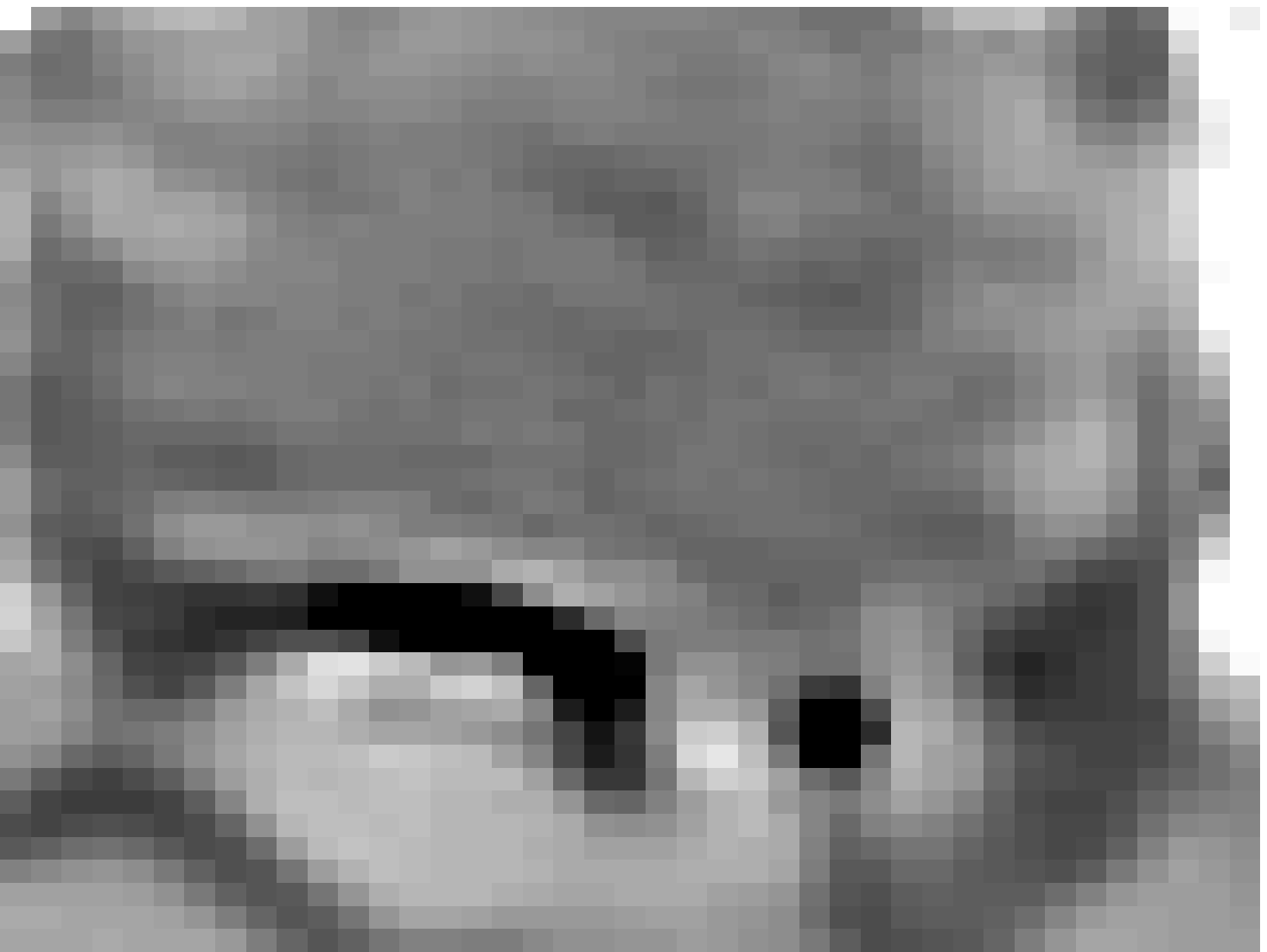}};
			\draw [-stealth, line width=2pt, cyan] (0.7,1.3) -- ++(-0.3,-0.3);
			\draw [-stealth, line width=2pt, cyan] (1.7,0.45) -- ++(-0.45,-0.0);
			\end{tikzpicture}&
			\begin{tikzpicture}
			\node[anchor=south west,inner sep=0] (image) at (0,0) {\includegraphics[width=.2\linewidth,height=.2\linewidth]{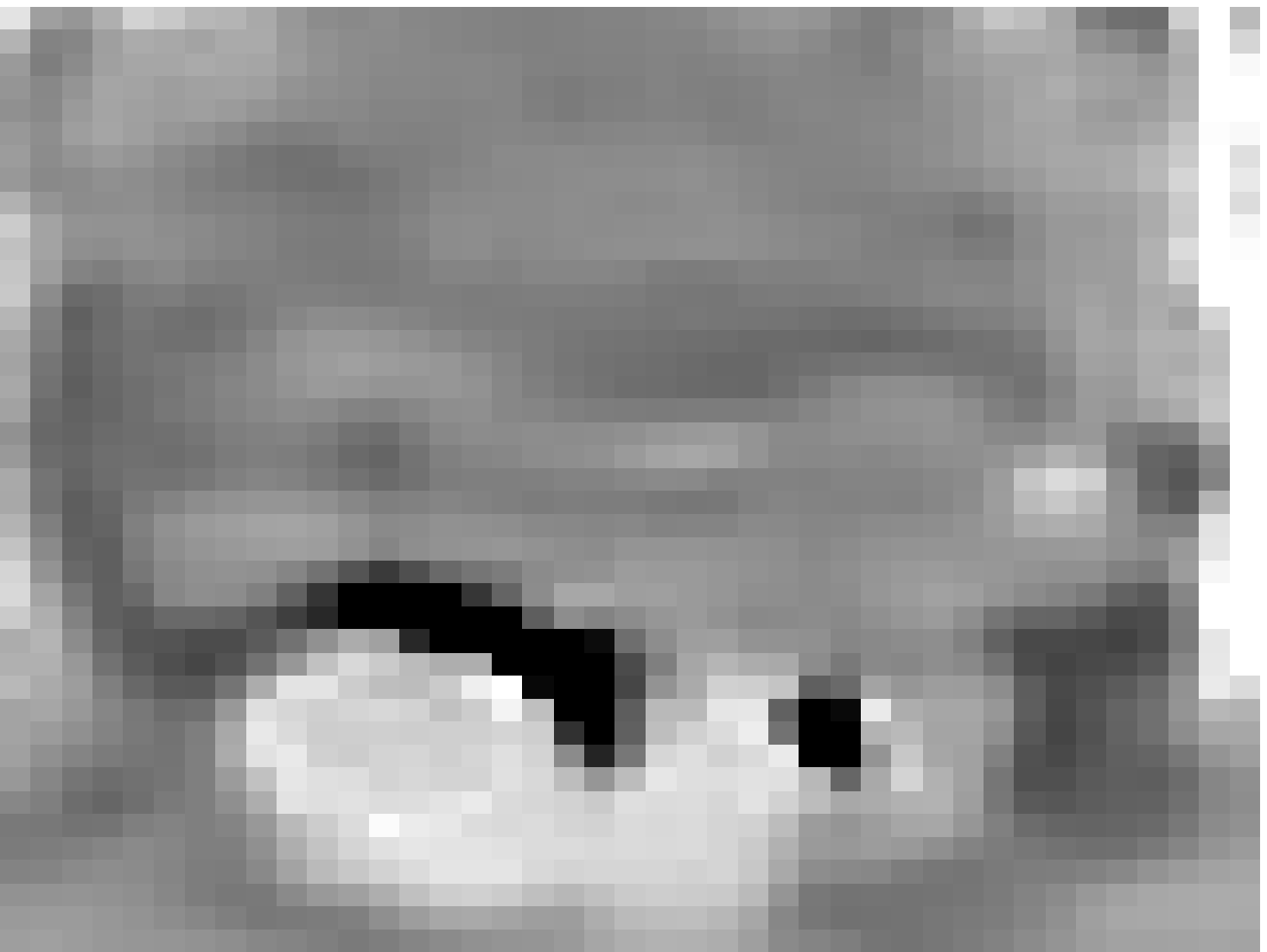}};
			\draw [-stealth, line width=2pt, cyan] (0.7,1.3) -- ++(-0.3,-0.3);
			\draw [-stealth, line width=2pt, cyan] (1.7,0.45) -- ++(-0.45,-0.0);
			\end{tikzpicture}&		
			\begin{tikzpicture}
			\node[anchor=south west,inner sep=0] (image) at (0,0) {\includegraphics[width=.2\linewidth,height=.2\linewidth]{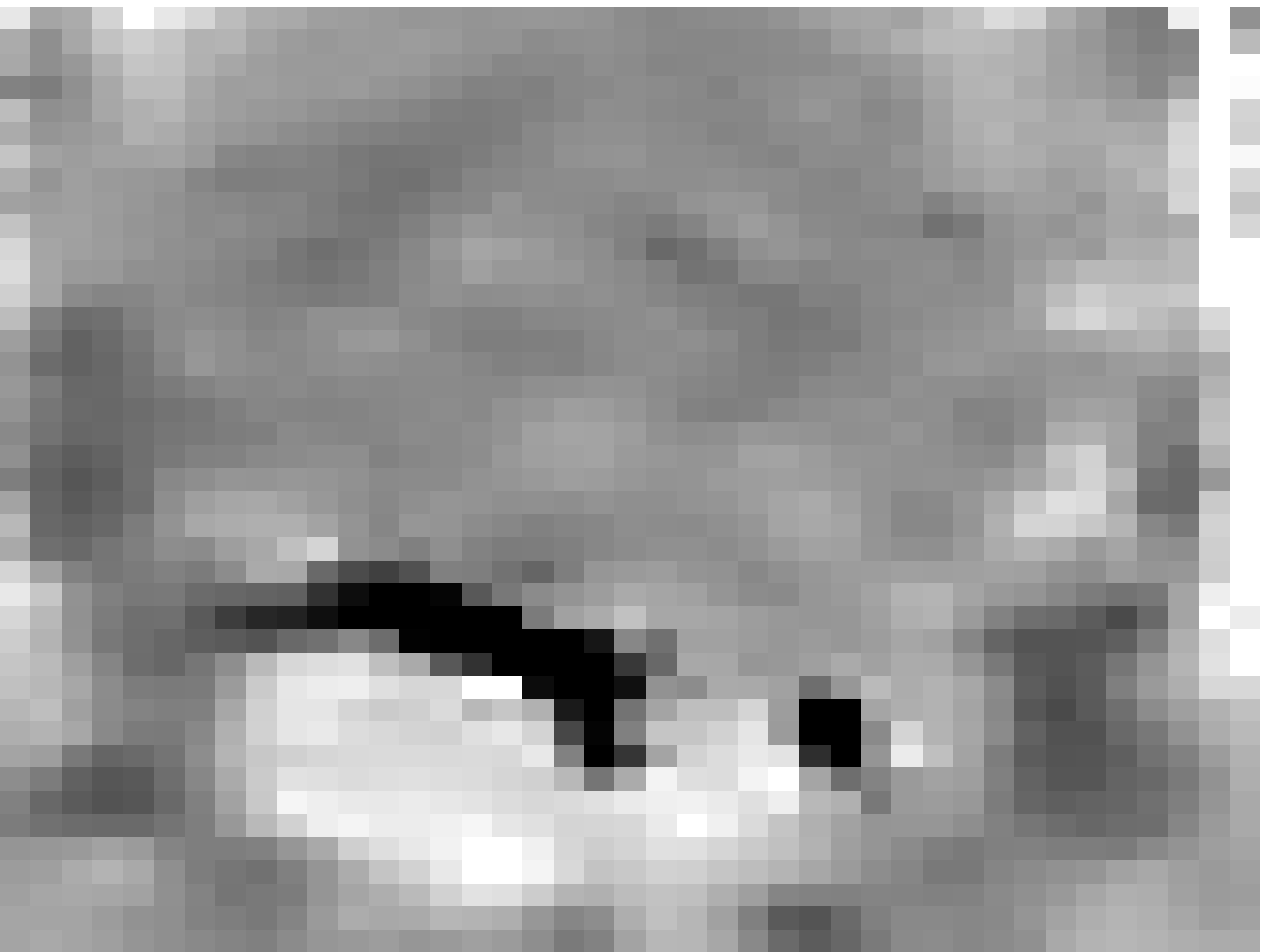}};
			\draw [-stealth, line width=2pt, cyan] (0.7,1.3) -- ++(-0.3,-0.3);
			\draw [-stealth, line width=2pt, cyan] (1.7,0.45) -- ++(-0.45,-0.0);
			\end{tikzpicture}&
			\begin{tikzpicture}
			\node[anchor=south west,inner sep=0] (image) at (0,0) {\includegraphics[width=.2\linewidth,height=.2\linewidth]{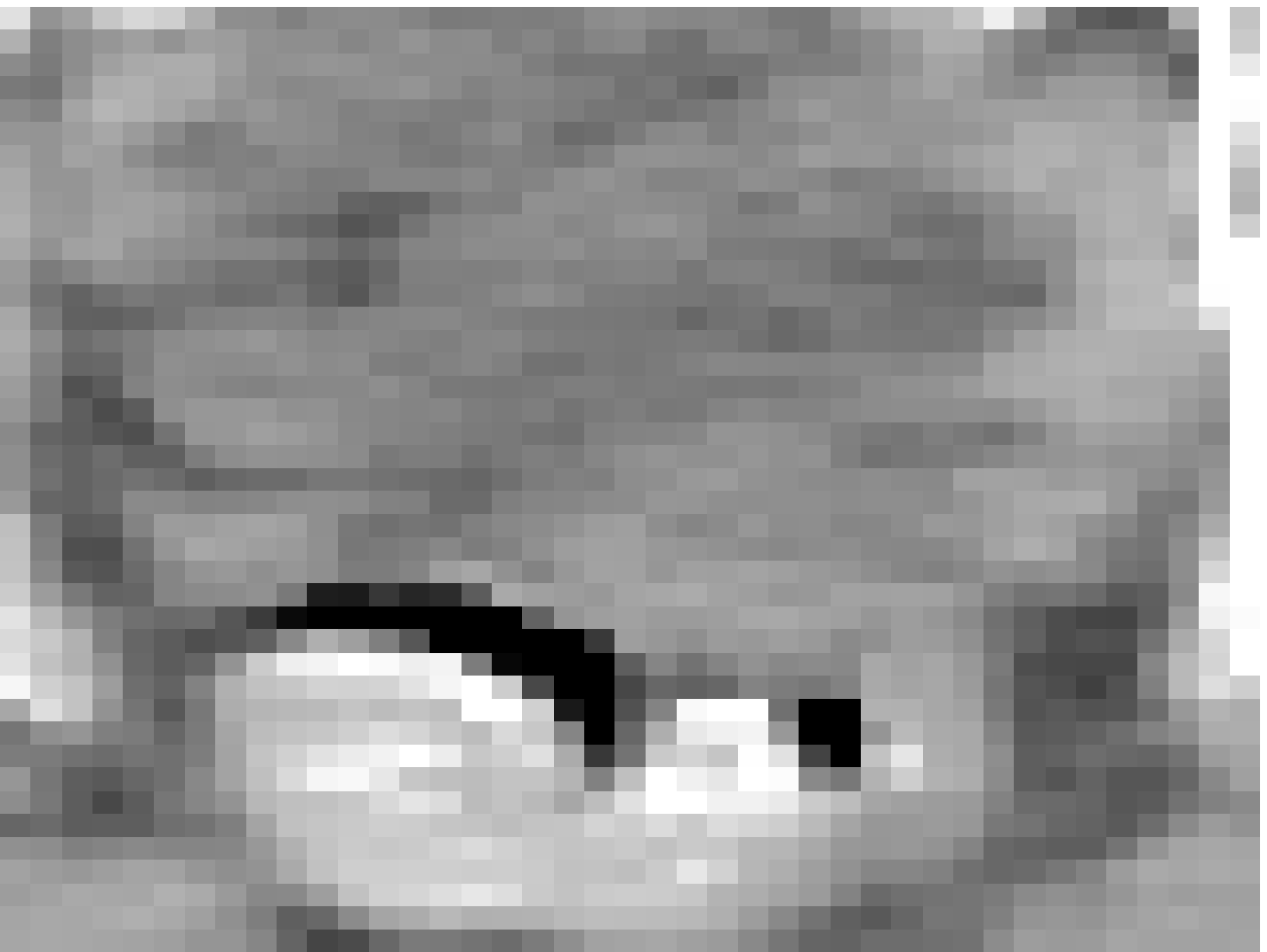}};
			\draw [-stealth, line width=2pt, cyan] (0.7,1.3) -- ++(-0.3,-0.3);
			\draw [-stealth, line width=2pt, cyan] (1.7,0.45) -- ++(-0.45,-0.0);
			\end{tikzpicture}\\
			MoDL&
			Neumann-Net&	
			PGD&
			Learned-PD&	
			AHP-Net
		\end{tabular}
		\caption{Zoom-in results of Fig.~\ref{sliceU_5000}.
		}
		\label{sliceZoomU_5000}
\end{figure}

Fig. \ref{Truth} demonstrates the selected  normal dose CT (NDCT) image and the zoomed region of interesting (ROI). For the image reconstructed by the non-learning methods and the NN models trained by deep learning methods, specifically for the dose level $I_i=5\times10^3$,
Fig. \ref{sliceU_5000} shows the  images reconstructed by different methods, and their zoomed-in images of  boxes in Fig. \ref{Truth}   are displayed in Fig.~\ref{sliceZoomU_5000}. 
As shown by the cyan arrows,  in comparison to he zoomed-in NDCT image, the results of FBPConvNet, MoDL, Neumann-Net, PGD and Learned-PD  were more blurred than that  of the proposed method.
See Table \ref{SNRRMSEU} for  quantitative comparison of the results shown in  Fig.~\ref{sliceU_5000}.

\begin{figure*}
	\begin{center}
		\begin{tabular}{c@{\hspace{0pt}}c@{\hspace{0pt}}c@{\hspace{0pt}}c@{\hspace{0pt}}c@{\hspace{0pt}}c@{\hspace{0pt}}c}
			\includegraphics[width=.198\linewidth,height=.132\linewidth]{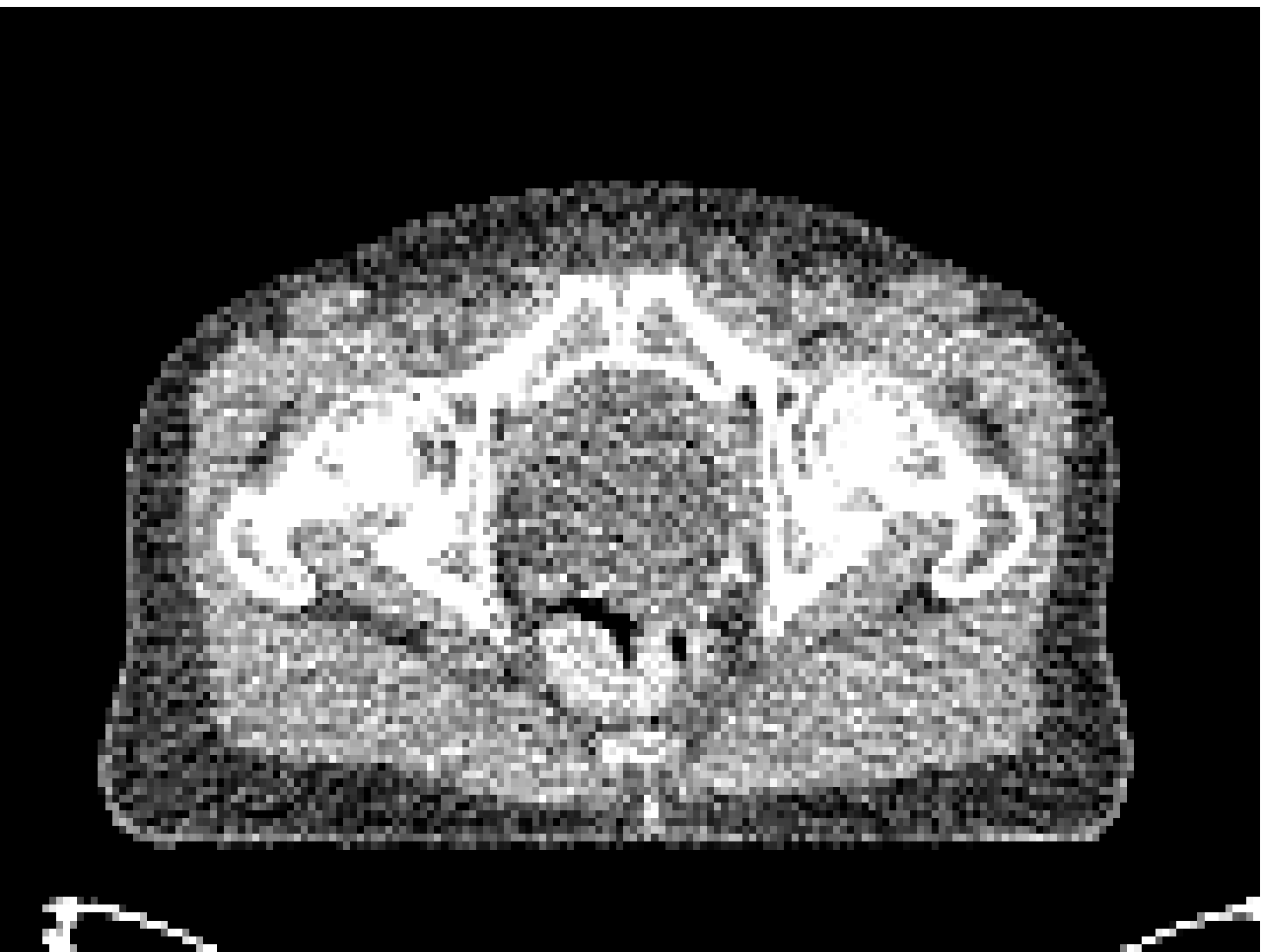}&		
			\includegraphics[width=.198\linewidth,height=.132\linewidth]{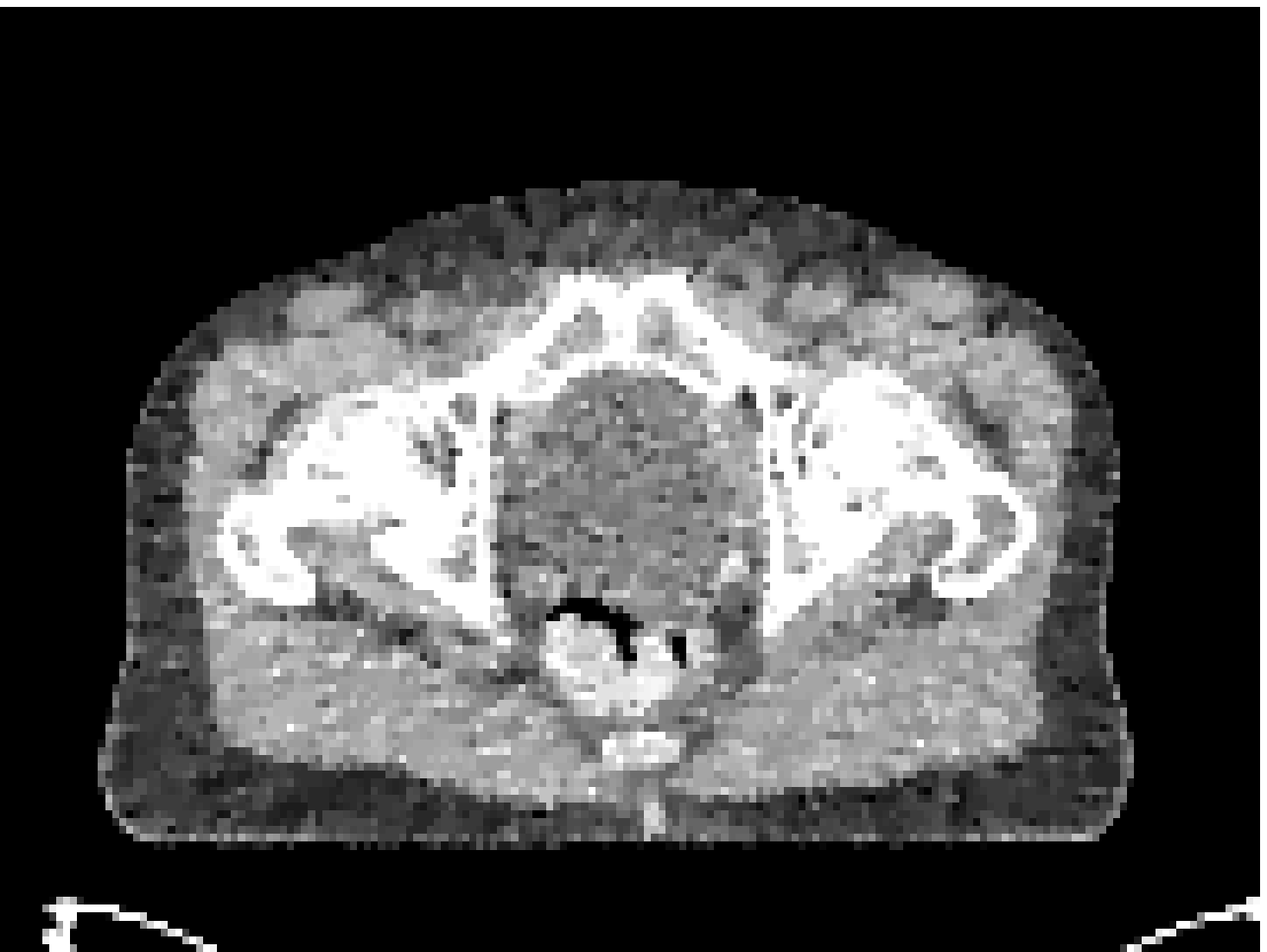}&		
			\includegraphics[width=.198\linewidth,height=.132\linewidth]{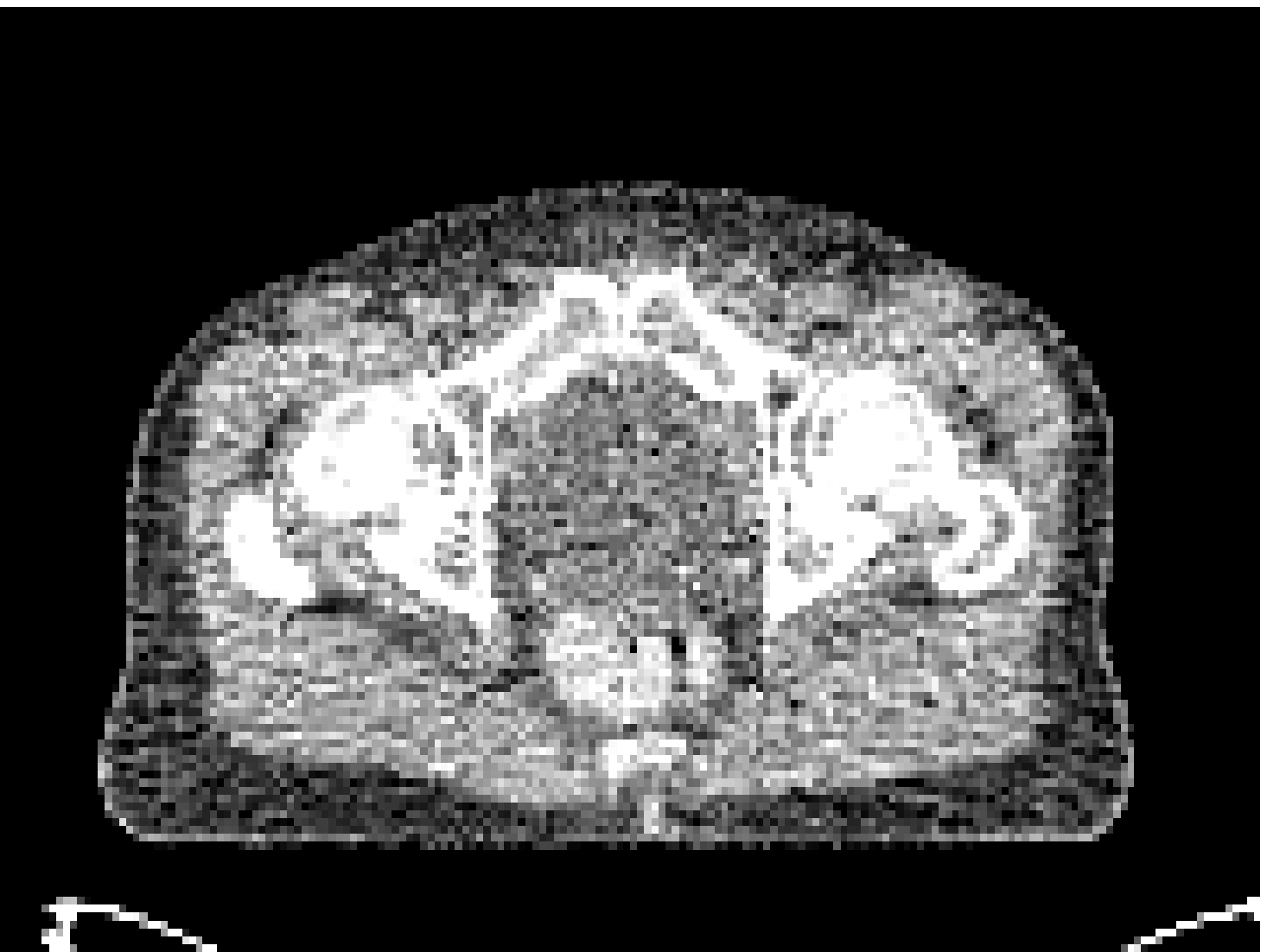}&
			\includegraphics[width=.198\linewidth,height=.132\linewidth]{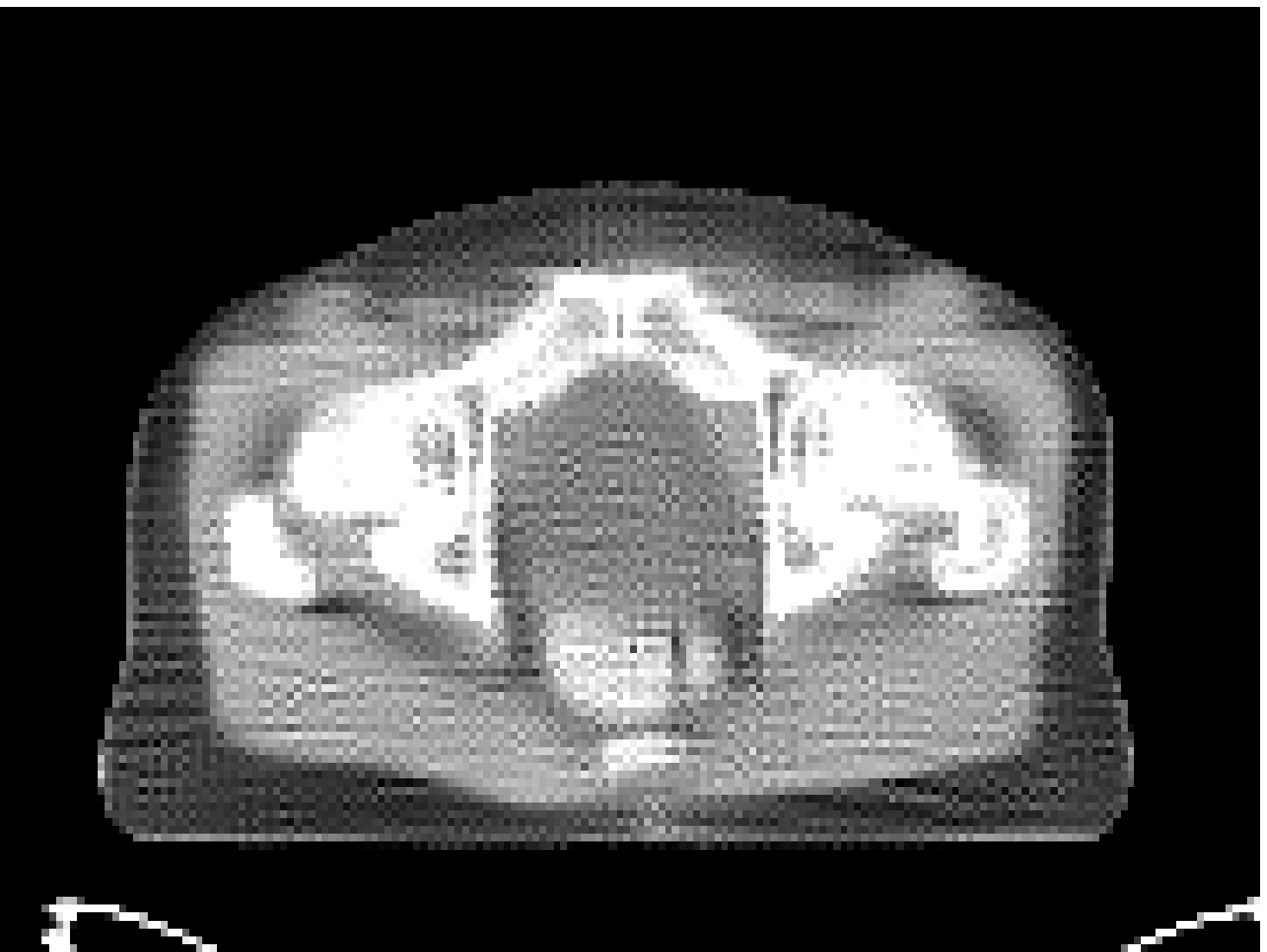}&
			\includegraphics[width=.198\linewidth,height=.132\linewidth]{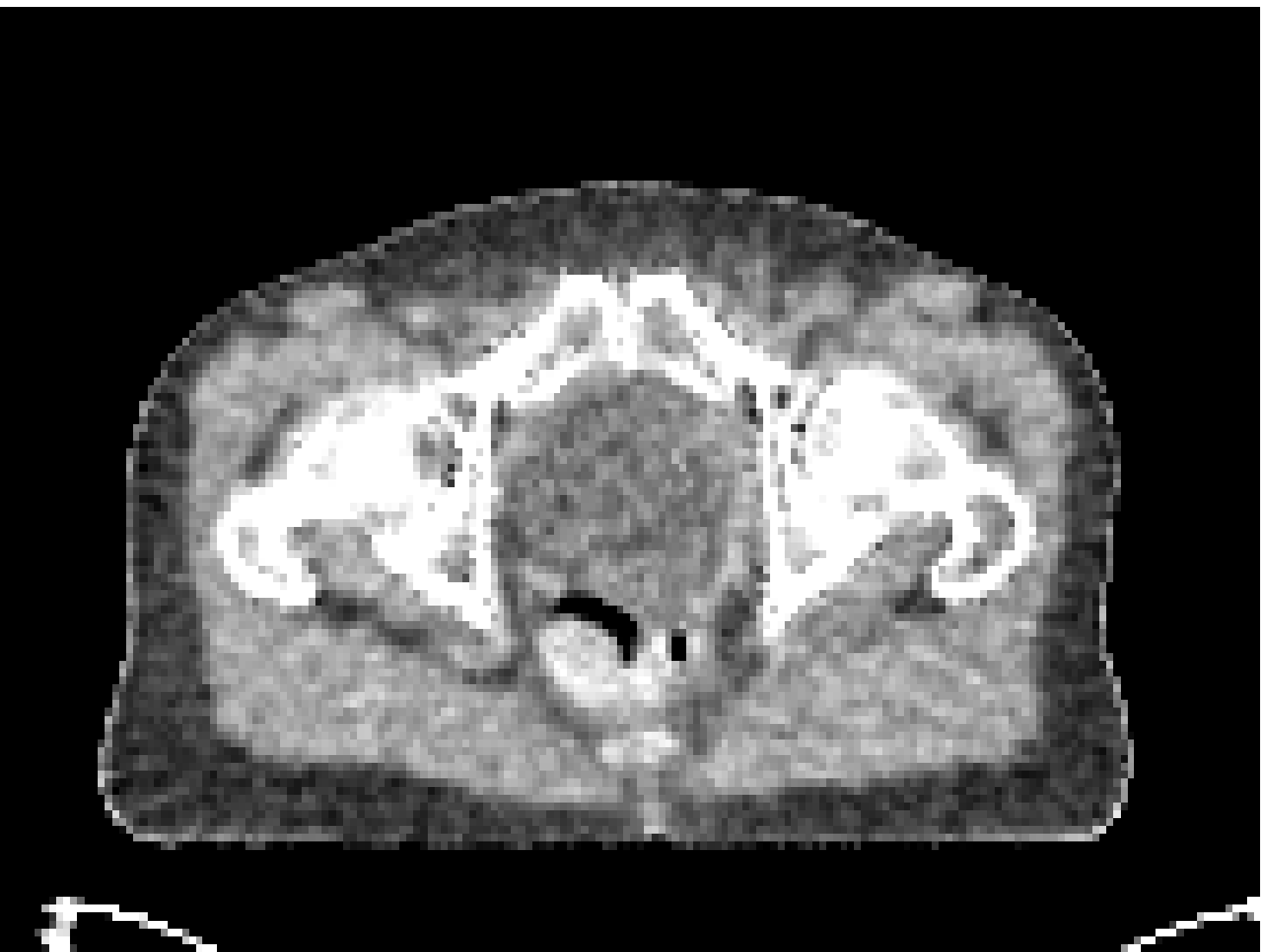}\\
			FBP&
			TV&
			KSVD&
			BM3D&
			FBPConvNet\\			
			\includegraphics[width=.198\linewidth,height=.132\linewidth]{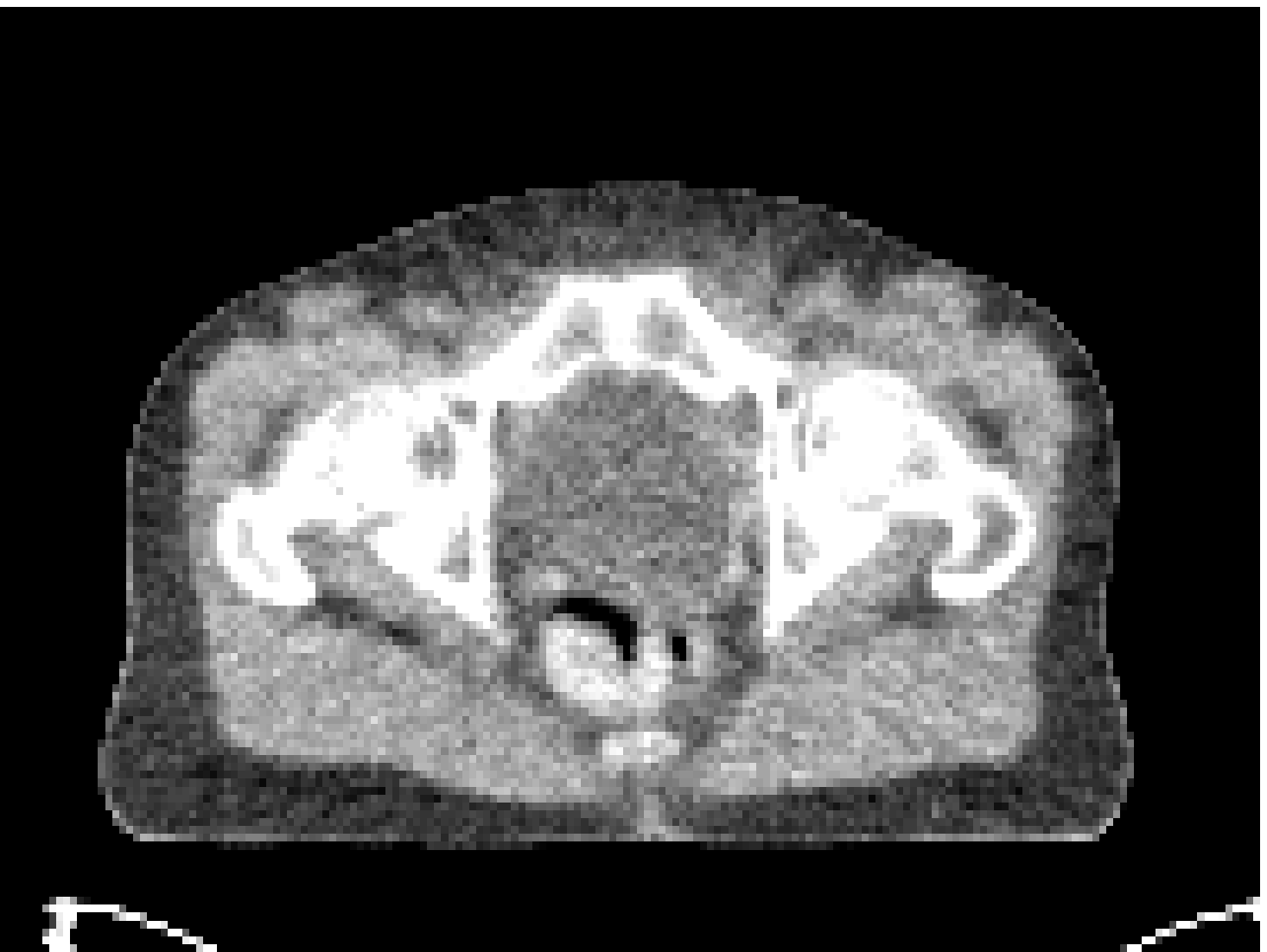}&
			\includegraphics[width=.198\linewidth,height=.132\linewidth]{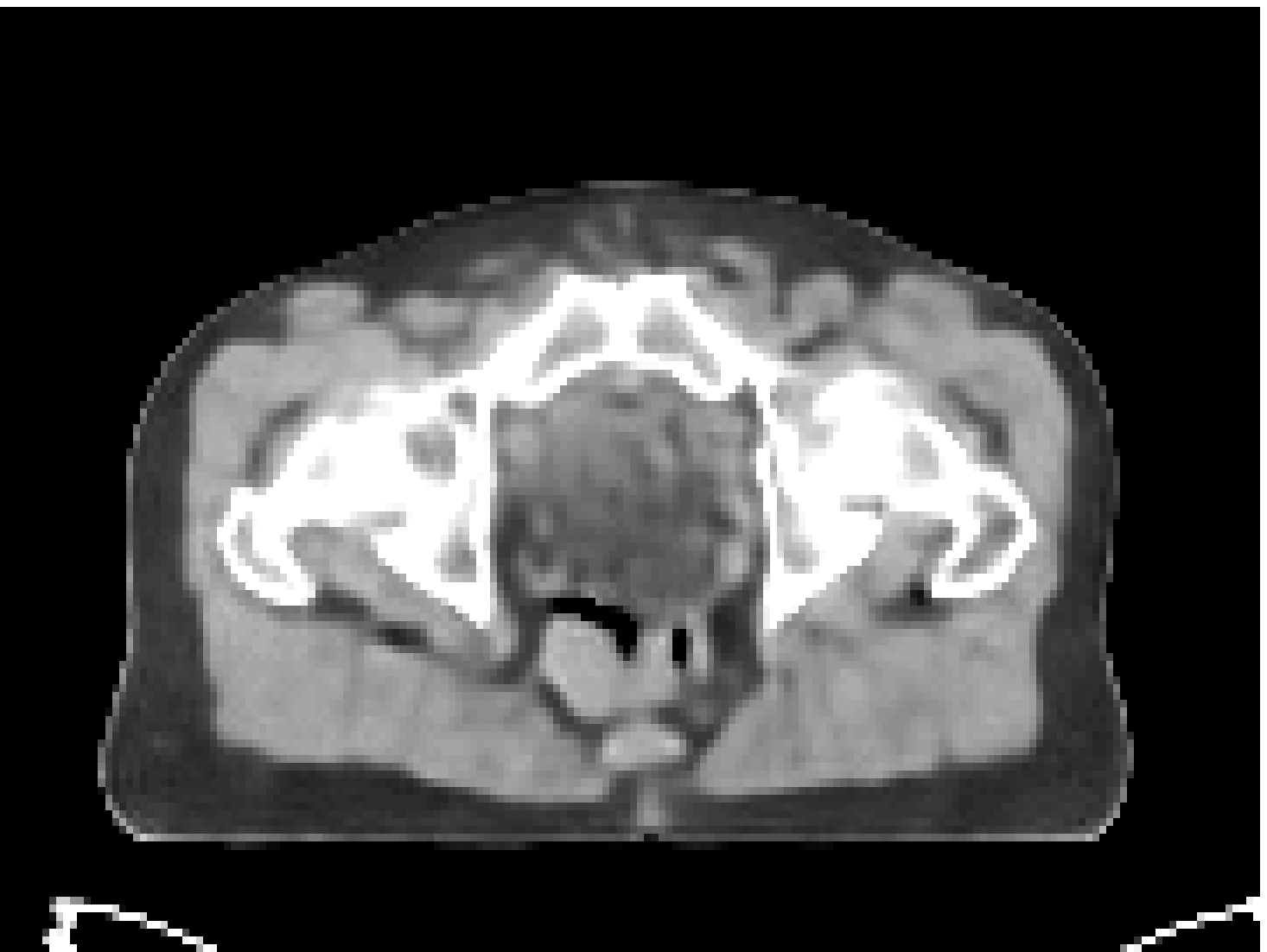}&
			\includegraphics[width=.198\linewidth,height=.132\linewidth]{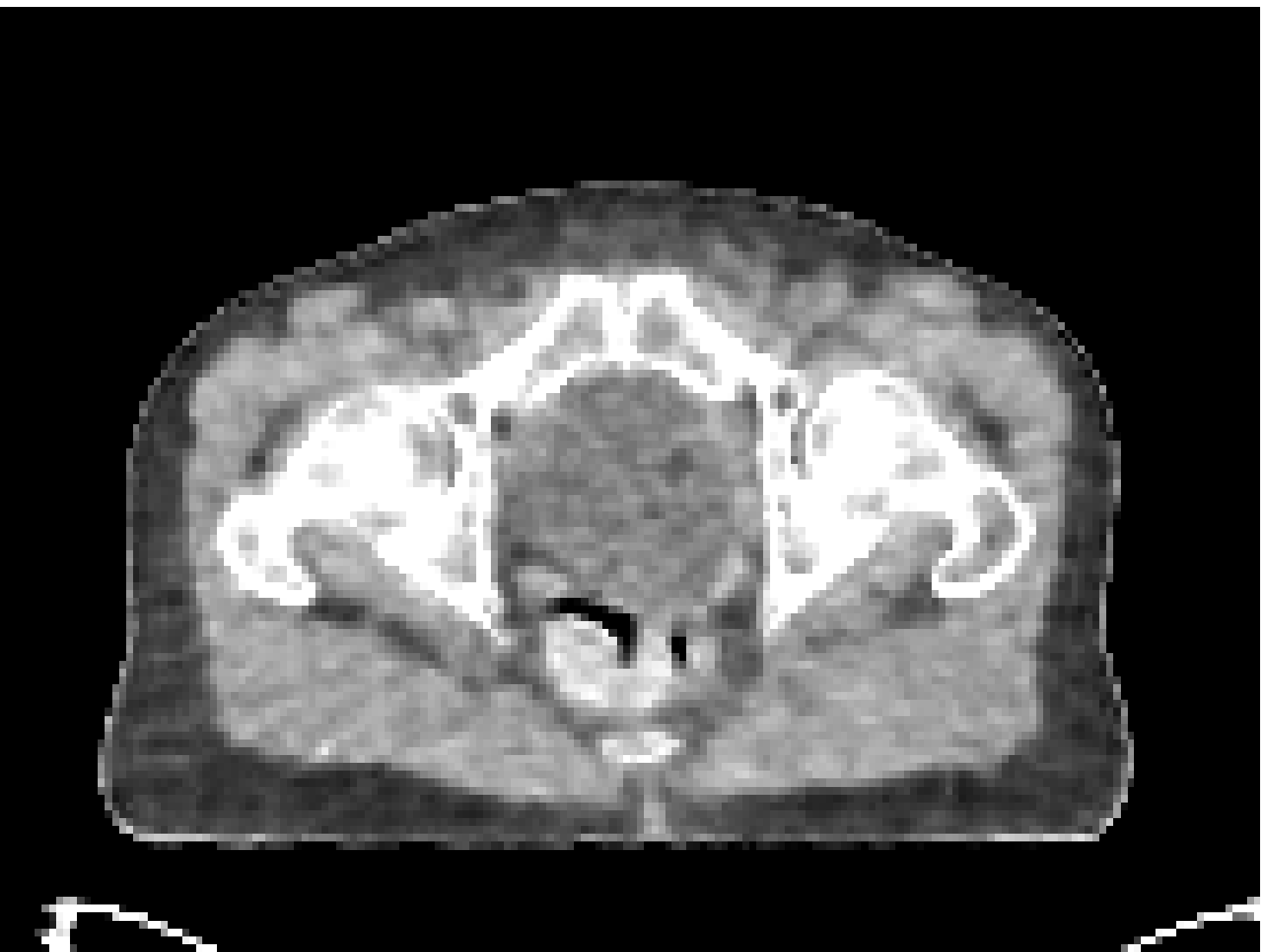}&				
			\includegraphics[width=.198\linewidth,height=.132\linewidth]{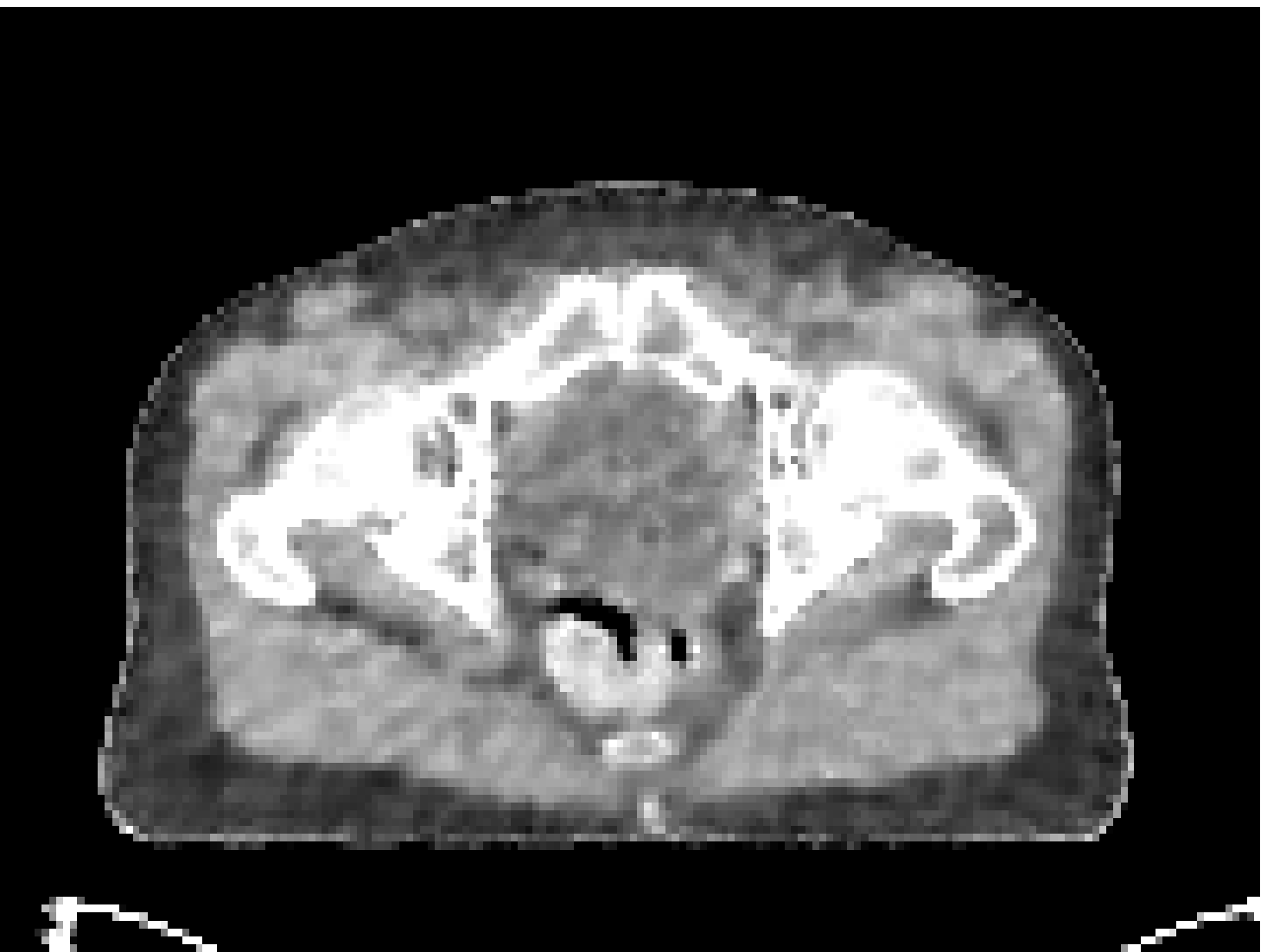}&		
			\includegraphics[width=.198\linewidth,height=.132\linewidth]{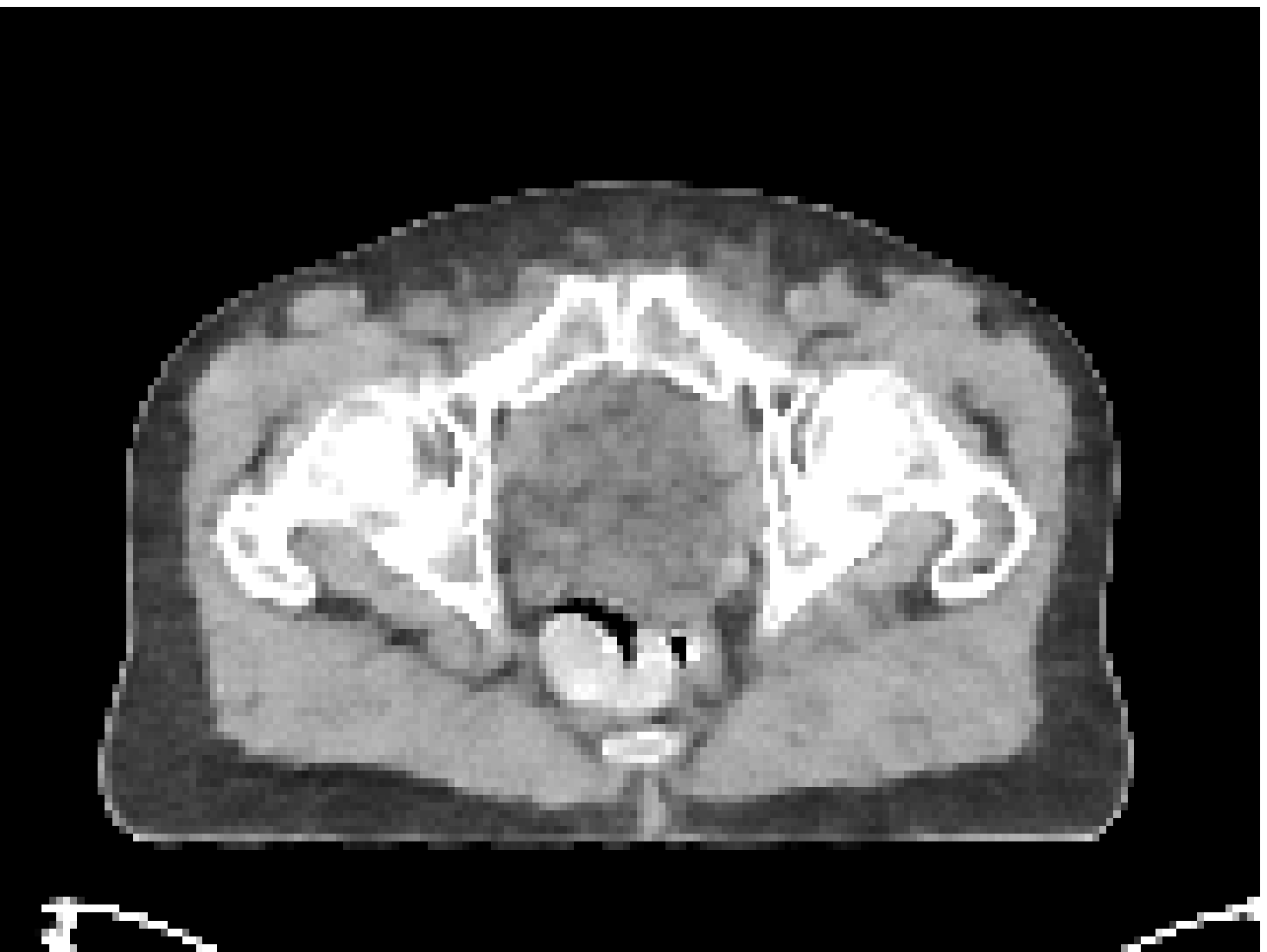}\\
			MoDL&
			Neumann-Net&
			PGD&
			Learn-PD&
			AHP-Net
		\end{tabular}
		\caption{Reconstruction results at dose level $I_i=1 \times 10^4$ by the universal models trained for varying dose levels.
		}
		\label{slice_10000}
	\end{center}
\end{figure*}

\begin{figure}
	\begin{center}
		\begin{tabular}{c@{\hspace{-2pt}}c@{\hspace{-2pt}}c@{\hspace{-2pt}}c@{\hspace{-2pt}}c@{\hspace{-2pt}}c@{\hspace{-2pt}}c}
			\begin{tikzpicture}
			\node[anchor=south west,inner sep=0] (image) at (0,0) {\includegraphics[width=.2\linewidth,height=.2\linewidth]{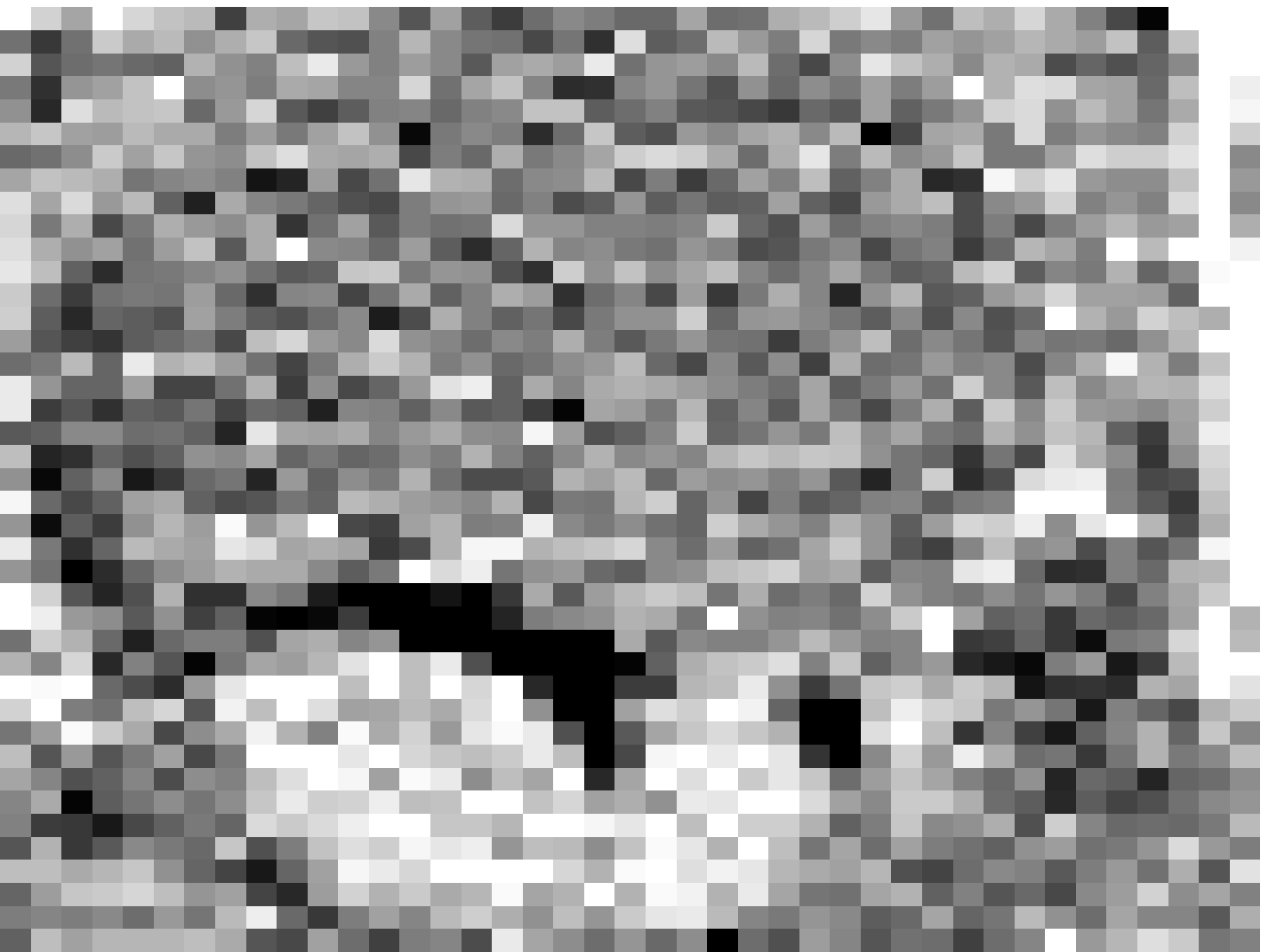}};
			\draw [-stealth, line width=2pt, cyan] (0.7,1.3) -- ++(-0.3,-0.3);
			\draw [-stealth, line width=2pt, cyan] (1.7,0.45) -- ++(-0.45,-0.0);
			\end{tikzpicture}&
			\begin{tikzpicture}
			\node[anchor=south west,inner sep=0] (image) at (0,0) {\includegraphics[width=.2\linewidth,height=.2\linewidth]{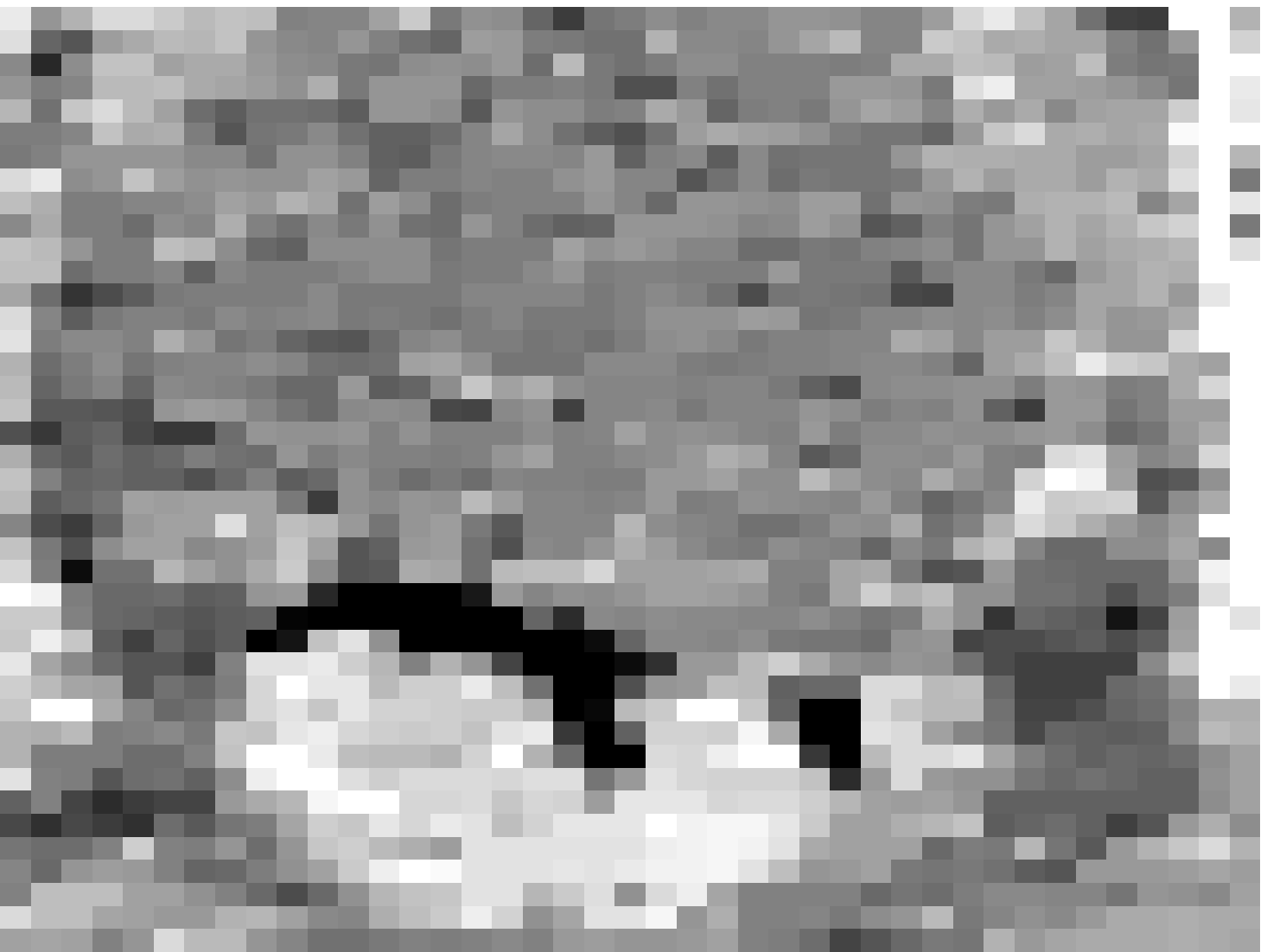}};
			\draw [-stealth, line width=2pt, cyan] (0.7,1.3) -- ++(-0.3,-0.3);
			\draw [-stealth, line width=2pt, cyan] (1.7,0.45) -- ++(-0.45,-0.0);
			\end{tikzpicture}&
			\begin{tikzpicture}
			\node[anchor=south west,inner sep=0] (image) at (0,0) {\includegraphics[width=.2\linewidth,height=.2\linewidth]{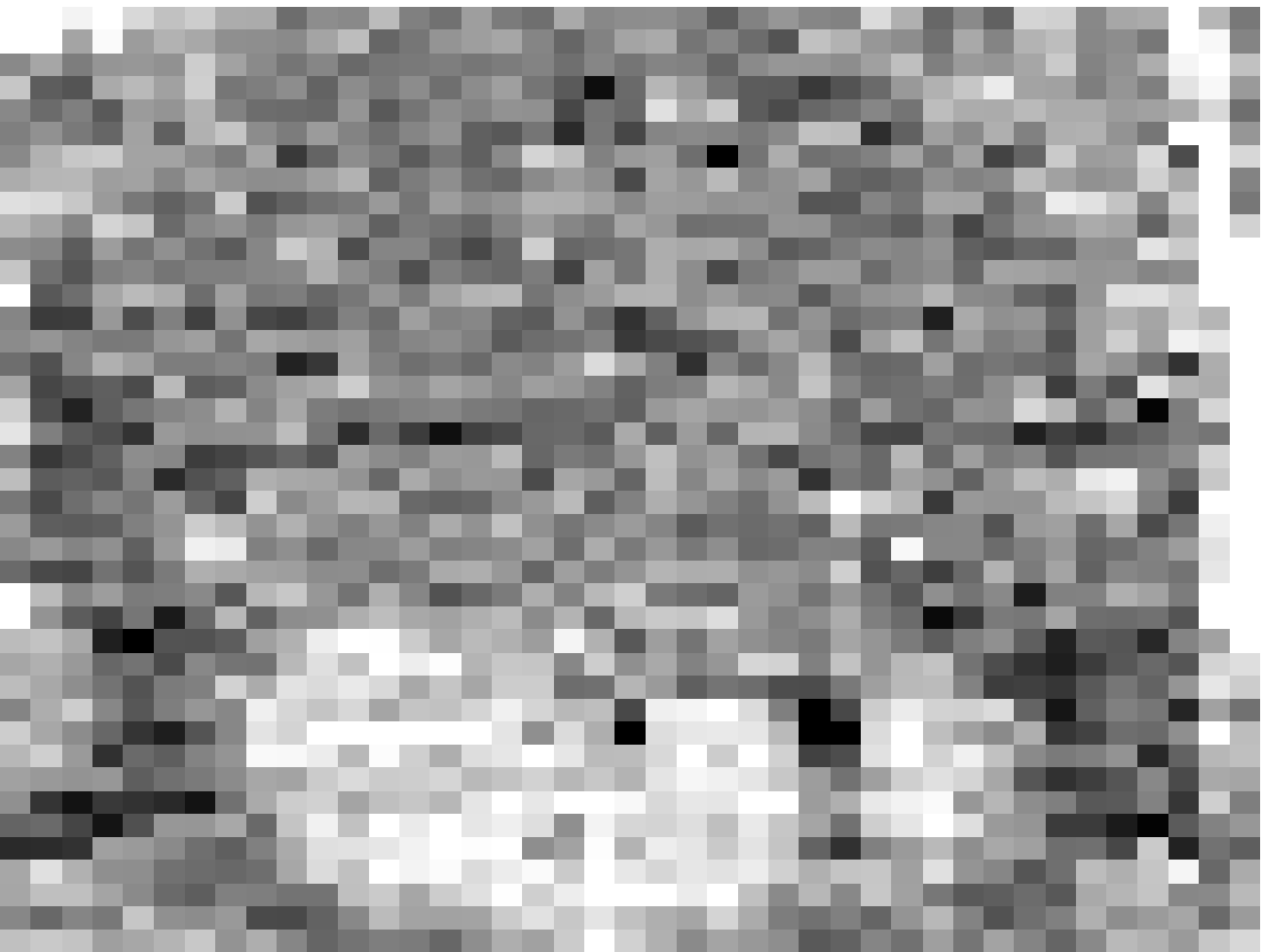}};
			\draw [-stealth, line width=2pt, cyan] (0.7,1.3) -- ++(-0.3,-0.3);
			\draw [-stealth, line width=2pt, cyan] (1.7,0.45) -- ++(-0.45,-0.0);
			\end{tikzpicture}&
			\begin{tikzpicture}
			\node[anchor=south west,inner sep=0] (image) at (0,0) {\includegraphics[width=.2\linewidth,height=.2\linewidth]{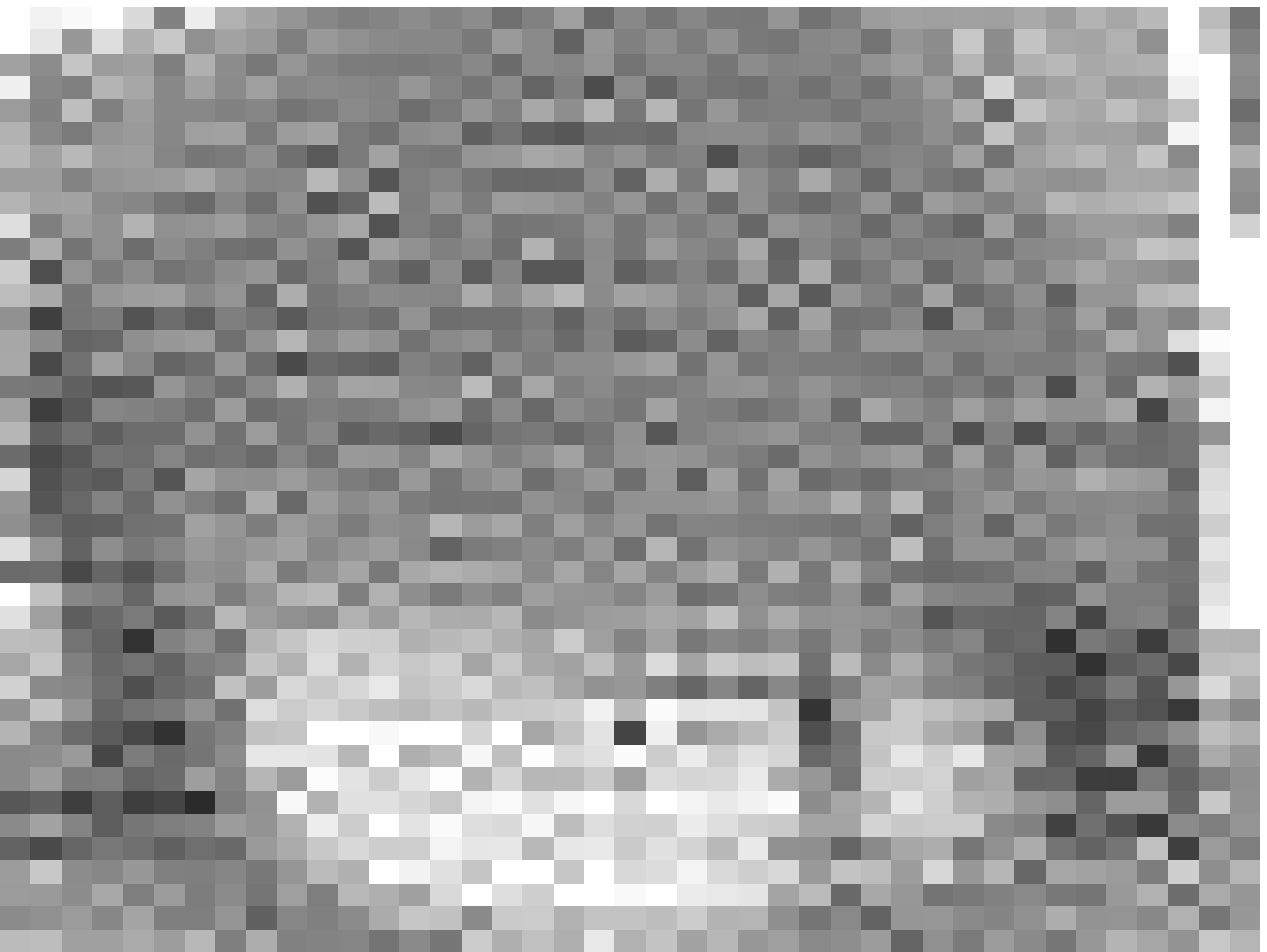}};
			\draw [-stealth, line width=2pt, cyan] (0.7,1.3) -- ++(-0.3,-0.3);
			\draw [-stealth, line width=2pt, cyan] (1.7,0.45) -- ++(-0.45,-0.0);
			\end{tikzpicture}&
			\begin{tikzpicture}
			\node[anchor=south west,inner sep=0] (image) at (0,0) {\includegraphics[width=.2\linewidth,height=.2\linewidth]{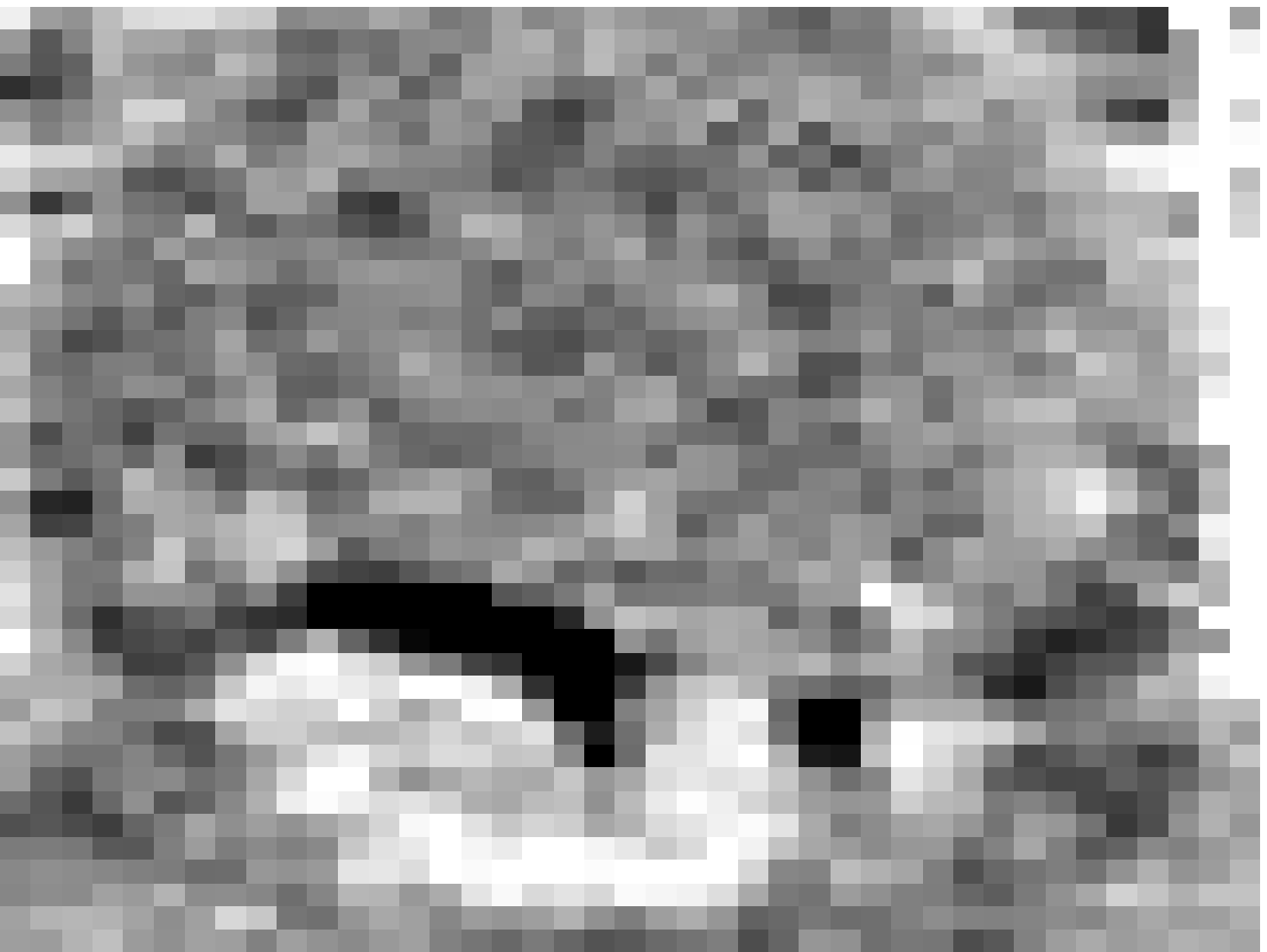}};
			\draw [-stealth, line width=2pt, cyan] (0.7,1.3) -- ++(-0.3,-0.3);
			\draw [-stealth, line width=2pt, cyan] (1.7,0.45) -- ++(-0.45,-0.0);
			\end{tikzpicture}\\
			FBP&		
			TV&
			KSVD&
			BM3D&
			FBPConvNet\\
			\begin{tikzpicture}
			\node[anchor=south west,inner sep=0] (image) at (0,0) {\includegraphics[width=.2\linewidth,height=.2\linewidth]{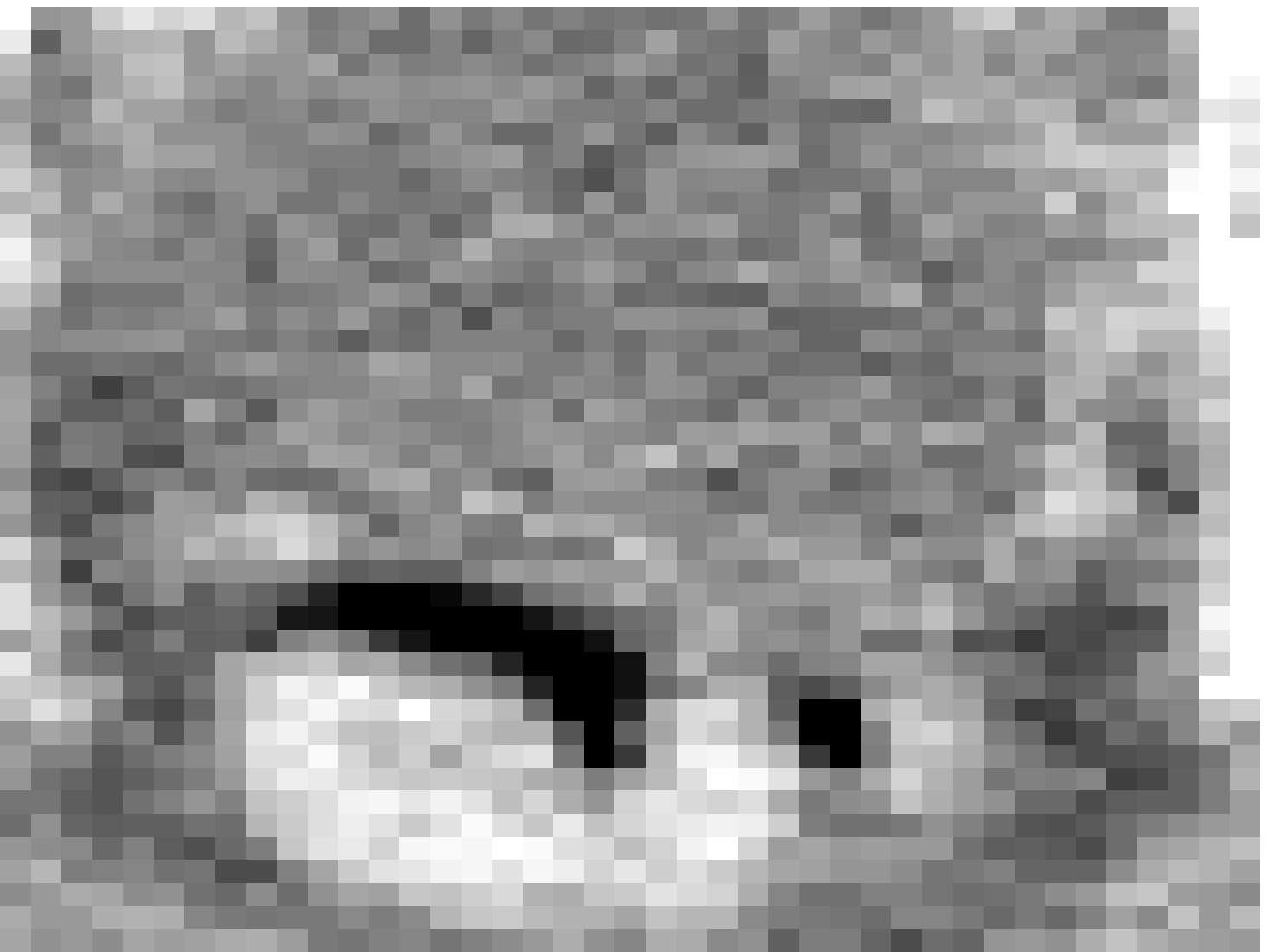}};
			\draw [-stealth, line width=2pt, cyan] (0.7,1.3) -- ++(-0.3,-0.3);
			\draw [-stealth, line width=2pt, cyan] (1.7,0.45) -- ++(-0.45,-0.0);
			\end{tikzpicture}&		
			\begin{tikzpicture}
			\node[anchor=south west,inner sep=0] (image) at (0,0) {\includegraphics[width=.2\linewidth,height=.2\linewidth]{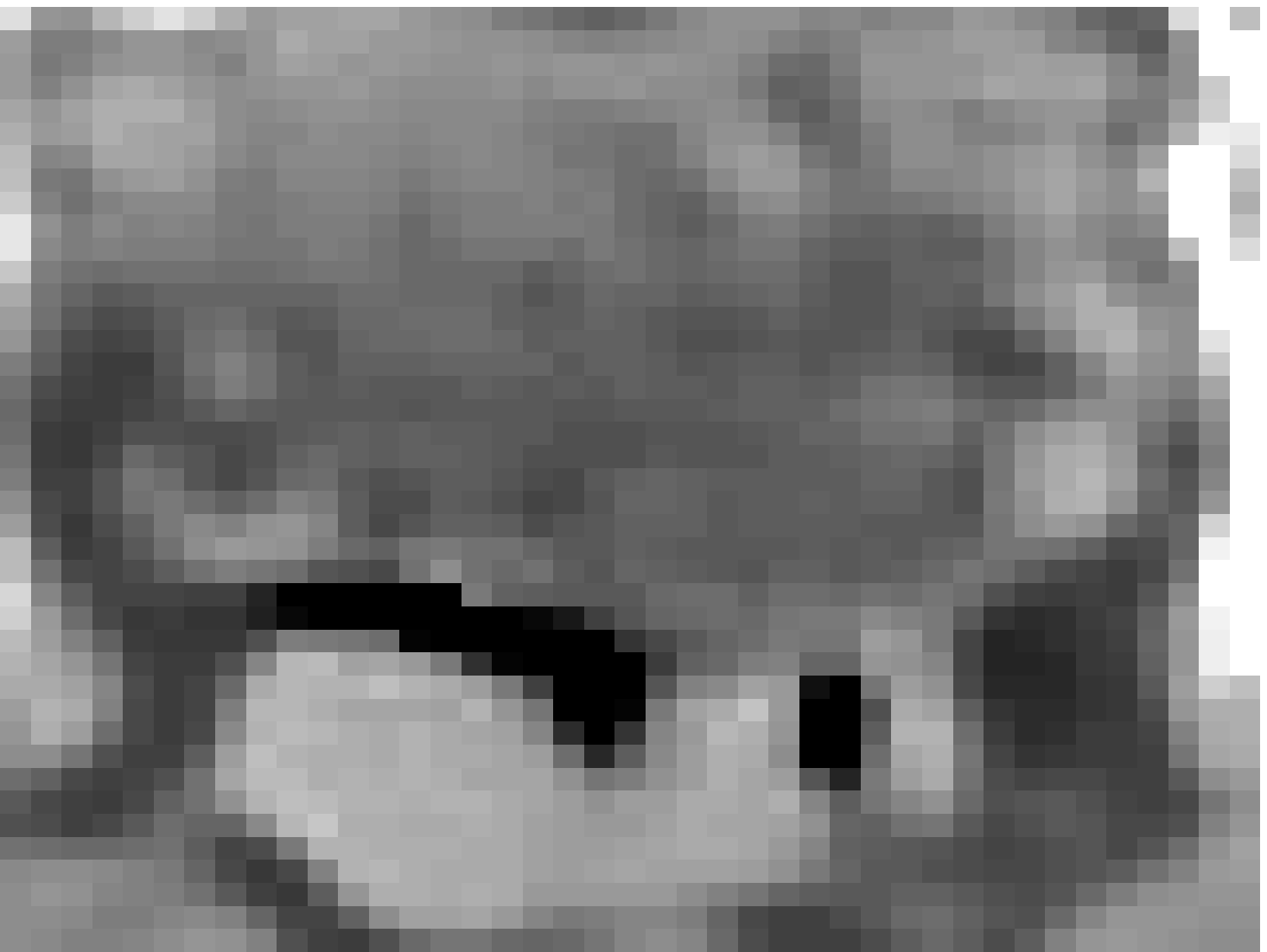}};
			\draw [-stealth, line width=2pt, cyan] (0.7,1.3) -- ++(-0.3,-0.3);
			\draw [-stealth, line width=2pt, cyan] (1.7,0.45) -- ++(-0.45,-0.0);
			\end{tikzpicture}&
			\begin{tikzpicture}
			\node[anchor=south west,inner sep=0] (image) at (0,0) {\includegraphics[width=.2\linewidth,height=.2\linewidth]{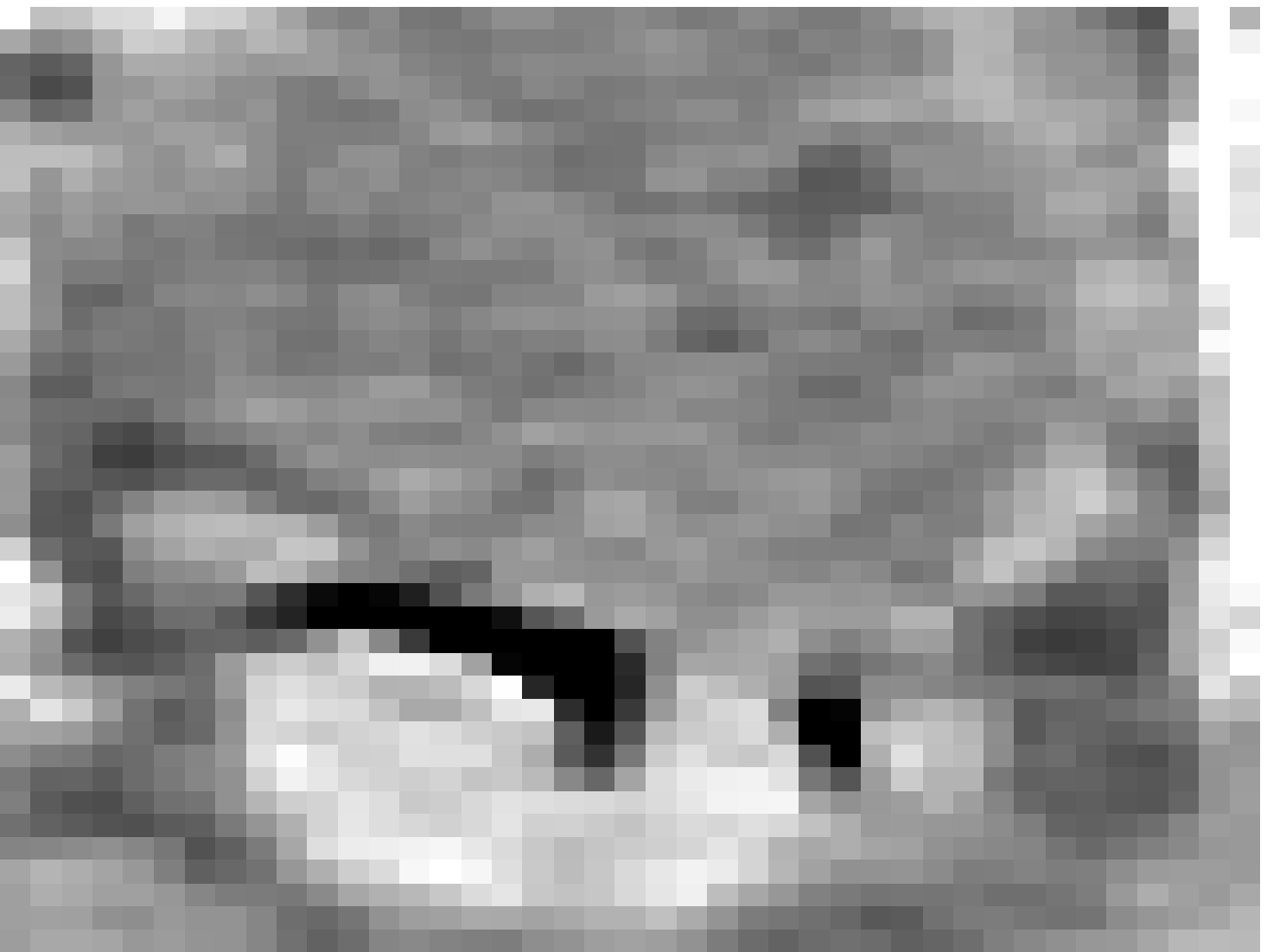}};
			\draw [-stealth, line width=2pt, cyan] (0.7,1.3) -- ++(-0.3,-0.3);
			\draw [-stealth, line width=2pt, cyan] (1.7,0.45) -- ++(-0.45,-0.0);
			\end{tikzpicture}&		
			\begin{tikzpicture}
			\node[anchor=south west,inner sep=0] (image) at (0,0) {\includegraphics[width=.2\linewidth,height=.2\linewidth]{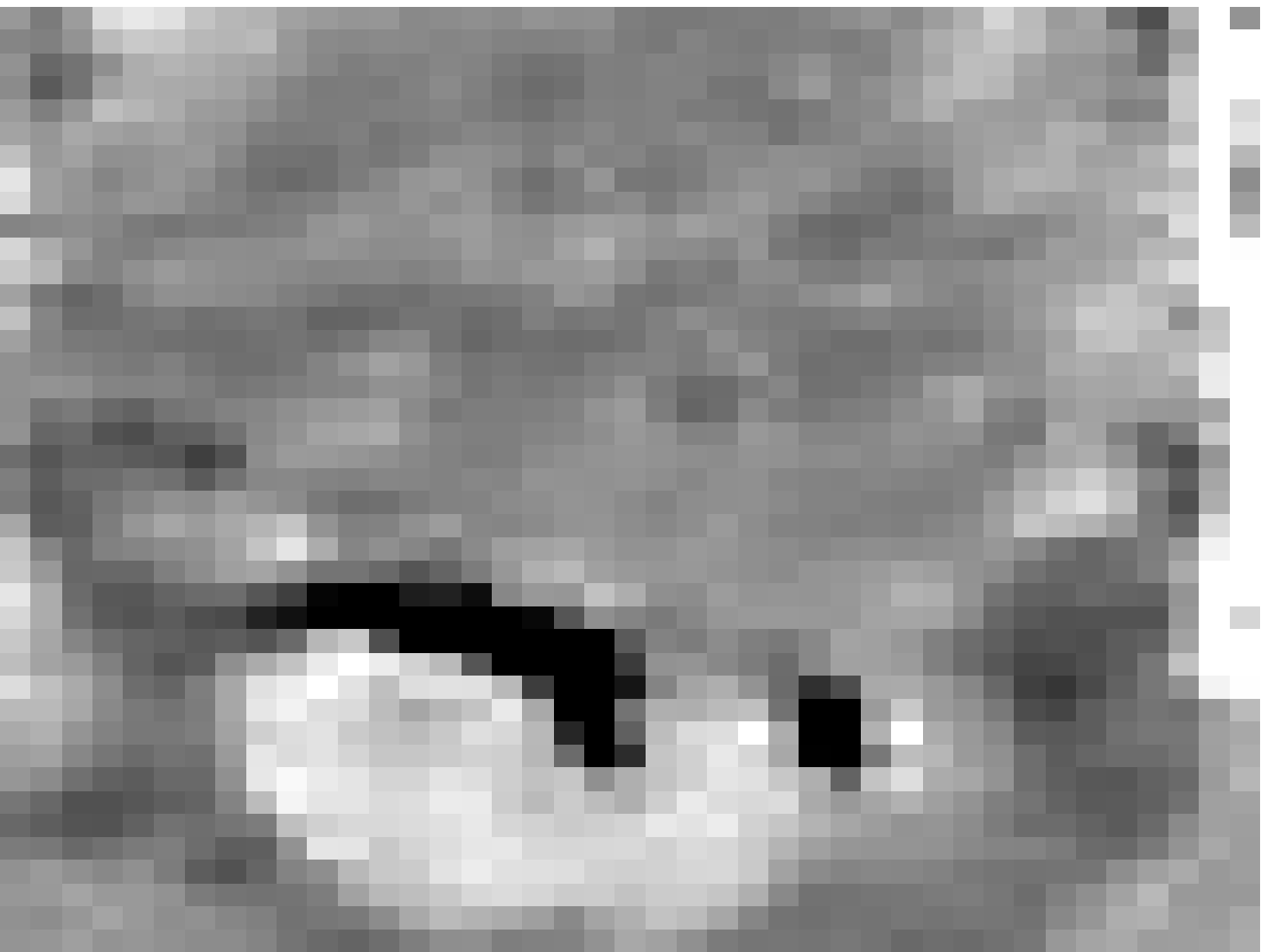}};
			\draw [-stealth, line width=2pt, cyan] (0.7,1.3) -- ++(-0.3,-0.3);
			\draw [-stealth, line width=2pt, cyan] (1.7,0.45) -- ++(-0.45,-0.0);
			\end{tikzpicture}&
			\begin{tikzpicture}
			\node[anchor=south west,inner sep=0] (image) at (0,0) {\includegraphics[width=.2\linewidth,height=.2\linewidth]{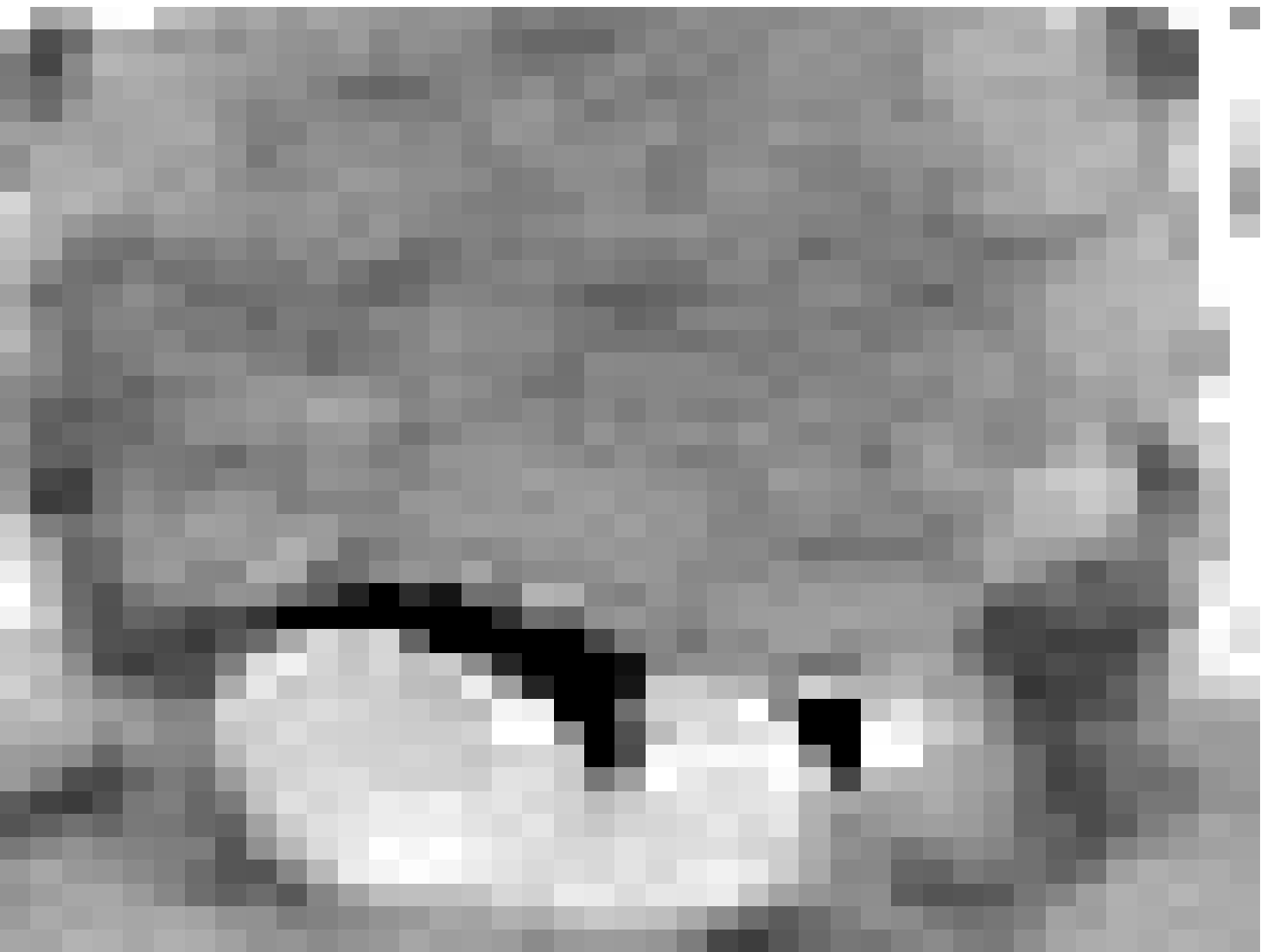}};
			\draw [-stealth, line width=2pt, cyan] (0.7,1.3) -- ++(-0.3,-0.3);
			\draw [-stealth, line width=2pt, cyan] (1.7,0.45) -- ++(-0.45,-0.0);
			\end{tikzpicture}\\
			MoDL&
			Neumann-Net&	
			PGD&
			Learned-PD&	
			AHP-Net
		\end{tabular}
		\caption{Zoom-in results of Fig.~\ref{slice_10000}.
		}
		\label{sliceZoom_10000}
	\end{center}
\end{figure}

For the image reconstructed  by the non-learning methods and the universal NN model trained by deep learning methods for varying dose levels, specifically for the dose level $I_i= 1\times10^4$,
Fig. \ref{slice_10000} shows the  images reconstructed by different methods, and their zoomed-in images of  boxes in Fig. \ref{Truth}   are displayed in Fig.~\ref{sliceZoom_10000}. 
As shown by the cyan arrows,  in comparison to he zoomed-in NDCT image, the results of FBPConvNet, MoDL, Neumann-Net, PGD and Learned-PD were more blurred than that  of the proposed method.
See Table \ref{SNRRMSEU} for  quantitative comparison for the results shown in  Fig.~\ref{slice_10000}.

\begin{figure}
	\begin{center}
		\includegraphics[width=.8\linewidth]{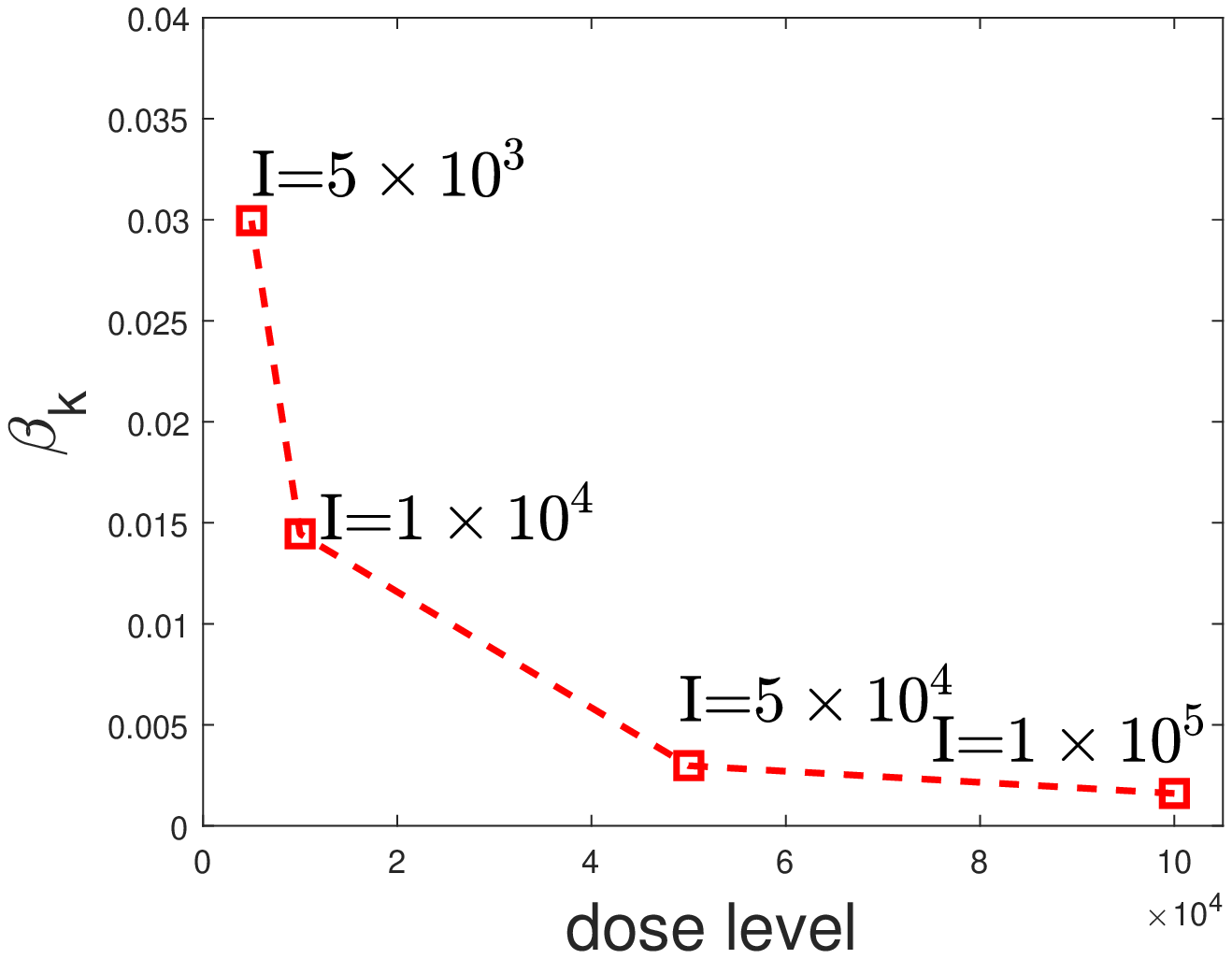}	
		\caption{Predicted hyper-parameter at different dose levels  by the universal models trained for varying dose levels.}
		\label{Param}
	\end{center}
\end{figure}

For the  hyper-parameters set in MBIR,
Fig. \ref{Param} shows the predicted hyper-parameters at different dose levels  by the universal models trained for varying dose levels.
It can be seen that
the predicted  hyper-parameter adapts well with inputs of different noise levels, where a lower-dose-level input requires larger hyper-parameter to incorporate more prior knowledge, and vice versa.

\begin{table}
	\centering
	\caption{Quantitative metrics on the  reconstruction results for the image slice shown in Fig. \ref{sliceU_5000} and Fig. \ref{slice_10000}.}
		\scalebox{0.78}{
	\begin{tabular}{@{\hspace{0pt}}c@{\hspace{2.5pt}}c@{\hspace{2.5pt}}c@{\hspace{2.5pt}}c@{\hspace{2.5pt}}c@{\hspace{2.5pt}}c@{\hspace{2.5pt}}c@{\hspace{2.5pt}}c@{\hspace{2.5pt}}c@{\hspace{2.5pt}}c@{\hspace{2.5pt}}c@{\hspace{2.5pt}}c}
		\hline
		\multirow{2}{*}{Method}&\multirow{2}{*}{Index} &\multirow{2}{*}{FBP} &\multirow{2}{*}{TV} &\multirow{2}{*}{KSVD}&\multirow{2}{*}{BM3D} &FBPConv  &\multirow{2}{*}{~MoDL}&Neumann &\multirow{2}{*}{~PGD} &Learned  &AHP            \\
		&&&&&&Net&& Net&&-PD& Net\\
		\hline
		\multirow{3}{*}{Fig. \ref{sliceU_5000}} &PSNR  &25.12	&28.09	&25.53	&27.42	&30.07	&31.83	&28.85 &32.03	&31.65	&$\bm{33.71}$\\
		                                        &RMSE  &57.78	&41.06	&54.90	&44.17	&32.68	&26.70	&37.60 &26.09	&27.25 	&$\bm{21.49}$\\
		                                        &SSIM  &0.7278	&0.8509 &0.6124	&0.7267	&0.9222	&0.9377	&0.9042&0.9412	&0.9425  &$\bm{0.9522}$\\
		\hline
		\multirow{3}{*}{Fig. \ref{slice_10000}} &PSNR &28.31	&31.65	&28.37	&29.46	&31.39	&31.34	&29.50	&32.90	&32.56	&$\bm{34.89}$\\
		                                        &RMSE &40.03	&27.24	&39.58	&34.91	&28.06	&28.22	&34.91	&23.59	&24.55	&$\bm{18.77}$\\
		                                        &SSIM &0.8430	&0.9355	&0.7351	&0.8046	&0.9283	&0.9356	&0.923	&0.9505	&0.9504	&$\bm{0.9617}$\\
		\hline                                                                                                                       
	\end{tabular}
			}
	\label{SNRRMSEU}
\end{table}

\subsection{Ablation study}
The ablation
study conducted in this section focuses on how the following two parts impact the performance of image reconstruction: (1)
the introduction of an MLP for adaptive prediction of the hyper-parameters and (2) the usages of 8 high-pass filters from B-spline framelet. Through the whole ablation study, all different versions of the AHP-Net are trained by the same procedure on the same training/validation dataset and tested on the same testing dataset. Different dose levels are trained with different models.
See  Table \ref{ablationComparisionU} for the results from the different versions.
\begin{itemize}
	\item 	\emph{no filter vs spline framelet filter bank}. \,\,
	To verify the effectiveness of the spline framelet filter bank of the proposed AHP-Net, the same NN but there is no filter was adopted.
	We set $\bm{z}^k=\bm{x}^k$, which the method is the same as MoDL. See the comparison of the  column titled ``No Filter" vs ``AHP-Net'' in Table~\ref{ablationComparisionU}. The results under different dose level showed the advantage of 8 filters over  no filter,  about $1.5-2.2$ dB in PSNR, which indicates the joint benefits of framelet filter bank and adaptive hyper-parameters.
	
	\item \emph{$\nabla$ vs spline framelet filter bank}. \,\,
    The ablation study was performed on  AHP-Net and the same NN but whose 8 framelet filters are replaced by 2 filters of $\nabla$ ($[1,-1], [1,-1]^\top$) seen in TV regularization.
	See the comparison of the  column titled as ``Using $\nabla$" vs ``AHP-Net'' in Table~\ref{ablationComparisionU}. The results under different dose level showed the advantage of 8 filters over  gradient operator $\nabla$,  about $1.1-1.9$ dB in PSNR. It is noted that in addition to performance gain by using the framelet filter bank, the performance benefit brought by MLP-based predictor is also fully exploited when using $\nabla$, as only 2 parameters are involved when using $\nabla$.
	
	\item \emph{Learnable filters vs Spline framelet filter bank}. \,\, It would be interesting to see whether a learnable filter outperformed our spline frame filter; see the column titled as ``Learnable filters" in Table~\ref{ablationComparisionU}. 
%
	In the case of $I_i=1\times10^5$, the performance gain of learning an MLP for adaptive prediction of hyper-parameters is  $0.5$dB in PSNR.
	In the case of lower dose levels, $I_i = 5\times 10^4,1 \times 10^4, 5\times 10^3$, there is a significant improvement of the proposed method over  ``Learnable filters",  about $1.3-1.7$dB in PSNR.
	One possible cause is that network training needs to solve a 
	highly non-convex optimization problem with millions of network parameters. The filter bank plays an important role in the performance of the NN. However, when they are treated as one part of network parameters, they do not receive sufficient attention in the training.
	\item \emph{Learning constant hyper-parameters as network weights vs Using an MLP for adaptive prediction of hyper-parameters}. \,\, The results from the proposed AHP-Net  are compared to that from the same NN but whose hyper-parameters $\{\beta_j^k\}$ are treated as network parameters learned on training dataset; see the column titled as ``Learnable HP" in Table~\ref{ablationComparisionU}.  In the case of higher dose levels, $I_i=1\times 10^5, 5\times 10^4$, the performance gain of learning an MLP for adaptive prediction of hyper-parameters is about $0.3-0.5$ dB in PSNR. In the case of lower dose levels, $I_i=1\times10^4, 5\times 10^3$, there is a significant improvement of MLP-based adaptive prediction over learnable constant hyper-parameters,  about $1.2-1.4$ dB in PSNR.
\end{itemize}

\begin{figure}
\centering
\begin{tabular}{c@{\hspace{0pt}}c}
	\includegraphics[width=.33\linewidth,height=.22\linewidth]{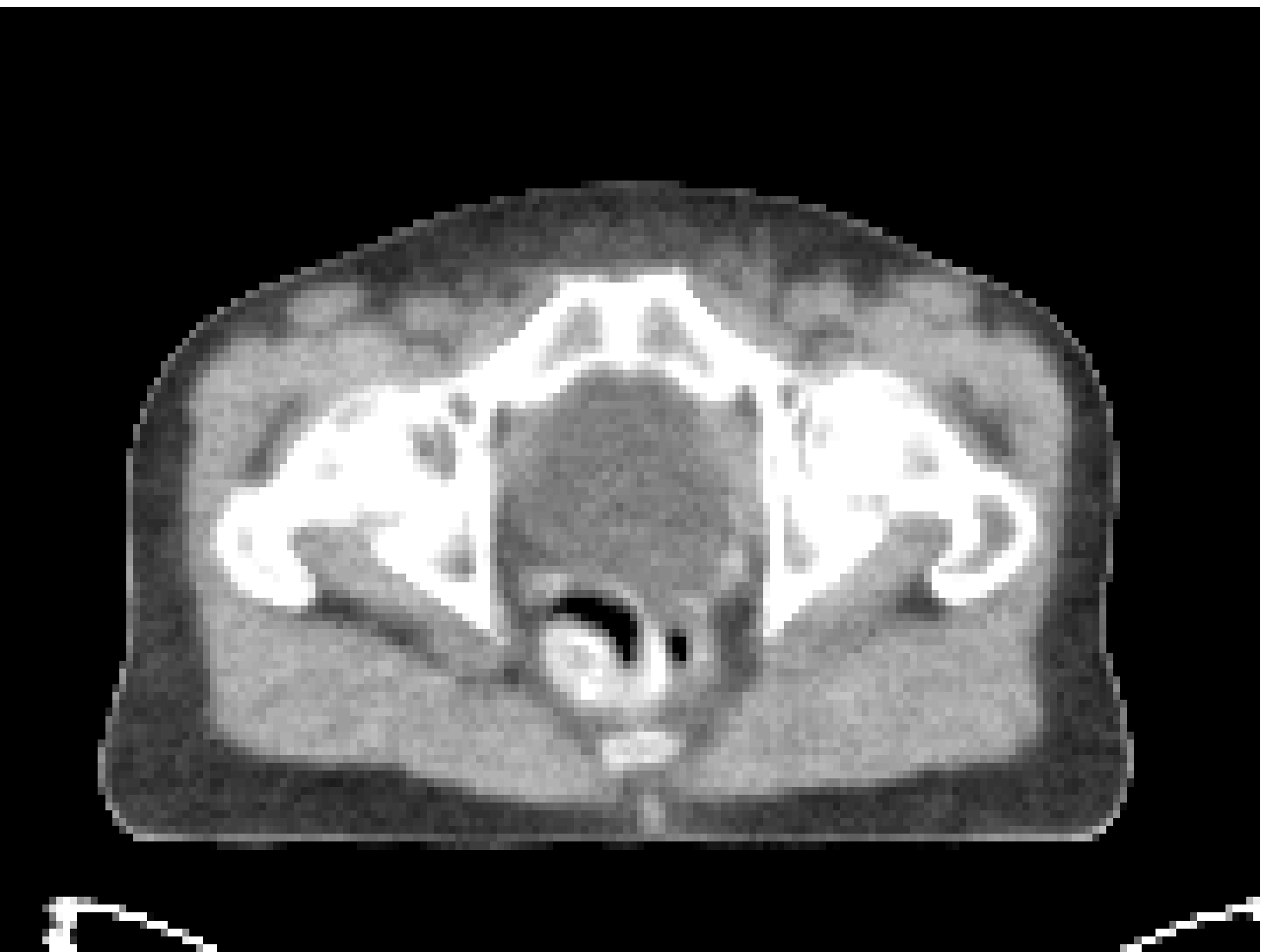}&
	\includegraphics[width=.33\linewidth,height=.22\linewidth]{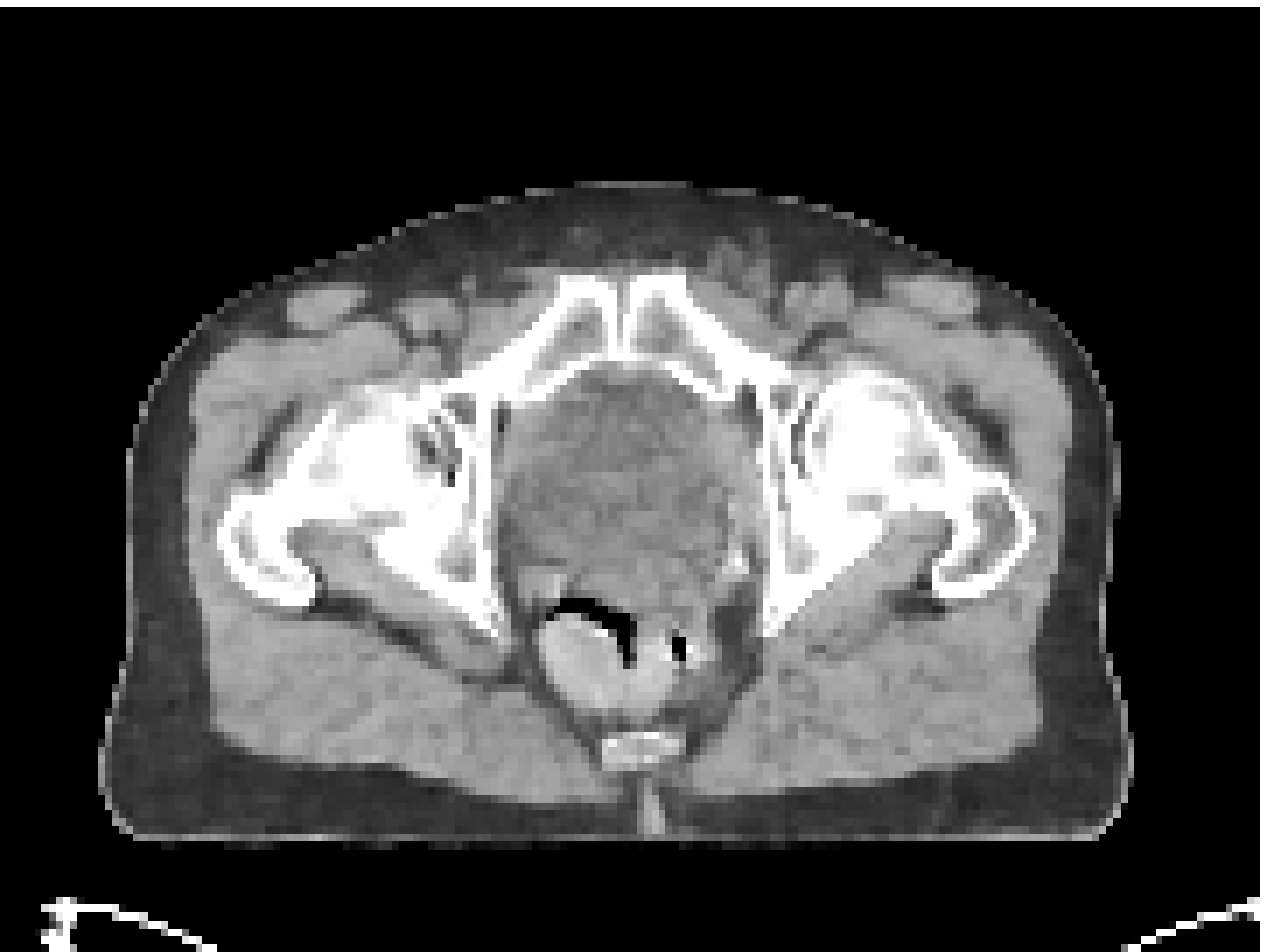}
	\\
	(a)&
	(b)
\end{tabular}
\begin{tabular}{c@{\hspace{0pt}}c@{\hspace{0pt}}c}

	\includegraphics[width=.33\linewidth,height=.22\linewidth]{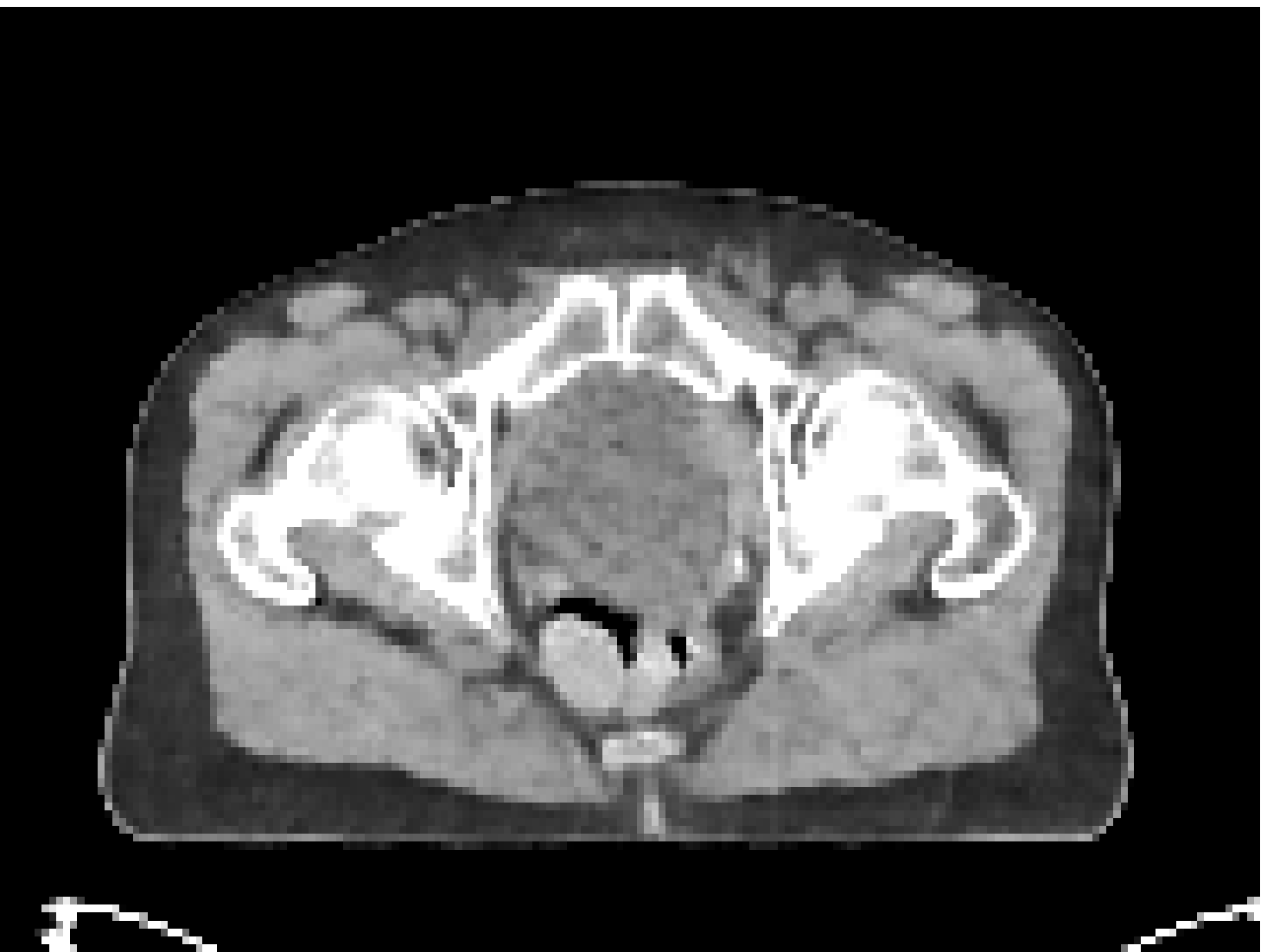}&
	\includegraphics[width=.33\linewidth,height=.22\linewidth]{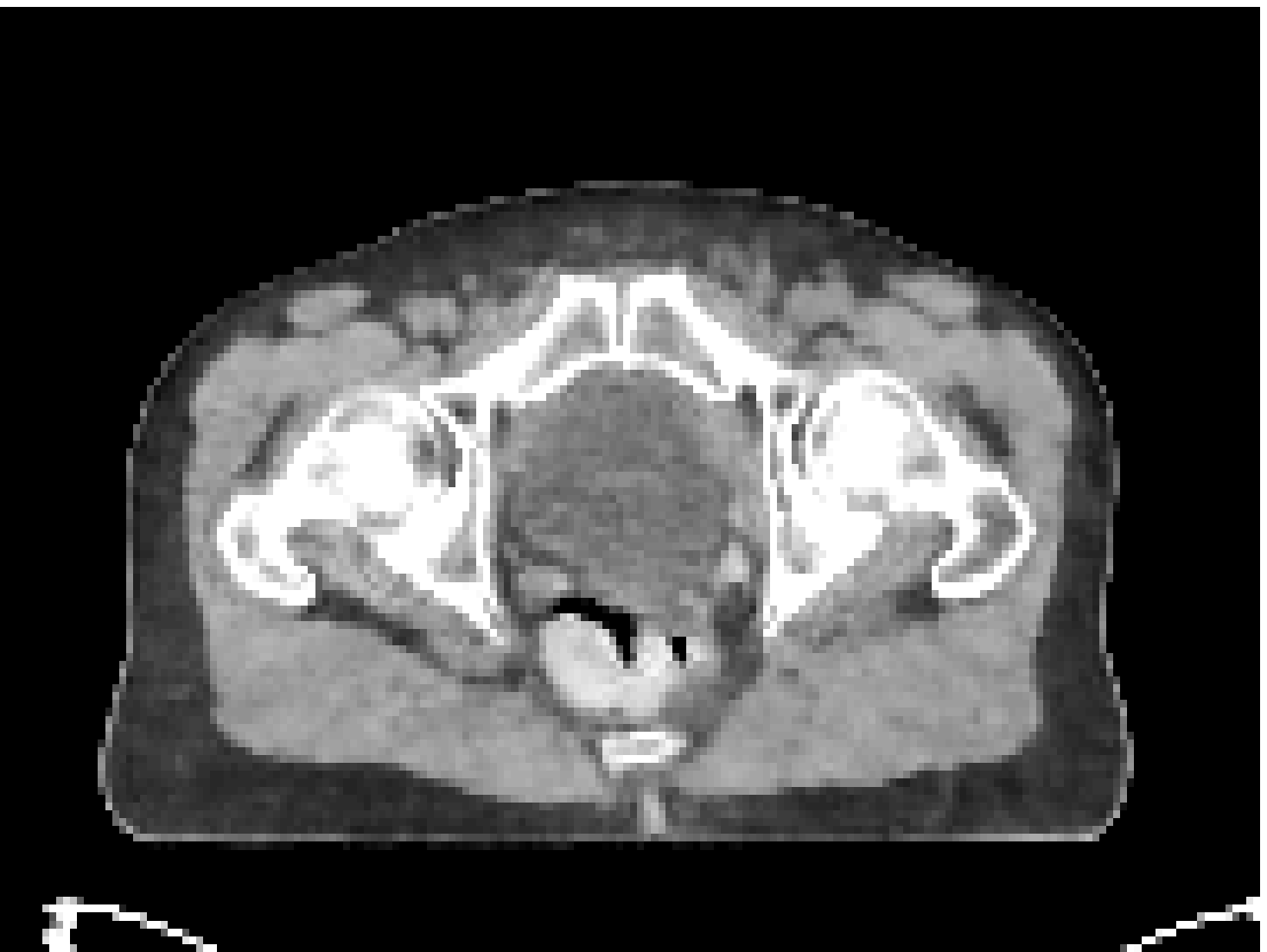}&
	\includegraphics[width=.33\linewidth,height=.22\linewidth]{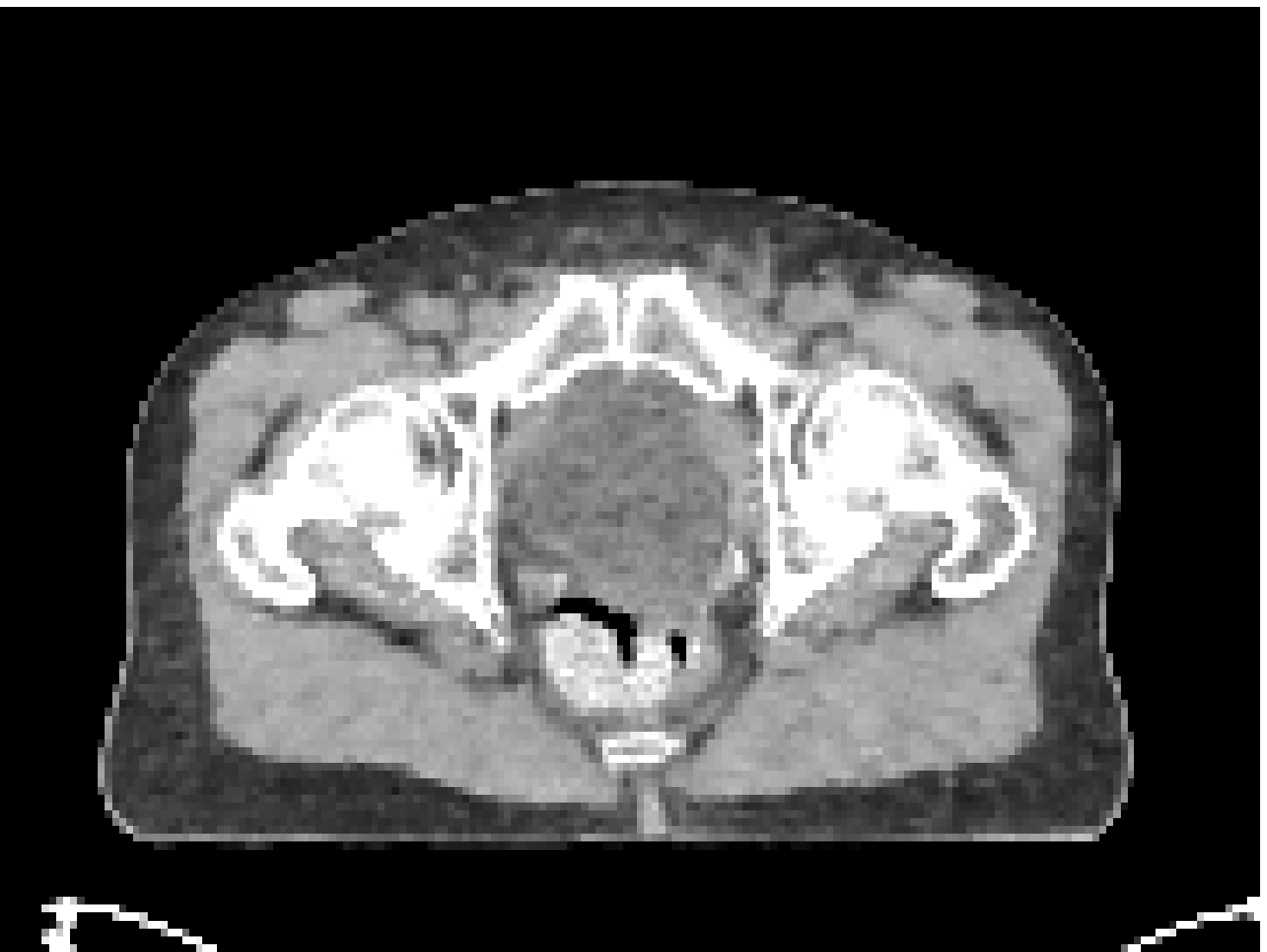}\\
	(d)&
	(e)&
	(c)\\
\end{tabular}
\caption{Reconstruction results at dose level $I_i=5\times10^4$  by the models trained under same dose level.
	(a) No filter; (b) Using $\nabla$;   (c) Learnable filters;  (d) Learnable HP; (e) AHP-net.
}
\label{ablationsliceU_50000}
\end{figure}

\begin{figure}
	\begin{center}
		\begin{tabular}{c@{\hspace{-1pt}}c@{\hspace{-1pt}}c@{\hspace{-1pt}}c@{\hspace{-1pt}}c@{\hspace{-1pt}}c@{\hspace{-1pt}}c}
			\begin{tikzpicture}
			\node[anchor=south west,inner sep=0] (image) at (0,0) {\includegraphics[width=.2\linewidth,height=.2\linewidth]{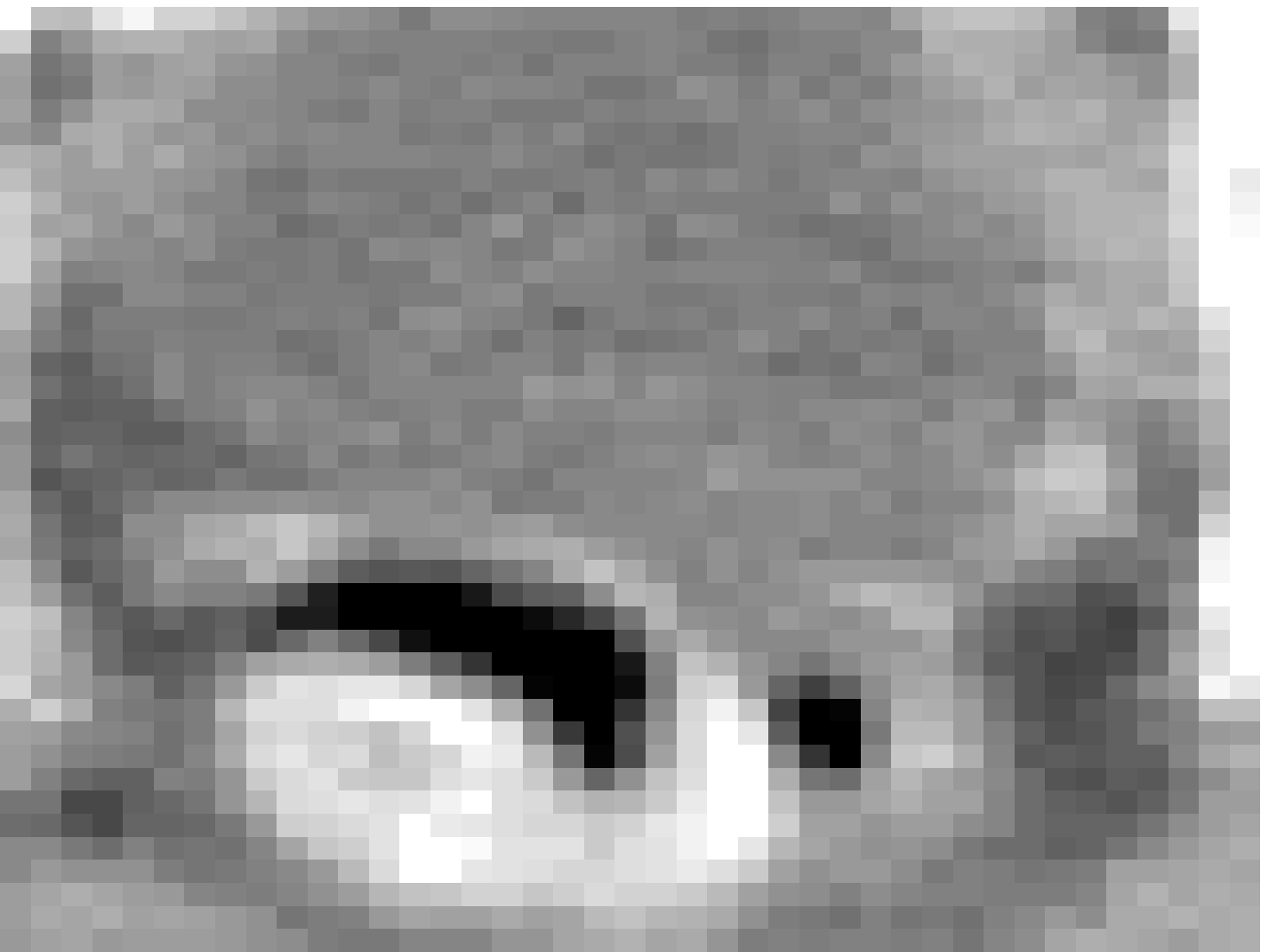}};
			\draw [-stealth, line width=2pt, cyan] (0.7,1.3) -- ++(-0.3,-0.3);
			\draw [-stealth, line width=2pt, cyan] (1.7,0.45) -- ++(-0.45,-0.0);
			\end{tikzpicture}&
			\begin{tikzpicture}
			\node[anchor=south west,inner sep=0] (image) at (0,0) {\includegraphics[width=.2\linewidth,height=.2\linewidth]{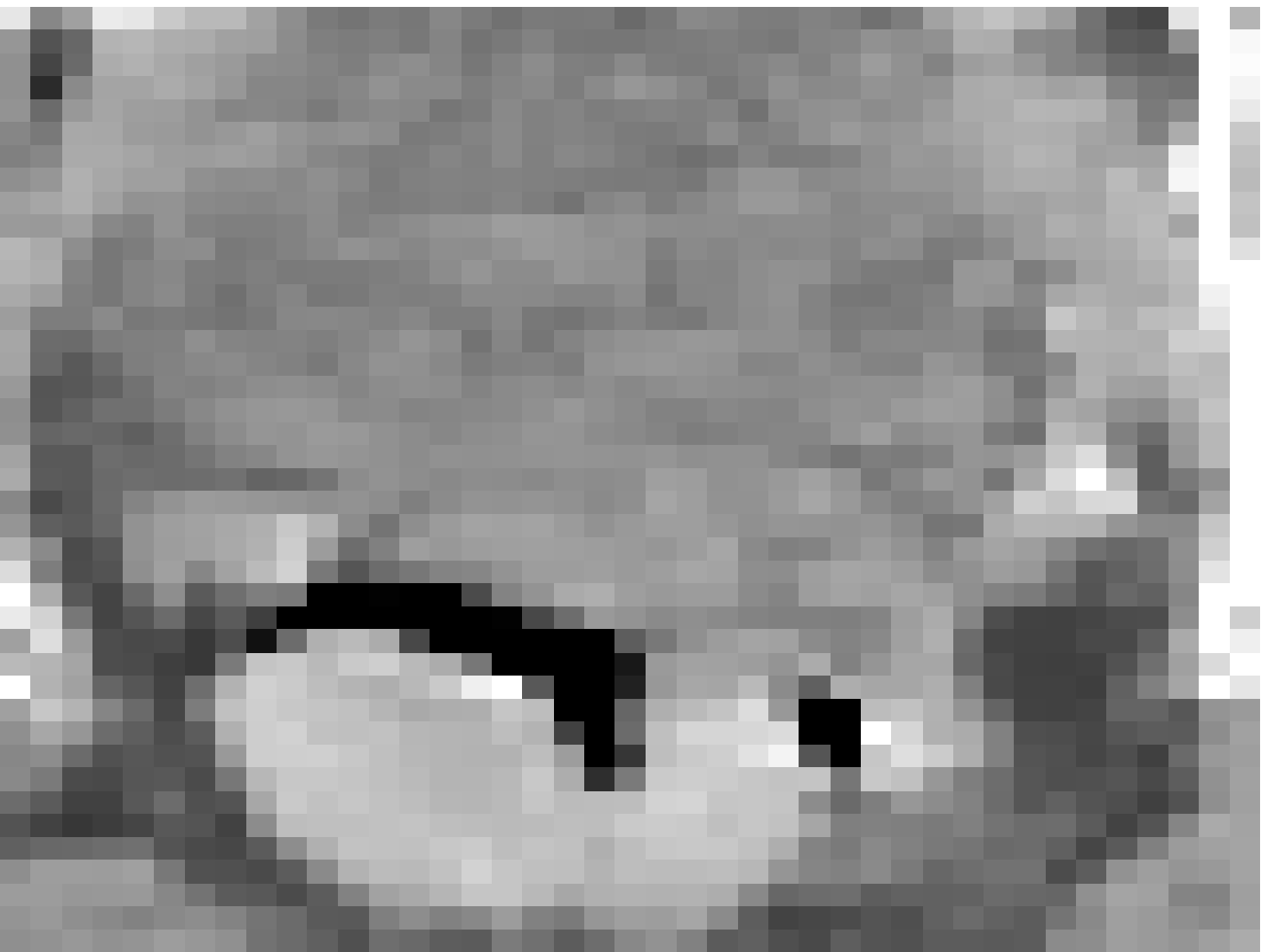}};
			\draw [-stealth, line width=2pt, cyan] (0.7,1.3) -- ++(-0.3,-0.3);
			\draw [-stealth, line width=2pt, cyan] (1.7,0.45) -- ++(-0.45,-0.0);
			\end{tikzpicture}
			&
			\begin{tikzpicture}
			\node[anchor=south west,inner sep=0] (image) at (0,0) {\includegraphics[width=.2\linewidth,height=.2\linewidth]{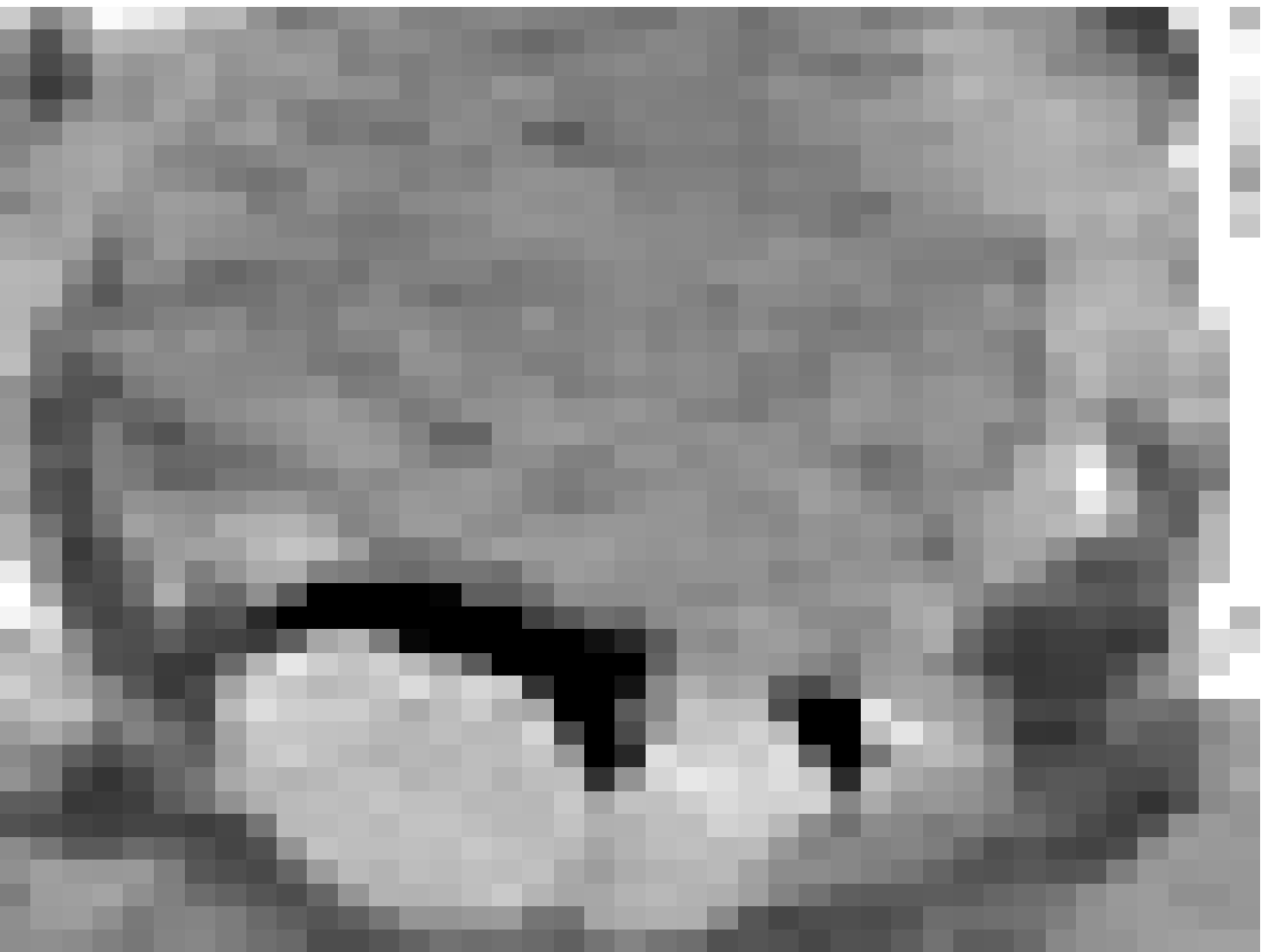}};
			\draw [-stealth, line width=2pt, cyan] (0.7,1.3) -- ++(-0.3,-0.3);
			\draw [-stealth, line width=2pt, cyan] (1.7,0.45) -- ++(-0.45,-0.0);
			\end{tikzpicture}&
			\begin{tikzpicture}
			\node[anchor=south west,inner sep=0] (image) at (0,0) {\includegraphics[width=.2\linewidth,height=.2\linewidth]{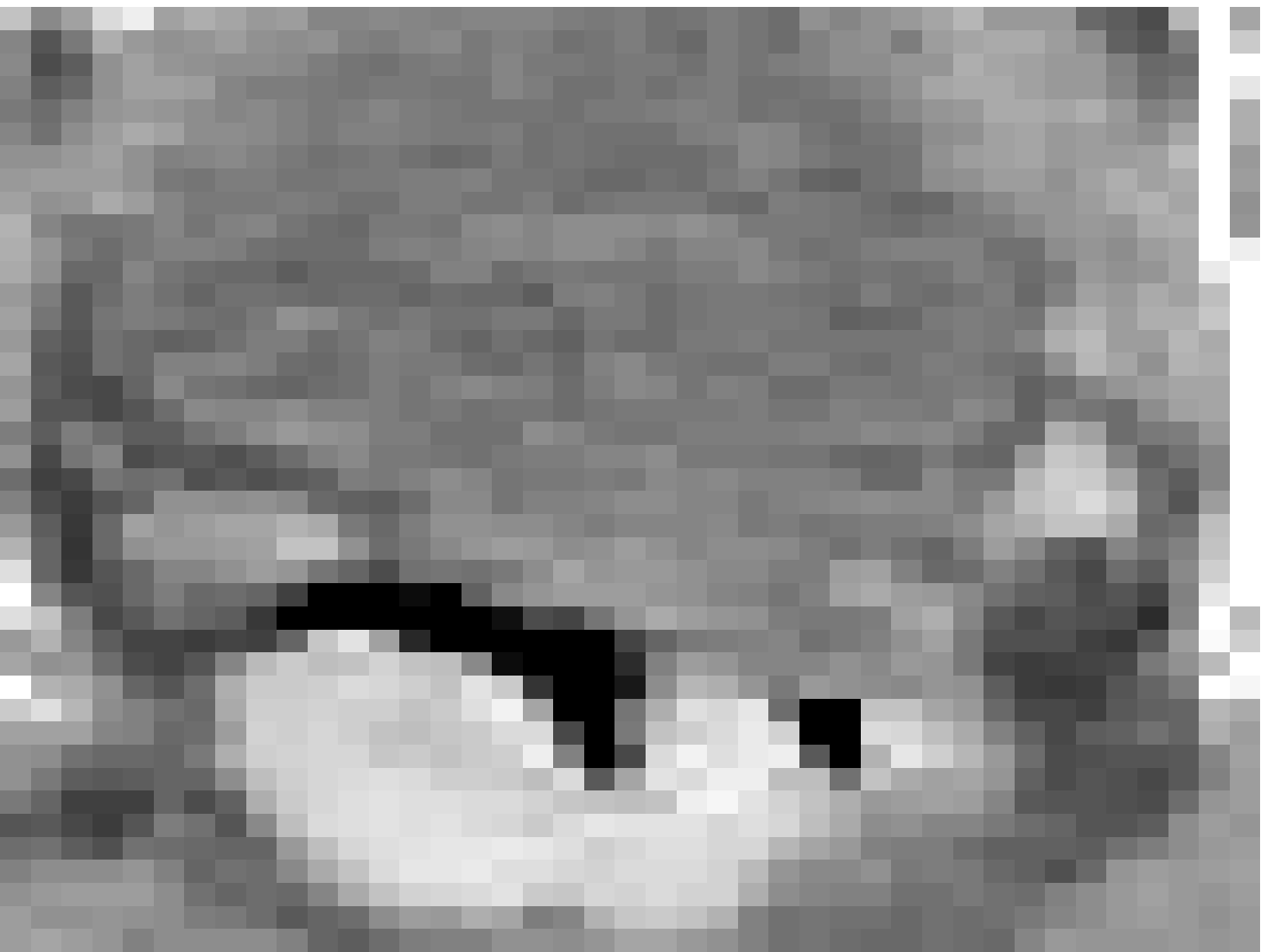}};
			\draw [-stealth, line width=2pt, cyan] (0.7,1.3) -- ++(-0.3,-0.3);
			\draw [-stealth, line width=2pt, cyan] (1.7,0.45) -- ++(-0.45,-0.0);
			\end{tikzpicture}&
			\begin{tikzpicture}
			\node[anchor=south west,inner sep=0] (image) at (0,0) {\includegraphics[width=.2\linewidth,height=.2\linewidth]{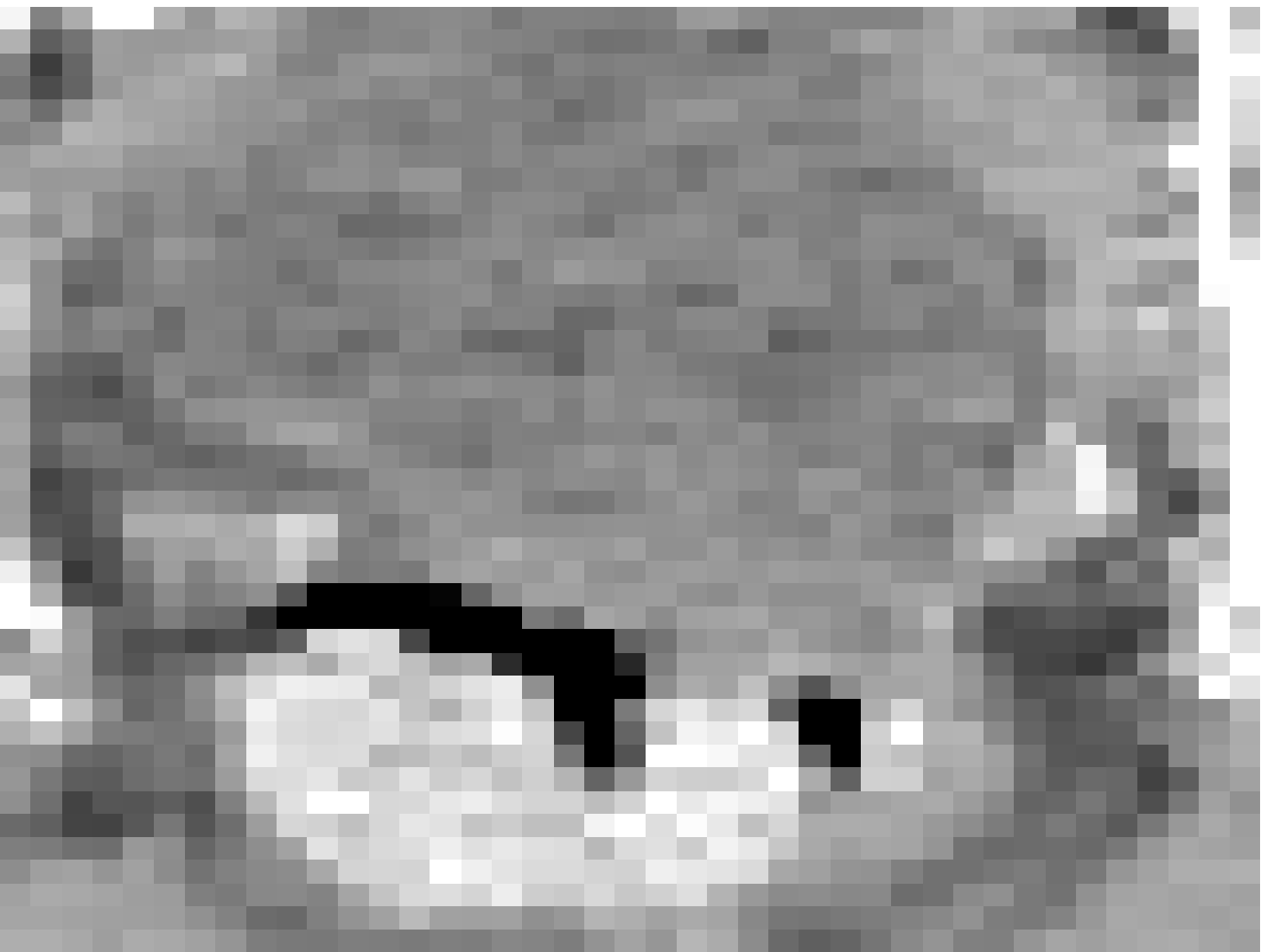}};
			\draw [-stealth, line width=2pt, cyan] (0.7,1.3) -- ++(-0.3,-0.3);
			\draw [-stealth, line width=2pt, cyan] (1.7,0.45) -- ++(-0.45,-0.0);
			\end{tikzpicture}\\		
			(a)&
			(b)&
			(c)&
			(d)&
			(e)
		\end{tabular}
		\caption{Zoom-in results of Fig.~\ref{ablationsliceU_50000} corresponding to the red boxes in Fig. \ref{Truth}.
			(a) No Filter; (b) Using $\nabla$;  (c) Learnable filters;  (d) Learnable HP; (e) AHP-net.
		}
		\label{ablationsliceZoomU_50000}
	\end{center}
\end{figure}

Fig. \ref{ablationsliceU_50000} visualized a comparison result  from different versions of the AHP-Net with dose level $I_i=5\times10^4$, and Fig.~\ref{ablationsliceZoomU_50000} shows their zoomed-in regions of  the boxes in Fig. \ref{Truth}. It can be seen that these figures correspond well with the results shown in Table~\ref{ablationComparisionU}; see more comparison in supplementary file.
	\begin{table}
	\caption{Quantitative comparison (Mean$\pm$STD) of  the results reconstructed by the different versions of the AHP-Net for ablation study}
	\scalebox{0.78}{
		\begin{tabular}{c@{\hspace{3pt}}c@{\hspace{3pt}}c@{\hspace{3pt}}c@{\hspace{3pt}}c@{\hspace{3pt}}c@{\hspace{3pt}}c@{\hspace{3pt}}c@{\hspace{3pt}}c}
			\hline
			{Dose }                    &\multirow{2}{*}{Index}       &\multirow{2}{*}{~~No Filter}  &\multirow{2}{*}{~Using $\nabla$}     &Learnable   &Learnable        &\multirow{2}{*}{~~AHP-Net}         \\
			{level}                   & &&                                                                                                               & filters     &HP     &            \\
			\hline
			\multirow{3}{*}{$1 \times 10^5$}         &PSNR       &$38.94\pm3.33$   &$39.26\pm2.86$ & $39.62\pm2.50$      &$40.86\pm2.54$        &$\bm{41.12\pm2.69}$\\	                                    
			                                &RMSE       &$16.88\pm14.58$  &$15.46\pm4.62$ & $14.67\pm3.66$      &$12.80\pm4.10$        &$\bm{12.41\pm3.31}$\\
			                                &SSIM       &$0.9841\pm0.02$&$0.9656\pm0.01$  &	$0.9856\pm0.01$      &$0.9686\pm0.02$       &$\bm{0.9875\pm0.01}$\\
			\hline
			\multirow{3}{*}{$5\times10^4$}    &PSNR     &$38.09\pm2.80$  &$38.41\pm2.98$  & $38.27\pm2.50$       &$39.03\pm3.11$        &$\bm{39.55\pm3.08}$\\ 
			                                  &RMSE     &$17.80\pm6.79$  &$17.25\pm7.16$  & $17.13\pm4.43$       &$16.14\pm7.24$        &$\bm{14.98\pm5.06}$\\ 
			                                  &SSIM     &$0.9796\pm0.01$&$0.9566\pm 0.02$ & $0.9812\pm0.01$       &$0.9618\pm0.02$       &$\bm{0.9819\pm0.01}$\\
			\hline
			\multirow{3}{*}{$1 \times 10^4$}         &PSNR      &$36.20\pm2.55$   &$36.91\pm2.49$  & $36.46\pm2.49$       &$36.78\pm2.42$        &$\bm{38.20\pm2.50}$\\
			                                &RMSE      &$21.65\pm5.21$   &$19.98\pm 4.51$ &	$20.97\pm4.09$       &$20.14\pm3.59$        &$\bm{17.26\pm4.60}$\\
			                                &SSIM      &$0.9656\pm0.01$&$0.9383\pm0.02$	  & $0.9712\pm0.01$       &$0.9368\pm0.02$       &$\bm{0.9738\pm 0.01}$\\      
			\hline
			\multirow{3}{*}{$5\times10^3$}   &PSNR     &$35.07\pm2.81$   &$35.08\pm3.31$  & $35.31\pm2.45$      &$35.63\pm2.49$        &$\bm{36.85\pm1.85}$\\
		 	                                 &RMSE     &$25.23\pm13.45$  &$26.21\pm19.65$ & $23.88\pm4.45$      &$23.08\pm5.19$        &$\bm{36.93\pm2.42}$\\
			                                 &SSIM     &$0.9605\pm0.02$ &$0.9191\pm0.05$  &	$0.9624\pm0.01$      &$0.9229\pm0.03$       &$\bm{0.9673\pm 0.01}$\\
			\hline                                                                                                                       
		\end{tabular} 
	}
	\label{ablationComparisionU}
\end{table}

\subsection{Cross-validation}
In this section, a $k$-fold cross-validation is  conducted on FBPConvNet and the proposed AHP-Net to give a comprehensive measure of its model's performance throughout the whole dataset. More specifically, 100 scans are equally partitioned into $5$ folders. Then, for each folder, the images associated with the scans from this fold are used for testing, and all others are used for training.  One single model is trained for different noise levels.
See Table \ref{CrossValid} for the list of average quantitative metric values of the results from  each test dataset.
The mean of PSNR for the results from the proposed AHP-Net is about  $2.4-3.5$dB higher than that of FBPConvNet.
However, the difference of the STD of PSNR is no more than $0.25$dB.

	\begin{table}
	\caption{  Cross-validation Quantitative  results of one single model  trained for different dose levels.}
	
	\scalebox{0.78}{
		\begin{tabular}{c@{\hspace{3pt}}|c@{\hspace{3pt}}|c@{\hspace{3pt}}|c@{\hspace{4pt}}c@{\hspace{4pt}}c@{\hspace{4pt}}c@{\hspace{4pt}}c@{\hspace{4pt}}|c}
			\hline
			Dose  &\multirow{2}{*}{Method}  & \multirow{2}{*}{Index}       &Fold  &Fold   &Fold  &Fold        &Fold         & \multirow{2}{*}{Mean$\pm$STD} \\
			level&&&1&2&3&4&5&\\
			\hline
\multirow{6}{*}{$1 \times 10^5$}&\multirow{3}{*}{FBPConvNet}                 &PSNR        &37.81    &37.69    &38.15    &37.06    &37.44   &$ 37.63\pm0.41$ \\	                                    
			                       &                                &RMSE        &17.24    &17.03    &16.36    &18.33    &18.49   &$17.49\pm0.90$\\
			                       &                                &SSIM        &0.9741   &0.9746   &0.9755   &0.9736   &0.9855  &$ 0.9739\pm0.00$\\
			 \cline{2-9}
			                       &\multirow{3}{*}{AHP-Net}        &PSNR        &39.68  &40.24   &39.40       &40.06       &$40.54$  &$\bm{39.98\pm0.45}$\\	                                    
			                       &                                &RMSE        &15.57  &13.24   &14.43       &13.19       &$13.15$  &$\bm{13.92\pm1.07}$\\
			                       &                                &SSIM        &0.9834 &0.9778  &0.9821      &0.9769      &$0.9703$ &$\bm{0.9811\pm0.01}$\\                                  
			\hline
\multirow{6}{*}{$5\times10^4$}&\multirow{3}{*}{FBPConvNet}          &PSNR        &37.26   &37.20  &37.46  &36.73  &37.06   &$37.14\pm0.27$\\ 
			                       &                                &RMSE        &18.35   &18.00  &17.69  &19.04  &19.32   &$18.48\pm 0.69$ \\ 
			                       &                                &SSIM        &0.9717  & 0.9723 &0.9724&0.9718 &0.9695   &$0.9715\pm0.00$\\
			\cline{2-9} 
	                               &\multirow{3}{*}{AHP-Net}        &PSNR        &39.30  &38.29   &39.02       &39.72       &$40.07$  &$\bm{39.58\pm0.42}$\\ 
			                       &                                &RMSE        &16.15  &13.87   &15.02       &13.69       &$13.85$  &$\bm{14.52\pm1.06}$\\ 
			                       &                                &SSIM        &0.9811 &0.9757  &0.9800      &0.9748      &$0.9832$ &$\bm{0.9790\pm0.00}$\\                                
			\hline
\multirow{6}{*}{$1 \times 10^4$}&\multirow{3}{*}{FBPConvNet}                 &PSNR        &34.63    &34.68    &34.69    &34.53    &34.76  &$34.66\pm0.09$\\
			                       &                                &RMSE        &24.83    &24.10    &24.36    &24.58   & 25.19  &$24.61\pm 0.42$ \\
			                       &                                &SSIM        &0.9562   &0.9548   & 0.9566  &0.9552    &0.9508 &$0.9547\pm0.01$\\
			\cline{2-9}       
			                       &\multirow{3}{*}{AHP-Net}        &PSNR        &37.31  &37.59   &37.21       &37.95       &$37.92$ &$\bm{37.59\pm0.34}$\\
			                       &                                &RMSE        &19.72  &17.60   &18.38       &16.70       &$17.65$ &$\bm{18.01\pm1.13}$\\
			                       &                                &SSIM        &0.9684 &0.9634  &0.9685      &0.9634      &$0.9713$&$\bm{0.9670\pm0.00}$\\                                      
			\hline
\multirow{6}{*}{$5\times10^3$}&\multirow{3}{*}{FBPConvNet}          &PSNR        &32.60    &32.68    &32.78    &32.30    &32.71   &$32.61\pm0.19$ \\
		 	                       &                                &RMSE        &31.39    &30.42    &30.40    &31.91    &32.02   &$31.23\pm0.78$ \\
			                       &                                &SSIM        &0.9355   &0.9310   &0.9375   &0.9266    &0.9230  &$0.9307\pm0.01$ \\
			\cline{2-9}
			                       &\multirow{3}{*}{AHP-Net}        &PSNR        &35.72  &36.03   &35.87       &36.57       &$36.39$  &$\bm{36.12\pm0.36}$ \\
			                       &                                &RMSE        &23.57  &20.97   &21.40       &19.54       &$21.03$  &$\bm{21.30\pm1.45}$\\
			                       &                                &SSIM        &0.9563 &0.9522  &0.9587      &0.9536      &$0.9603$ &$\bm{0.9562\pm0.00}$\\                                
			\hline                                                                                                                    
		\end{tabular} 
	}
	\label{CrossValid}
\end{table}
\section{Discussion and Conclusion}
\label{conclusion}
Similarly to many optimization-unrolling-based deep learning methods, this paper also proposed a deep learning method for LDCT image reconstruction, the AHP-Net, that unrolls the half-quadratic splitting scheme for MBIR. Each stage of the AHP-Net contains one inversion block and a denoising block, where the denoising block is built-on a CNN. The main difference of the proposed AHP-Net from other deep learning solutions lies the design of the inversion block.

In the proposed inversion block, we  replaced the often-used gradient operator $\nabla$ by the filter banks with 8 high-pass filters from linear spline framelet transform, motivated by its success in $\ell_1$-norm relating regularization in image recovery. Moreover, we proposed to
pay special attention to the hyper-parameters involved in the inversion block, and presented a MLP-based NN to predict hyper-parameters that adaptive to both dose level and image content.

The experiments showed the advantage of the proposed method over classic non-learning methods and some representative deep learning based methods for LDCT reconstruction. Also, another advantage is that the proposed method can only train a single model with  competitive performance to process measurement data with varying dose levels.

As different dose levels implicitly refers to different SNRs. The proposed methods indeed allows to train a universal model with competitive performance to process the data with unknown SNRs, which can be valuable to practical applications in medical CT imaging. In practical medical CT scans, the dose level is not fixed and the SNR of the data is also different for different people under the same dose level, \emph{e.g.} adult vs child. The adaptivity of the proposed method to different SNRs of data can be welcomed in practice.

There still exists some issues in the proposed method. One is that instead of reconstructing the 3D volume as a whole, the proposed method separately reconstructs image slices. The main reason we take such a reconstruction scheme is that
the number of  network parameters of the NN adopted in this paper will increase exponentially when constructing the whole 3D volume. The resulting memory usage in GPU will be too excessive to make the computation feasible under current available commodity GPUs. 

In future work, we will study how to design compact
deep NN that can directly reconstruct 3D volume in LDCT image reconstruction. Also, we will investigate the applications of the proposed method in other medical image reconstruction problems, \emph{e.g.}
sparse-view CT reconstruction and  image reconstruction from sparse samples in MRI.
\IEEEtriggeratref{100}
\bibliographystyle{ieeetr}

\begin{thebibliography}{10}

\bibitem{sidky2006accurate}
E.~Y. Sidky, C.-M. Kao, and X.~Pan, ``{Accurate image reconstruction from
  few-views and limited-angle data in divergent-beam CT},'' {\em Journal of
  X-ray Science and Technology}, vol.~14, no.~2, pp.~119--139, 2006.

\bibitem{whiting06pop}
B.~R. Whiting, P.~Massoumzadeh, O.~A. Earl, J.~A. O~Sullivan, D.~L. Snyder, and
  J.~F. Williamson, ``Properties of preprocessed sinogram data in {X}-ray
  computed tomography,'' {\em Medical physics}, vol.~33, no.~9, pp.~3290--3303,
  2006.

\bibitem{thibault2007three}
J.-B. Thibault, K.~D. Sauer, C.~A. Bouman, and J.~Hsieh, ``A three-dimensional
  statistical approach to improved image quality for multislice helical {CT},''
  {\em Medical physics}, vol.~34, no.~11, pp.~4526--4544, 2007.

\bibitem{nuyts13mtp}
J.~Nuyts, B.~De~Man, J.~A. Fessler, W.~Zbijewski, and F.~J. Beekman,
  ``Modelling the physics in the iterative reconstruction for transmission
  computed tomography,'' {\em Physics in medicine and biology}, vol.~58,
  no.~12, p.~R63, 2013.

\bibitem{zhang2005total}
X.~Zhang and J.~Froment, ``Total variation based fourier reconstruction and
  regularization for computer tomography,'' in {\em Nuclear Science Symposium
  Conference Record, 2005 IEEE}, vol.~4, pp.~2332--2336, IEEE, 2005.

\bibitem{sidky2008image}
E.~Y. Sidky and X.~Pan, ``Image reconstruction in circular cone-beam computed
  tomography by constrained, total-variation minimization,'' {\em Physics in
  Medicine \& Biology}, vol.~53, no.~17, p.~4777, 2008.

\bibitem{chen2008prior}
G.~Chen, J.~Tang, and S.~Leng, ``{Prior image constrained compressed sensing
  (PICCS): a method to accurately reconstruct dynamic CT images from highly
  undersampled projection data sets},'' {\em Medical physics}, vol.~35, no.~2,
  pp.~660--663, 2008.

\bibitem{jia2011gpu}
X.~Jia, B.~Dong, Y.~Lou, and S.~B. Jiang, ``{GPU}-based iterative cone-beam
  {CT} reconstruction using tight frame regularization,'' {\em Physics in
  Medicine \& Biology}, vol.~56, no.~13, p.~3787, 2011.

\bibitem{gao20124d}
H.~Gao, R.~Li, Y.~Lin, and L.~Xing, ``4{D} cone beam {CT} via spatiotemporal
  tensor framelet,'' {\em Medical physics}, vol.~39, no.~11, pp.~6943--6946,
  2012.

\bibitem{jia20104d}
X.~Jia, Y.~Lou, B.~Dong, Z.~Tian, and S.~Jiang, ``{4D computed tomography
  reconstruction from few-projection data via temporal non-local
  regularization},'' in {\em International Conference on Medical Image
  Computing and Computer-Assisted Intervention}, pp.~143--150, Springer, 2010.

\bibitem{gao2011robust}
H.~Gao, J.~Cai, Z.~Shen, and H.~Zhao, ``Robust principal component
  analysis-based four-dimensional computed tomography,'' {\em Physics in
  Medicine \& Biology}, vol.~56, no.~11, p.~3181, 2011.

\bibitem{gao2011multi}
H.~Gao, H.~Yu, S.~Osher, and G.~Wang, ``{Multi-energy CT based on a prior rank,
  intensity and sparsity model (PRISM)},'' {\em Inverse problems}, vol.~27,
  no.~11, p.~115012, 2011.

\bibitem{cai2014cine}
J.~Cai, X.~Jia, H.~Gao, S.~B. Jiang, Z.~Shen, and H.~Zhao, ``{Cine cone beam CT
  reconstruction using low-rank matrix factorization: algorithm and a
  proof-of-principle study},'' {\em IEEE transactions on medical imaging},
  vol.~33, no.~8, pp.~1581--1591, 2014.

\bibitem{chen2015synchronized}
G.~Chen and Y.~Li, ``{Synchronized multiartifact reduction with tomographic
  reconstruction (SMART-RECON): A statistical model based iterative image
  reconstruction method to eliminate limited-view artifacts and to mitigate the
  temporal-average artifacts in time-resolved CT},'' {\em Medical physics},
  vol.~42, no.~8, pp.~4698--4707, 2015.

\bibitem{xu2012low}
Q.~Xu, H.~Yu, X.~Mou, L.~Zhang, J.~Hsieh, and G.~Wang, ``Low-dose {X}-ray {CT}
  reconstruction via dictionary learning,'' {\em IEEE transactions on medical
  imaging}, vol.~31, no.~9, pp.~1682--1697, 2012.

\bibitem{bai2017z}
T.~Bai, H.~Yan, X.~Jia, S.~Jiang, G.~Wang, and X.~Mou, ``{Z-index
  parameterization for volumetric CT image reconstruction via 3-D dictionary
  learning},'' {\em IEEE transactions on medical imaging}, vol.~36, no.~12,
  pp.~2466--2478, 2017.

\bibitem{gupta2018cnn}
H.~Gupta, K.~H. Jin, H.~Q. Nguyen, M.~T. McCann, and M.~Unser, ``{CNN}-based
  projected gradient descent for consistent {CT} image reconstruction,'' {\em
  IEEE transactions on medical imaging}, vol.~37, no.~6, pp.~1440--1453, 2018.

\bibitem{he2018optimizing}
J.~He, Y.~Yang, Y.~Wang, D.~Zeng, Z.~Bian, H.~Zhang, J.~Sun, Z.~Xu, and J.~Ma,
  ``{Optimizing a parameterized plug-and-play ADMM for iterative low-dose CT
  reconstruction},'' {\em IEEE transactions on medical imaging}, vol.~38,
  no.~2, pp.~371--382, 2018.

\bibitem{adler2018learned}
J.~Adler and O.~{\"O}ktem, ``Learned primal-dual reconstruction,'' {\em IEEE
  transactions on medical imaging}, vol.~37, no.~6, pp.~1322--1332, 2018.

\bibitem{ramani2012regularization}
S.~Ramani, Z.~Liu, J.~Rosen, J.-F. Nielsen, and J.~A. Fessler, ``Regularization
  parameter selection for nonlinear iterative image restoration and {MRI}
  reconstruction using {GCV} and {SURE}-based methods,'' {\em IEEE Transactions
  on Image Processing}, vol.~21, no.~8, pp.~3659--3672, 2012.

\bibitem{hansen1992analysis}
P.~C. Hansen, ``Analysis of discrete ill-posed problems by means of the
  {L}-curve,'' {\em SIAM review}, vol.~34, no.~4, pp.~561--580, 1992.

\bibitem{conn1991globally}
A.~R. Conn, N.~I. Gould, and P.~Toint, ``A globally convergent augmented
  lagrangian algorithm for optimization with general constraints and simple
  bounds,'' {\em SIAM Journal on Numerical Analysis}, vol.~28, no.~2,
  pp.~545--572, 1991.

\bibitem{jin2012iterative}
B.~Jin, Y.~Zhao, and J.~Zou, ``Iterative parameter choice by discrepancy
  principle,'' {\em IMA Journal of Numerical Analysis}, vol.~32, no.~4,
  pp.~1714--1732, 2012.

\bibitem{golub1979generalized}
G.~H. Golub, M.~Heath, and G.~Wahba, ``Generalized cross-validation as a method
  for choosing a good ridge parameter,'' {\em Technometrics}, vol.~21, no.~2,
  pp.~215--223, 1979.

\bibitem{wang2019global}
Y.~Wang, W.~Yin, and J.~Zeng, ``{Global convergence of ADMM in nonconvex
  nonsmooth optimization},'' {\em Journal of Scientific Computing}, vol.~78,
  no.~1, pp.~29--63, 2019.

\bibitem{mardani2017recurrent}
M.~Mardani, H.~Monajemi, V.~Papyan, S.~Vasanawala, D.~Donoho, and J.~Pauly,
  ``Recurrent generative adversarial networks for proximal learning and
  automated compressive image recovery,'' {\em arXiv preprint
  arXiv:1711.10046}, 2017.

\bibitem{sun2016deep}
J.~Sun, H.~Li, Z.~Xu, {\em et~al.}, ``Deep {ADMM}-{N}et for compressive sensing
  {MRI},'' in {\em Advances in neural information processing systems},
  pp.~10--18, 2016.

\bibitem{wu2017iterative}
D.~Wu, K.~Kim, G.~El~Fakhri, and Q.~Li, ``Iterative low-dose {CT}
  reconstruction with priors trained by artificial neural network,'' {\em IEEE
  transactions on medical imaging}, vol.~36, no.~12, pp.~2479--2486, 2017.

\bibitem{geman1995nonlinear}
D.~Geman and C.~Yang, ``Nonlinear image recovery with half-quadratic
  regularization,'' {\em IEEE transactions on Image Processing}, vol.~4, no.~7,
  pp.~932--946, 1995.

\bibitem{chen2017lowS}
H.~Chen, Y.~Zhang, W.~Zhang, P.~Liao, K.~Li, J.~Zhou, and G.~Wang, ``Low-dose
  {CT} via convolutional neural network,'' {\em Biomedical optics express},
  vol.~8, no.~2, pp.~679--694, 2017.

\bibitem{chen2017low}
H.~Chen, Y.~Zhang, M.~K. Kalra, F.~Lin, Y.~Chen, P.~Liao, J.~Zhou, and G.~Wang,
  ``Low-dose {CT} with a residual encoder-decoder convolutional neural
  network,'' {\em IEEE transactions on medical imaging}, vol.~36, no.~12,
  pp.~2524--2535, 2017.

\bibitem{han2016deep}
Y.~S. Han, J.~Yoo, and J.~C. Ye, ``Deep residual learning for compressed
  sensing {CT} reconstruction via persistent homology analysis,'' {\em arXiv
  preprint arXiv:1611.06391}, 2016.

\bibitem{li2017low}
H.~Li and K.~Mueller, ``Low-dose {CT} streak artifacts removal using deep
  residual neural network,'' in {\em Proc. Fully Three-Dimensional Image
  Reconstruction Radiol. Nucl. Med.(Fully3D)}, pp.~191--194, 2017.

\bibitem{jin2017deep}
K.~H. Jin, M.~T. McCann, E.~Froustey, and M.~Unser, ``Deep convolutional neural
  network for inverse problems in imaging,'' {\em IEEE Transactions on Image
  Processing}, vol.~26, no.~9, pp.~4509--4522, 2017.

\bibitem{wolterink2017generative}
J.~M. Wolterink, T.~Leiner, M.~A. Viergever, and I.~I{\v{s}}gum, ``Generative
  adversarial networks for noise reduction in low-dose {CT},'' {\em IEEE
  transactions on medical imaging}, vol.~36, no.~12, pp.~2536--2545, 2017.

\bibitem{yang2018low}
Q.~Yang, P.~Yan, Y.~Zhang, H.~Yu, Y.~Shi, X.~Mou, M.~K. Kalra, Y.~Zhang,
  L.~Sun, and G.~Wang, ``Low-dose {CT} image denoising using a generative
  adversarial network with wasserstein distance and perceptual loss,'' {\em
  IEEE transactions on medical imaging}, vol.~37, no.~6, pp.~1348--1357, 2018.

\bibitem{ye2018deep}
J.~C. Ye, Y.~Han, and E.~Cha, ``{Deep convolutional framelets: A general deep
  learning framework for inverse problems},'' {\em SIAM Journal on Imaging
  Sciences}, vol.~11, no.~2, pp.~991--1048, 2018.

\bibitem{kang2017deep}
E.~Kang, J.~Min, and J.~C. Ye, ``A deep convolutional neural network using
  directional wavelets for low-dose {X}-ray {CT} reconstruction,'' {\em Medical
  physics}, vol.~44, no.~10, pp.~e360--e375, 2017.

\bibitem{kang2017wavelet}
E.~Kang, J.~C. Ye, {\em et~al.}, ``{Wavelet domain residual network (WavResNet)
  for low-dose X-ray CT reconstruction},'' {\em arXiv preprint
  arXiv:1703.01383}, 2017.

\bibitem{chen2018learn}
H.~Chen, Y.~Zhang, Y.~Chen, J.~Zhang, W.~Zhang, H.~Sun, Y.~Lv, P.~Liao,
  J.~Zhou, and G.~Wang, ``{LEARN}: Learned experts' assessment-based
  reconstruction network for sparse-data {CT},'' {\em IEEE transactions on
  medical imaging}, vol.~37, no.~6, pp.~1333--1347, 2018.

\bibitem{pelt2013fast}
D.~M. Pelt and K.~J. Batenburg, ``{Fast tomographic reconstruction from limited
  data using artificial neural networks},'' {\em IEEE Transactions on Image
  Processing}, vol.~22, no.~12, pp.~5238--5251, 2013.

\bibitem{hauptmann2018model}
A.~Hauptmann, F.~Lucka, M.~Betcke, N.~Huynh, J.~Adler, B.~Cox, P.~Beard,
  S.~Ourselin, and S.~Arridge, ``{Model-based learning for accelerated,
  limited-view 3-D photoacoustic tomography},'' {\em IEEE transactions on
  medical imaging}, vol.~37, no.~6, pp.~1382--1393, 2018.

\bibitem{bazrafkan2019deep}
S.~Bazrafkan, V.~Van~Nieuwenhove, J.~Soons, J.~De~Beenhouwer, and J.~Sijbers,
  ``{Deep neural network assisted iterative reconstruction method for low dose
  CT},'' {\em arXiv preprint arXiv:1906.00650}, 2019.

\bibitem{aggarwal2018modl}
H.~K. Aggarwal, M.~P. Mani, and M.~Jacob, ``{MoDL: Model-based deep learning
  architecture for inverse problems},'' {\em IEEE transactions on medical
  imaging}, vol.~38, no.~2, pp.~394--405, 2018.

\bibitem{gilton2019neumann}
D.~Gilton, G.~Ongie, and R.~Willett, ``Neumann networks for inverse problems in
  imaging,'' {\em arXiv preprint arXiv:1901.03707}, 2019.

\bibitem{ding2018statistical}
Q.~Ding, Y.~Long, X.~Zhang, and J.~A. Fessler, ``{Statistical Image
  Reconstruction Using Mixed Poisson-Gaussian Noise Model for X-Ray CT},'' {\em
  arXiv preprint arXiv:1801.09533}, 2018.

\bibitem{LinComparison}
L.~Fu, T.~Lee, S.~M. Kim, A.~M. Alessio, P.~E. Kinahan, Z.~Chang, K.~D. Sauer,
  M.~K. Kalra, and B.~De~Man, ``{Comparison Between Pre-Log and Post-Log
  Statistical Models in Ultra-Low-Dose CT Reconstruction},'' {\em IEEE
  Transactions on Medical Imaging}, vol.~36, no.~3, pp.~707--720, 2017.

\bibitem{sauer1993a}
K.~D. Sauer and C.~A. Bouman, ``A local update strategy for iterative
  reconstruction from projections,'' {\em IEEE Transactions on Signal
  Processing}, vol.~41, no.~2, pp.~534--548, 1993.

\bibitem{fessler2000statistical}
J.~A. Fessler, ``Statistical image reconstruction methods for transmission
  tomography,'' {\em Handbook of medical imaging}, vol.~2, pp.~1--70, 2000.

\bibitem{dong2010mra}
B.~Dong, Z.~Shen, {\em et~al.}, ``{MRA} based wavelet frames and
  applications,'' {\em IAS Lecture Notes Series, Summer Program on "The
  Mathematics of Image Processing", Park City Mathematics Institute}, vol.~19,
  2010.

\bibitem{zheng2017sparse}
X.~Zheng, I.~Y. Chun, Z.~Li, Y.~Long, and J.~A. Fessler, ``{Sparse-View X-Ray
  CT Reconstruction Using $\ell_1$ Prior with Learned Transform},'' {\em arXiv
  preprint arXiv:1711.00905}, 2017.

\bibitem{huang2017densely}
G.~Huang, Z.~Liu, L.~Van Der~Maaten, and K.~Q. Weinberger, ``Densely connected
  convolutional networks,'' in {\em Proceedings of the IEEE conference on
  computer vision and pattern recognition}, pp.~4700--4708, 2017.

\bibitem{paszke2017automatic}
A.~Paszke, S.~Gross, S.~Chintala, G.~Chanan, E.~Yang, Z.~DeVito, Z.~Lin,
  A.~Desmaison, L.~Antiga, and A.~Lerer, ``Automatic differentiation in
  pytorch,'' 2017.

\bibitem{zhang2010discriminative}
Q.~Zhang and B.~Li, ``Discriminative k-svd for dictionary learning in face
  recognition,'' in {\em 2010 IEEE computer society conference on computer
  vision and pattern recognition}, pp.~2691--2698, IEEE, 2010.

\bibitem{dabov2007image}
K.~Dabov, A.~Foi, V.~Katkovnik, and K.~Egiazarian, ``Image denoising by sparse
  3-d transform-domain collaborative filtering,'' {\em IEEE Transactions on
  image processing}, vol.~16, no.~8, pp.~2080--2095, 2007.

\bibitem{wang2004image}
Z.~Wang, A.~C. Bovik, H.~R. Sheikh, E.~P. Simoncelli, {\em et~al.}, ``Image
  quality assessment: from error visibility to structural similarity,'' {\em
  IEEE transactions on image processing}, vol.~13, no.~4, pp.~600--612, 2004.

\end{thebibliography}

\clearpage
\appendix

\begin{itemize}
	
	\item  The details of calculating  the gradients of the loss function w.r.t. all parameters, using  back-propagation  over the deep architecture in Fig. 3.
	\item The details of the implementation and training of comparison methods
	\item Visual comparison on more examples.
\end{itemize}
\section{Back-propagation }
\label{APP:BP}
During the train process, we need to calculate gradients about  $\Theta_k=\{\theta_\mathcal{D}^{k-1}, \theta_\mathcal{P}^k\}, k=1,\cdots K$.
The loss of the proposed neural network is as follows,
\begin{equation}\label{eqn:loss1}
\mathcal{L}(\Theta)=\frac{1}{J}\sum_{j=1}^{J}\left(\|\bm{x}_{j}^{{{K}}}-\bm{x}_{j} \|_2^2+\sum_{k=1}^{K-1}\mu_k\|\bm{x}^k_{j}-\bm{x}_{j} \|_2^2\right),
\end{equation}
We give the details of the gradient calculation procedure.
Recall that the iteration of the proposed method is
as follows
\begin{subequations}
	\label{backpros}
	\begin{align}
		 &\tilde{\bm x}^{k-1}= 	\mathcal{D}^{k-1}_{\textrm{cnn}}(\cdot,\theta_\mathcal{D}^{k-1}) \\
		&\beta^k=\mathcal{P}_{\textrm{mlp}}^k(\cdot,\theta_{\mathcal{P}}^k)\\
		&\bm z_i^k=\bm F_i  \tilde{\bm x}^{k-1} \\
		&\bm{x}^{k}=\left(\bm{A}^\top\bm{A}+\sum_{i=1}^L\beta_i^k\bm{F}_i^\top\bm{F}_i\right)^{-1}\left(\bm{A}^\top\bm{y}+\sum_{i=1}^{L}\beta_i^k\bm{F}_i^\top\bm{z}_i^k\right),\nonumber\\
	\end{align}
\end{subequations}
For simplify, we omit the data index $j$,
\begin{eqnarray}
\frac{\partial \mathcal{L}(\Theta)}{\partial \theta_\mathcal{P}^k}&=& \frac{\partial L(\Theta)}{\partial \bm{x}^{k}}\cdot\frac{\partial \bm{x}^{k}}{\partial \theta_\mathcal{P}^k},\\
\frac{\partial \mathcal{L}(\Theta)}{\partial \theta_\mathcal{D}^{k-1}}&=&\frac{\partial L(\Theta)}{\partial \bm{x}^{k}}\cdot \frac{\partial \bm{x}^{k}}{\partial \theta_\mathcal{D}^{k-1}},
\end{eqnarray}
where

\begin{eqnarray}
\frac{\partial \mathcal{L}(\Theta)}{\partial \bm{x}^{k}}
&=&\mu_k\frac{\partial \| \bm{x}^k- \bm{x}\|_2^2}{\partial \bm{x}^{k}} +\mu_{k+1}\frac{\partial  \| \bm{x}^{k+1}- \bm{x}\|_2^2}{\partial \bm{x}^{k+1}}\cdot\frac{\partial \bm{x}^{k+1}}{\partial \bm{x}^{k}}  \nonumber\\
&+&\cdots  +\mu_{{K}-1} \frac{\partial \| \bm{x}^{{K}-1}- \bm{x}\|_2^2}{\bm{x}^{{K}-1}}  \cdots \frac{\partial \bm{x}^{{K}-1}}{\partial \bm{x}^{{K}-2}} \cdot \frac{\partial \bm{x}^{k+1}}{\partial \bm{x}^{k}} ,\nonumber\\
&+&   \frac{\partial \| \bm{x}^{{K}}- \bm{x}\|_2^2}{\partial \bm{x}^{{K}}} \cdot \frac{\partial \bm{x}^{{K}}}{\partial \bm{x}^{{K}-1}} \cdots \frac{\partial \bm{x}^{k+1}}{\partial \bm{x}^{k}} 
,
\end{eqnarray}

and

\begin{subequations}
	\label{backproS}
	\begin{numcases}
	{} \frac{\partial \| \bm{x}^{k}- \bm{x}\|_2^2}{\partial \bm{x}^{k}}=2(\bm{x}^{k}-\bm{x}) \\
	\frac{\partial \bm{x}^{k}}{\partial \bm{x}^{k-1}}=\sum_{i=1}^L\frac{\partial \bm{x}^{k}}{\partial \bm{z}_{i}^{k}}\frac{\partial \bm{z}_{i}^k}{\partial \tilde{\bm{x}}^{k-1}}\frac{\partial \tilde{\bm{x}}^{k-1}}{\partial \bm{x}^{k-1}},\\
	\frac{\partial \bm{x}^{k}}{\partial \theta_{\mathcal{D}}^{k-1}}=\sum_{i=1}^L \frac{\partial \bm{x}^{k}}{\partial \bm{z}_{i}^{k}} \frac{\partial \bm{z}_{i}^k}{\partial \tilde{\bm{x}}^{k-1}}\frac{\partial \tilde{\bm{x}}^{k-1}}{\partial \theta_{\mathcal{D}}^{k-1}},\\
	\frac{\partial \bm{x}^{k}}{\partial \theta_{\mathcal{P}}^{k}}= \frac{\partial \bm{x}^{k}}{\partial \beta^{k}}\frac{\partial \beta^{k}}{\partial \theta_{\mathcal{P}}^{k}}
	\end{numcases}
\end{subequations}

\begin{subequations}
	\label{backproSS}
	\begin{numcases}
	{}\frac{\partial\bm x^k}{\partial \bm z_i^k}= \beta_i^k \bm{F}_i(\bm{A}^\top\bm{A}+\sum_{i=1}^L\beta_i^k\bm{F}_i^\top\bm{F}_i)^{-1}\\
	\frac{\partial \bm x^k}{\partial \beta_i^k}=(\bm{F}^\top_i\bm z_i^k - \bm{F}^\top_i\bm{F}_i\bm x^k )^T(\bm{A}^\top\bm{A}+\sum_{i=1}^L\beta_i^k\bm{F}_i^\top\bm{F}_i)^{-1}\\
	\frac{\partial \bm{z}_i^k}{\partial \tilde{\bm{x}}^{k-1}}=\bm{F}^T_i.\\
	\frac{\partial \bm{\beta}^k}{\partial \theta_{\mathcal{P}}^k}=\frac{\mathcal{P}_{fcn}^k(\bm{d}^k,\theta_{\mathcal{P}}^k)}{\partial \theta_{\mathcal{P}}^k}.\\
	\frac{\partial \tilde{\bm{x}}^{k-1}}{\partial \theta_{\mathcal{D}}^k}=\frac{\mathcal{D}_{cnn}^k([\bm{x}^0,\bm{x}^1,\cdots \bm{x}^k],\theta_\mathcal{D}^k)}{\partial \theta_{\mathcal{D}}^k},
	\end{numcases}
\end{subequations}

	\section{Experiments}
\label{APP:com}
\subsection{Comparison Method}
\subsubsection{PWLS-TV}
The regularization parameter $\lambda$  is set to $0.01$ for $I_i=1\times 10^5,5\times10^4$; $0.02$ for $I_i=1\times 10^4$; and  $0.03$ for $I_i=5\times 10^3$. The parameter $\mu$ is set to 10. 
\subsubsection{MoDL}
was proposed in \cite{aggarwal2018modl} for MRI reconstruction. We applied it in CT. The details of unrolling stage of MoDL is as follows.
\begin{eqnarray}
\bm{x}^{k}&=&(\bm{A}^T\bm{A}+\lambda\bm{I})^{-1}(\bm{A}^T\bm{y}+\lambda \tilde{\bm{x}}^{k-1}), \nonumber\\
\tilde{\bm{x}}^{k}&=&\mathcal{D}_{\textrm{cnn}}(\bm{x}^k;\theta^k) \nonumber
\end{eqnarray}
where $\mathcal{D}_{cnn}$ is the same neural network as AHP-Net and $\lambda$ is set as learnable parameter.
The loss of MoDL is
\begin{equation}
\mathcal{L}(\Theta)=\frac{1}{J}\sum_{j=1}^{J}\left(\|\bm{x}_{j}^{{{K}}}-\bm{x}_{j} \|_2^2+\sum_{k=1}^{K-1}\mu_k\|\bm{x}^k_{j}-\bm{x}_{j} \|_2^2\right),
\end{equation}
where $\Theta=(\theta^1,\cdots \theta^{{K}},\lambda)$.
For consistency, we set ${{K}}=3$. 
Adam optimizer was used with the momentum parameter $\beta=0.9$, mini-batch size to be $4$, and the learning rate to be $10^{-4}$. 
The model was trained for $50$ epochs. 
\subsubsection{Neumann-Net}
Neumann Network was proposed in \cite{gilton2019neumann} for linear inverse problem in imaging. With step size $\eta>0$, 
the inverse problem can be solved by $$ \bm{x}^k=\eta\sum_{k=0}^{{K}}(\bm{I}-\eta\bm{A}^T\bm{A}-\eta\bm{R})^k\bm{A}^\top\bm{y}.$$
The details of unrolling stage of Neumann-Net is as follows.
$$ \bm{x}^k=  \sum_{i=0}^{k-1}\bm{x}^{i-1}+\eta(\bm{I}-\eta\bm{A}^\top\bm{A})\bm{x}^{k-1} -\eta\mathcal{D}_{cnn}( \bm{x}^{k-1},\theta^k).$$

The loss of Neumann-Net is
\begin{equation}
\mathcal{L}(\Theta)=\frac{1}{J}\sum_{j=1}^{J}\left(\|\bm{x}_{j}^{{{K}}}-\bm{x}_{j} \|_2^2+\sum_{k=1}^{K-1}\mu_k\|\bm{x}^k_{j}-\bm{x}_{j} \|_2^2\right),
\end{equation}
where $\Theta=(\theta^1,\cdots \theta^{{K}},\eta)$.
For consistency, we set ${{K}}=3$. 
Adam optimizer was used with the momentum parameter $\beta=0.9$, mini-batch size to be $4$, and the learning rate to be $10^{-4}$. 
The model was trained for $50$ epochs. 		
\section{Result}
\subsection{Visual comparison}
\label{APP:VC}
\begin{figure*}
	\begin{center}
		\begin{tabular}{c@{\hspace{0pt}}c@{\hspace{0pt}}c@{\hspace{0pt}}c@{\hspace{0pt}}c@{\hspace{0pt}}c@{\hspace{0pt}}c}
			\includegraphics[width=.198\linewidth,height=.132\linewidth]{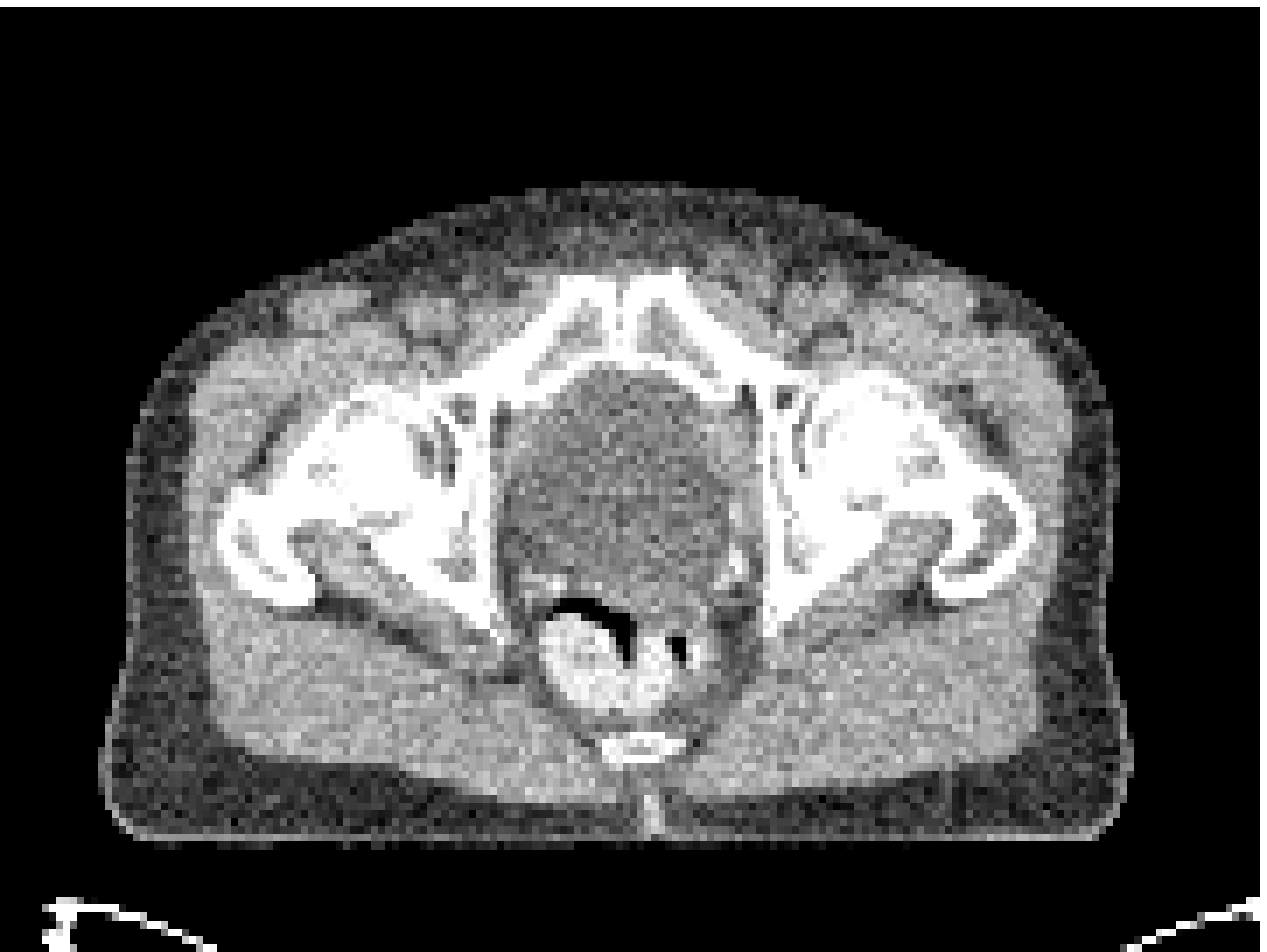}&
			\includegraphics[width=.198\linewidth,height=.132\linewidth]{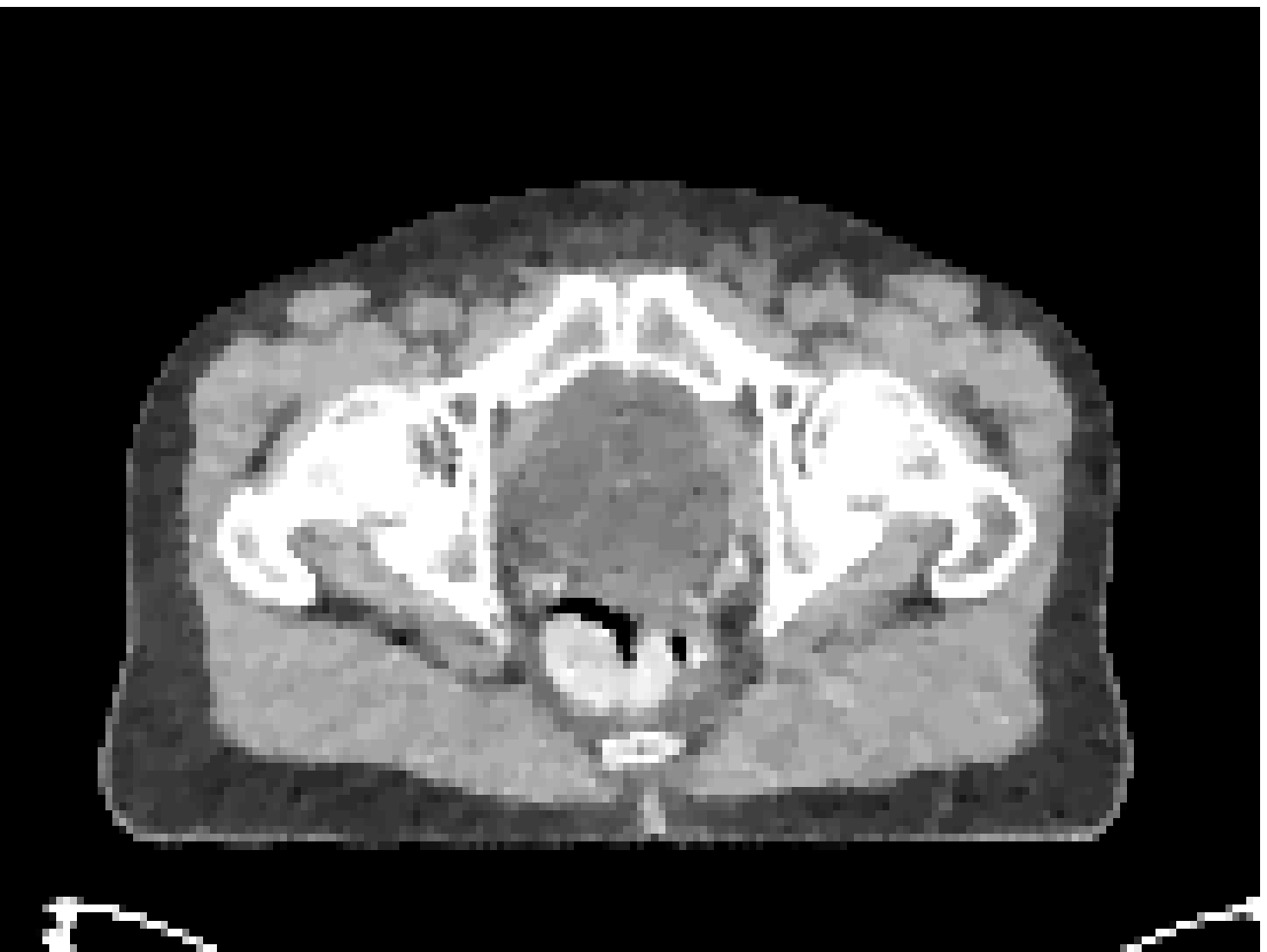}&			
			\includegraphics[width=.198\linewidth,height=.132\linewidth]{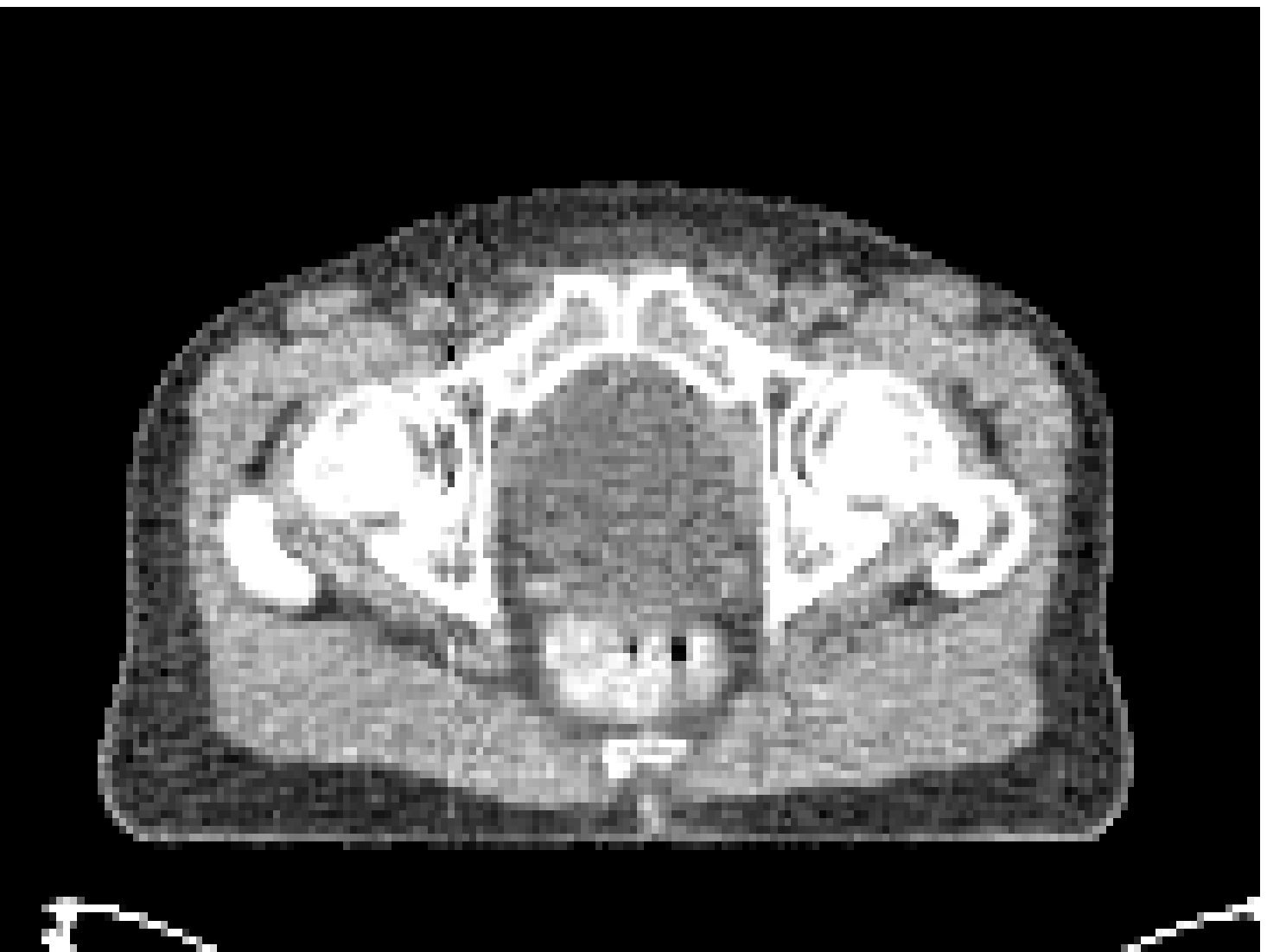}&
			\includegraphics[width=.198\linewidth,height=.132\linewidth]{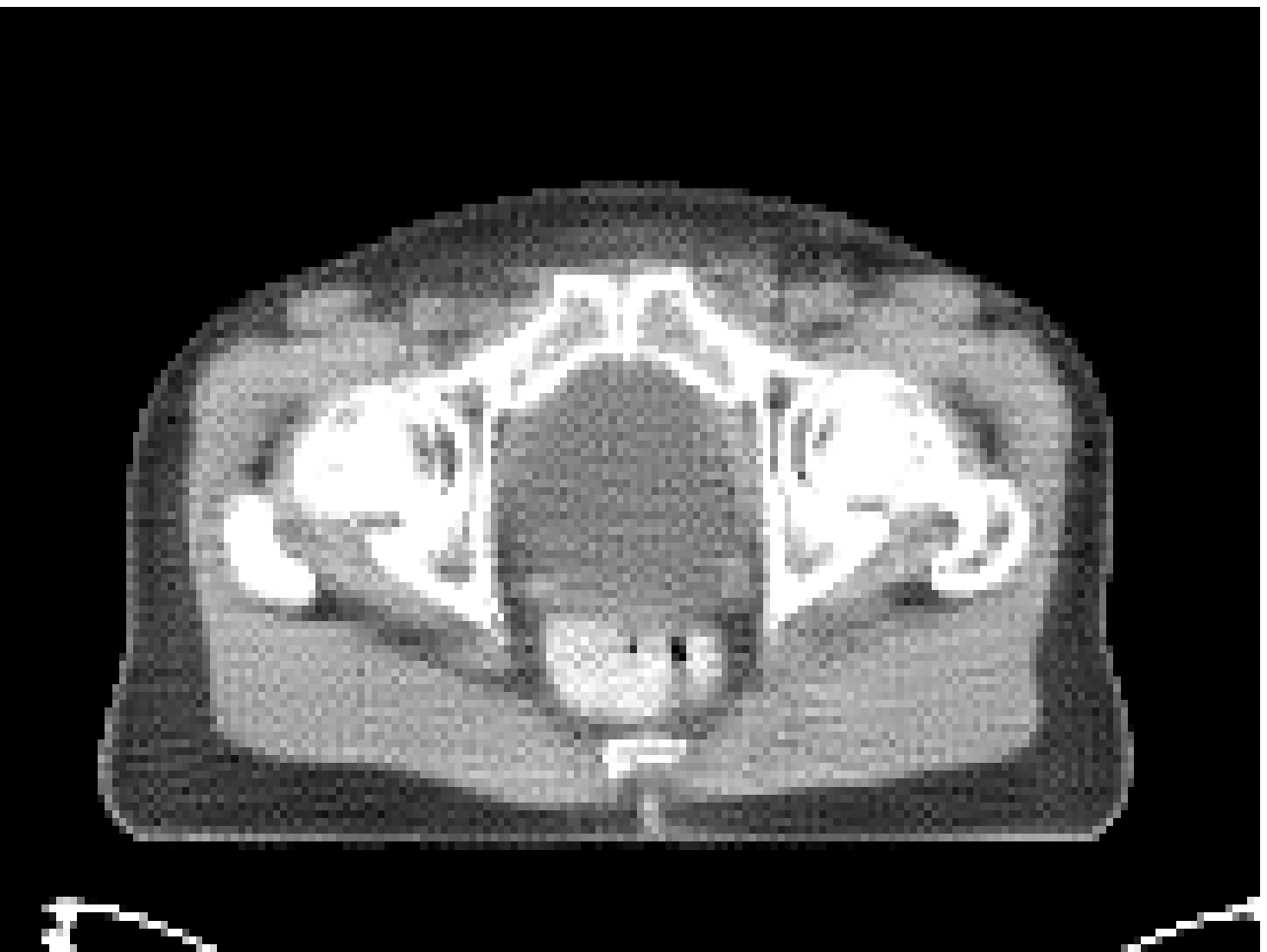}&
			\includegraphics[width=.198\linewidth,height=.132\linewidth]{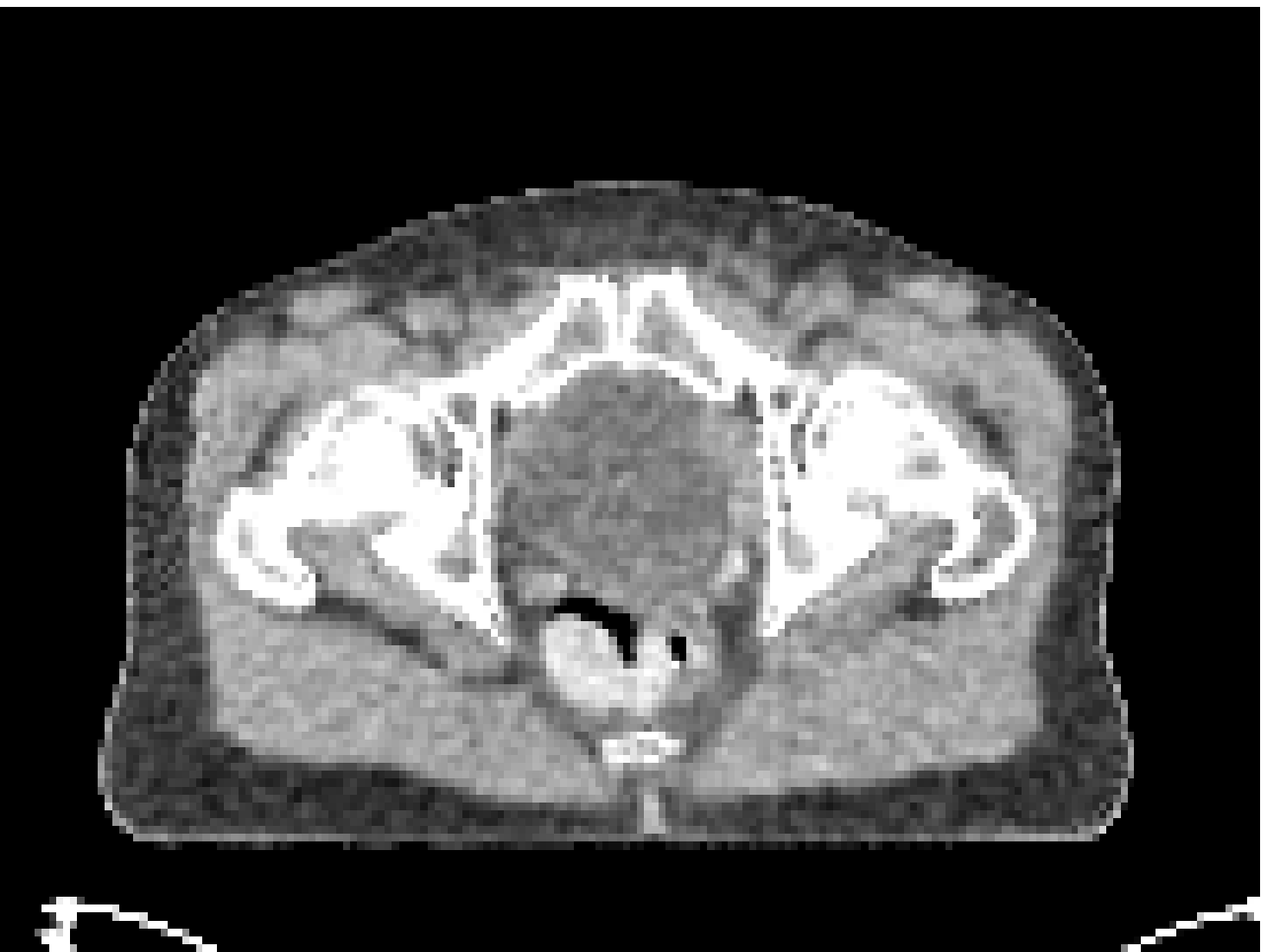}\\
			FBP&
			TV&
			KSVD&
			BM3D&									
			FBPConvNet\\			
			\includegraphics[width=.198\linewidth,height=.132\linewidth]{MoDLHU_50000_50000_2_38.eps}&
			\includegraphics[width=.198\linewidth,height=.132\linewidth]{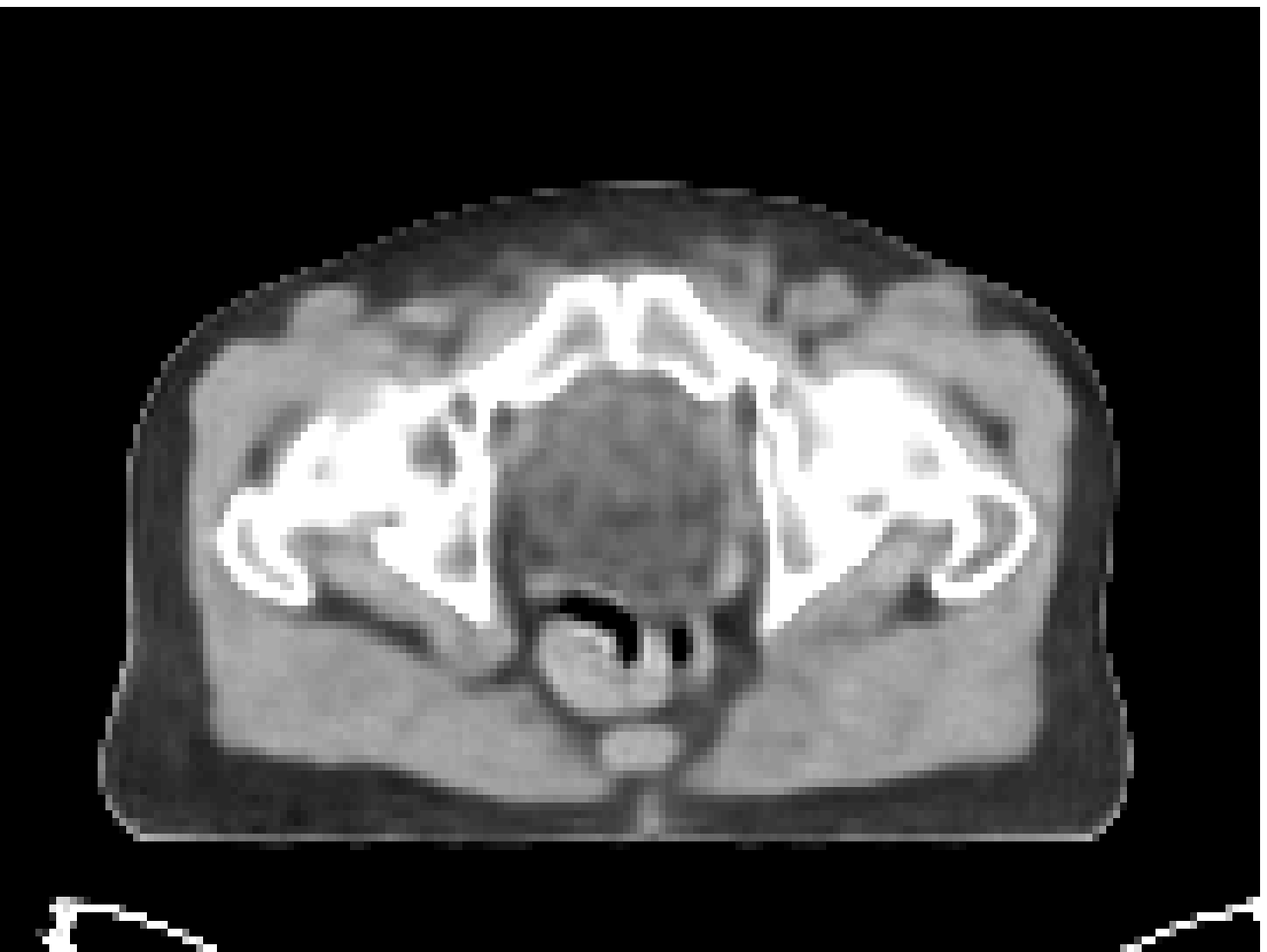}&	
			\includegraphics[width=.198\linewidth,height=.132\linewidth]{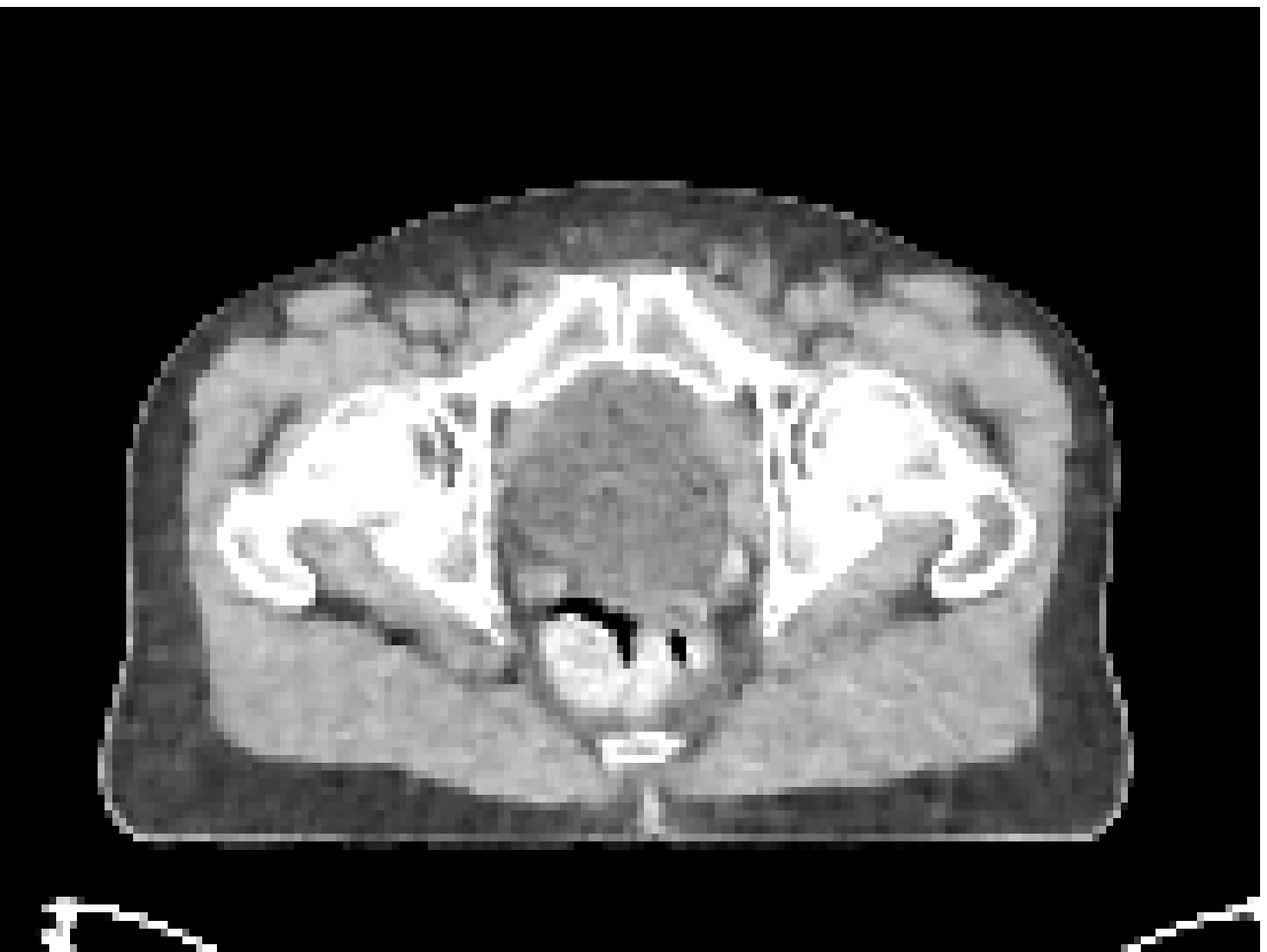}&
			\includegraphics[width=.198\linewidth,height=.132\linewidth]{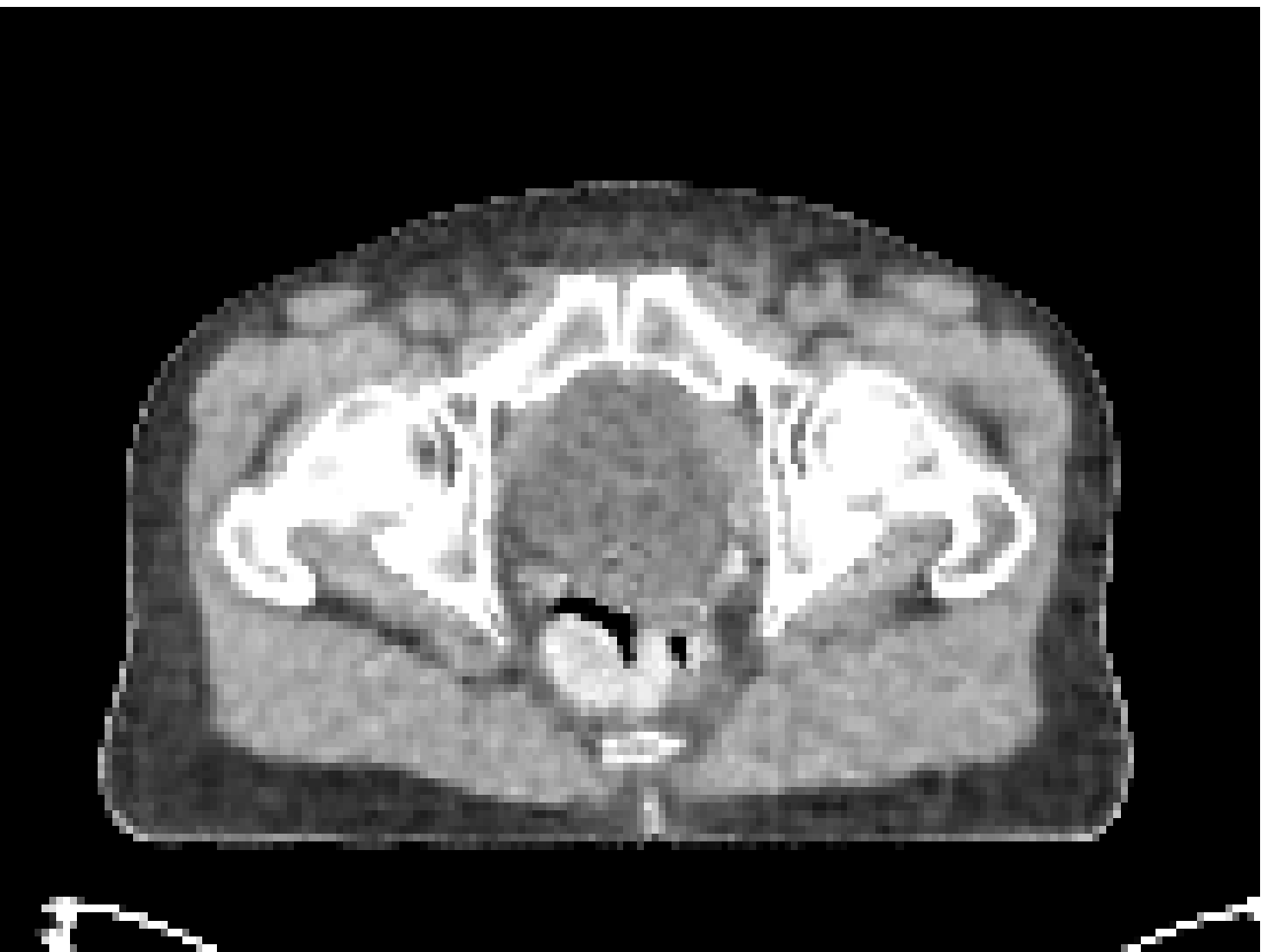}&								
			\includegraphics[width=.198\linewidth,height=.132\linewidth]{AHPHU_50000_50000_2_38.eps}\\
			MoDL&
			Neumann-Net&
			PGD&
			Learn-PD&
			AHP-Net
		\end{tabular}
		\caption{Reconstruction results at dose level $I_i=5\times10^4$ by the models trained under same dose level.}
		\label{sliceU_50000}
	\end{center}
\end{figure*}

\begin{figure*}
	\begin{center}
		\begin{tabular}{c@{\hspace{0pt}}c@{\hspace{0pt}}c@{\hspace{0pt}}c@{\hspace{0pt}}c@{\hspace{0pt}}c@{\hspace{0pt}}c}
			\includegraphics[width=.198\linewidth,height=.132\linewidth]{FBPHU_10000_2_38.eps}&
			\includegraphics[width=.198\linewidth,height=.132\linewidth]{TVHU_10000_2_38.eps}&
			\includegraphics[width=.198\linewidth,height=.132\linewidth]{KSVD_ProstateMix_I_10000_sigma_100.eps}&
			\includegraphics[width=.198\linewidth,height=.132\linewidth]{BM3D_ProstateMix_I_10000_sigma_100.eps}&
			\includegraphics[width=.198\linewidth,height=.132\linewidth]{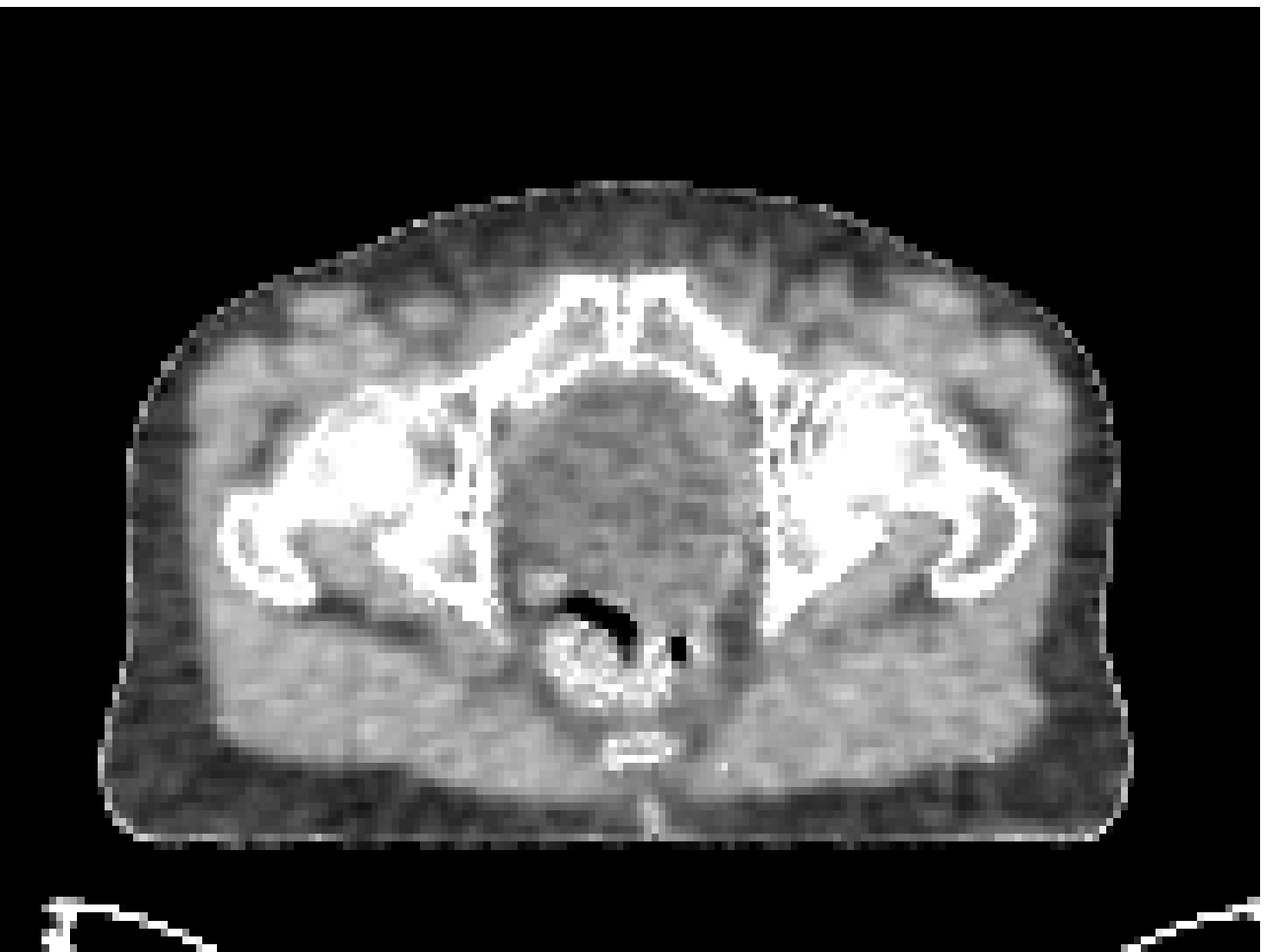}\\	
			FBP&
			TV&
			KSVD&
			BM3D&							
			FBPConvNet\\
			\includegraphics[width=.198\linewidth,height=.132\linewidth]{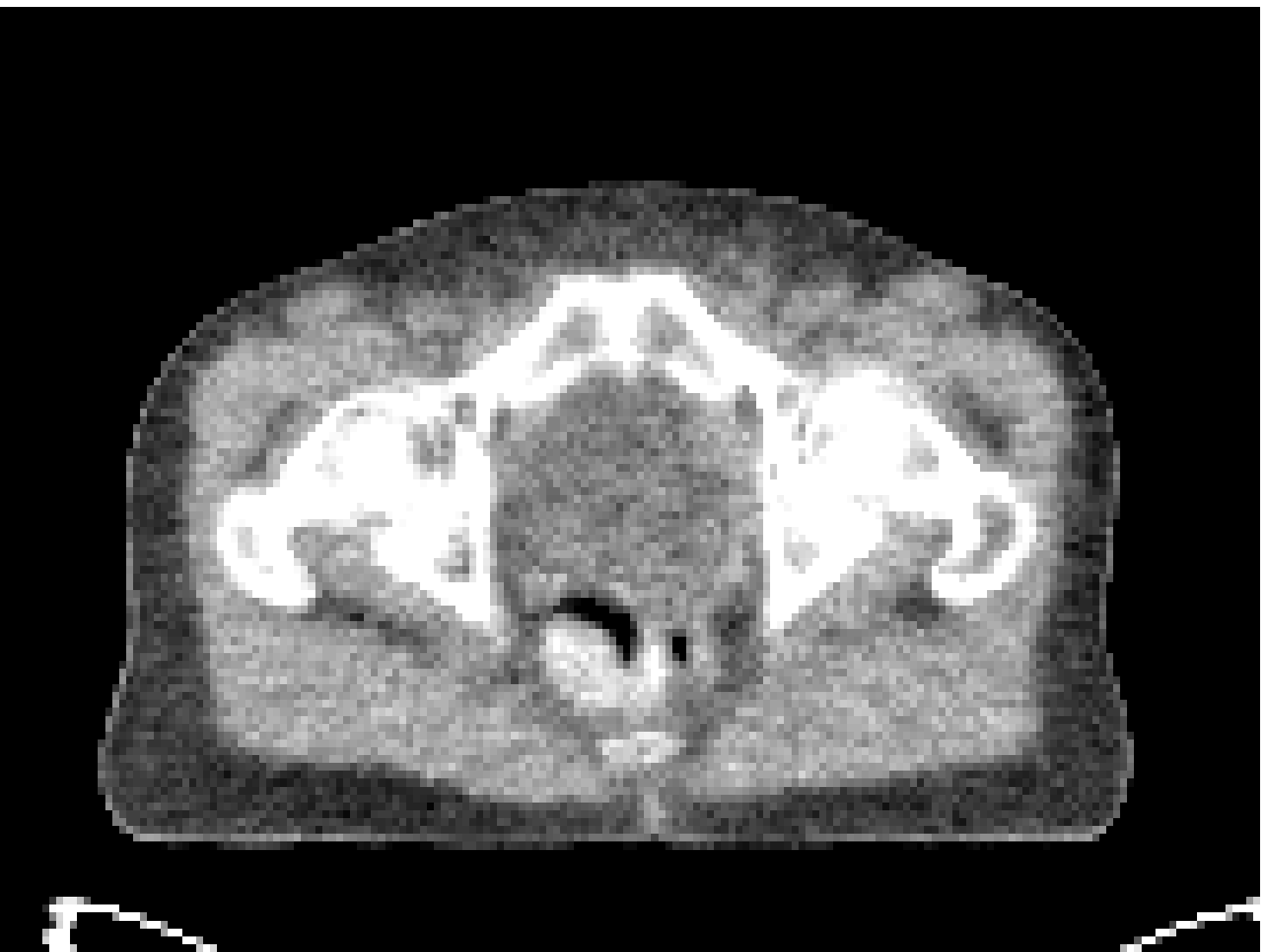}&
			\includegraphics[width=.198\linewidth,height=.132\linewidth]{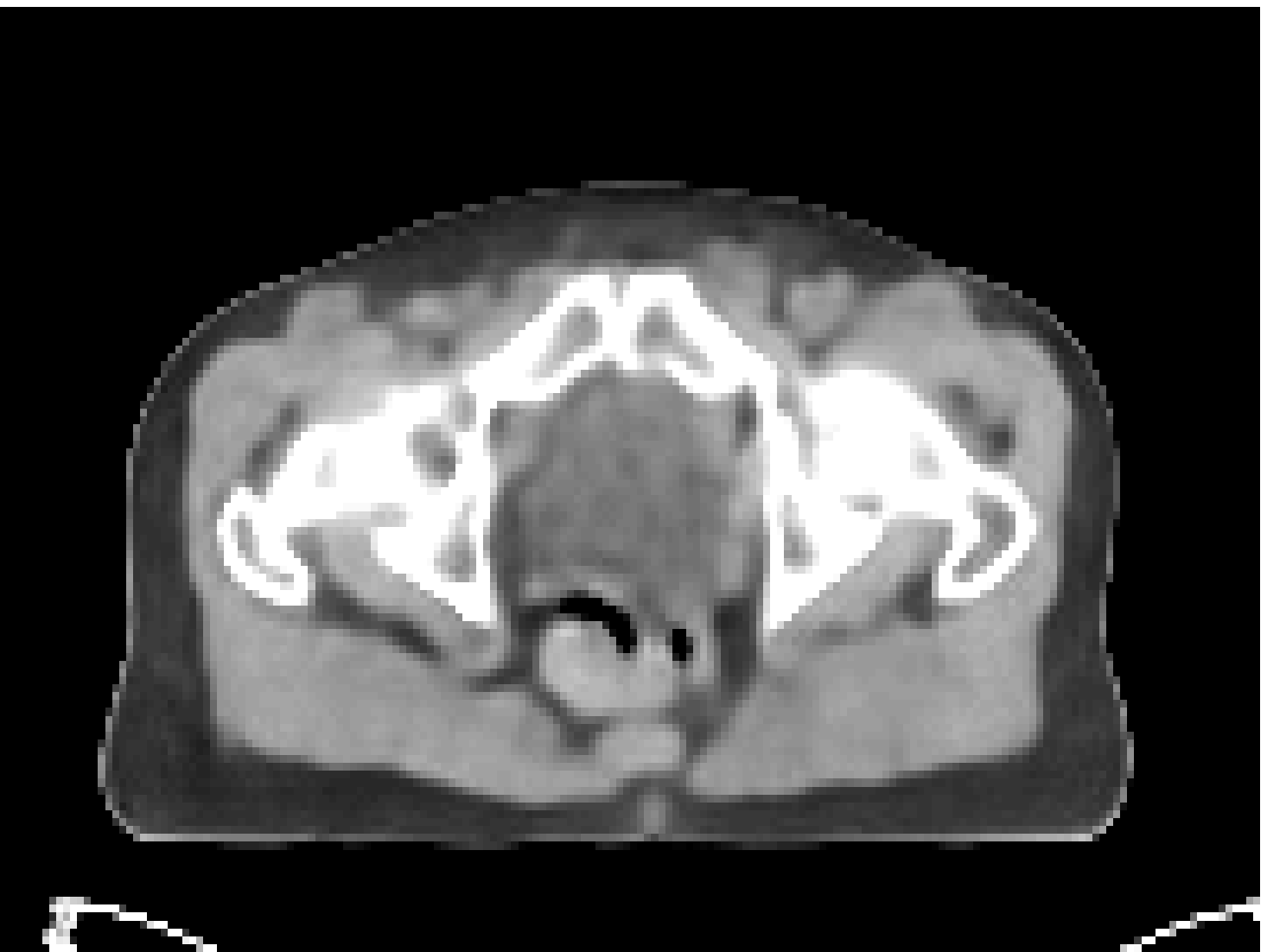}&	
			\includegraphics[width=.198\linewidth,height=.132\linewidth]{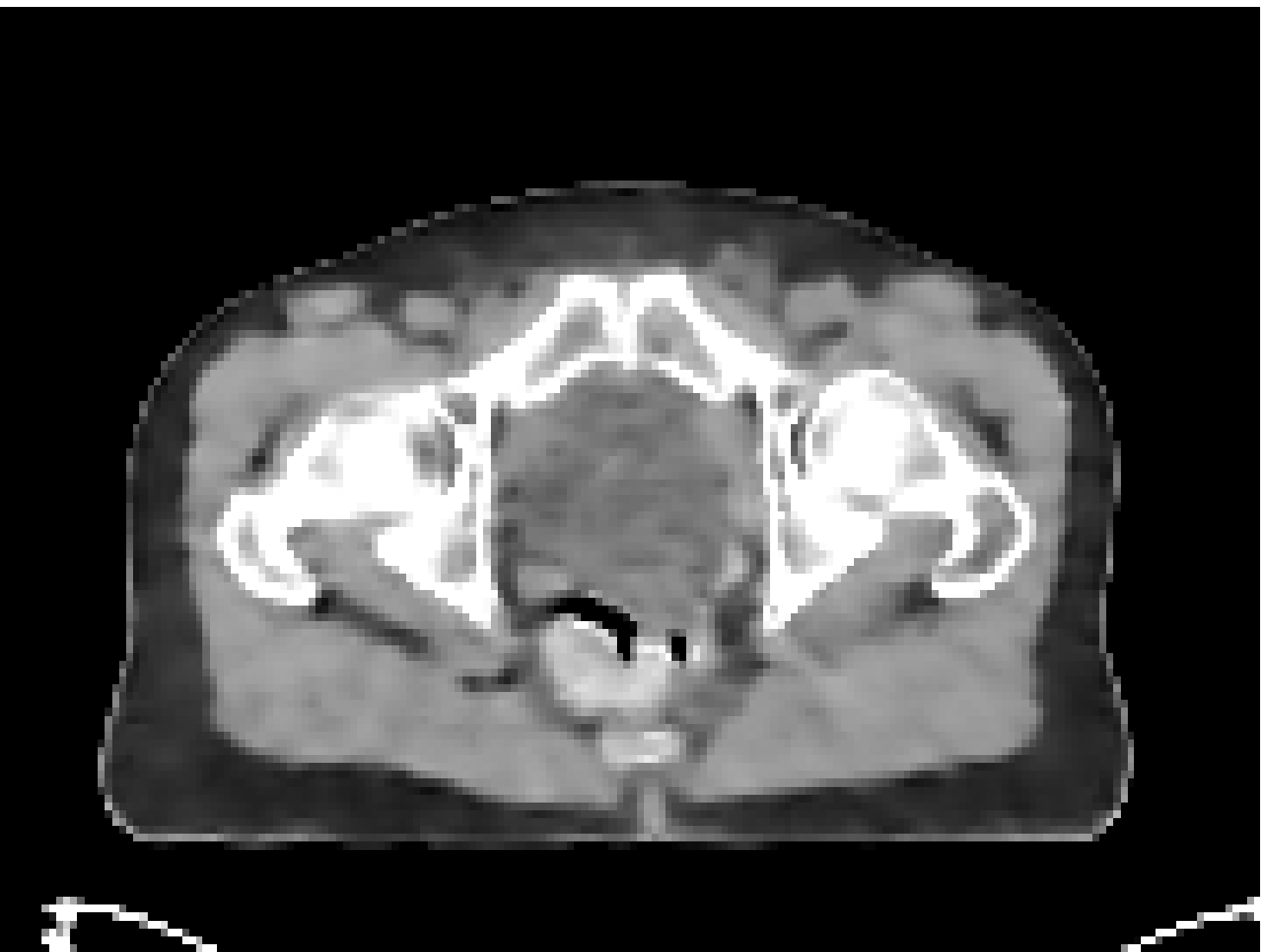}&
			\includegraphics[width=.198\linewidth,height=.132\linewidth]{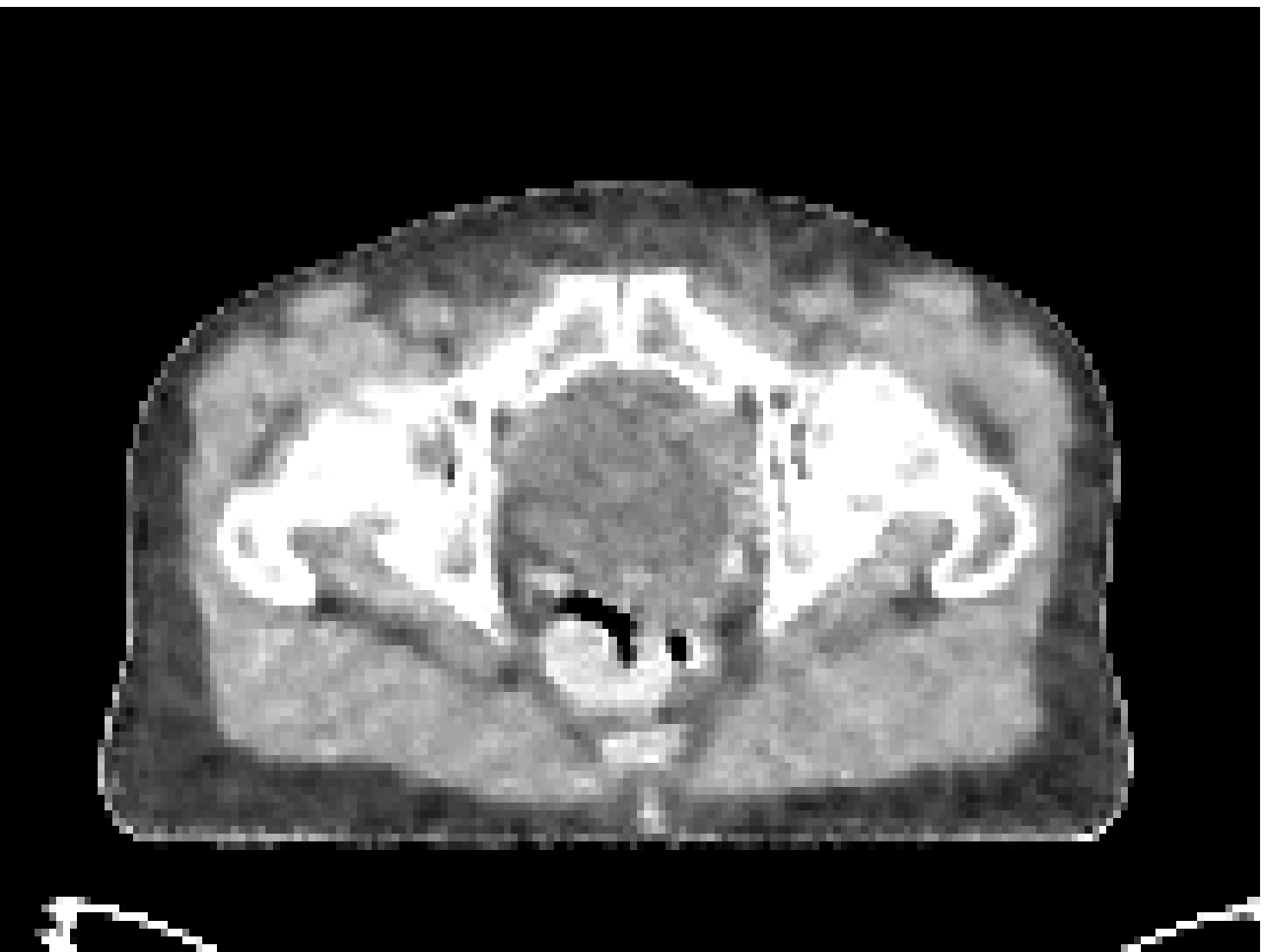}&								
			\includegraphics[width=.198\linewidth,height=.132\linewidth]{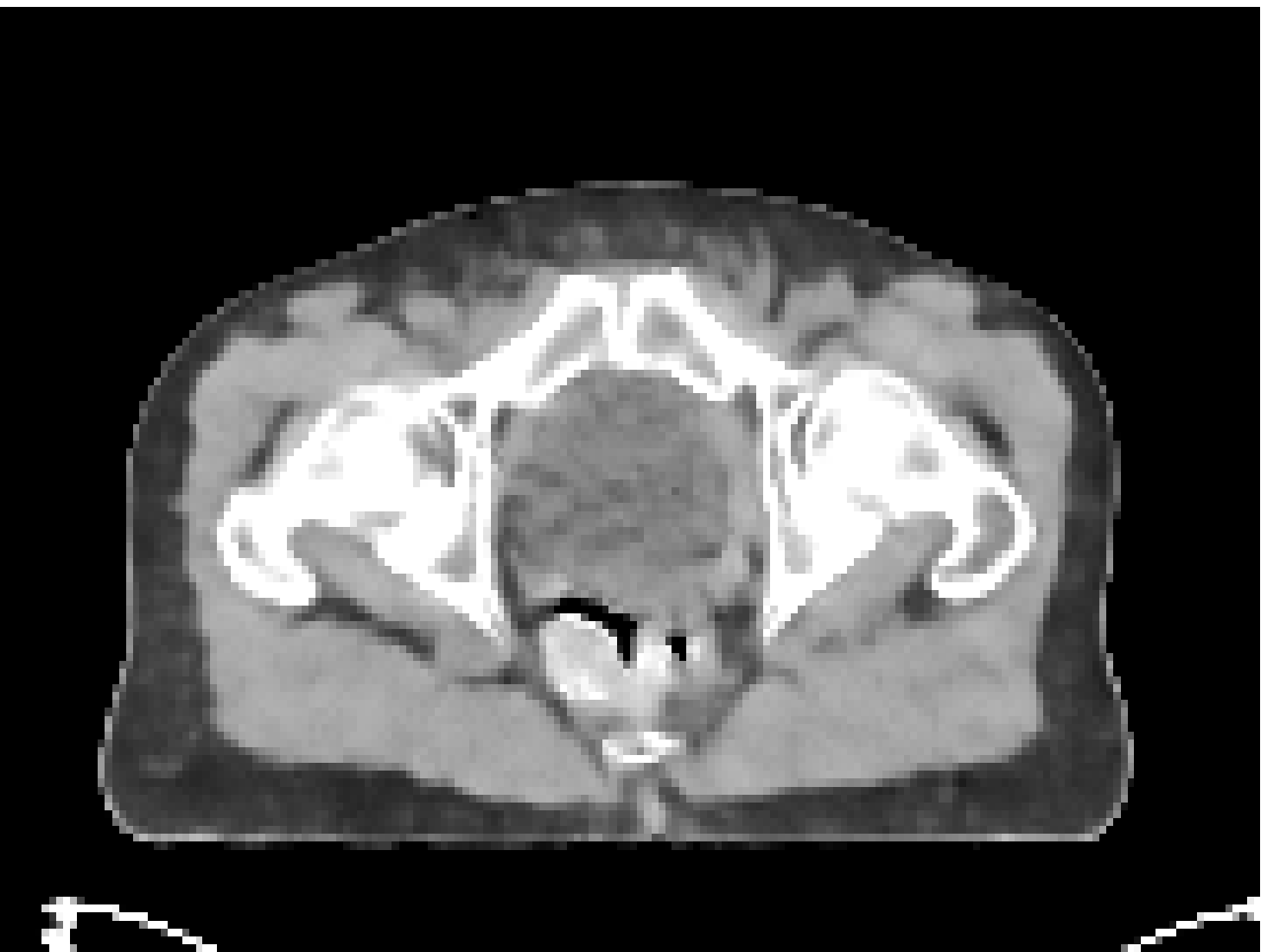}\\
			MoDL&
			Neumann-Net&
			PGD&
			Learn-PD&
			AHP-Net
		\end{tabular}
		\caption{Reconstruction results at dose level $I_i=10^4$ by the models trained under same dose level.}
		\label{sliceU_10000}
	\end{center}
\end{figure*}

\begin{figure}
	\begin{center}
		\begin{tabular}{c@{\hspace{-1pt}}c@{\hspace{-1pt}}c@{\hspace{-1pt}}c@{\hspace{-1pt}}c@{\hspace{-1pt}}c@{\hspace{-1pt}}c}
			\begin{tikzpicture}
			\node[anchor=south west,inner sep=0] (image) at (0,0) {\includegraphics[width=.2\linewidth,height=.2\linewidth]{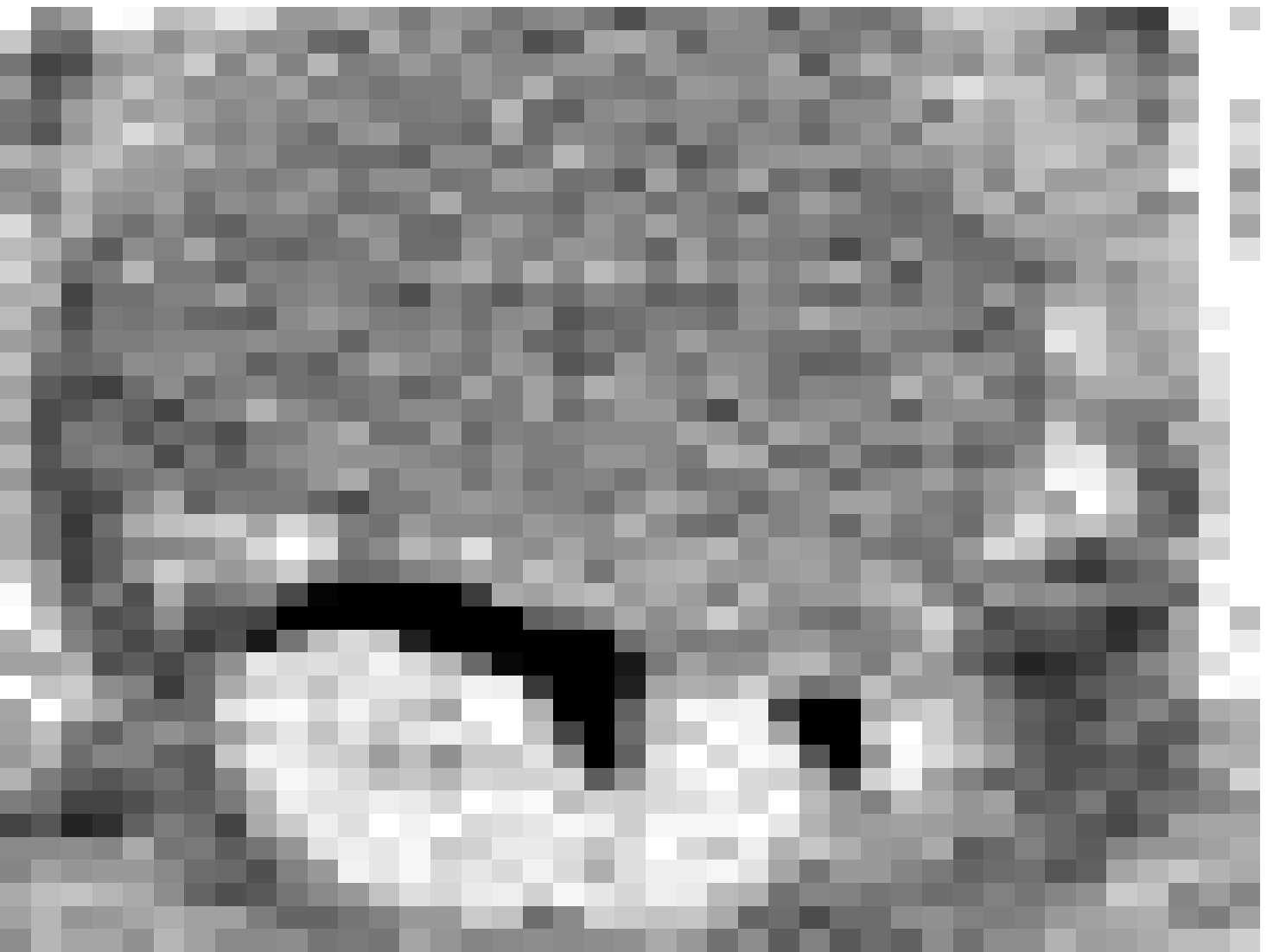}};
			\draw [-stealth, line width=2pt, cyan] (0.7,1.3) -- ++(-0.3,-0.3);
			\draw [-stealth, line width=2pt, cyan] (1.7,0.45) -- ++(-0.45,-0.0);
			\end{tikzpicture}&
			\begin{tikzpicture}
			\node[anchor=south west,inner sep=0] (image) at (0,0) {\includegraphics[width=.2\linewidth,height=.2\linewidth]{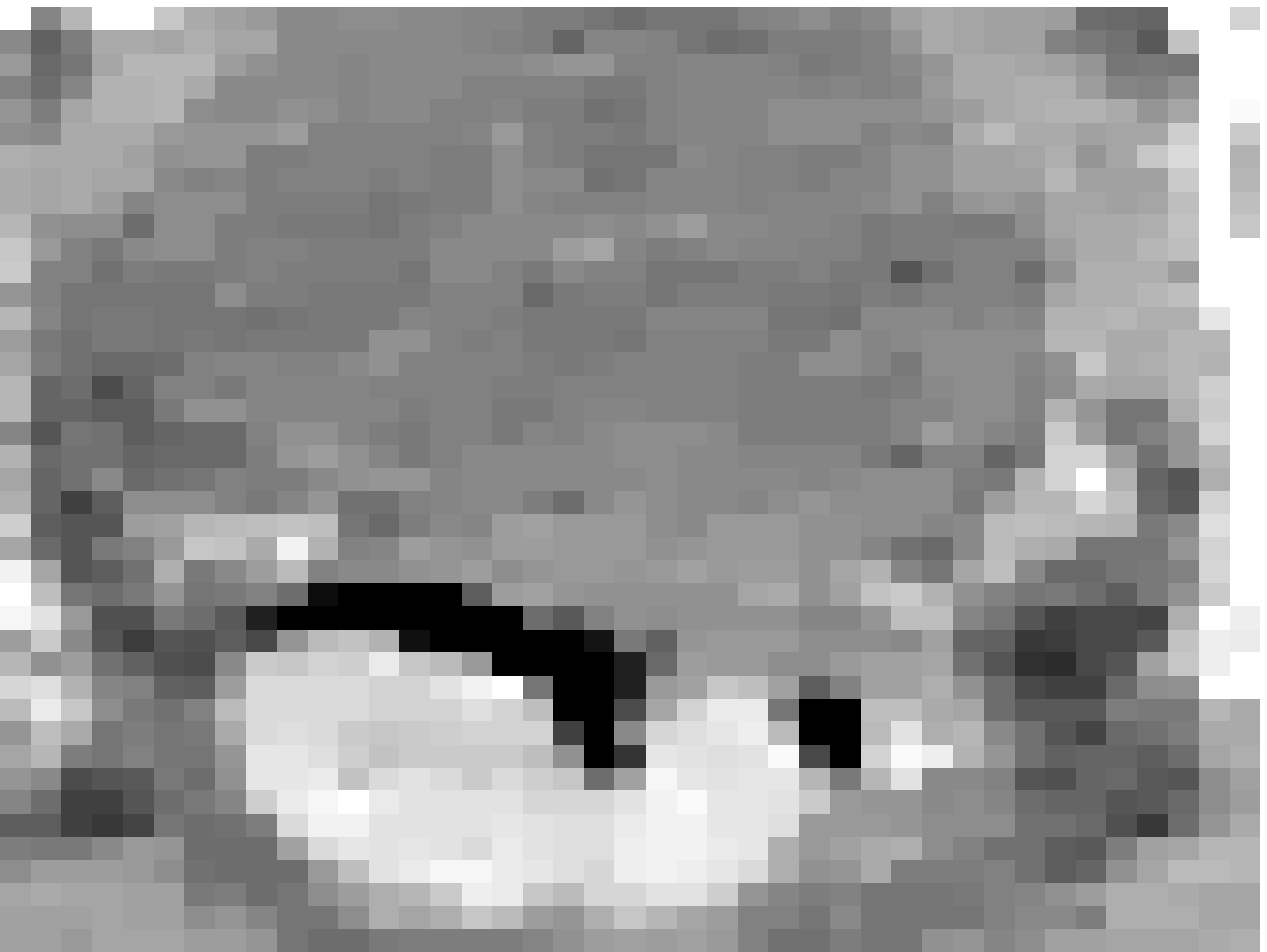}};
			\draw [-stealth, line width=2pt, cyan] (0.7,1.3) -- ++(-0.3,-0.3);
			\draw [-stealth, line width=2pt, cyan] (1.7,0.45) -- ++(-0.45,-0.0);
			\end{tikzpicture}&
			\begin{tikzpicture}
			\node[anchor=south west,inner sep=0] (image) at (0,0) {\includegraphics[width=.2\linewidth,height=.2\linewidth]{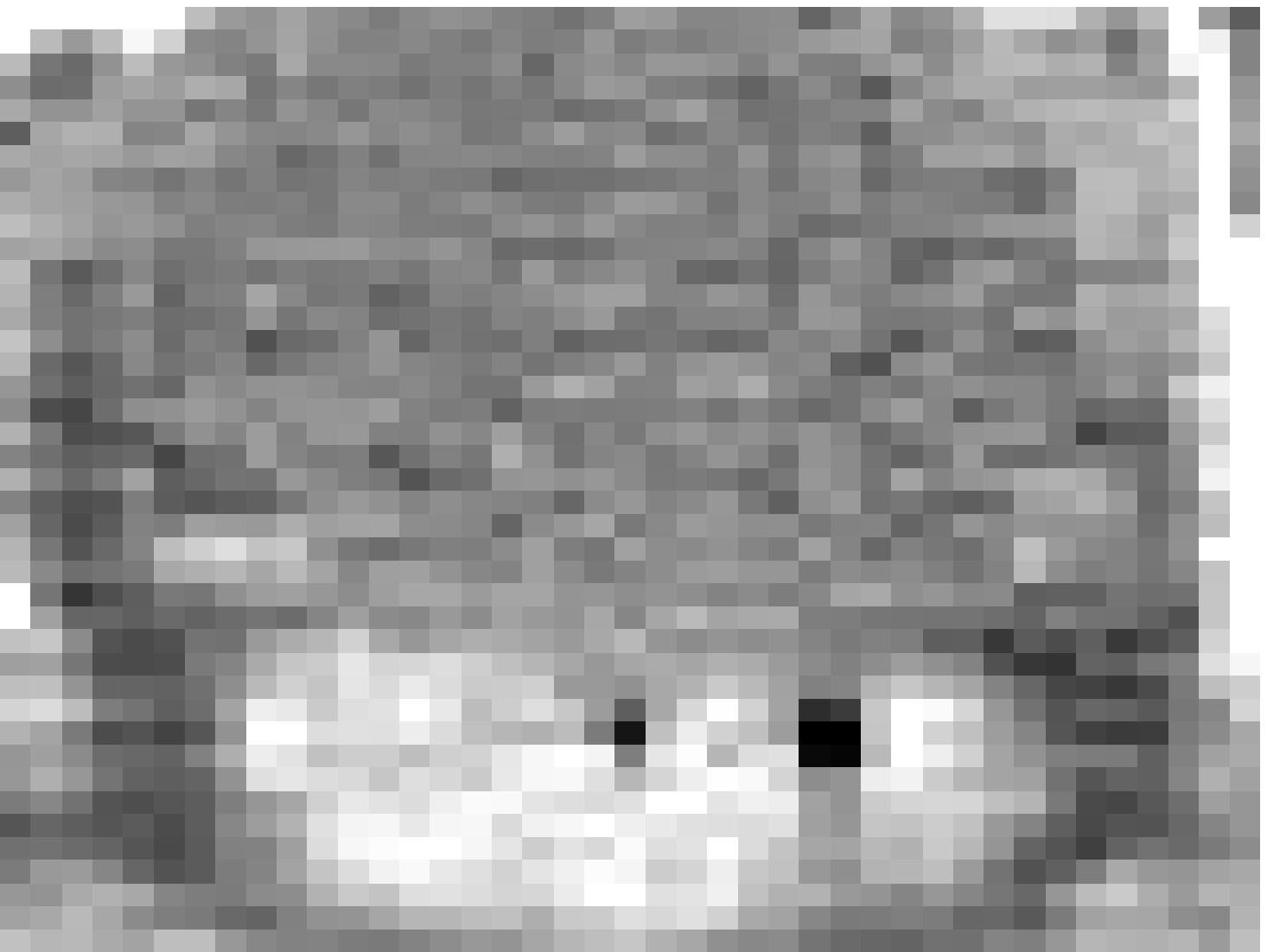}};
			\draw [-stealth, line width=2pt, cyan] (0.7,1.3) -- ++(-0.3,-0.3);
			\draw [-stealth, line width=2pt, cyan] (1.7,0.45) -- ++(-0.45,-0.0);
			\end{tikzpicture}&
			\begin{tikzpicture}
			\node[anchor=south west,inner sep=0] (image) at (0,0) {\includegraphics[width=.2\linewidth,height=.2\linewidth]{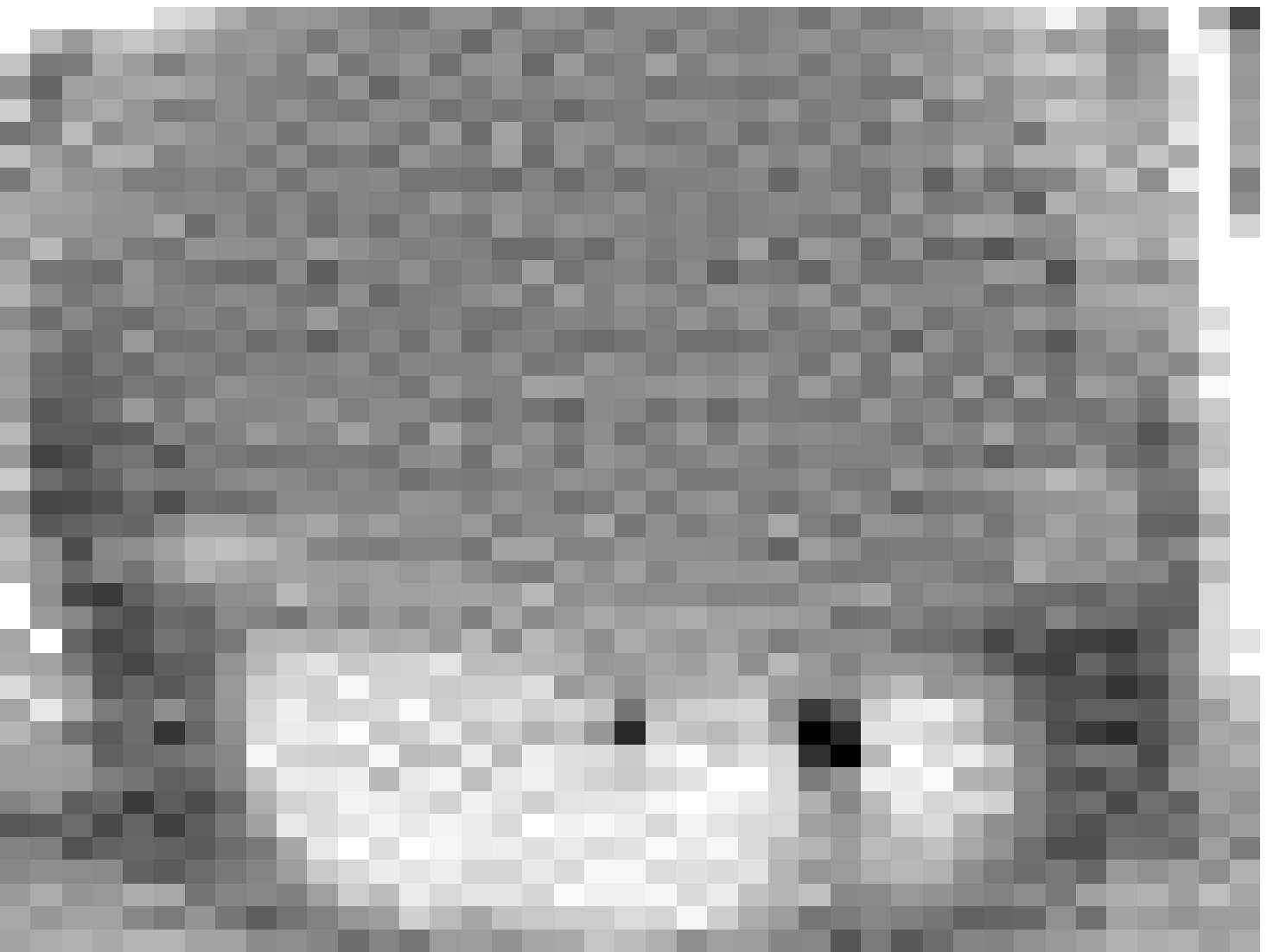}};
			\draw [-stealth, line width=2pt, cyan] (0.7,1.3) -- ++(-0.3,-0.3);
			\draw [-stealth, line width=2pt, cyan] (1.7,0.45) -- ++(-0.45,-0.0);
			\end{tikzpicture}&
			\begin{tikzpicture}
			\node[anchor=south west,inner sep=0] (image) at (0,0) {\includegraphics[width=.2\linewidth,height=.2\linewidth]{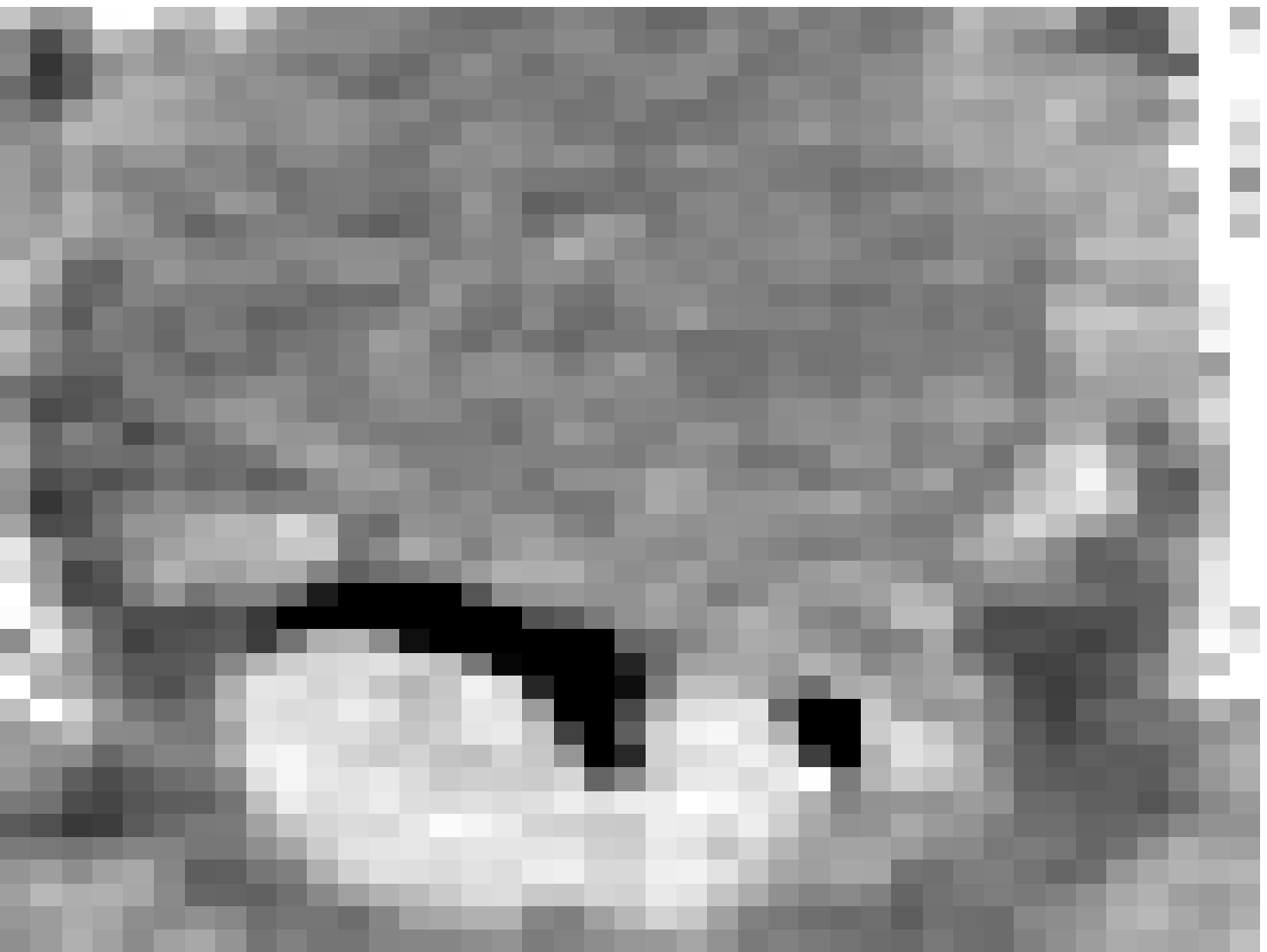}};
			\draw [-stealth, line width=2pt, cyan] (0.7,1.3) -- ++(-0.3,-0.3);
			\draw [-stealth, line width=2pt, cyan] (1.7,0.45) -- ++(-0.45,-0.0);
			\end{tikzpicture}\\
			FBP&		
			TV&
			KSVD&
			BM3D&
			FBPConvNet\\
			\begin{tikzpicture}
			\node[anchor=south west,inner sep=0] (image) at (0,0) {\includegraphics[width=.2\linewidth,height=.2\linewidth]{MoDLHUZoom_50000_50000_2_38.eps}};
			\draw [-stealth, line width=2pt, cyan] (0.7,1.3) -- ++(-0.3,-0.3);
			\draw [-stealth, line width=2pt, cyan] (1.7,0.45) -- ++(-0.45,-0.0);
			\end{tikzpicture}&		
			\begin{tikzpicture}
			\node[anchor=south west,inner sep=0] (image) at (0,0) {\includegraphics[width=.2\linewidth,height=.2\linewidth]{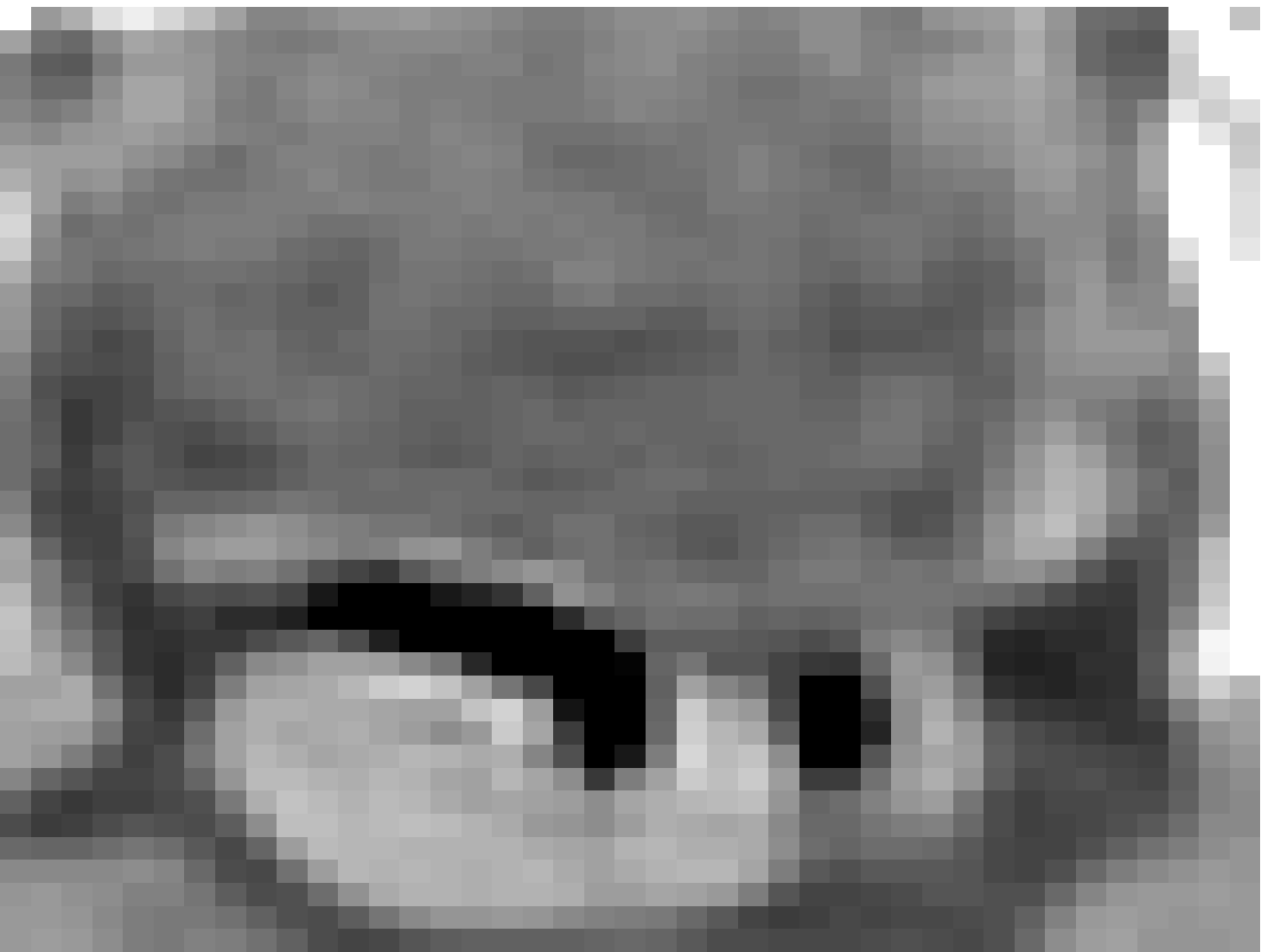}};
			\draw [-stealth, line width=2pt, cyan] (0.7,1.3) -- ++(-0.3,-0.3);
			\draw [-stealth, line width=2pt, cyan] (1.7,0.45) -- ++(-0.45,-0.0);
			\end{tikzpicture}&
			\begin{tikzpicture}
			\node[anchor=south west,inner sep=0] (image) at (0,0) {\includegraphics[width=.2\linewidth,height=.2\linewidth]{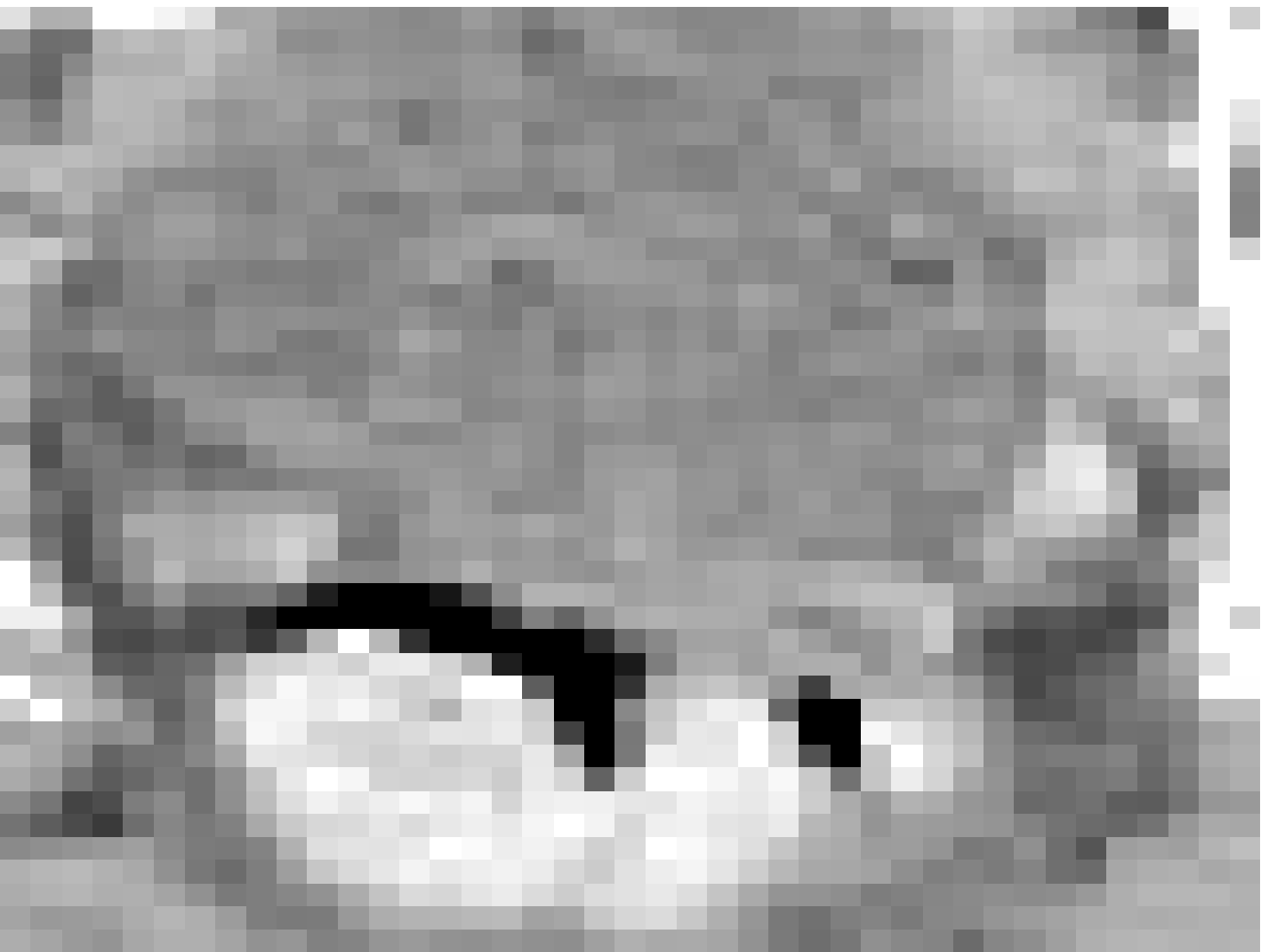}};
			\draw [-stealth, line width=2pt, cyan] (0.7,1.3) -- ++(-0.3,-0.3);
			\draw [-stealth, line width=2pt, cyan] (1.7,0.45) -- ++(-0.45,-0.0);
			\end{tikzpicture}&		
			\begin{tikzpicture}
			\node[anchor=south west,inner sep=0] (image) at (0,0) {\includegraphics[width=.2\linewidth,height=.2\linewidth]{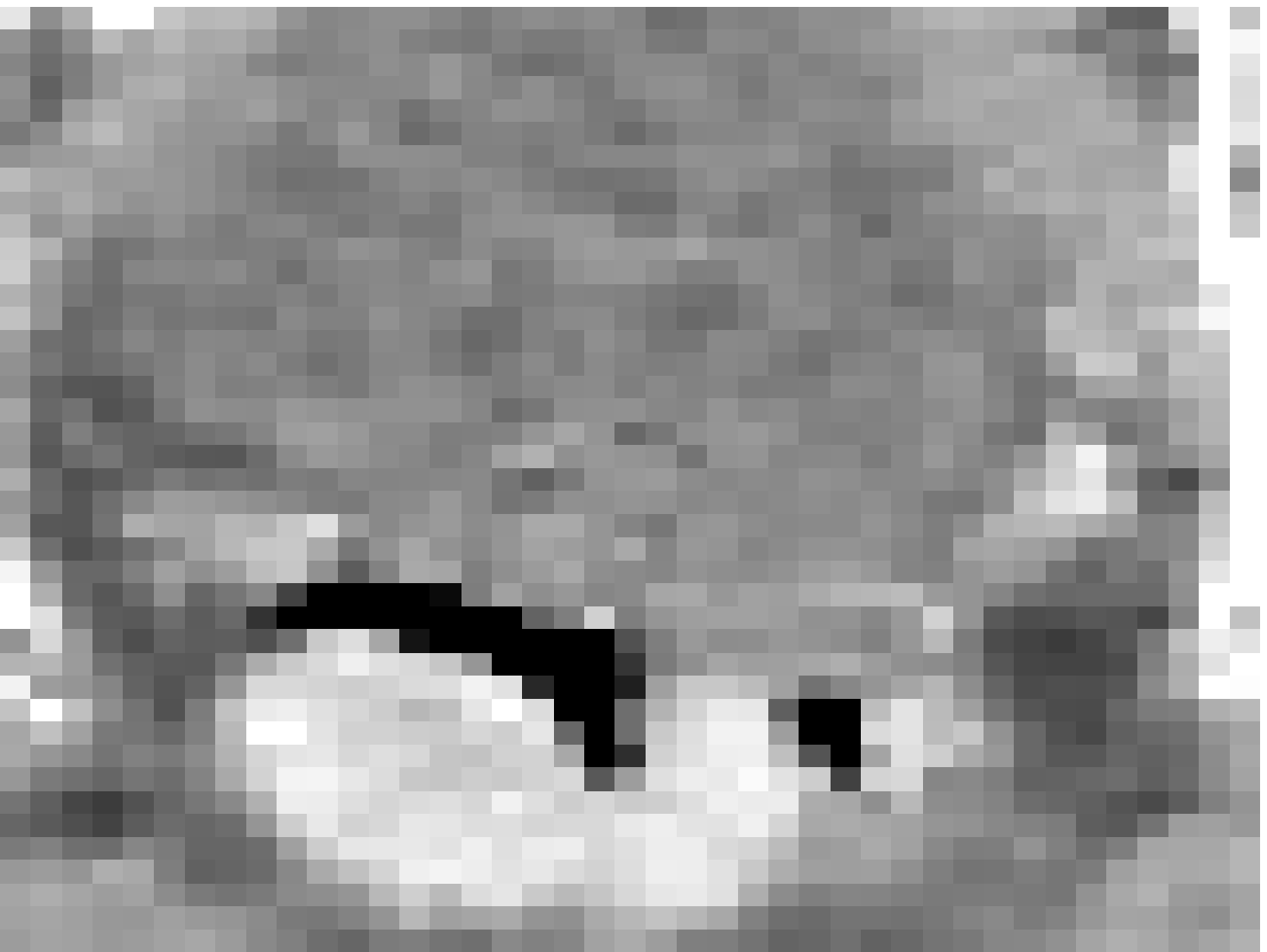}};
			\draw [-stealth, line width=2pt, cyan] (0.7,1.3) -- ++(-0.3,-0.3);
			\draw [-stealth, line width=2pt, cyan] (1.7,0.45) -- ++(-0.45,-0.0);
			\end{tikzpicture}&
			\begin{tikzpicture}
			\node[anchor=south west,inner sep=0] (image) at (0,0) {\includegraphics[width=.2\linewidth,height=.2\linewidth]{AHPHUZoom_50000_50000_2_38.eps}};
			\draw [-stealth, line width=2pt, cyan] (0.7,1.3) -- ++(-0.3,-0.3);
			\draw [-stealth, line width=2pt, cyan] (1.7,0.45) -- ++(-0.45,-0.0);
			\end{tikzpicture}\\
			MoDL&
			Neumann-Net&	
			PGD&
			Learned-PD&	
			AHP-Net
		\end{tabular}
		\caption{Zoom-in results of Fig.~\ref{sliceU_50000}.
		}
		\label{sliceZoomU_50000}
	\end{center}
\end{figure}

\begin{figure}
		\begin{tabular}{c@{\hspace{-1pt}}c@{\hspace{-1pt}}c@{\hspace{-1pt}}c@{\hspace{-1pt}}c@{\hspace{-1pt}}c@{\hspace{-1pt}}c}
			\begin{tikzpicture}
			\node[anchor=south west,inner sep=0] (image) at (0,0) {\includegraphics[width=.2\linewidth,height=.2\linewidth]{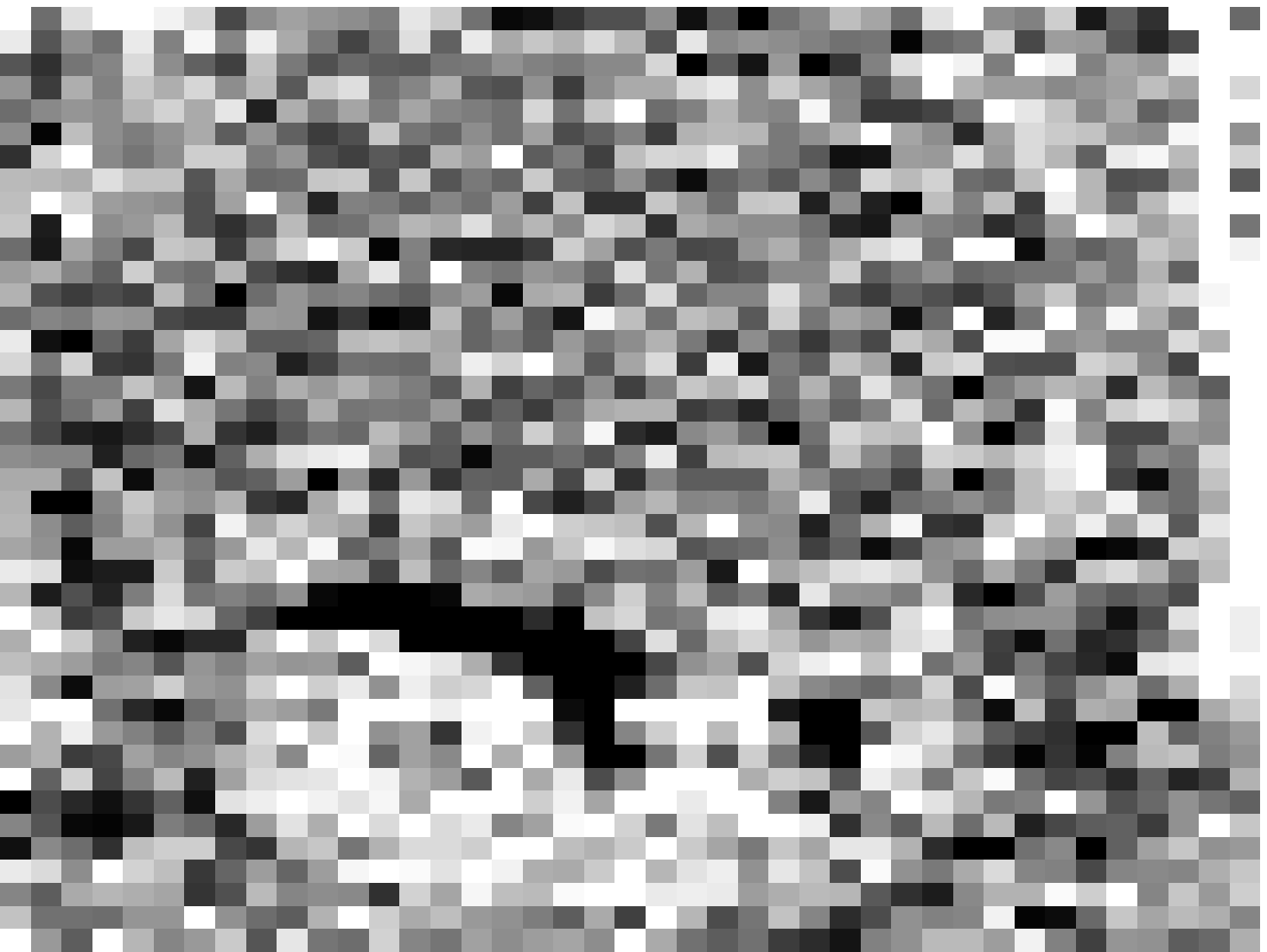}};
			\draw [-stealth, line width=2pt, cyan] (0.7,1.3) -- ++(-0.3,-0.3);
			\draw [-stealth, line width=2pt, cyan] (1.7,0.45) -- ++(-0.45,-0.0);
			\end{tikzpicture}&
			\begin{tikzpicture}
			\node[anchor=south west,inner sep=0] (image) at (0,0) {\includegraphics[width=.2\linewidth,height=.2\linewidth]{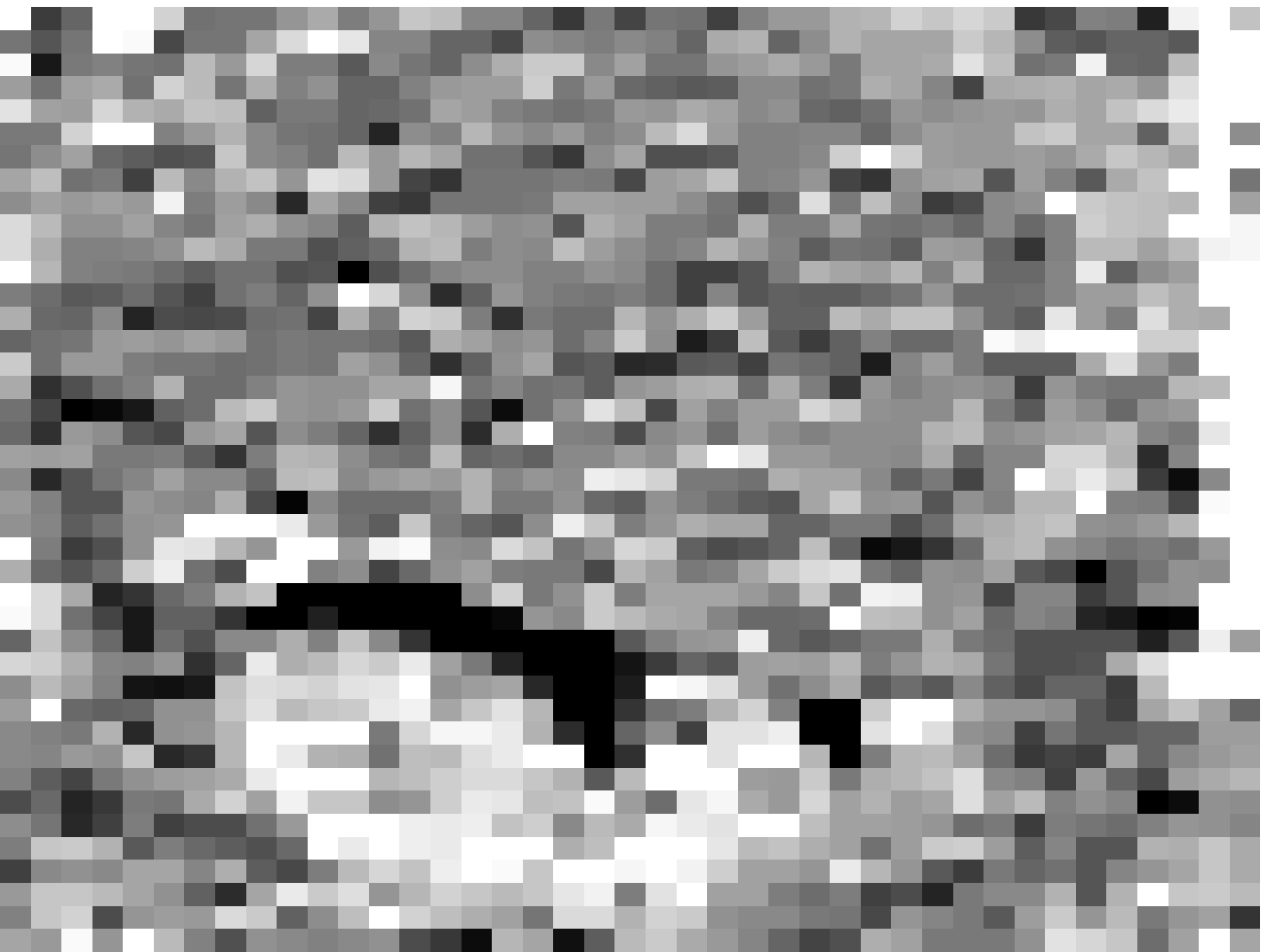}};
			\draw [-stealth, line width=2pt, cyan] (0.7,1.3) -- ++(-0.3,-0.3);
			\draw [-stealth, line width=2pt, cyan] (1.7,0.45) -- ++(-0.45,-0.0);
			\end{tikzpicture}&
			\begin{tikzpicture}
			\node[anchor=south west,inner sep=0] (image) at (0,0) {\includegraphics[width=.2\linewidth,height=.2\linewidth]{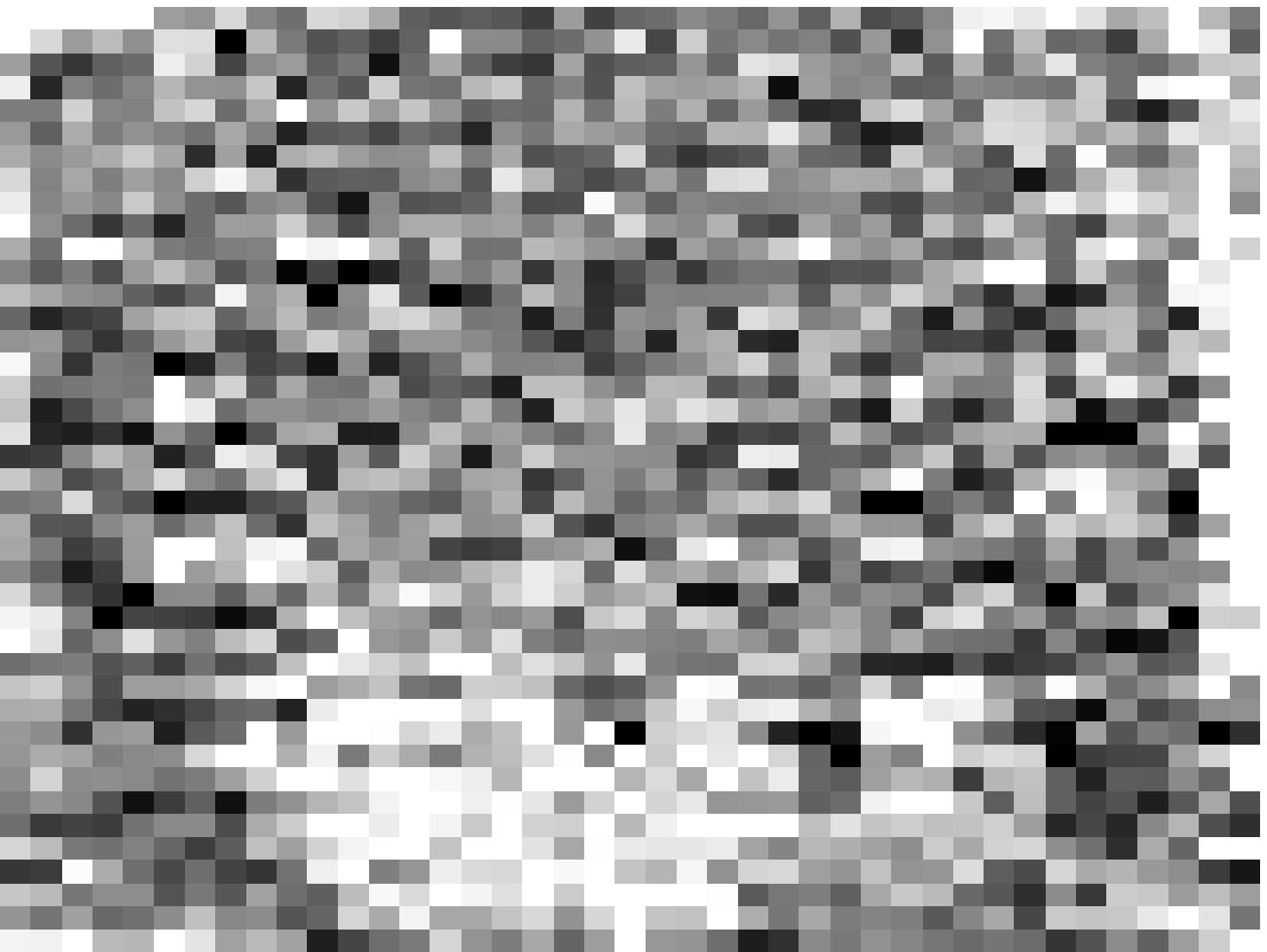}};
			\draw [-stealth, line width=2pt, cyan] (0.7,1.3) -- ++(-0.3,-0.3);
			\draw [-stealth, line width=2pt, cyan] (1.7,0.45) -- ++(-0.45,-0.0);
			\end{tikzpicture}&
			\begin{tikzpicture}
			\node[anchor=south west,inner sep=0] (image) at (0,0) {\includegraphics[width=.2\linewidth,height=.2\linewidth]{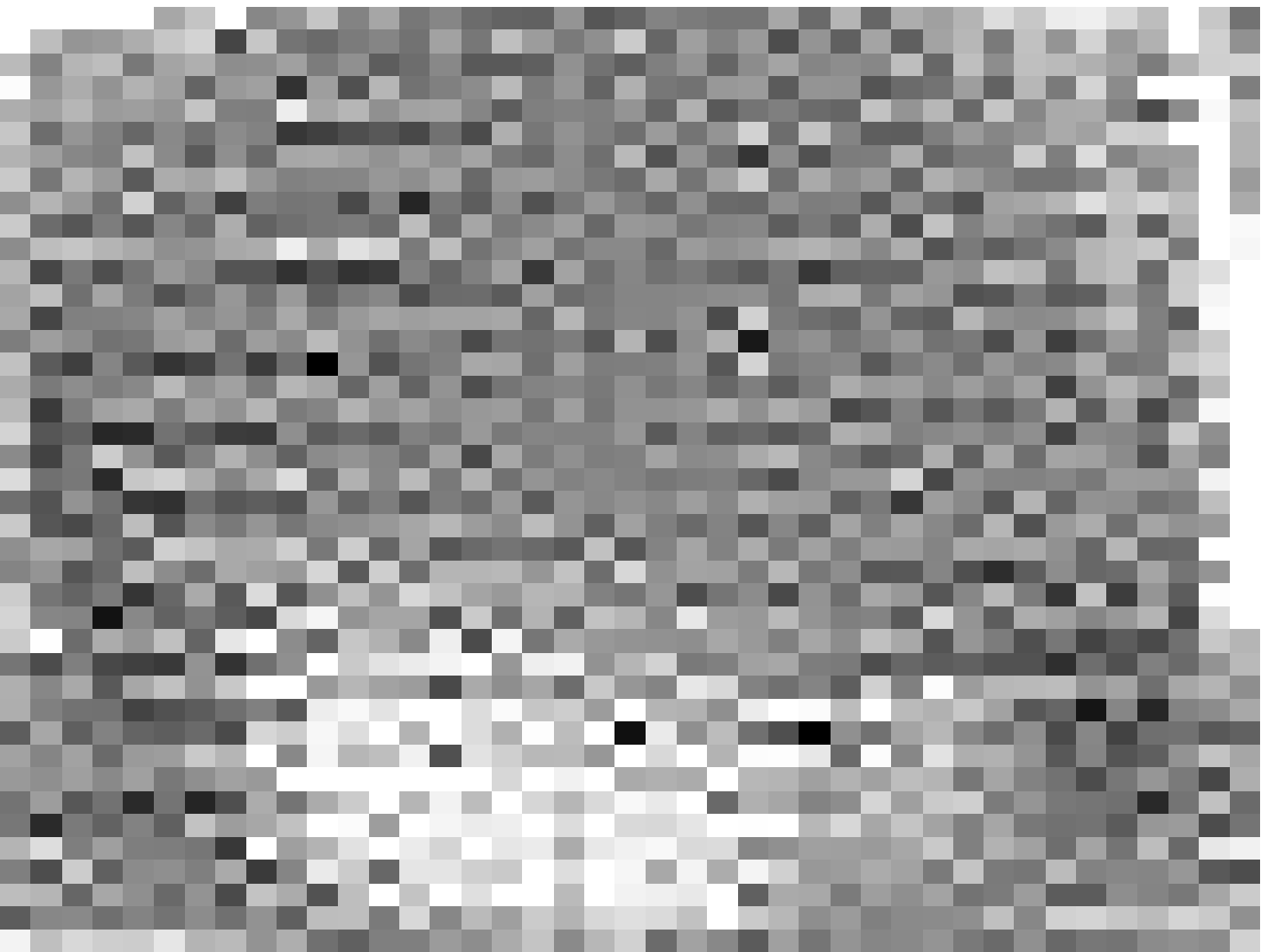}};
			\draw [-stealth, line width=2pt, cyan] (0.7,1.3) -- ++(-0.3,-0.3);
			\draw [-stealth, line width=2pt, cyan] (1.7,0.45) -- ++(-0.45,-0.0);
			\end{tikzpicture}&
			\begin{tikzpicture}
			\node[anchor=south west,inner sep=0] (image) at (0,0) {\includegraphics[width=.2\linewidth,height=.2\linewidth]{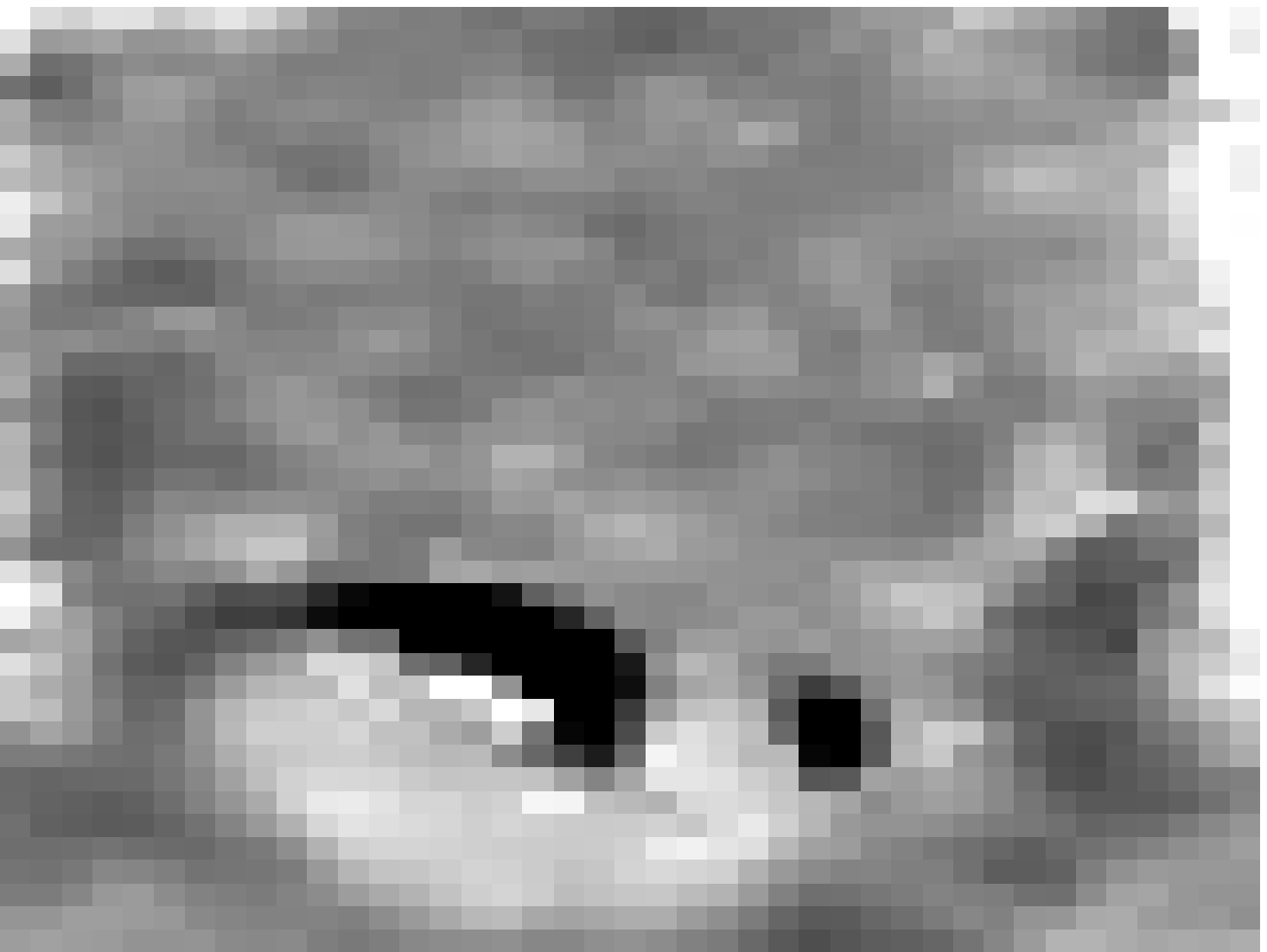}};
			\draw [-stealth, line width=2pt, cyan] (0.7,1.3) -- ++(-0.3,-0.3);
			\draw [-stealth, line width=2pt, cyan] (1.7,0.45) -- ++(-0.45,-0.0);
			\end{tikzpicture}\\
			FBP&		
			TV&
			KSVD&
			BM3D&
			FBPConvNet\\
			\begin{tikzpicture}
			\node[anchor=south west,inner sep=0] (image) at (0,0) {\includegraphics[width=.2\linewidth,height=.2\linewidth]{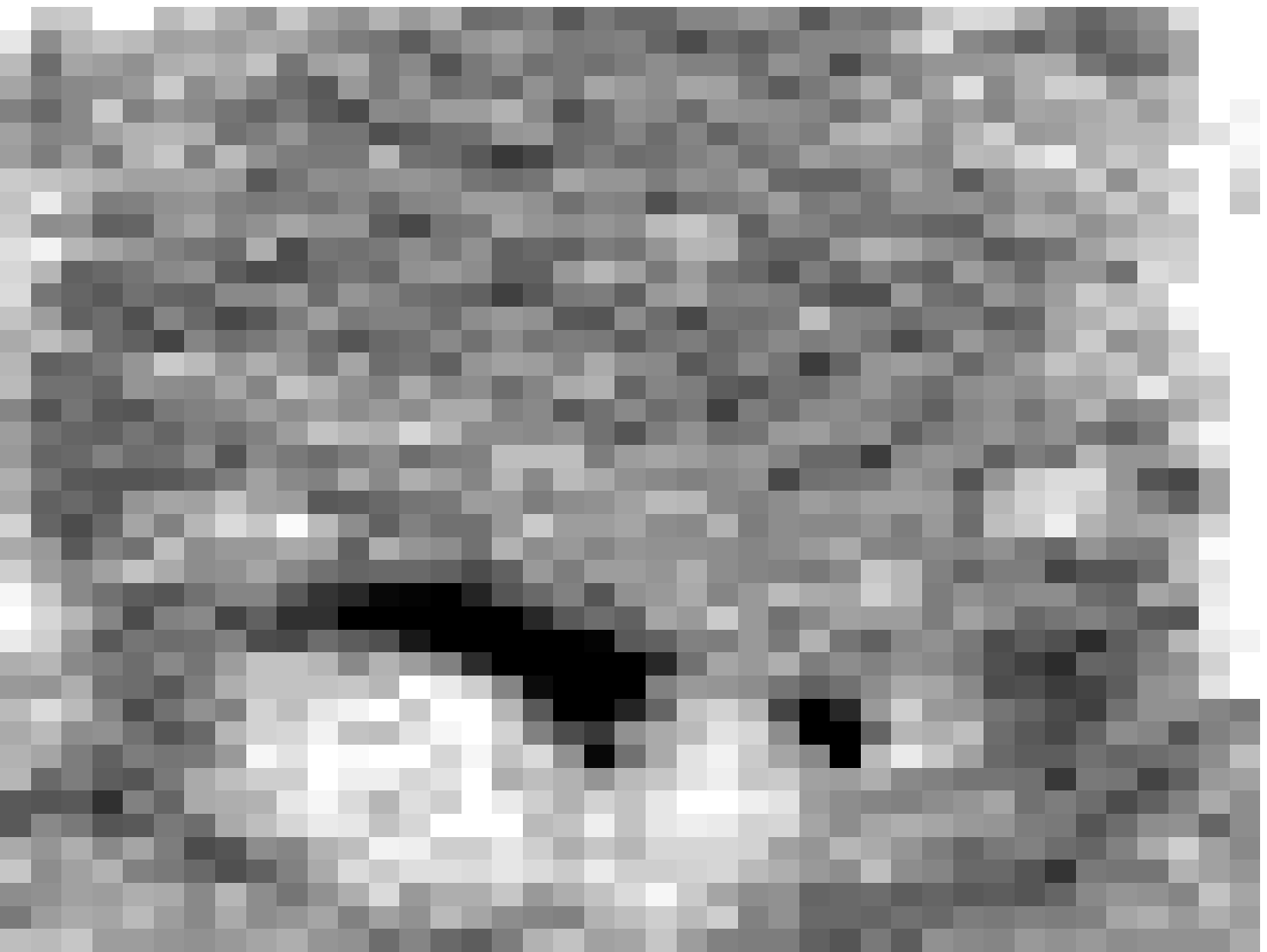}};
			\draw [-stealth, line width=2pt, cyan] (0.7,1.3) -- ++(-0.3,-0.3);
			\draw [-stealth, line width=2pt, cyan] (1.7,0.45) -- ++(-0.45,-0.0);
			\end{tikzpicture}&		
			\begin{tikzpicture}
			\node[anchor=south west,inner sep=0] (image) at (0,0) {\includegraphics[width=.2\linewidth,height=.2\linewidth]{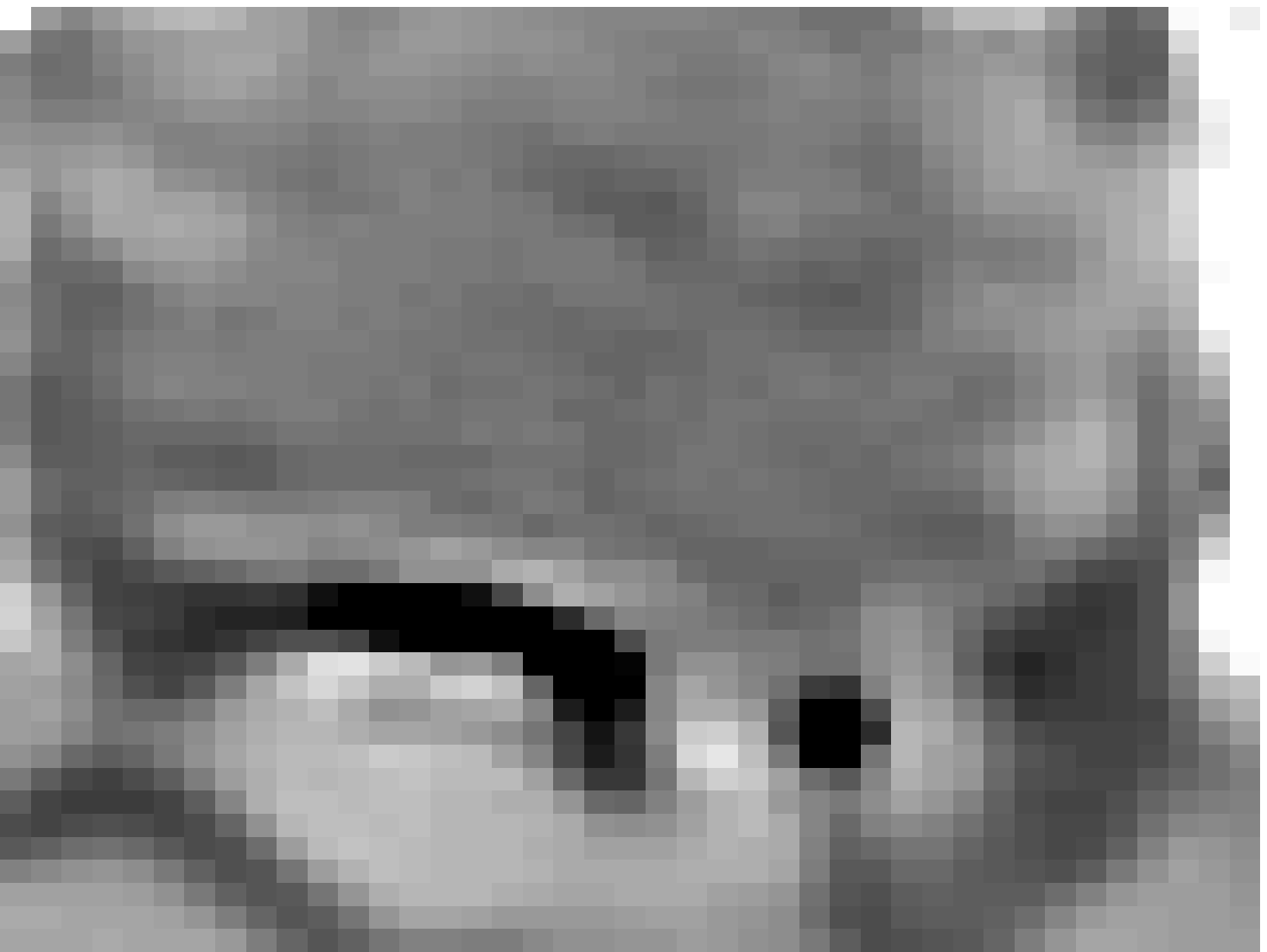}};
			\draw [-stealth, line width=2pt, cyan] (0.7,1.3) -- ++(-0.3,-0.3);
			\draw [-stealth, line width=2pt, cyan] (1.7,0.45) -- ++(-0.45,-0.0);
			\end{tikzpicture}&
			\begin{tikzpicture}
			\node[anchor=south west,inner sep=0] (image) at (0,0) {\includegraphics[width=.2\linewidth,height=.2\linewidth]{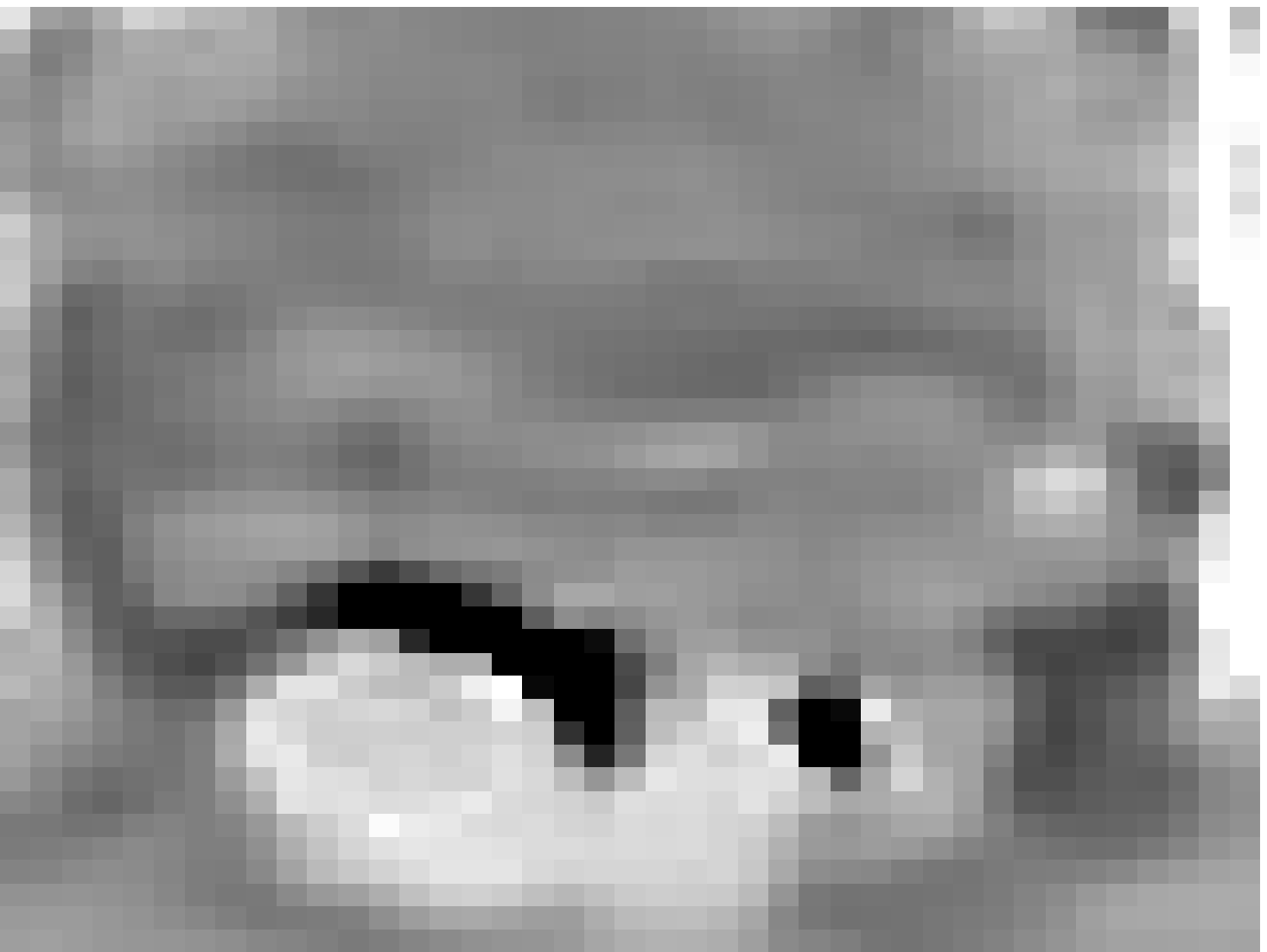}};
			\draw [-stealth, line width=2pt, cyan] (0.7,1.3) -- ++(-0.3,-0.3);
			\draw [-stealth, line width=2pt, cyan] (1.7,0.45) -- ++(-0.45,-0.0);
			\end{tikzpicture}&		
			\begin{tikzpicture}
			\node[anchor=south west,inner sep=0] (image) at (0,0) {\includegraphics[width=.2\linewidth,height=.2\linewidth]{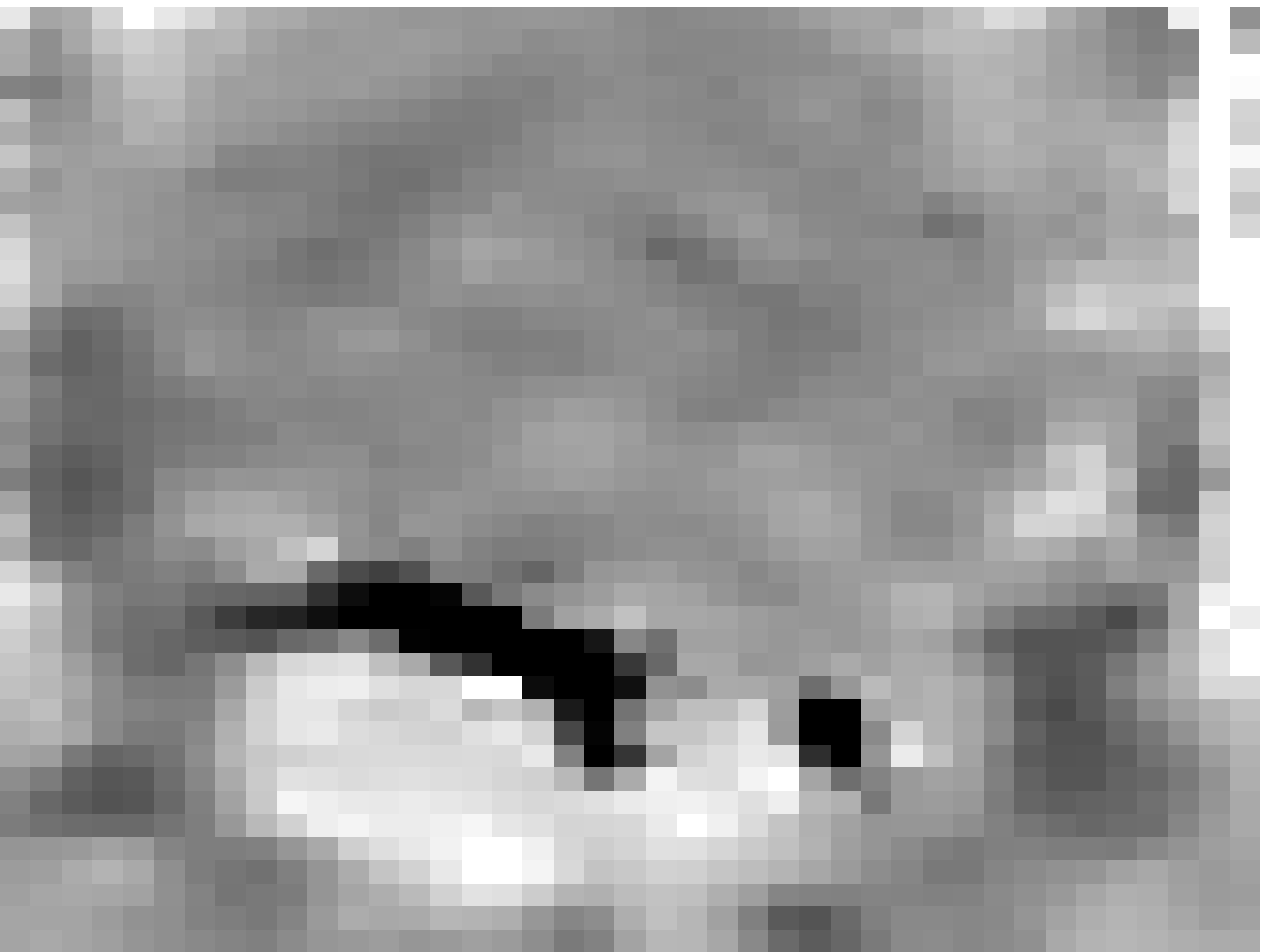}};
			\draw [-stealth, line width=2pt, cyan] (0.7,1.3) -- ++(-0.3,-0.3);
			\draw [-stealth, line width=2pt, cyan] (1.7,0.45) -- ++(-0.45,-0.0);
			\end{tikzpicture}&
			\begin{tikzpicture}
			\node[anchor=south west,inner sep=0] (image) at (0,0) {\includegraphics[width=.2\linewidth,height=.2\linewidth]{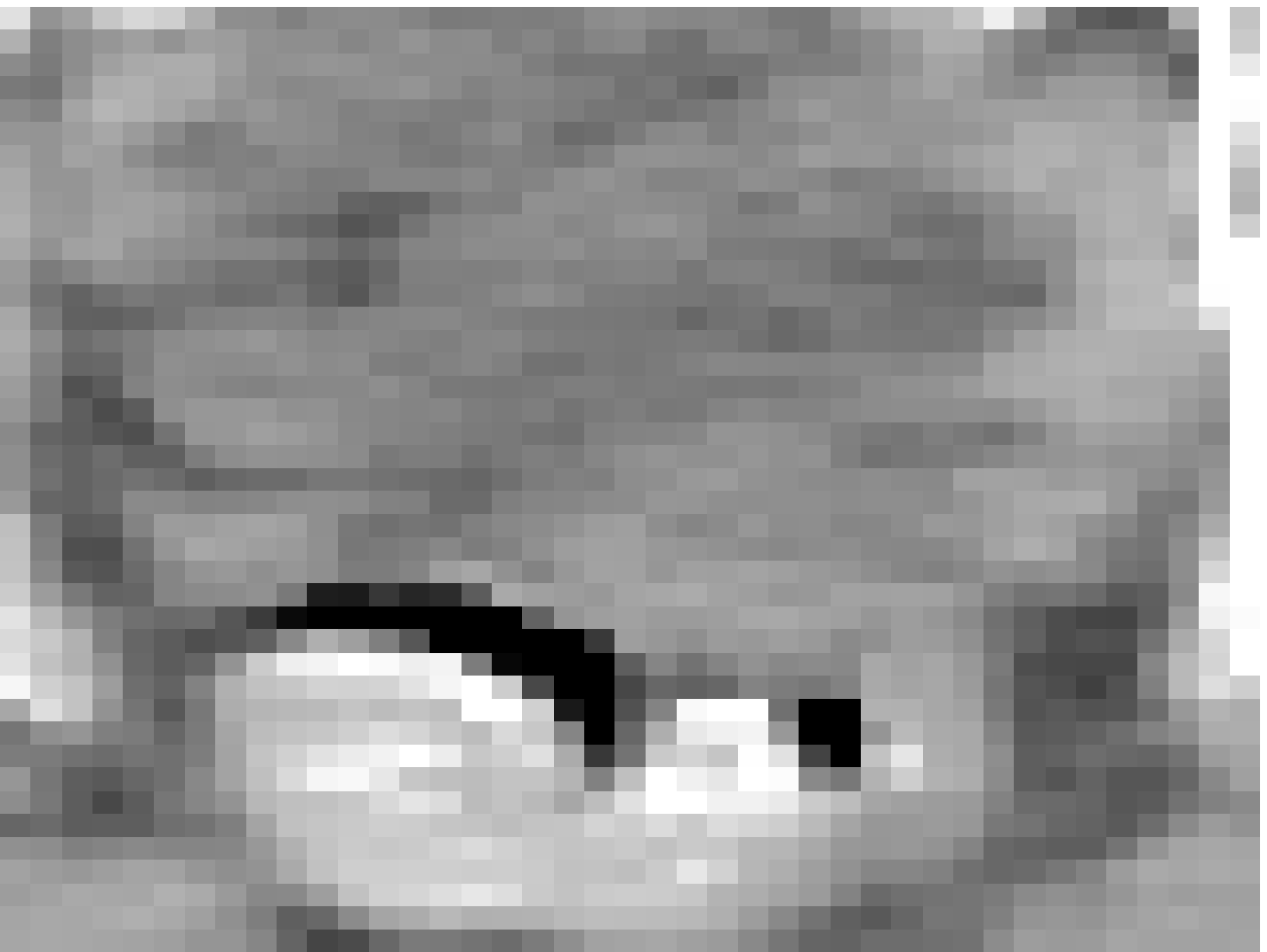}};
			\draw [-stealth, line width=2pt, cyan] (0.7,1.3) -- ++(-0.3,-0.3);
			\draw [-stealth, line width=2pt, cyan] (1.7,0.45) -- ++(-0.45,-0.0);
			\end{tikzpicture}\\
			MoDL&
			Neumann-Net&	
			PGD&
			Learned-PD&	
			AHP-Net
		\end{tabular}
		\caption{Zoom-in results of Fig.~\ref{sliceU_10000}.
		}
		\label{sliceZoomU_10000}
\end{figure}

Fig. \ref{sliceU_50000}  and Fig. \ref{sliceU_10000} shows the  images reconstructed by the models trained under same dose level.
And the corresponding zoomed-in images are displayed in Fig.~\ref{sliceZoomU_50000} and Fig.~\ref{sliceZoomU_10000}.

\begin{figure*}
	\begin{center}
		\begin{tabular}{c@{\hspace{0pt}}c@{\hspace{0pt}}c@{\hspace{0pt}}c@{\hspace{0pt}}c@{\hspace{0pt}}c@{\hspace{0pt}}c}
			\includegraphics[width=.198\linewidth,height=.132\linewidth]{FBPHU_50000_2_38.eps}&		
			\includegraphics[width=.198\linewidth,height=.132\linewidth]{TVHU_50000_2_38.eps}&			
			\includegraphics[width=.198\linewidth,height=.132\linewidth]{KSVD_ProstateMix_I_50000_sigma_100.eps}&
			\includegraphics[width=.198\linewidth,height=.132\linewidth]{BM3D_ProstateMix_I_50000_sigma_100.eps}&
			\includegraphics[width=.198\linewidth,height=.132\linewidth]{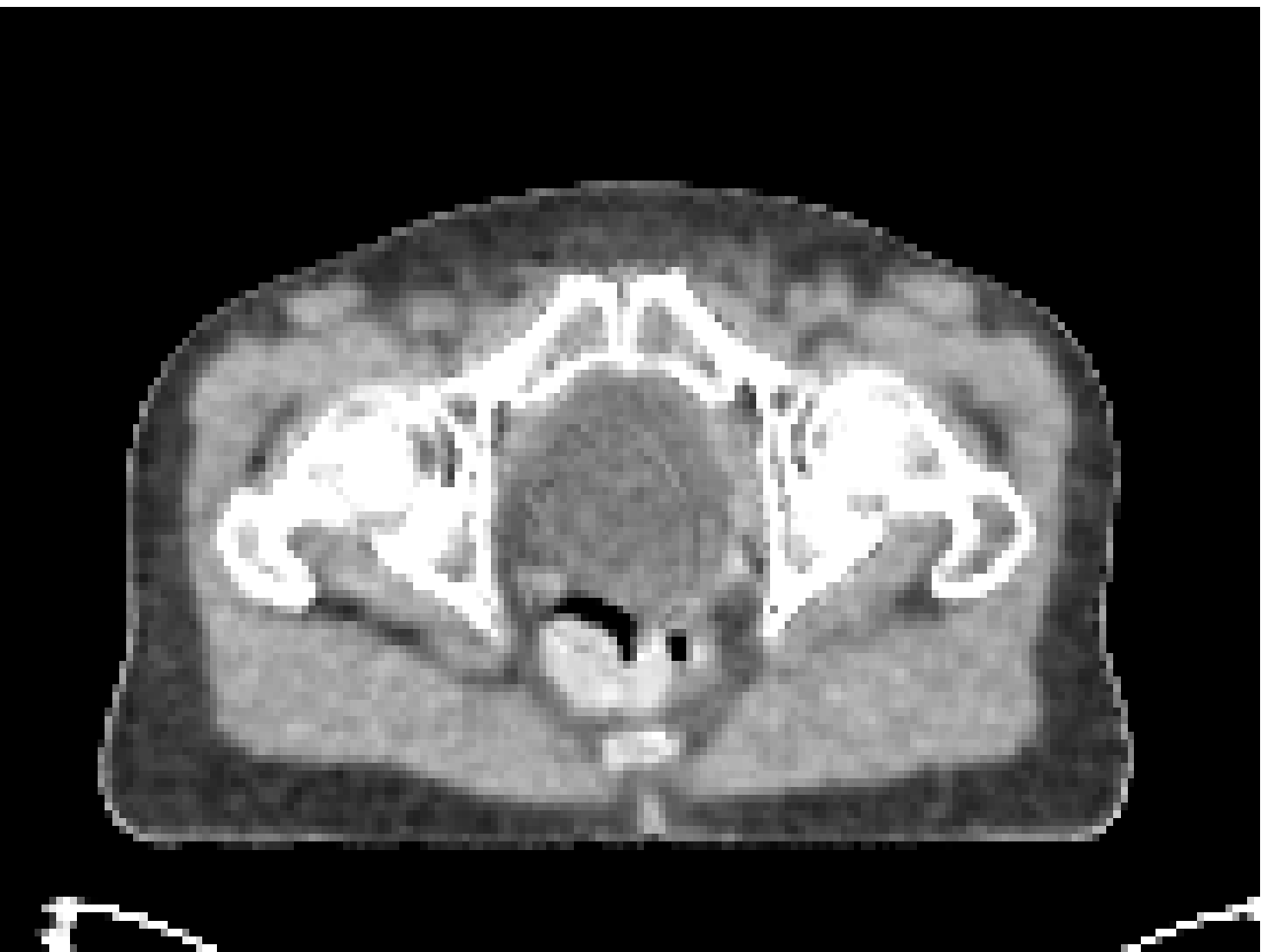}\\			
			FBP&
			TV&
			KSVD&
			BM3D&
			FBPConvNet\\		
			\includegraphics[width=.198\linewidth,height=.132\linewidth]{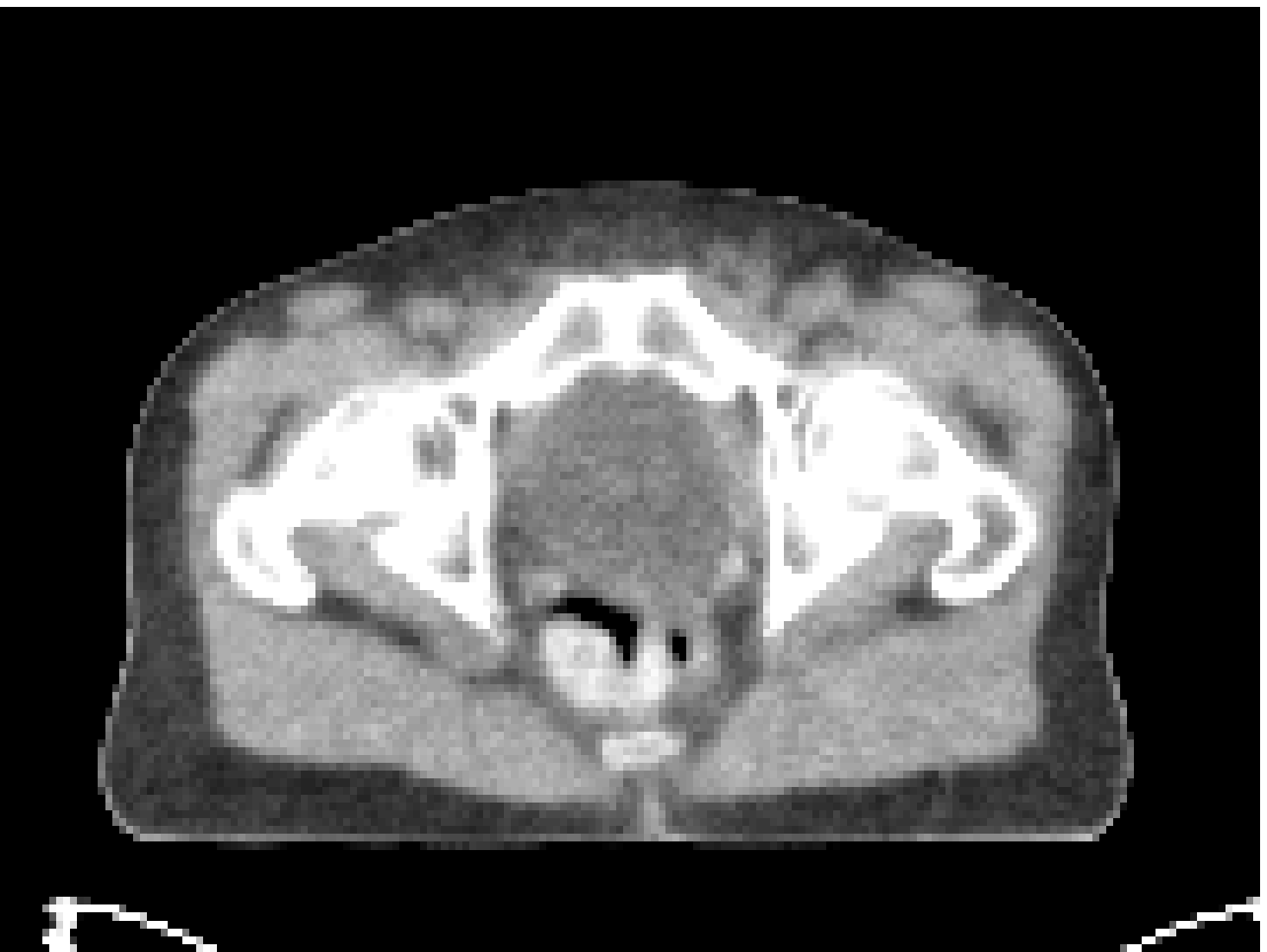}&
			\includegraphics[width=.198\linewidth,height=.132\linewidth]{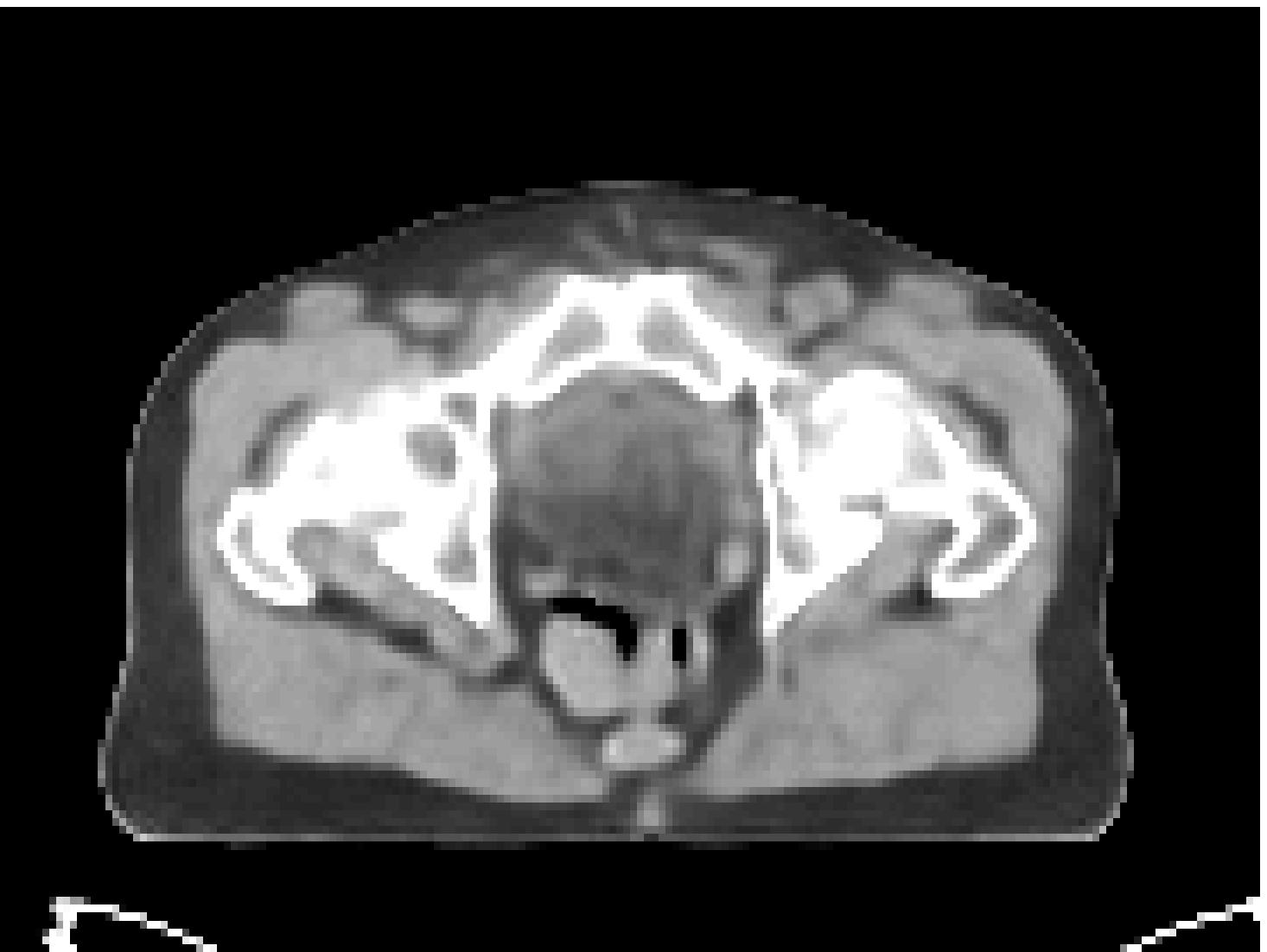}&
			\includegraphics[width=.198\linewidth,height=.132\linewidth]{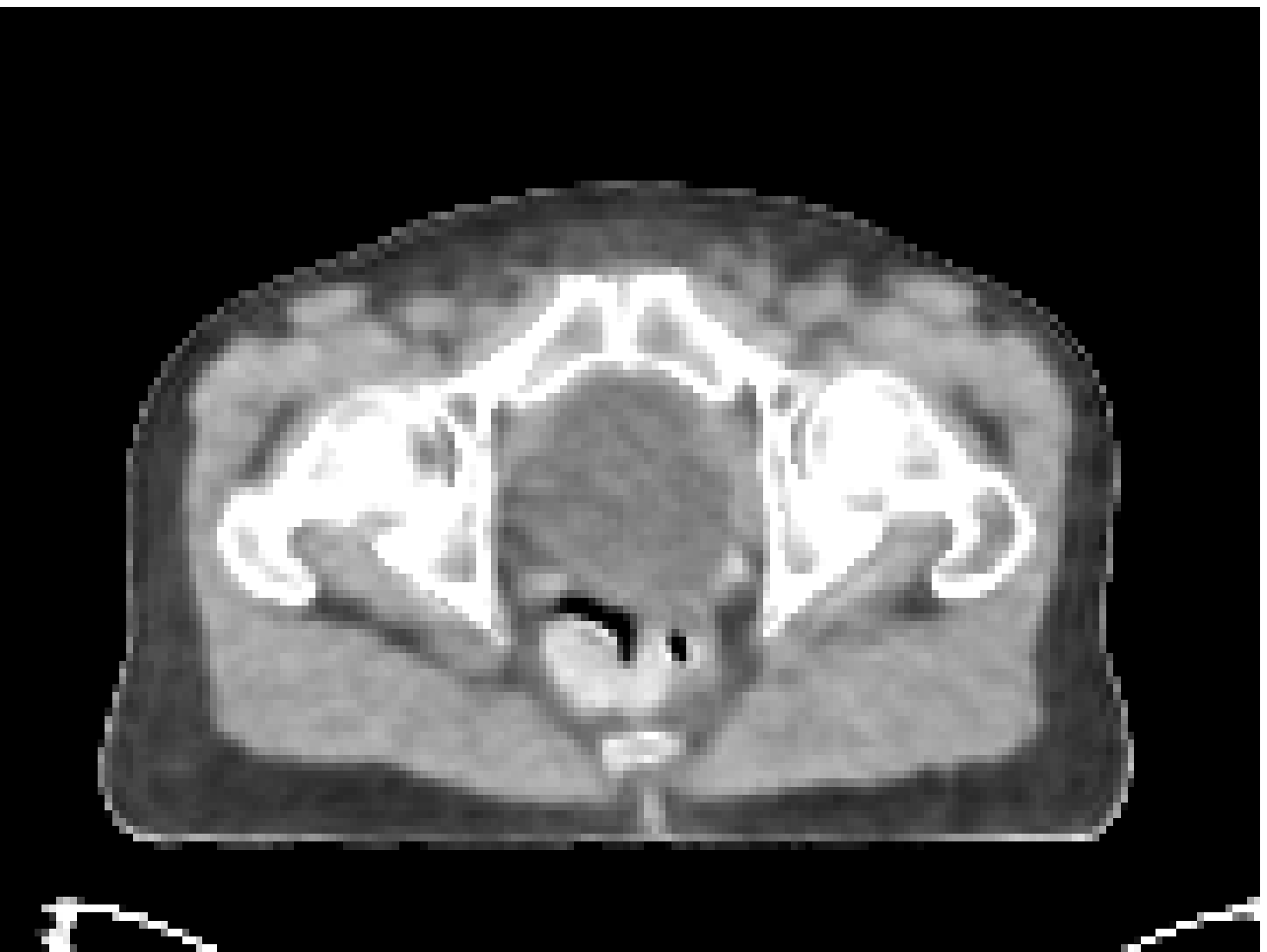}&		
			\includegraphics[width=.198\linewidth,height=.132\linewidth]{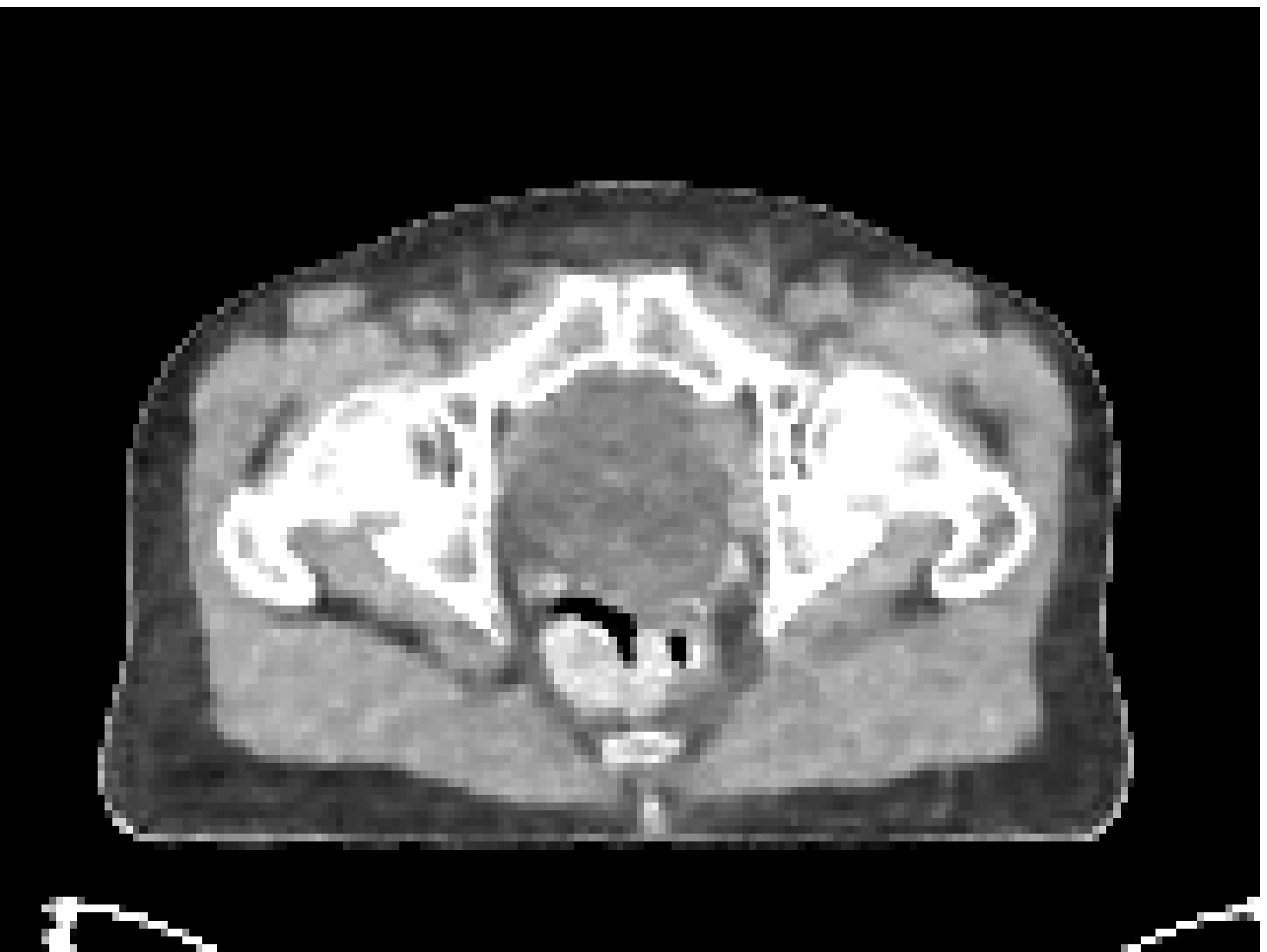}&		
			\includegraphics[width=.198\linewidth,height=.132\linewidth]{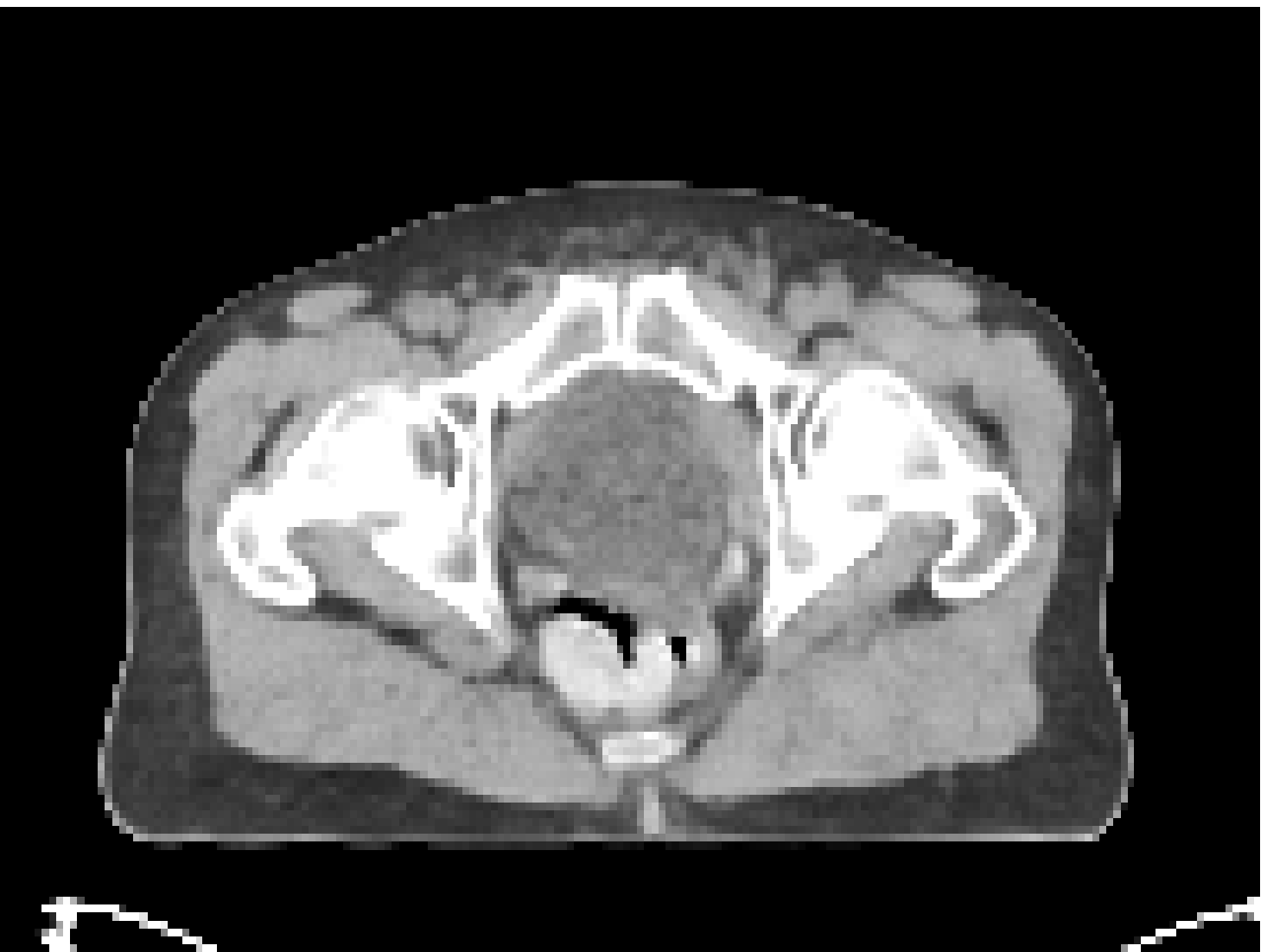}\\
			MoDL&
			Neumann-Net&
			PGD&
			Learn-PD&
			AHP-Net
		\end{tabular}
		\caption{Reconstruction results at dose level $I_i=5\times10^4$ by the universal models trained for varying dose levels.
		}
		\label{slice_50000}
	\end{center}
\end{figure*}

\begin{figure*}
	\begin{center}
		\begin{tabular}{c@{\hspace{0pt}}c@{\hspace{0pt}}c@{\hspace{0pt}}c@{\hspace{0pt}}c@{\hspace{0pt}}c@{\hspace{0pt}}c}
			\includegraphics[width=.198\linewidth,height=.132\linewidth]{FBPHU_5000_2_38.eps}&		
			\includegraphics[width=.198\linewidth,height=.132\linewidth]{TVHU_5000_2_38.eps}&			
			\includegraphics[width=.198\linewidth,height=.132\linewidth]{KSVD_ProstateMix_I_5000_sigma_100.eps}&
			\includegraphics[width=.198\linewidth,height=.132\linewidth]{BM3D_ProstateMix_I_5000_sigma_100.eps}&
			\includegraphics[width=.198\linewidth,height=.132\linewidth]{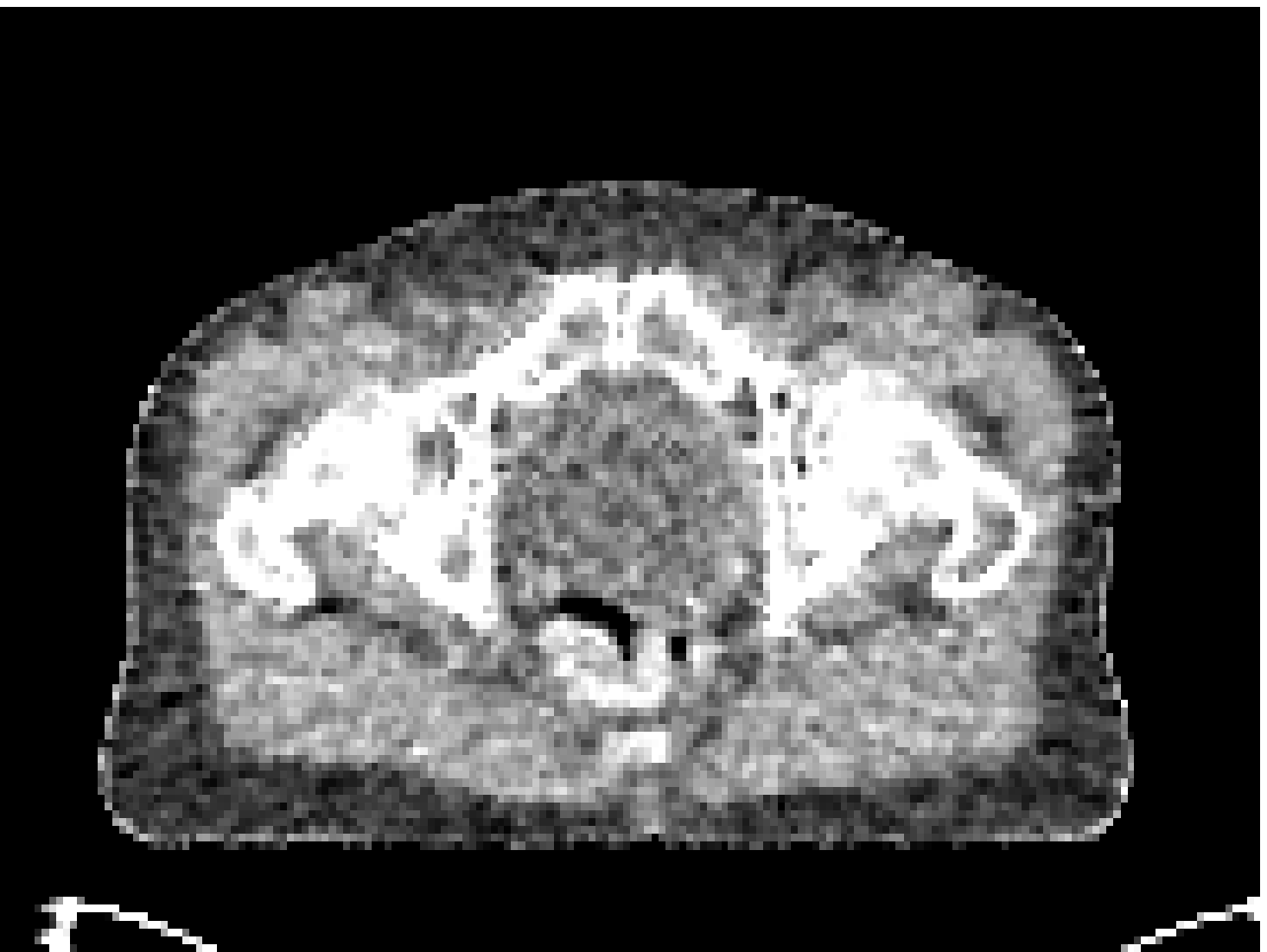}\\
			FBP&
			TV&
			KSVD&
			BM3D&			
			FBPConvNet\\				
			\includegraphics[width=.198\linewidth,height=.132\linewidth]{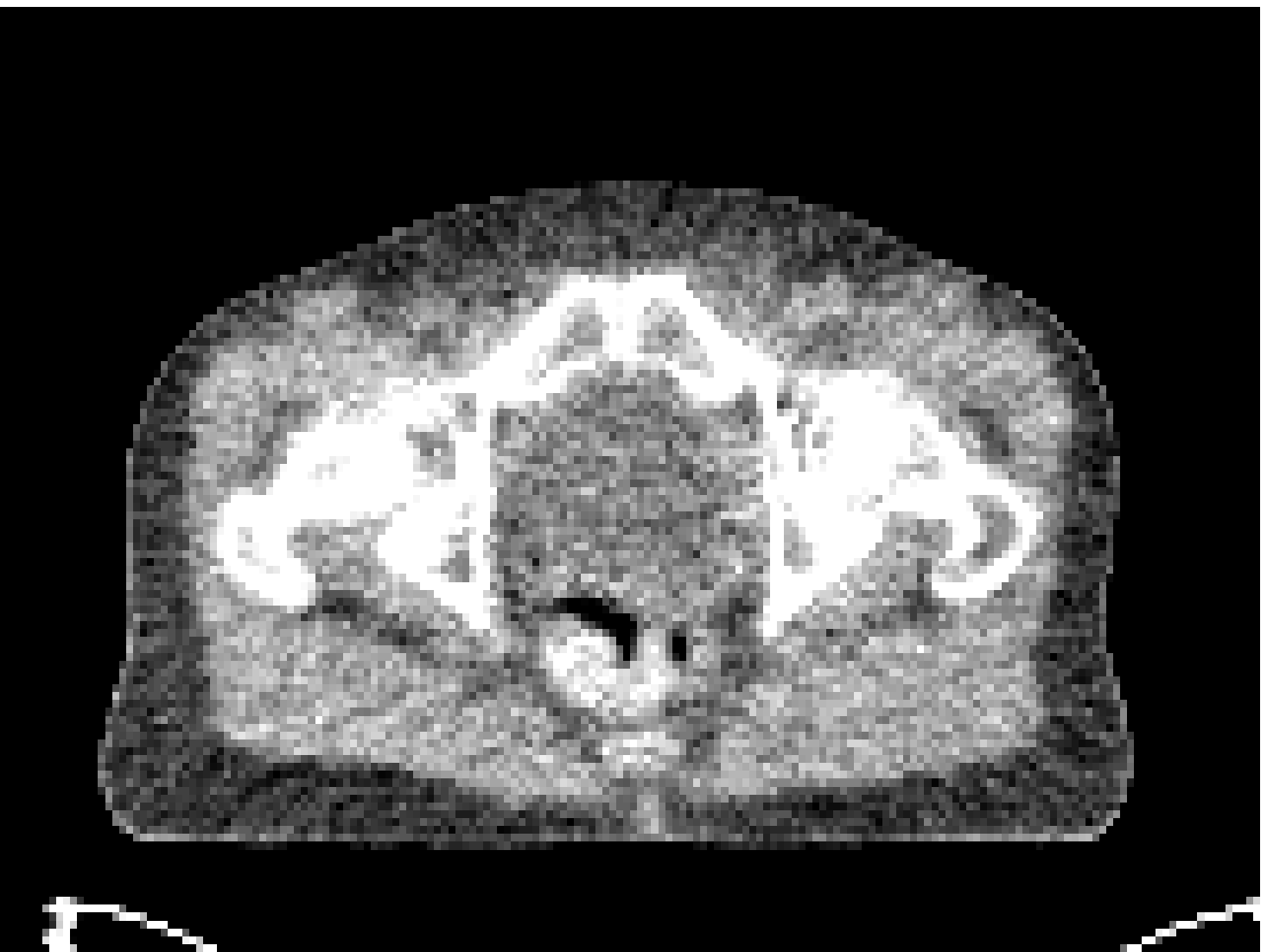}&
			\includegraphics[width=.198\linewidth,height=.132\linewidth]{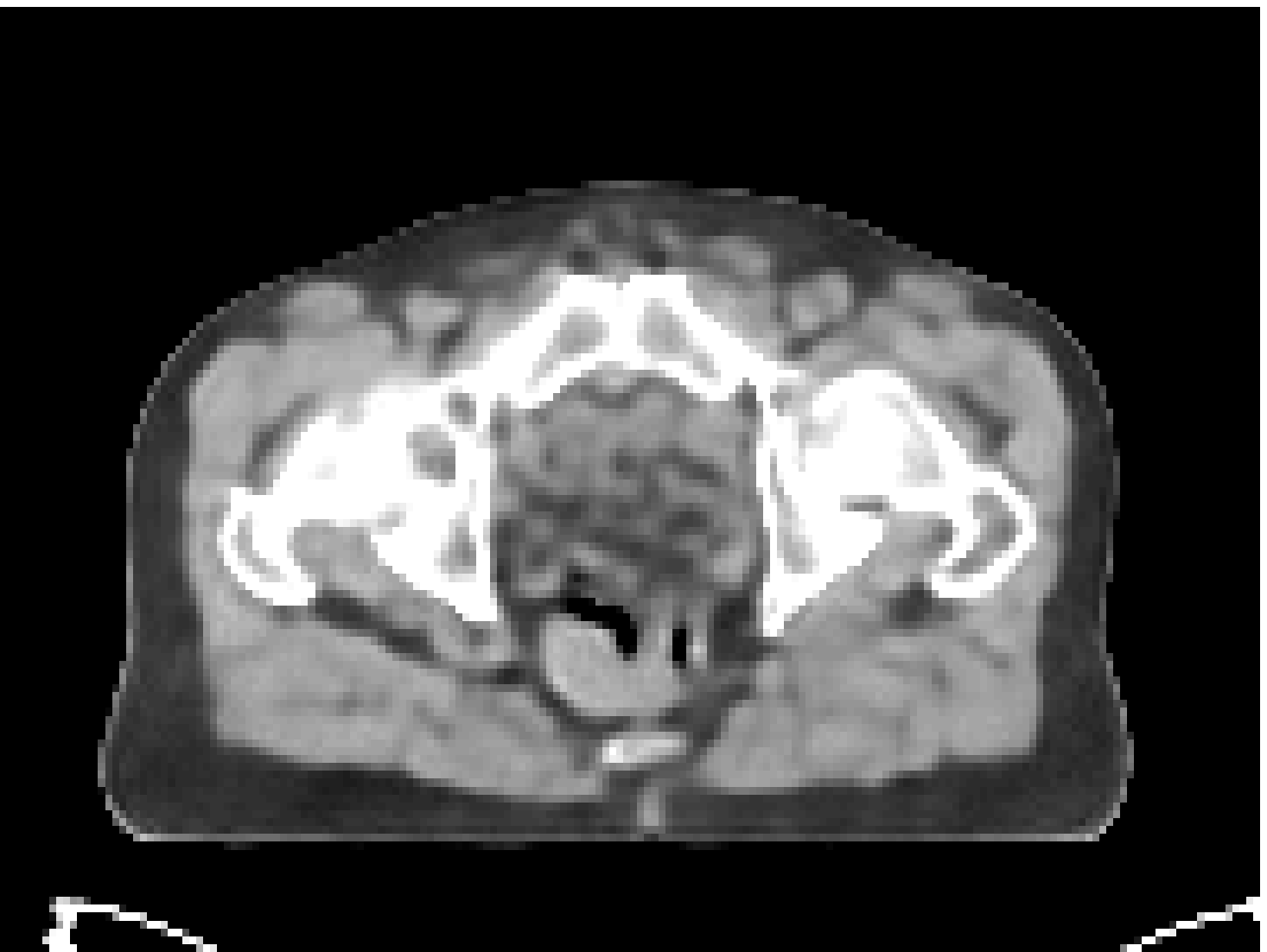}&
			\includegraphics[width=.198\linewidth,height=.132\linewidth]{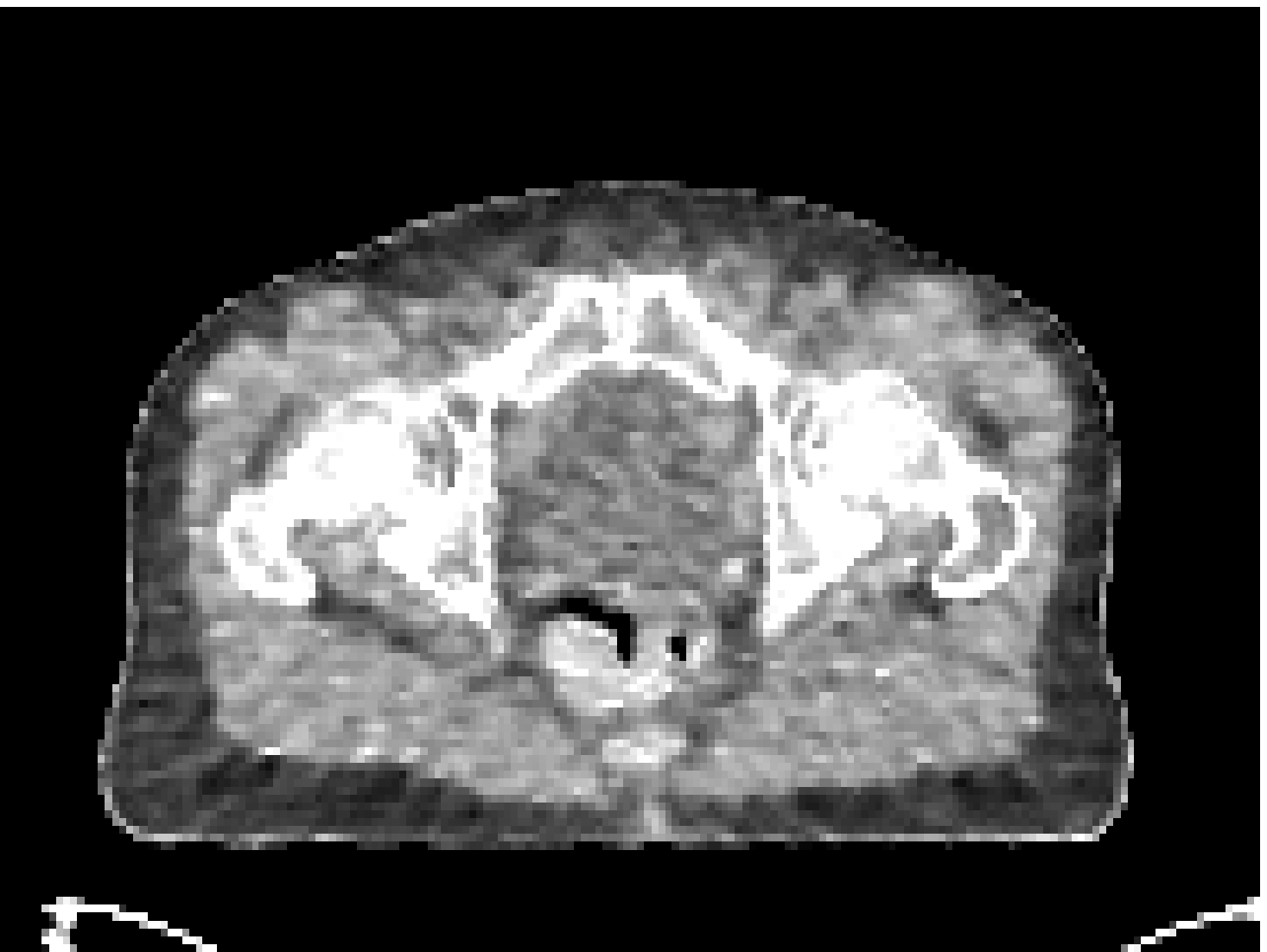}&			
			\includegraphics[width=.198\linewidth,height=.132\linewidth]{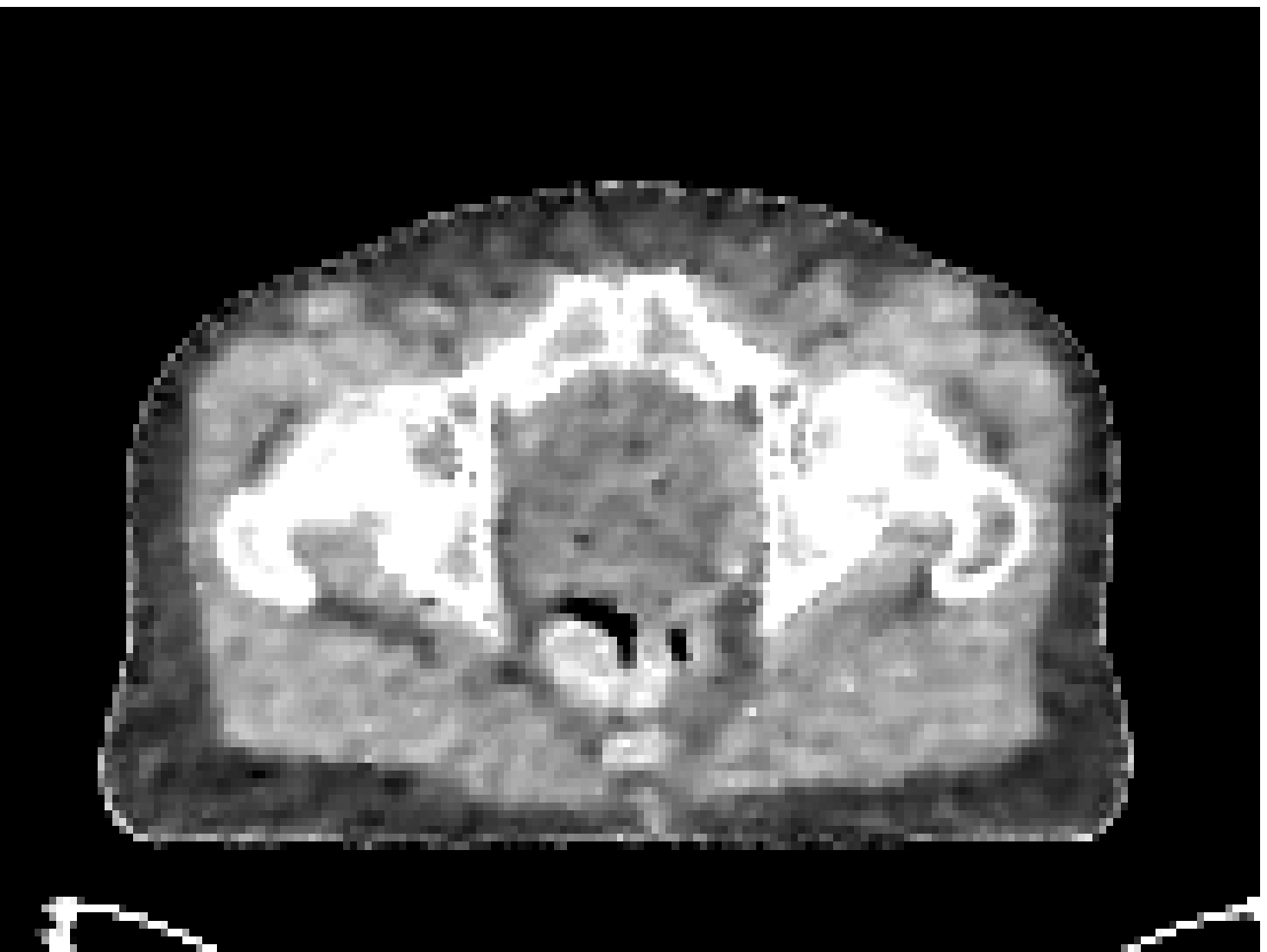}&		
			\includegraphics[width=.198\linewidth,height=.132\linewidth]{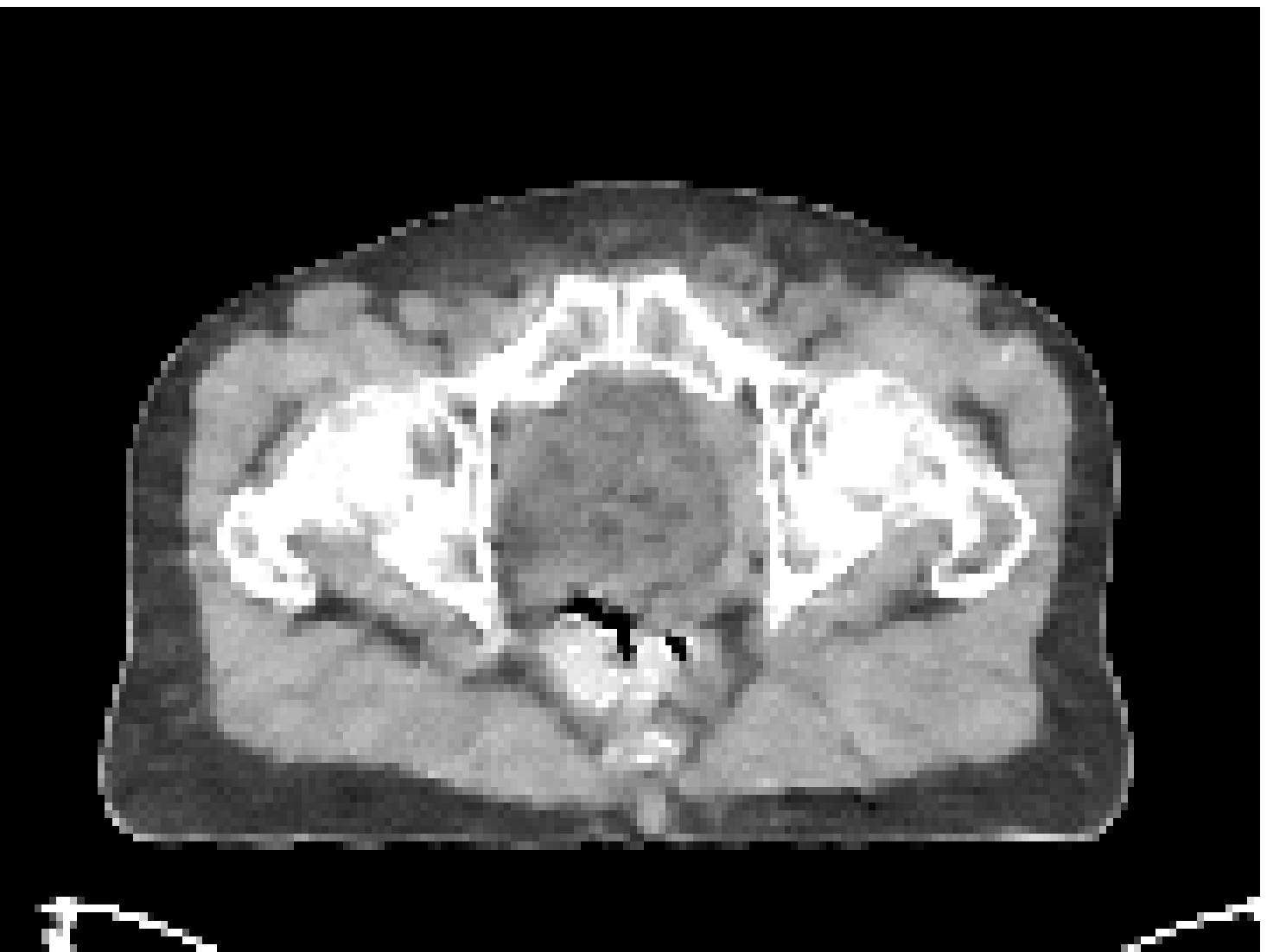}\\
			MoDL&
			Neumann-Net&
			PGD&
			Learn-PD&
			AHP-Net
		\end{tabular}
		\caption{Reconstruction results at dose level $I_i=5\times 10^3$ by the universal models trained for varying dose levels.
		}
		\label{slice_5000}
	\end{center}
\end{figure*}

\begin{figure}
	\begin{center}
		\begin{tabular}{c@{\hspace{-1pt}}c@{\hspace{-1pt}}c@{\hspace{-1pt}}c@{\hspace{-1pt}}c@{\hspace{-1pt}}c@{\hspace{-1pt}}c}
			\begin{tikzpicture}
			\node[anchor=south west,inner sep=0] (image) at (0,0) {\includegraphics[width=.2\linewidth,height=.2\linewidth]{FBPHUZoom_50000_2_38.eps}};
			\draw [-stealth, line width=2pt, cyan] (0.7,1.3) -- ++(-0.3,-0.3);
			\draw [-stealth, line width=2pt, cyan] (1.7,0.45) -- ++(-0.45,-0.0);
			\end{tikzpicture}&
			\begin{tikzpicture}
			\node[anchor=south west,inner sep=0] (image) at (0,0) {\includegraphics[width=.2\linewidth,height=.2\linewidth]{TVHUZoom_50000_2_38.eps}};
			\draw [-stealth, line width=2pt, cyan] (0.7,1.3) -- ++(-0.3,-0.3);
			\draw [-stealth, line width=2pt, cyan] (1.7,0.45) -- ++(-0.45,-0.0);
			\end{tikzpicture}&
			\begin{tikzpicture}
			\node[anchor=south west,inner sep=0] (image) at (0,0) {\includegraphics[width=.2\linewidth,height=.2\linewidth]{KSVD_ProstateMix_Zoom_I_50000_sigma_100.eps}};
			\draw [-stealth, line width=2pt, cyan] (0.7,1.3) -- ++(-0.3,-0.3);
			\draw [-stealth, line width=2pt, cyan] (1.7,0.45) -- ++(-0.45,-0.0);
			\end{tikzpicture}&
			\begin{tikzpicture}
			\node[anchor=south west,inner sep=0] (image) at (0,0) {\includegraphics[width=.2\linewidth,height=.2\linewidth]{BM3D_ProstateMix_Zoom_I_50000_sigma_100.eps}};
			\draw [-stealth, line width=2pt, cyan] (0.7,1.3) -- ++(-0.3,-0.3);
			\draw [-stealth, line width=2pt, cyan] (1.7,0.45) -- ++(-0.45,-0.0);
			\end{tikzpicture}&
			\begin{tikzpicture}
			\node[anchor=south west,inner sep=0] (image) at (0,0) {\includegraphics[width=.2\linewidth,height=.2\linewidth]{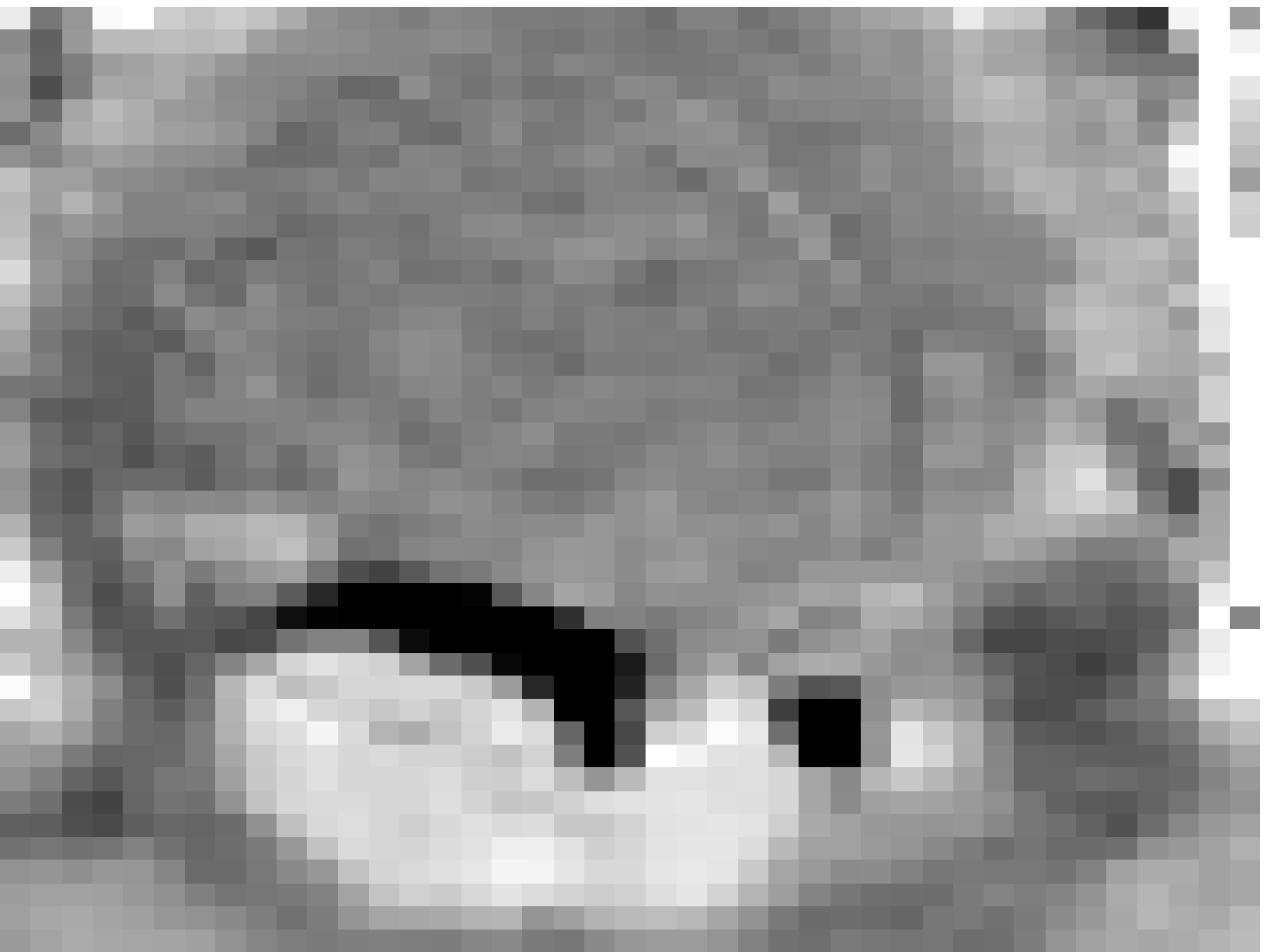}};
			\draw [-stealth, line width=2pt, cyan] (0.7,1.3) -- ++(-0.3,-0.3);
			\draw [-stealth, line width=2pt, cyan] (1.7,0.45) -- ++(-0.45,-0.0);
			\end{tikzpicture}\\
			FBP&		
			TV&
			KSVD&
			BM3D&
			FBPConvNet\\
			\begin{tikzpicture}
			\node[anchor=south west,inner sep=0] (image) at (0,0) {\includegraphics[width=.2\linewidth,height=.2\linewidth]{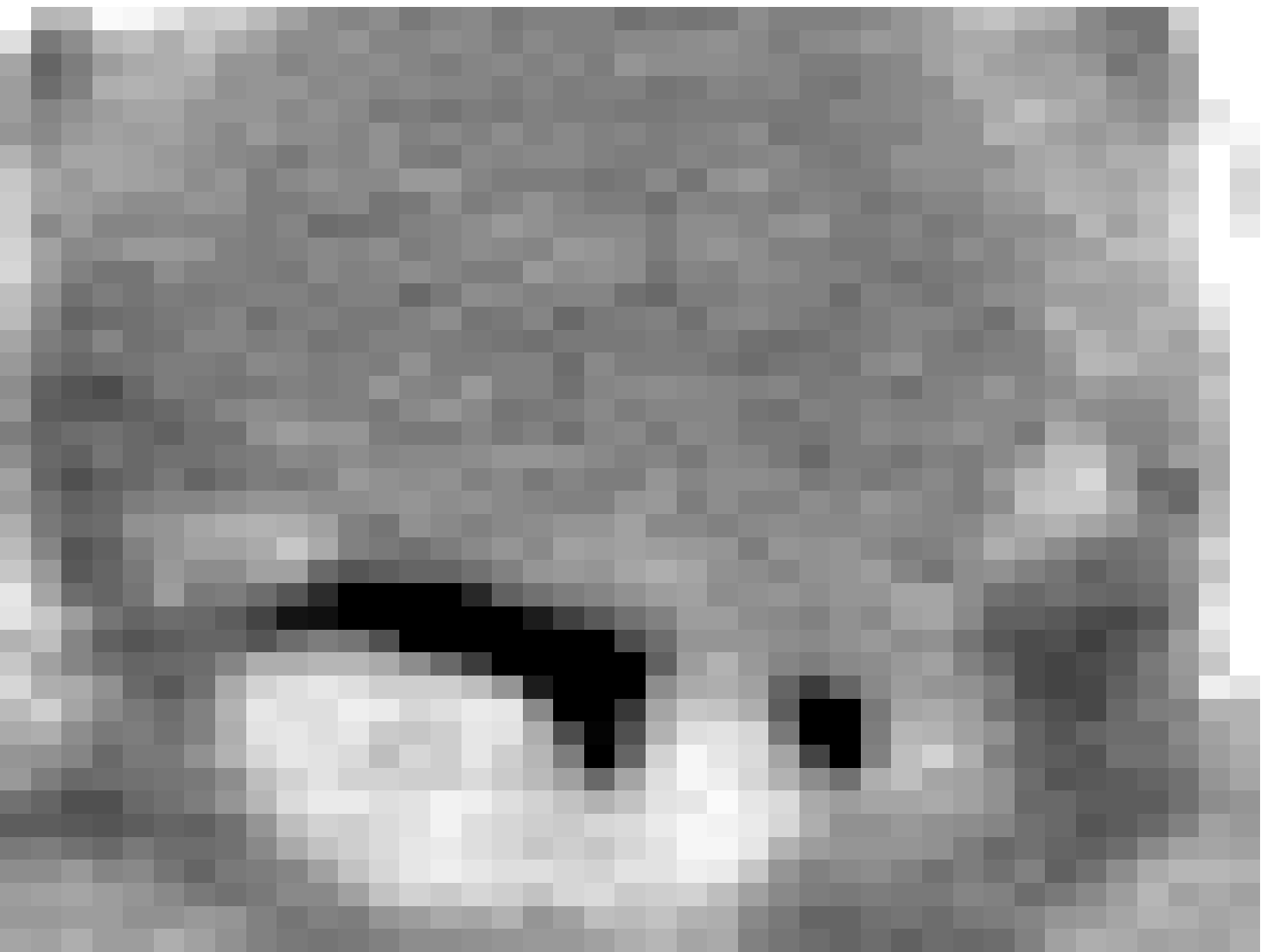}};
			\draw [-stealth, line width=2pt, cyan] (0.7,1.3) -- ++(-0.3,-0.3);
			\draw [-stealth, line width=2pt, cyan] (1.7,0.45) -- ++(-0.45,-0.0);
			\end{tikzpicture}&		
			\begin{tikzpicture}
			\node[anchor=south west,inner sep=0] (image) at (0,0) {\includegraphics[width=.2\linewidth,height=.2\linewidth]{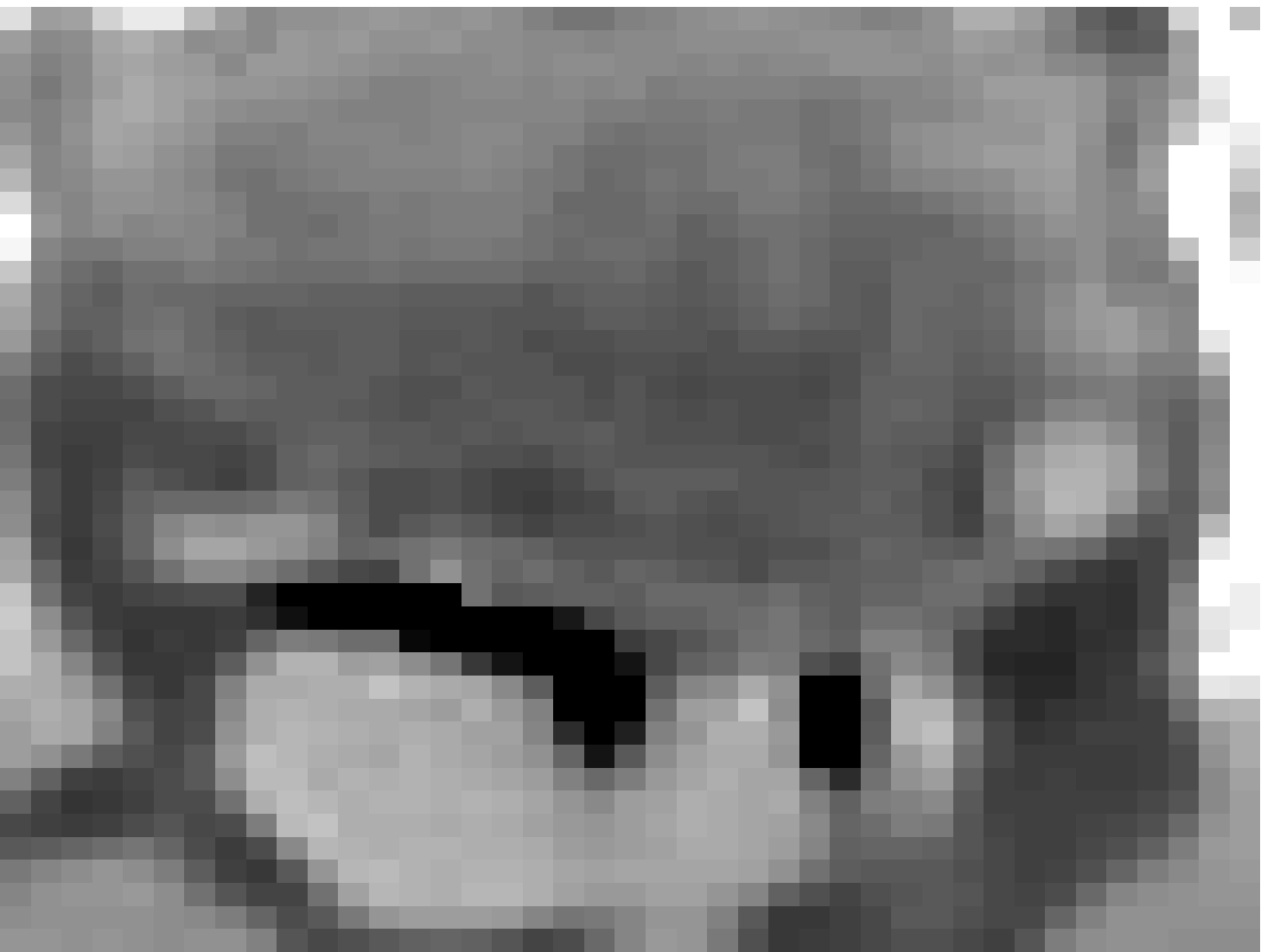}};
			\draw [-stealth, line width=2pt, cyan] (0.7,1.3) -- ++(-0.3,-0.3);
			\draw [-stealth, line width=2pt, cyan] (1.7,0.45) -- ++(-0.45,-0.0);
			\end{tikzpicture}&
			\begin{tikzpicture}
			\node[anchor=south west,inner sep=0] (image) at (0,0) {\includegraphics[width=.2\linewidth,height=.2\linewidth]{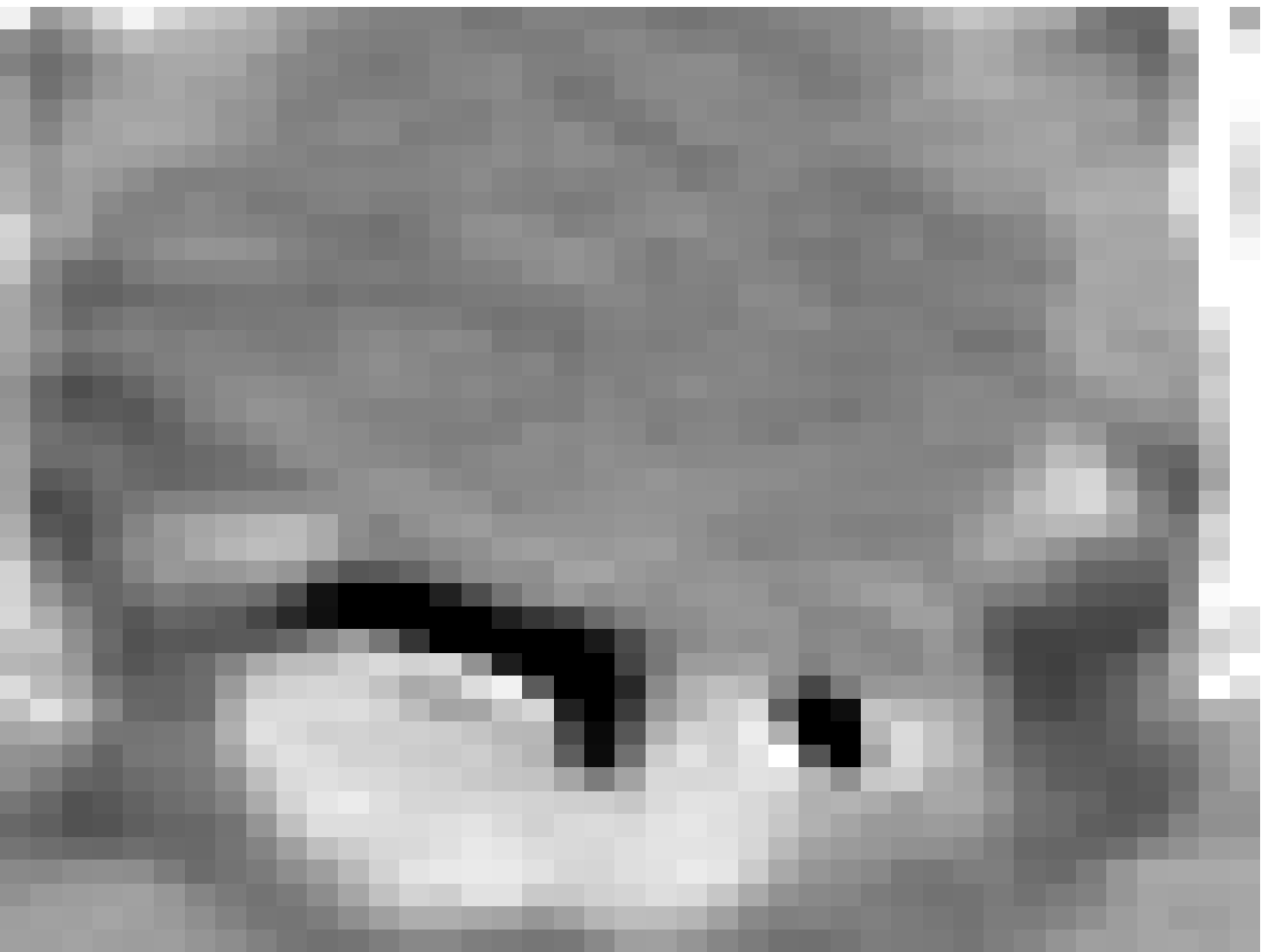}};
			\draw [-stealth, line width=2pt, cyan] (0.7,1.3) -- ++(-0.3,-0.3);
			\draw [-stealth, line width=2pt, cyan] (1.7,0.45) -- ++(-0.45,-0.0);
			\end{tikzpicture}&		
			\begin{tikzpicture}
			\node[anchor=south west,inner sep=0] (image) at (0,0) {\includegraphics[width=.2\linewidth,height=.2\linewidth]{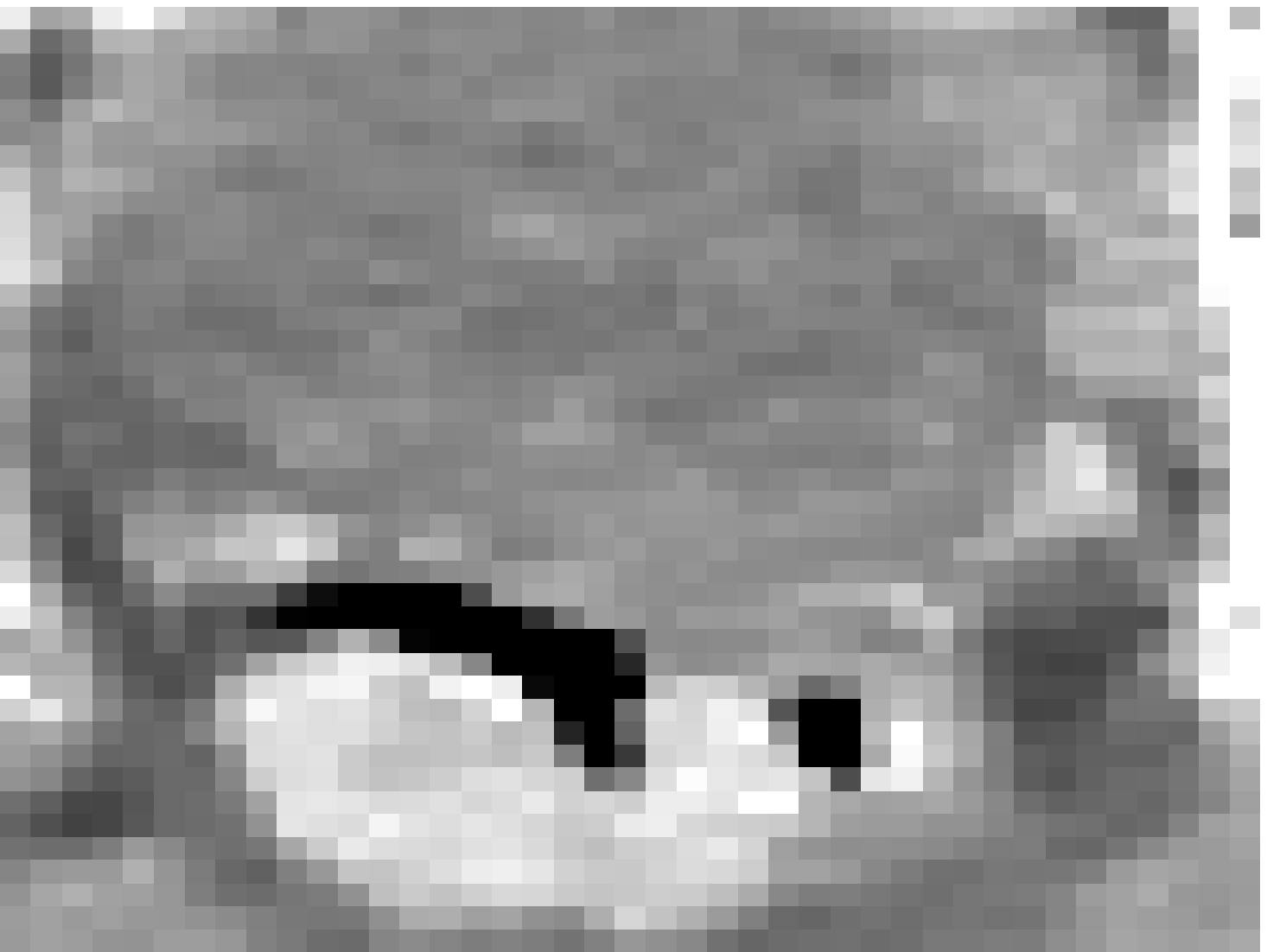}};
			\draw [-stealth, line width=2pt, cyan] (0.7,1.3) -- ++(-0.3,-0.3);
			\draw [-stealth, line width=2pt, cyan] (1.7,0.45) -- ++(-0.45,-0.0);
			\end{tikzpicture}&
			\begin{tikzpicture}
			\node[anchor=south west,inner sep=0] (image) at (0,0) {\includegraphics[width=.2\linewidth,height=.2\linewidth]{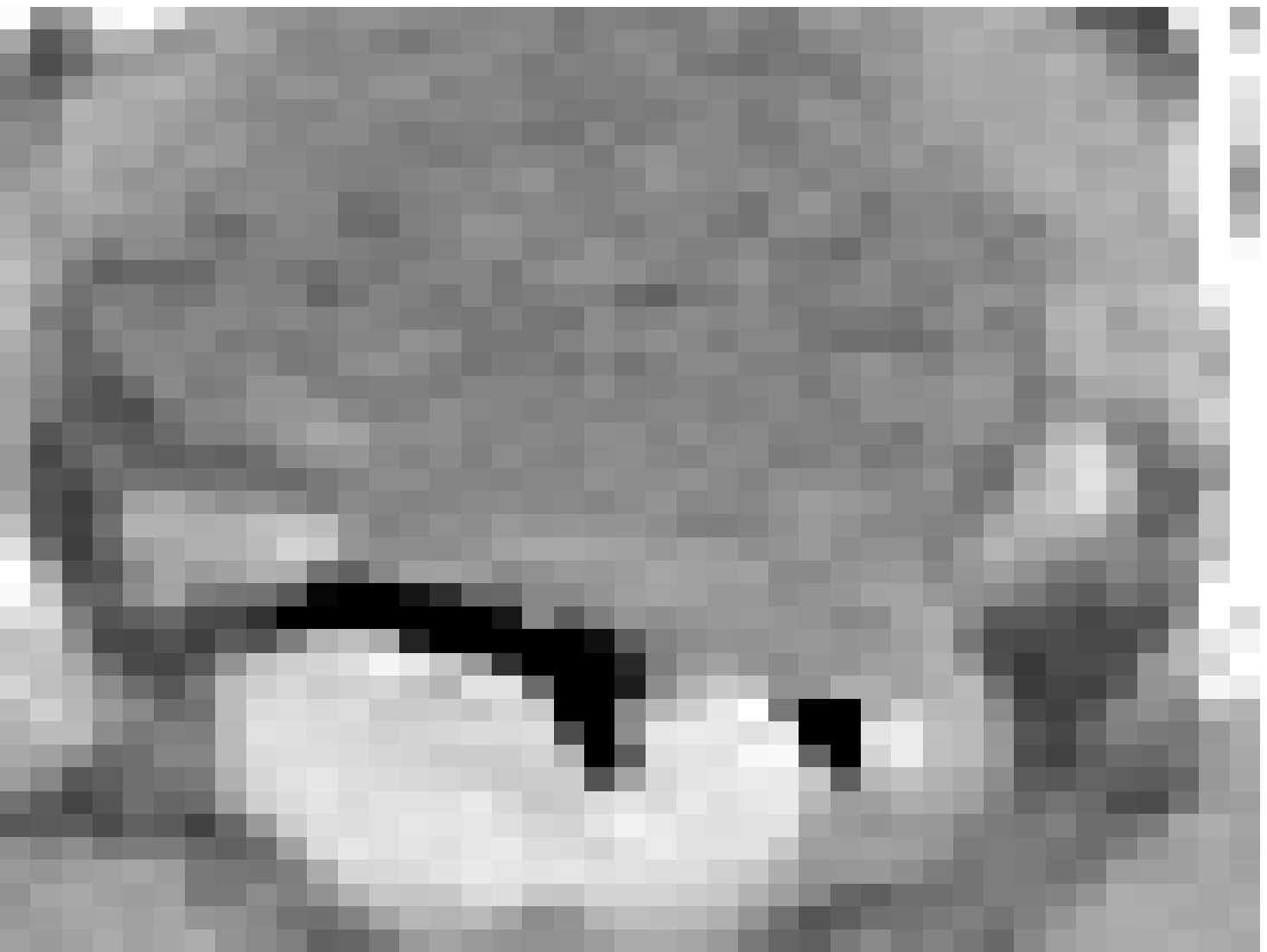}};
			\draw [-stealth, line width=2pt, cyan] (0.7,1.3) -- ++(-0.3,-0.3);
			\draw [-stealth, line width=2pt, cyan] (1.7,0.45) -- ++(-0.45,-0.0);
			\end{tikzpicture}\\
			MoDL&
			Neumann-Net&	
			PGD&
			Learned-PD&	
			AHP-Net
		\end{tabular}
		\caption{Zoom-in results of Fig.~\ref{slice_50000}.
		}
		\label{sliceZoom_50000}
	\end{center}
\end{figure}

\begin{figure}
	\begin{center}
		\begin{tabular}{c@{\hspace{-1pt}}c@{\hspace{-1pt}}c@{\hspace{-1pt}}c@{\hspace{-1pt}}c@{\hspace{-1pt}}c@{\hspace{-1pt}}c}
			\begin{tikzpicture}
			\node[anchor=south west,inner sep=0] (image) at (0,0) {\includegraphics[width=.2\linewidth,height=.2\linewidth]{FBPHUZoom_5000_2_38.eps}};
			\draw [-stealth, line width=2pt, cyan] (0.7,1.3) -- ++(-0.3,-0.3);
			\draw [-stealth, line width=2pt, cyan] (1.7,0.45) -- ++(-0.45,-0.0);
			\end{tikzpicture}&
			\begin{tikzpicture}
			\node[anchor=south west,inner sep=0] (image) at (0,0) {\includegraphics[width=.2\linewidth,height=.2\linewidth]{TVHUZoom_5000_2_38.eps}};
			\draw [-stealth, line width=2pt, cyan] (0.7,1.3) -- ++(-0.3,-0.3);
			\draw [-stealth, line width=2pt, cyan] (1.7,0.45) -- ++(-0.45,-0.0);
			\end{tikzpicture}&
			\begin{tikzpicture}
			\node[anchor=south west,inner sep=0] (image) at (0,0) {\includegraphics[width=.2\linewidth,height=.2\linewidth]{KSVD_ProstateMix_Zoom_I_5000_sigma_100.eps}};
			\draw [-stealth, line width=2pt, cyan] (0.7,1.3) -- ++(-0.3,-0.3);
			\draw [-stealth, line width=2pt, cyan] (1.7,0.45) -- ++(-0.45,-0.0);
			\end{tikzpicture}&
			\begin{tikzpicture}
			\node[anchor=south west,inner sep=0] (image) at (0,0) {\includegraphics[width=.2\linewidth,height=.2\linewidth]{BM3D_ProstateMix_Zoom_I_5000_sigma_100.eps}};
			\draw [-stealth, line width=2pt, cyan] (0.7,1.3) -- ++(-0.3,-0.3);
			\draw [-stealth, line width=2pt, cyan] (1.7,0.45) -- ++(-0.45,-0.0);
			\end{tikzpicture}&
			\begin{tikzpicture}
			\node[anchor=south west,inner sep=0] (image) at (0,0) {\includegraphics[width=.2\linewidth,height=.2\linewidth]{FBPConV_V_Zoom_5000_100_2_38.eps}};
			\draw [-stealth, line width=2pt, cyan] (0.7,1.3) -- ++(-0.3,-0.3);
			\draw [-stealth, line width=2pt, cyan] (1.7,0.45) -- ++(-0.45,-0.0);
			\end{tikzpicture}\\
			FBP&		
			TV&
			KSVD&
			BM3D&
			FBPConvNet\\
			\begin{tikzpicture}
			\node[anchor=south west,inner sep=0] (image) at (0,0) {\includegraphics[width=.2\linewidth,height=.2\linewidth]{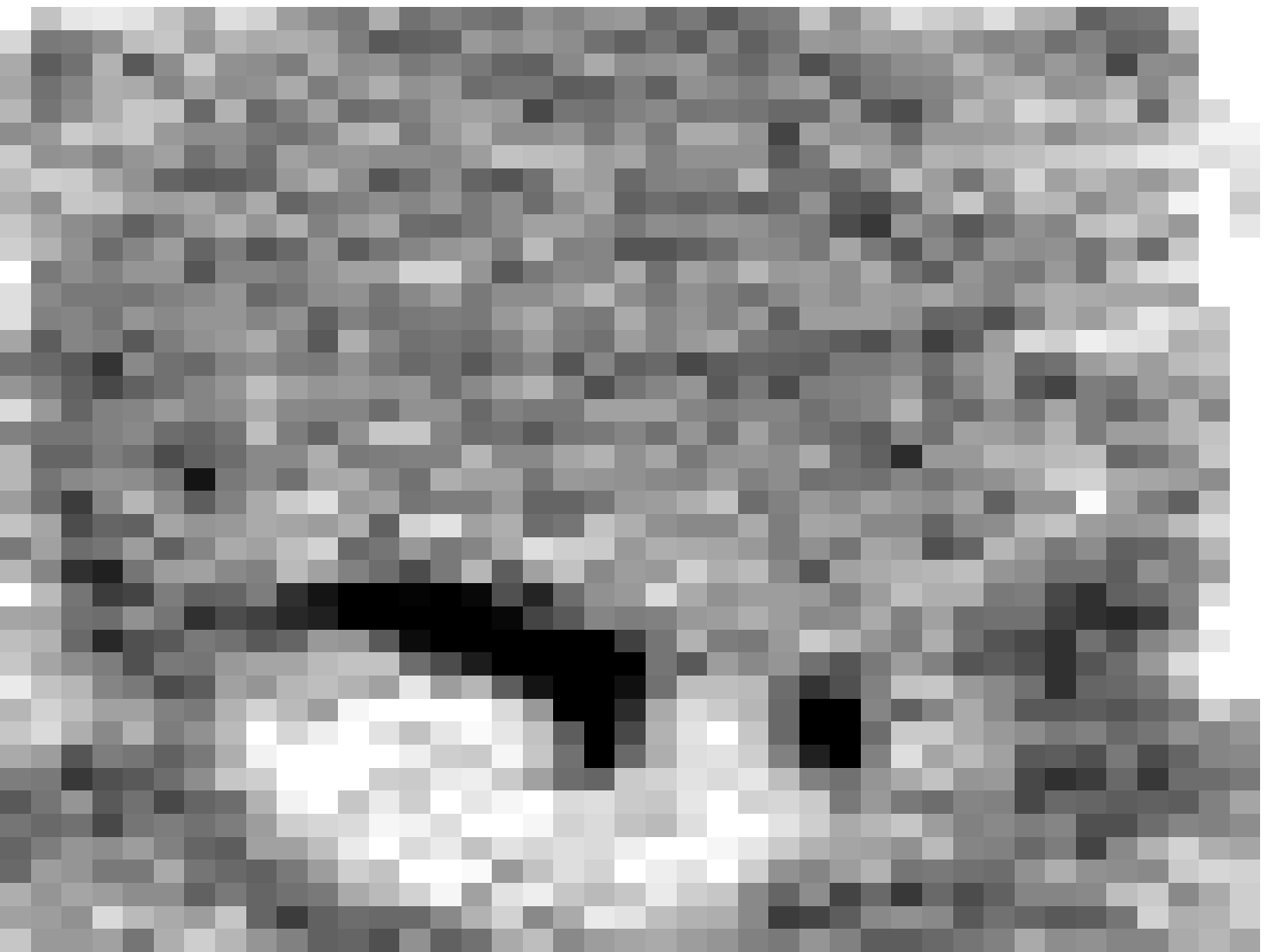}};
			\draw [-stealth, line width=2pt, cyan] (0.7,1.3) -- ++(-0.3,-0.3);
			\draw [-stealth, line width=2pt, cyan] (1.7,0.45) -- ++(-0.45,-0.0);
			\end{tikzpicture}&		
			\begin{tikzpicture}
			\node[anchor=south west,inner sep=0] (image) at (0,0) {\includegraphics[width=.2\linewidth,height=.2\linewidth]{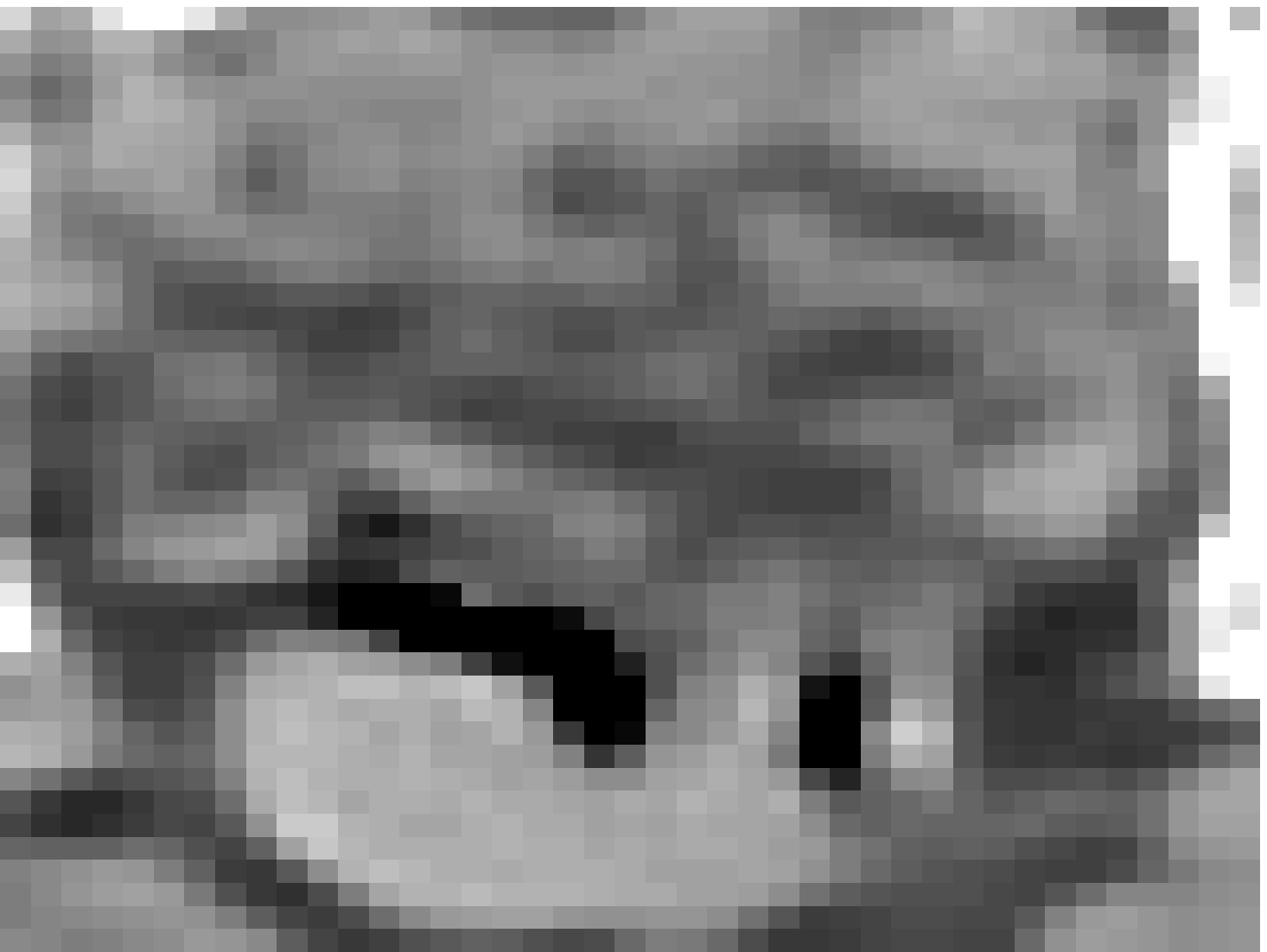}};
			\draw [-stealth, line width=2pt, cyan] (0.7,1.3) -- ++(-0.3,-0.3);
			\draw [-stealth, line width=2pt, cyan] (1.7,0.45) -- ++(-0.45,-0.0);
			\end{tikzpicture}&
			\begin{tikzpicture}
			\node[anchor=south west,inner sep=0] (image) at (0,0) {\includegraphics[width=.2\linewidth,height=.2\linewidth]{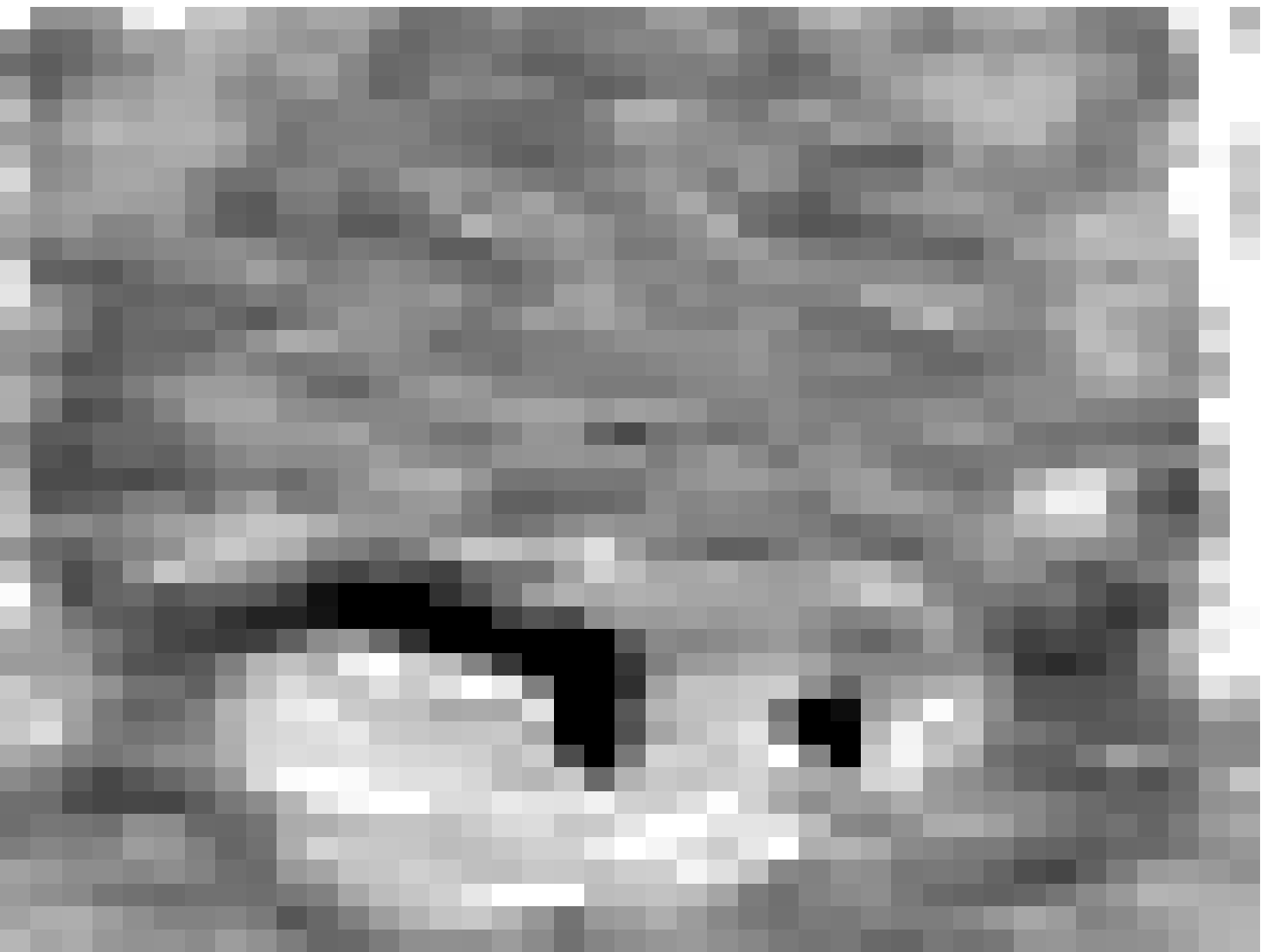}};
			\draw [-stealth, line width=2pt, cyan] (0.7,1.3) -- ++(-0.3,-0.3);
			\draw [-stealth, line width=2pt, cyan] (1.7,0.45) -- ++(-0.45,-0.0);
			\end{tikzpicture}&		
			\begin{tikzpicture}
			\node[anchor=south west,inner sep=0] (image) at (0,0) {\includegraphics[width=.2\linewidth,height=.2\linewidth]{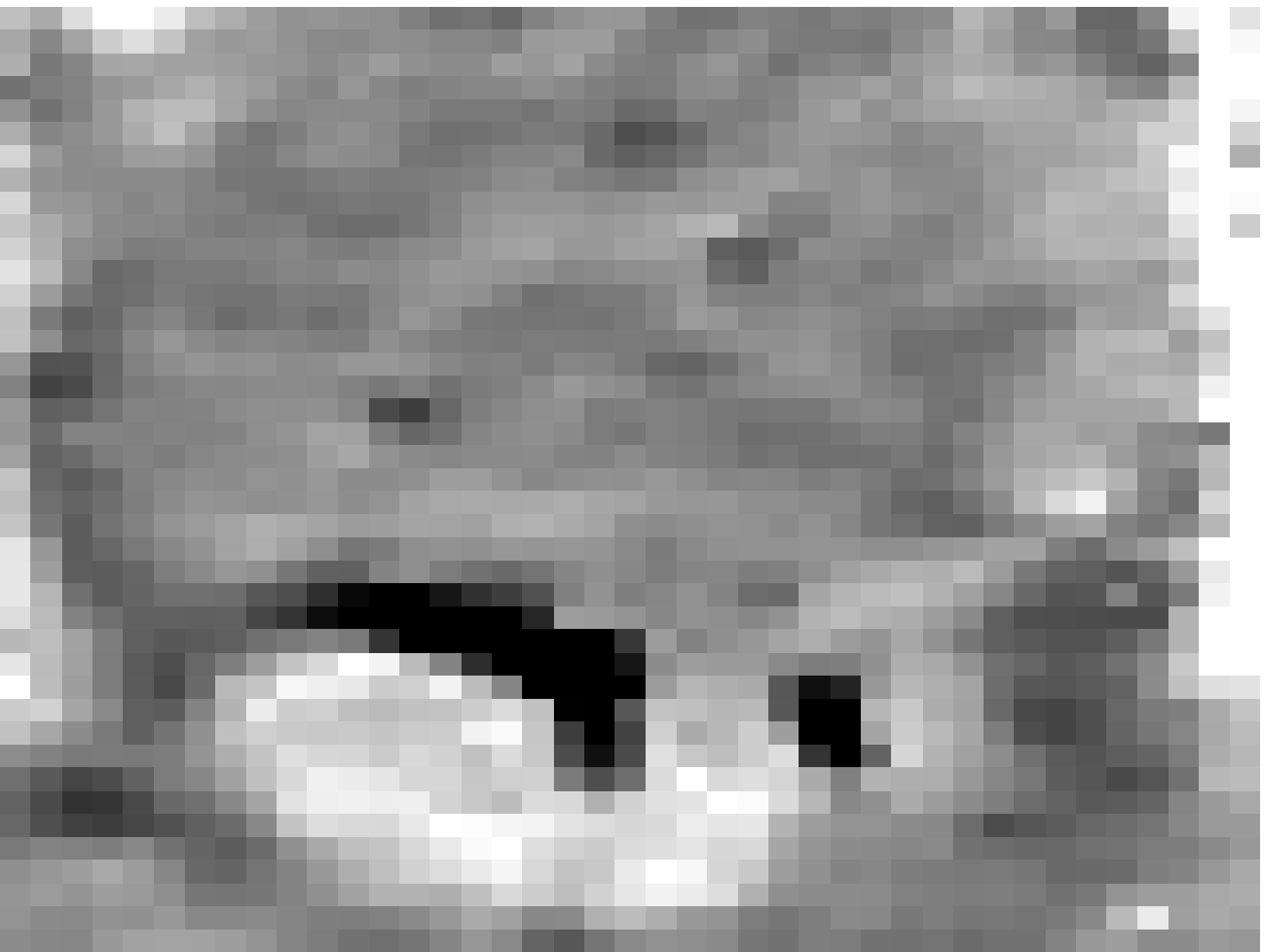}};
			\draw [-stealth, line width=2pt, cyan] (0.7,1.3) -- ++(-0.3,-0.3);
			\draw [-stealth, line width=2pt, cyan] (1.7,0.45) -- ++(-0.45,-0.0);
			\end{tikzpicture}&
			\begin{tikzpicture}
			\node[anchor=south west,inner sep=0] (image) at (0,0) {\includegraphics[width=.2\linewidth,height=.2\linewidth]{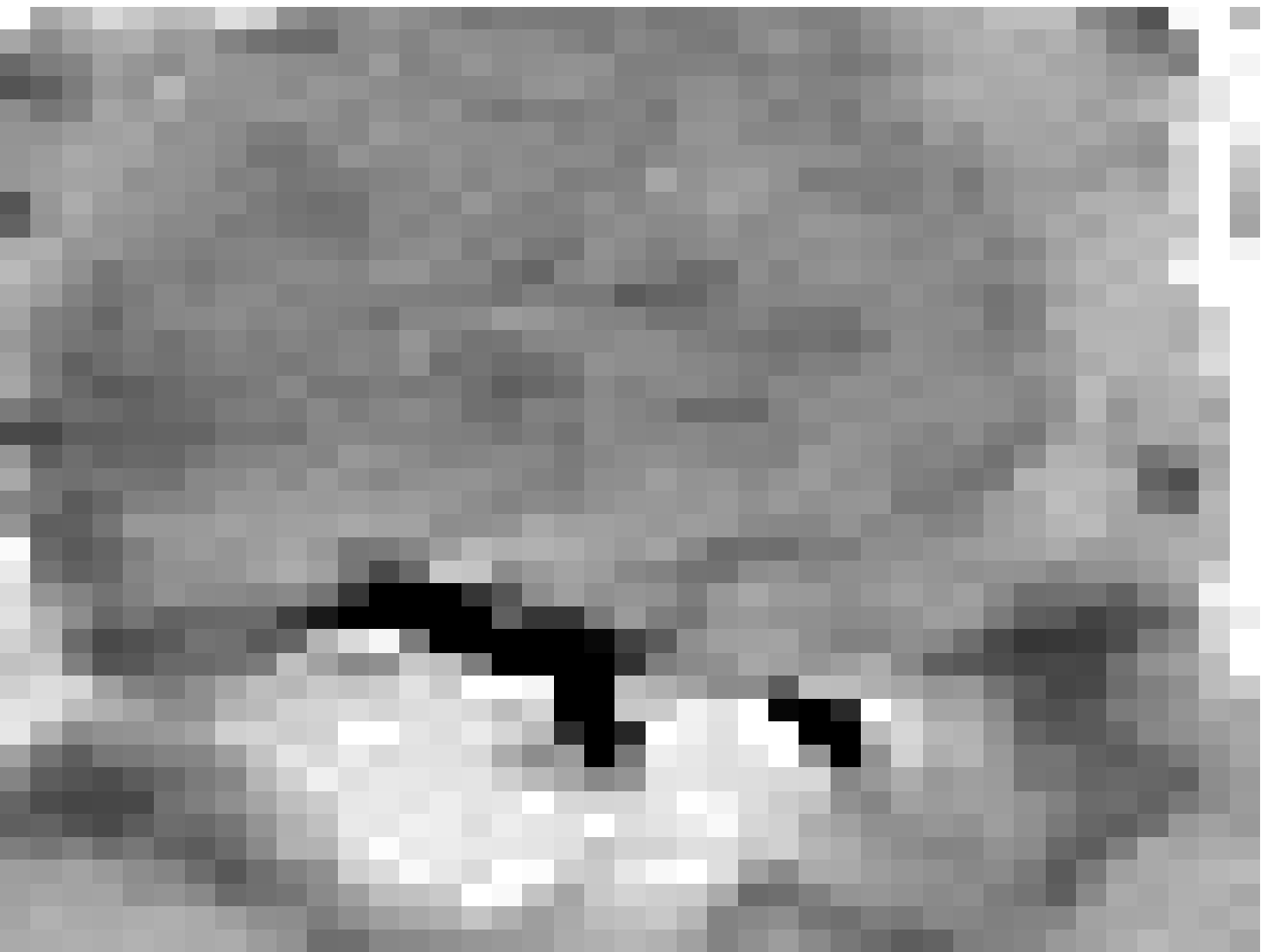}};
			\draw [-stealth, line width=2pt, cyan] (0.7,1.3) -- ++(-0.3,-0.3);
			\draw [-stealth, line width=2pt, cyan] (1.7,0.45) -- ++(-0.45,-0.0);
			\end{tikzpicture}\\
			MoDL&
			Neumann-Net&	
			PGD&
			Learned-PD&	
			AHP-Net
		\end{tabular}
		\caption{Zoom-in results of Fig.~\ref{slice_5000}.
		}
		\label{sliceZoom_5000}
	\end{center}
\end{figure}

Fig. \ref{slice_50000}  and Fig. \ref{slice_5000} shows the  images reconstructed by the universal models trained for varying dose levels.
And the corresponding zoomed-in images are displayed in Fig.~\ref{sliceZoom_50000} and Fig.~\ref{sliceZoom_5000}.

See Table \ref{SNRRMSE} for  quantitative comparison for the results shown in  Fig. \ref{sliceU_50000}, Fig. \ref{sliceU_10000}, Fig. \ref{slice_50000}  and Fig. \ref{slice_5000} .



	\begin{table*}
		\caption{Quantitative metrics on the  reconstruction results for the image slice shown in Fig. \ref{sliceU_50000}, Fig. \ref{sliceU_10000}, Fig. \ref{slice_50000} and Fig. \ref{slice_5000}.}
			\scalebox{1.0}{
    			\begin{tabular}{ccccccccccccc}
    		\hline\hline
    		{Method}                                 &Index      & FBP   &  TV &KSVD &BM3D     &FBPCo-nvNet  &MoDL   &Neuma-nnNet &PGD &Learned-PD  &AHP-Net            \\
			\hline
			\multirow{3}{*}{Fig. \ref{sliceU_50000}}  &PSNR&33.20	 &35.49	 &32.34	 &33.81	 &34.65	 &20.98	 &34.96	 &35.26	     &36.04	 &$\bm{36.05}$\\ 
			                                          &RMSE&22.79	 &17.50	 &25.07	 &21.16	 &19.29	 &93.07	 &18.61	 &17.99	     &16.45	 &$\bm{15.67}$\\ 
			                                          &SSIM&0.9559	 &0.9752 &0.8947 &0.9165 &0.9638 &0.8114 &0.9735 &0.9341	 &0.9750 &$\bm{0.9758}$\\ 
			 \hline                                         
			 \multirow{3}{*}{Fig. \ref{sliceU_10000}} &PSNR&28.31	 &31.65	 &28.37	 &29.46	 &32.00	 &32.88	 &31.35	 &31.83	 &32.48	 &$\bm{34.40}$\\ 
			                                          &RMSE&40.03	 &27.24	 &39.58	 &34.91	 &26.17	 &23.65	 &28.22	 &26.69	 &24.77	 &$\bm{19.85}$\\ 
			                                          &SSIM&0.843	 &0.9355 &0.7351 &0.8046 &0.9437 &0.9498 &0.9511 &0.9300 &0.9515 &$\bm{0.9651}$\\ 
			 \hline
			 \multirow{3}{*}{Fig. \ref{slice_50000}}  &PSNR&33.20	 &35.50	 &32.34	     &33.81	 &33.45	 &32.59	 &33.40	 &34.30	 &35.20	 &$\bm{37.24}$\\ 
			                                          &RMSE&22.79	 &17.50	 &25.07	     &21.16	 &22.14	 &24.46	 &22.28	 &20.08	 &18.10	 &$\bm{14.31}$\\ 
			                                          &SSIM&0.9559	 &0.9752 &0.8947	 &0.9165 &0.9497 &0.9356 &0.9685 &0.9641 &0.9714 &$\bm{0.9783}$\\ 
			 \hline                                         
			 \multirow{3}{*}{Fig. \ref{slice_5000}}   &PSNR&25.12	 &28.09	 &25.53	 &27.42	 &29.50	 &30.41	 &30.92	 &31.14	 &30.82	 &$\bm{33.34}$\\ 
			                                          &RMSE&57.78 	 &41.06	 &54.90	 &44.17	 &34.90	 &31.44	 &29.64	 &28.91	 &29.97	 &$\bm{22.44}$\\ 
			                                          &SSIM&0.7278	 &0.8509 &0.6124 &0.7267 &0.899	 &0.9143 &0.9363 &0.9324 &0.9322 &$\bm{0.9473}$\\ 
			\hline\hline                                                                                                                       
		\end{tabular}
		     	}
		\label{SNRRMSE}
	\end{table*}
	\subsection{Ablation study }
\begin{figure*}
	\begin{center}
		\begin{tabular}{c@{\hspace{0pt}}c@{\hspace{0pt}}c@{\hspace{0pt}}c@{\hspace{0pt}}c@{\hspace{0pt}}c@{\hspace{0pt}}c}
			\includegraphics[width=.198\linewidth,height=.132\linewidth]{MoDLHU_10000_10000_2_38.eps}&
			\includegraphics[width=.198\linewidth,height=.132\linewidth]{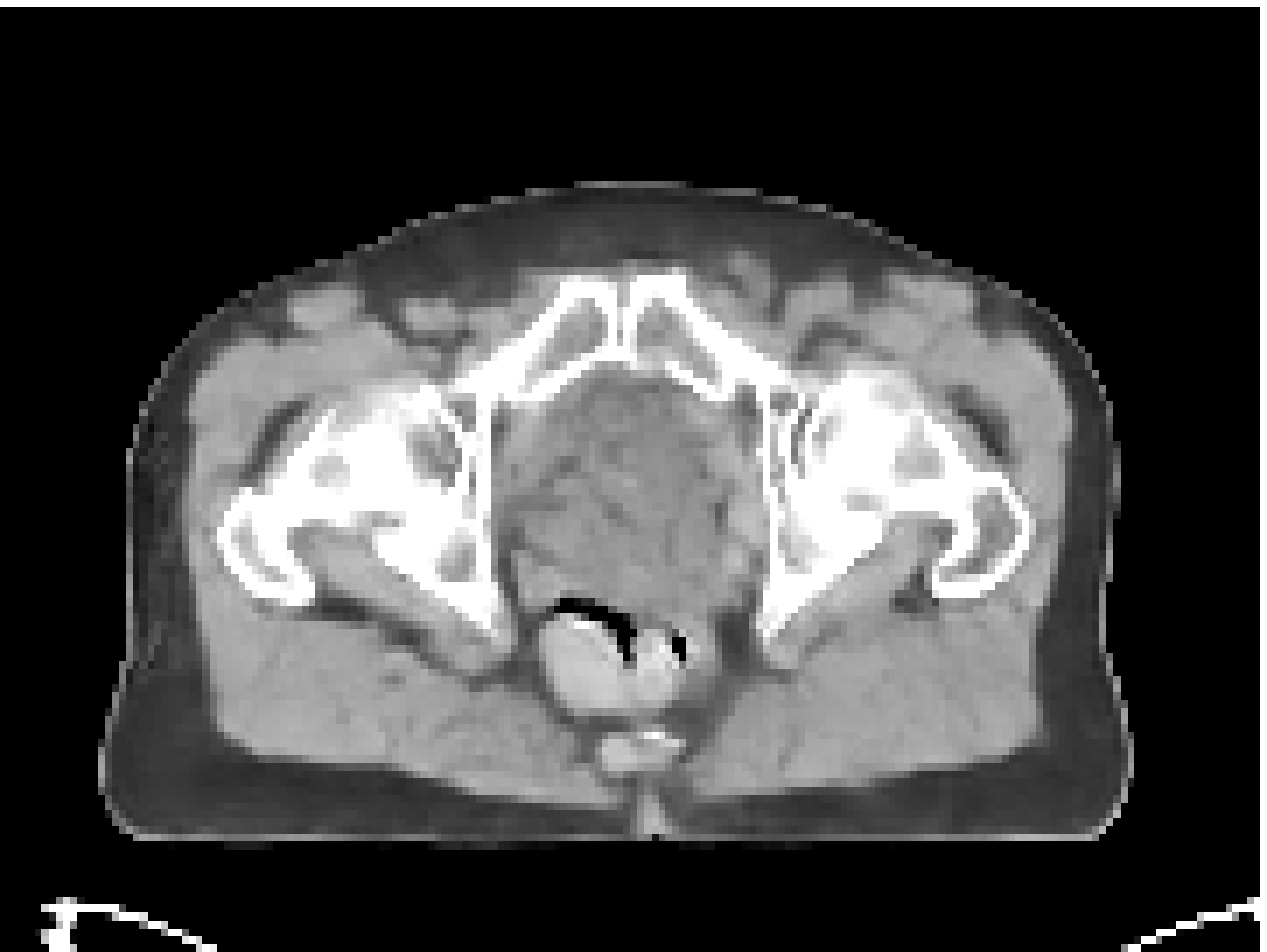}&
			\includegraphics[width=.198\linewidth,height=.132\linewidth]{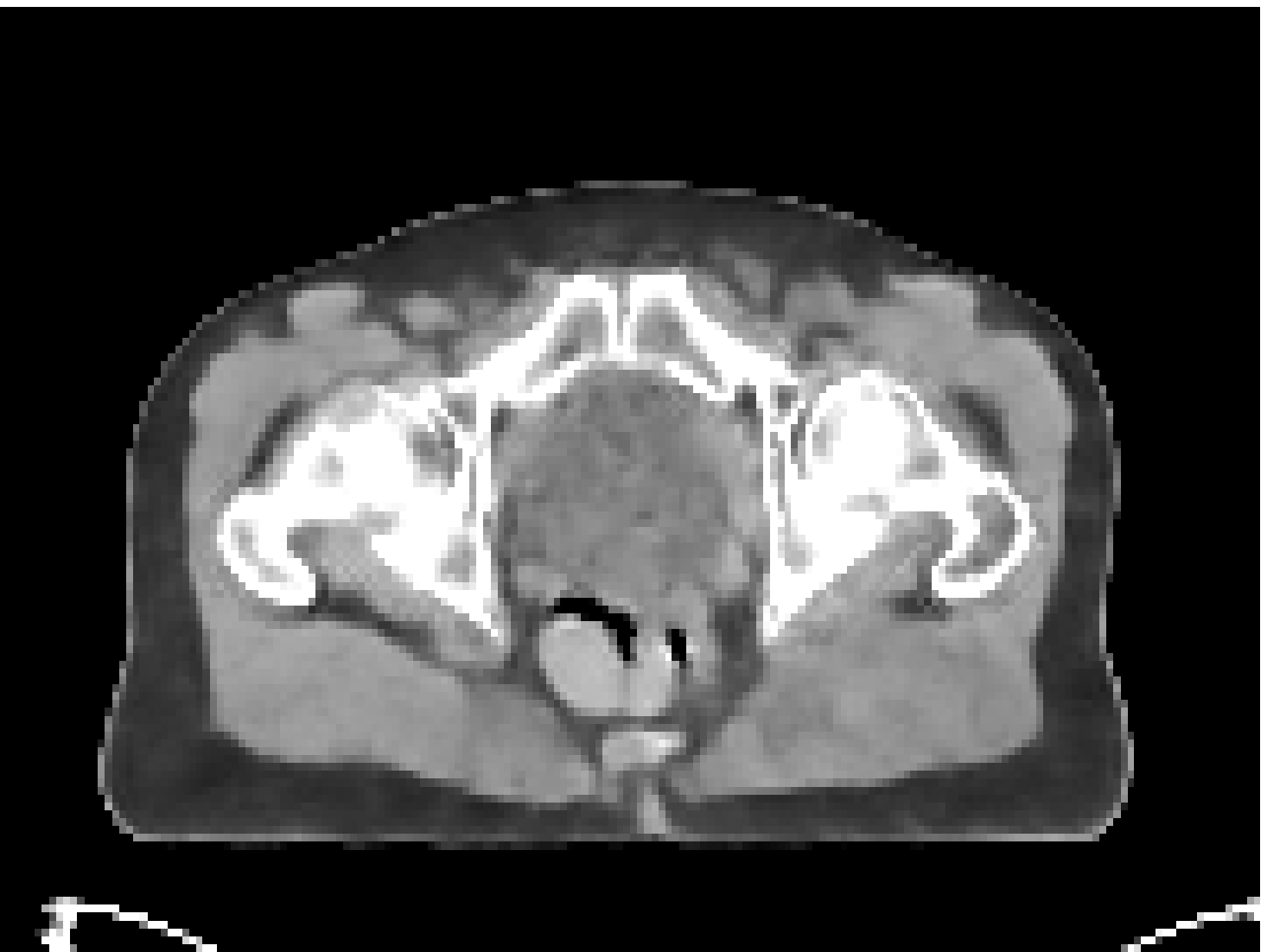}&								
			\includegraphics[width=.198\linewidth,height=.132\linewidth]{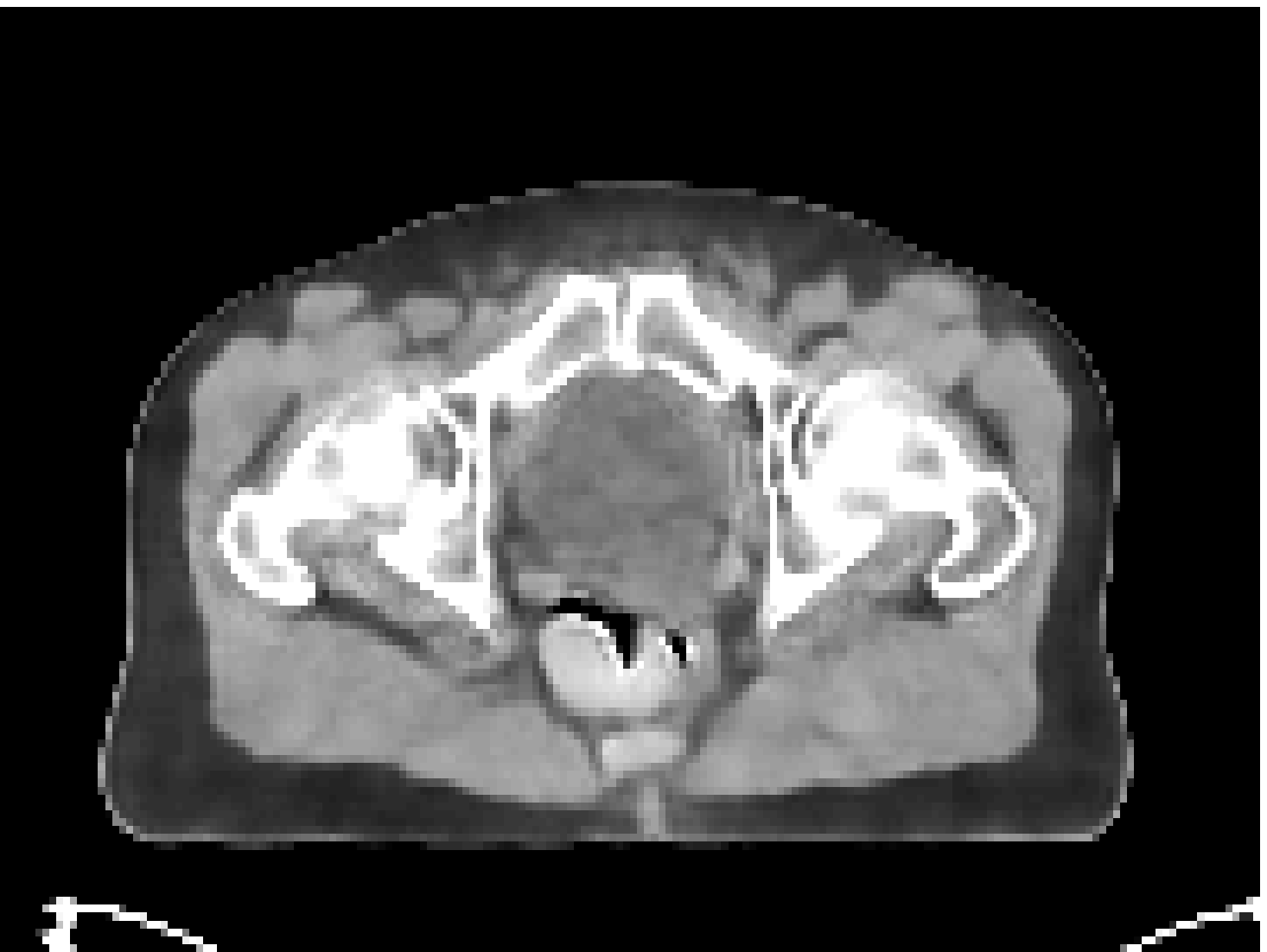}&
			\includegraphics[width=.198\linewidth,height=.132\linewidth]{AHPHU_10000_10000_2_38.eps}\\	
			(a)&
			(b)&		
			(c)&
			(d)&
			(e)
		\end{tabular}
		\caption{Reconstruction results at dose level $I_i=10^4$  by the models trained under same dose level.
			(a) No filter; (b) Using $\nabla$;   (c) Learnable filters;  (d) Learnable HP; (e) AHP-net.
		}
		\label{ablationsliceU_10000}
	\end{center}
\end{figure*}

\begin{figure*}
	\begin{center}
		\begin{tabular}{c@{\hspace{0pt}}c@{\hspace{0pt}}c@{\hspace{0pt}}c@{\hspace{0pt}}c@{\hspace{0pt}}c@{\hspace{0pt}}c}
			\includegraphics[width=.198\linewidth,height=.132\linewidth]{MoDLHU_5000_5000_2_38.eps}&
			\includegraphics[width=.198\linewidth,height=.132\linewidth]{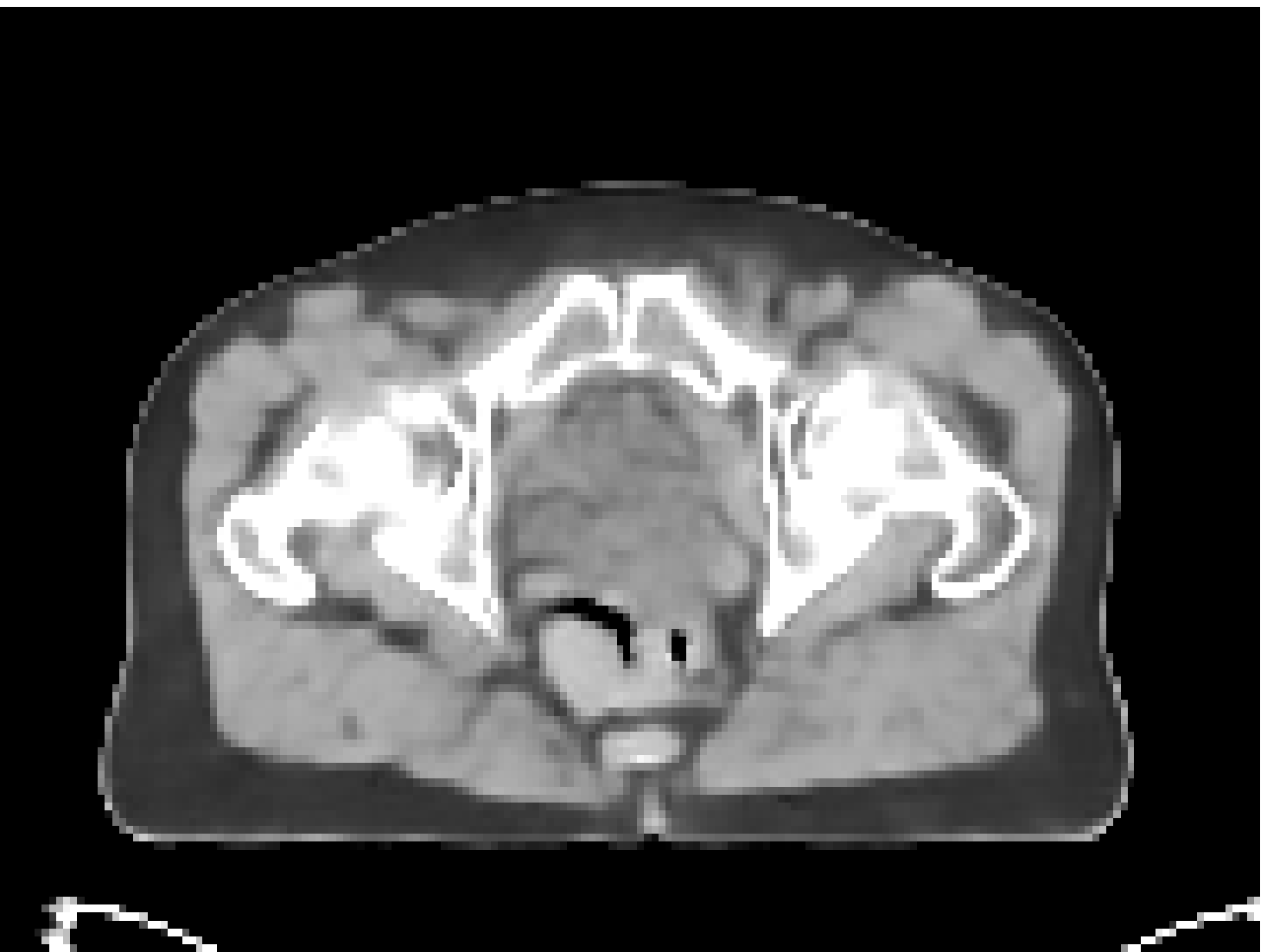}&
			\includegraphics[width=.198\linewidth,height=.132\linewidth]{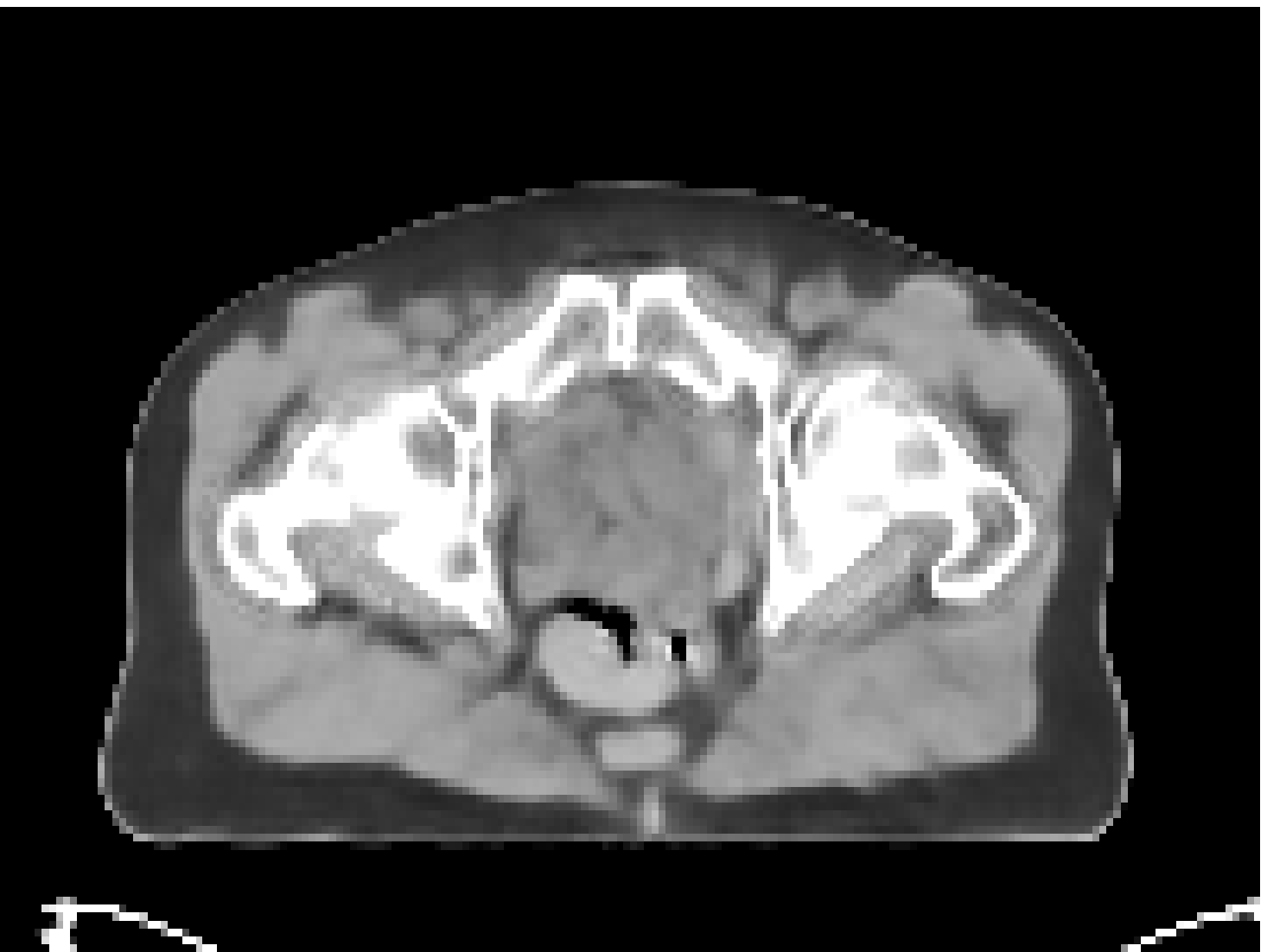}&								
			\includegraphics[width=.198\linewidth,height=.132\linewidth]{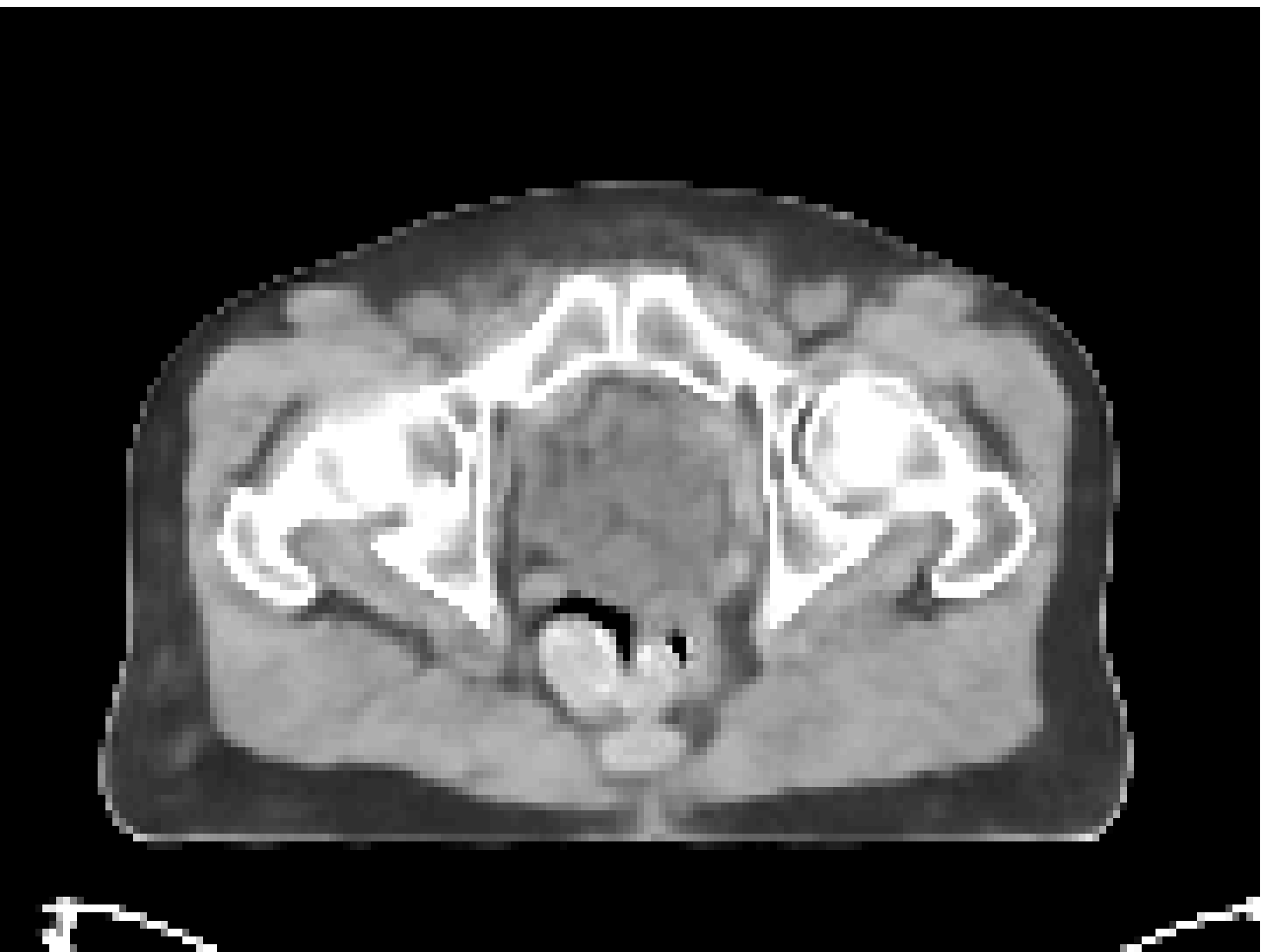}&
			\includegraphics[width=.198\linewidth,height=.132\linewidth]{AHPHU_5000_5000_2_38.eps}\\	
			(a)&
			(b)&		
			(c)&
			(d)&
			(e)
		\end{tabular}
		\caption{Reconstruction results at dose level $I_i=5\times10^3$  by the models trained under same dose level.
			(a) No filter; (b) Using $\nabla$;   (c) Learnable filters;  (d) Learnable HP; (e) AHP-net.
		}
		\label{ablationsliceU_5000}
	\end{center}
\end{figure*}

\begin{figure}
	\begin{center}
		\begin{tabular}{c@{\hspace{-1pt}}c@{\hspace{-1pt}}c@{\hspace{-1pt}}c@{\hspace{-1pt}}c@{\hspace{-1pt}}c@{\hspace{-1pt}}c}
			\begin{tikzpicture}
			\node[anchor=south west,inner sep=0] (image) at (0,0) {\includegraphics[width=.2\linewidth,height=.2\linewidth]{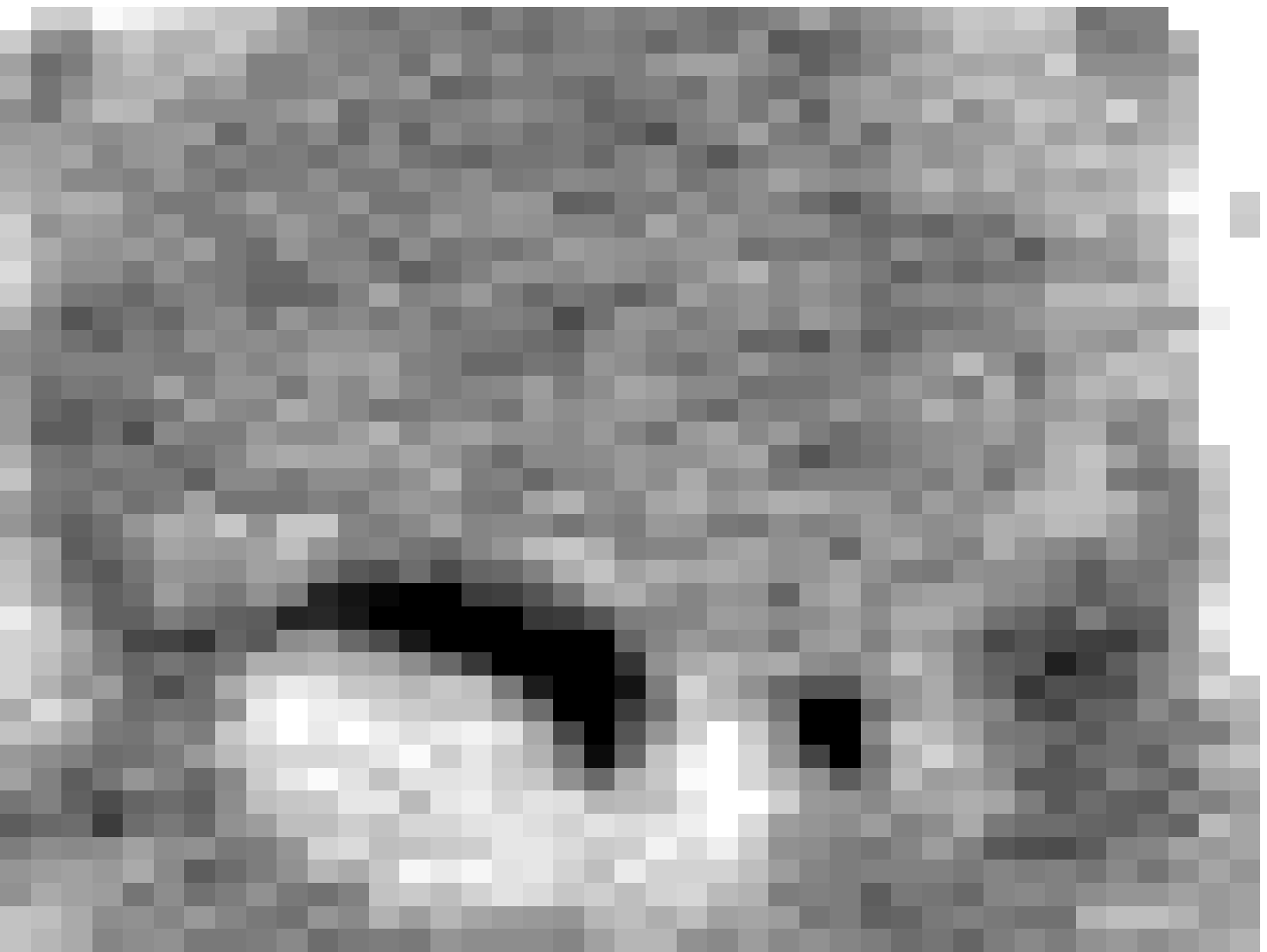}};
			\draw [-stealth, line width=2pt, cyan] (0.7,1.3) -- ++(-0.3,-0.3);
			\draw [-stealth, line width=2pt, cyan] (1.7,0.45) -- ++(-0.45,-0.0);
			\end{tikzpicture}&
			\begin{tikzpicture}
			\node[anchor=south west,inner sep=0] (image) at (0,0) {\includegraphics[width=.2\linewidth,height=.2\linewidth]{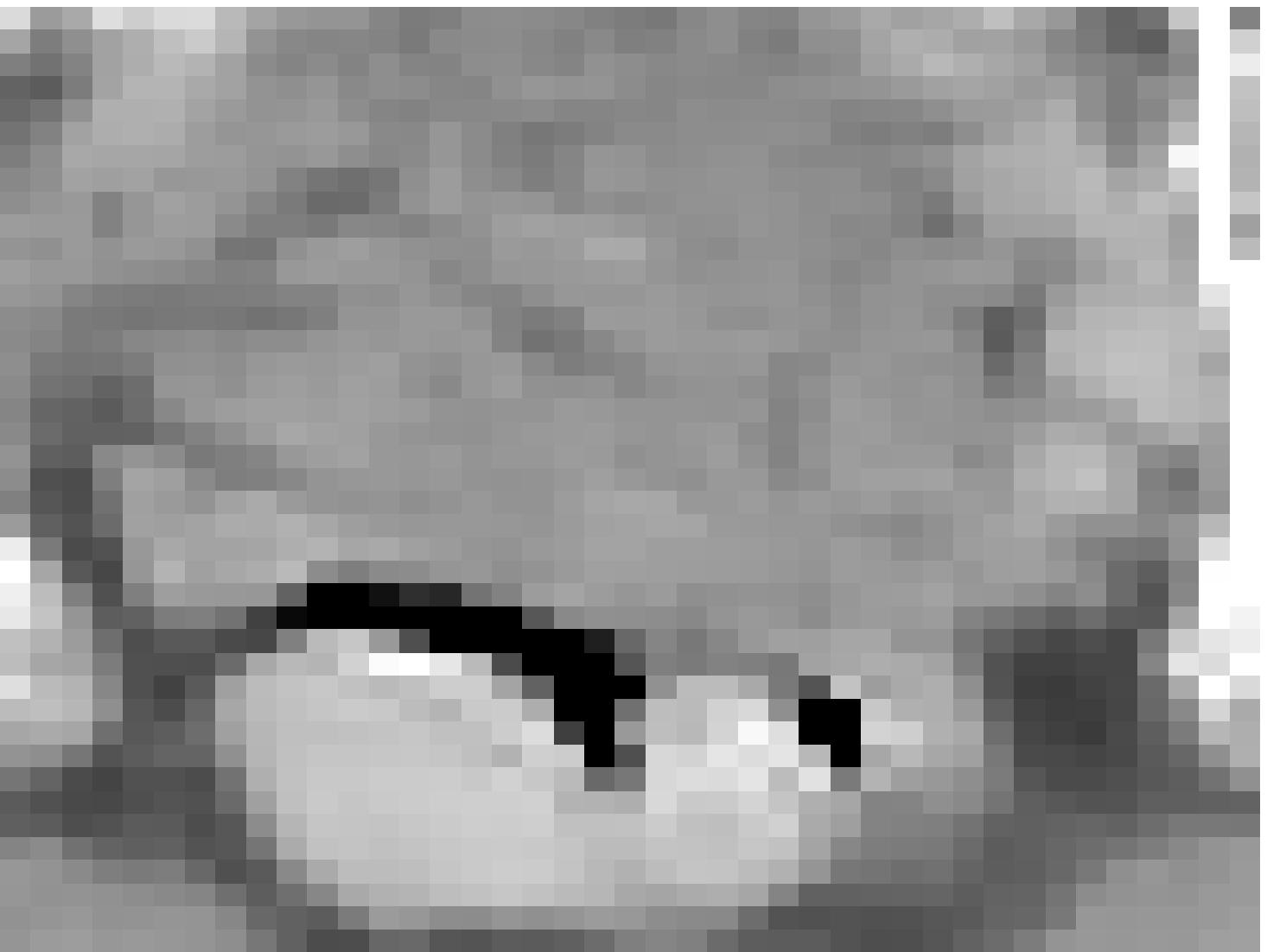}};
			\draw [-stealth, line width=2pt, cyan] (0.7,1.3) -- ++(-0.3,-0.3);
			\draw [-stealth, line width=2pt, cyan] (1.7,0.45) -- ++(-0.45,-0.0);
			\end{tikzpicture}
			&
			\begin{tikzpicture}
			\node[anchor=south west,inner sep=0] (image) at (0,0) {\includegraphics[width=.2\linewidth,height=.2\linewidth]{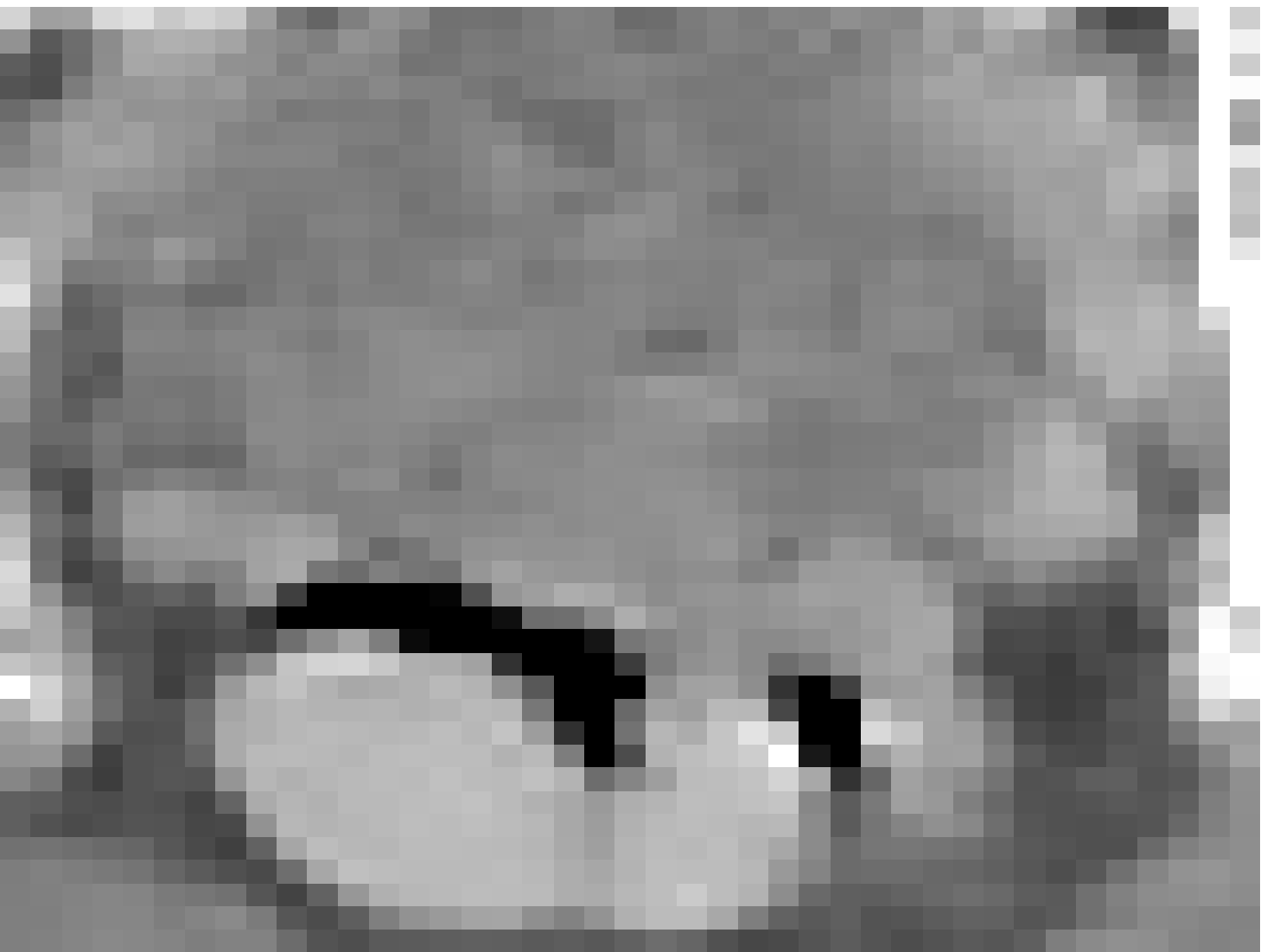}};
			\draw [-stealth, line width=2pt, cyan] (0.7,1.3) -- ++(-0.3,-0.3);
			\draw [-stealth, line width=2pt, cyan] (1.7,0.45) -- ++(-0.45,-0.0);
			\end{tikzpicture}&
			\begin{tikzpicture}
			\node[anchor=south west,inner sep=0] (image) at (0,0) {\includegraphics[width=.2\linewidth,height=.2\linewidth]{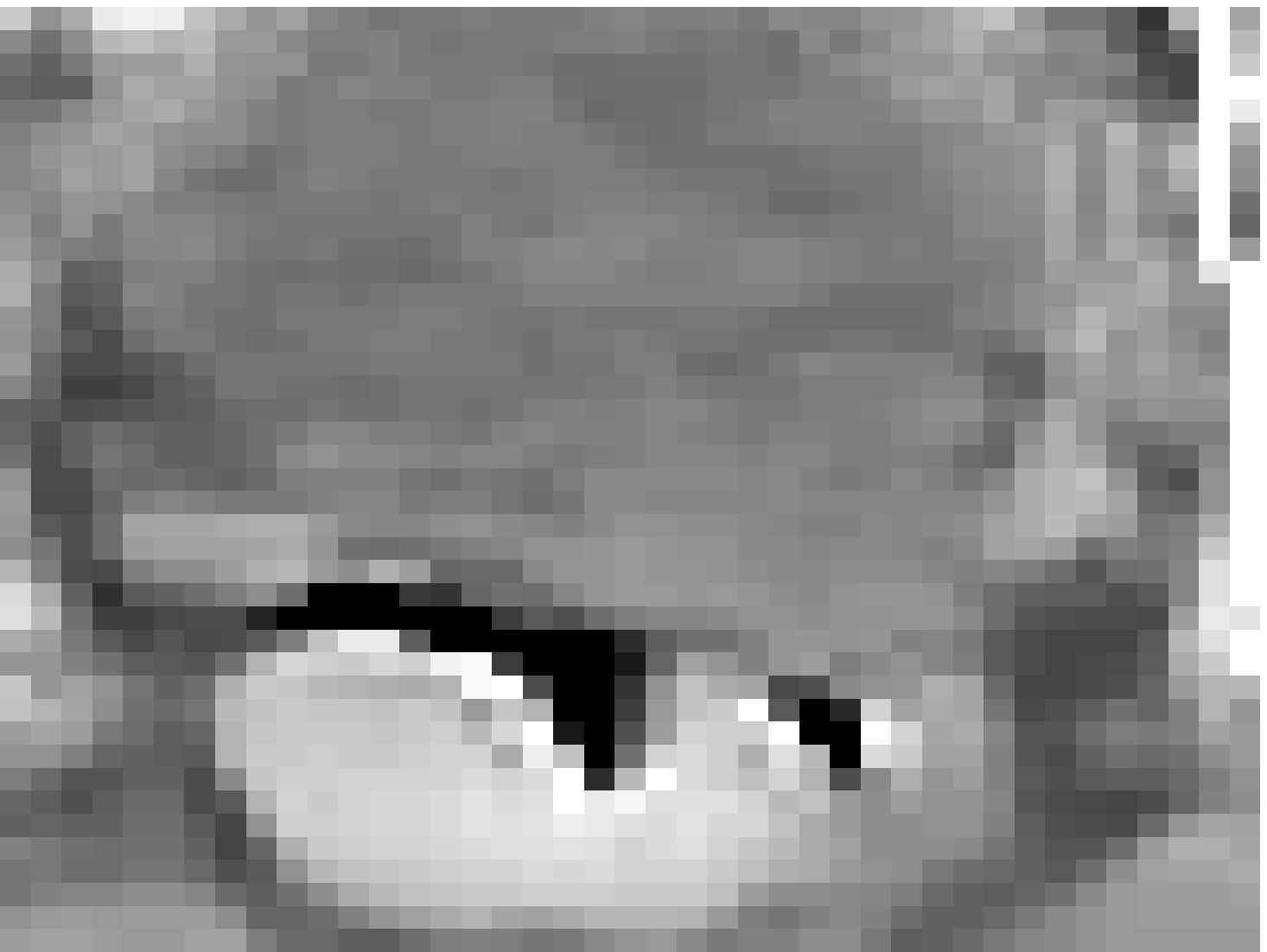}};
			\draw [-stealth, line width=2pt, cyan] (0.7,1.3) -- ++(-0.3,-0.3);
			\draw [-stealth, line width=2pt, cyan] (1.7,0.45) -- ++(-0.45,-0.0);
			\end{tikzpicture}&
			\begin{tikzpicture}
			\node[anchor=south west,inner sep=0] (image) at (0,0) {\includegraphics[width=.2\linewidth,height=.2\linewidth]{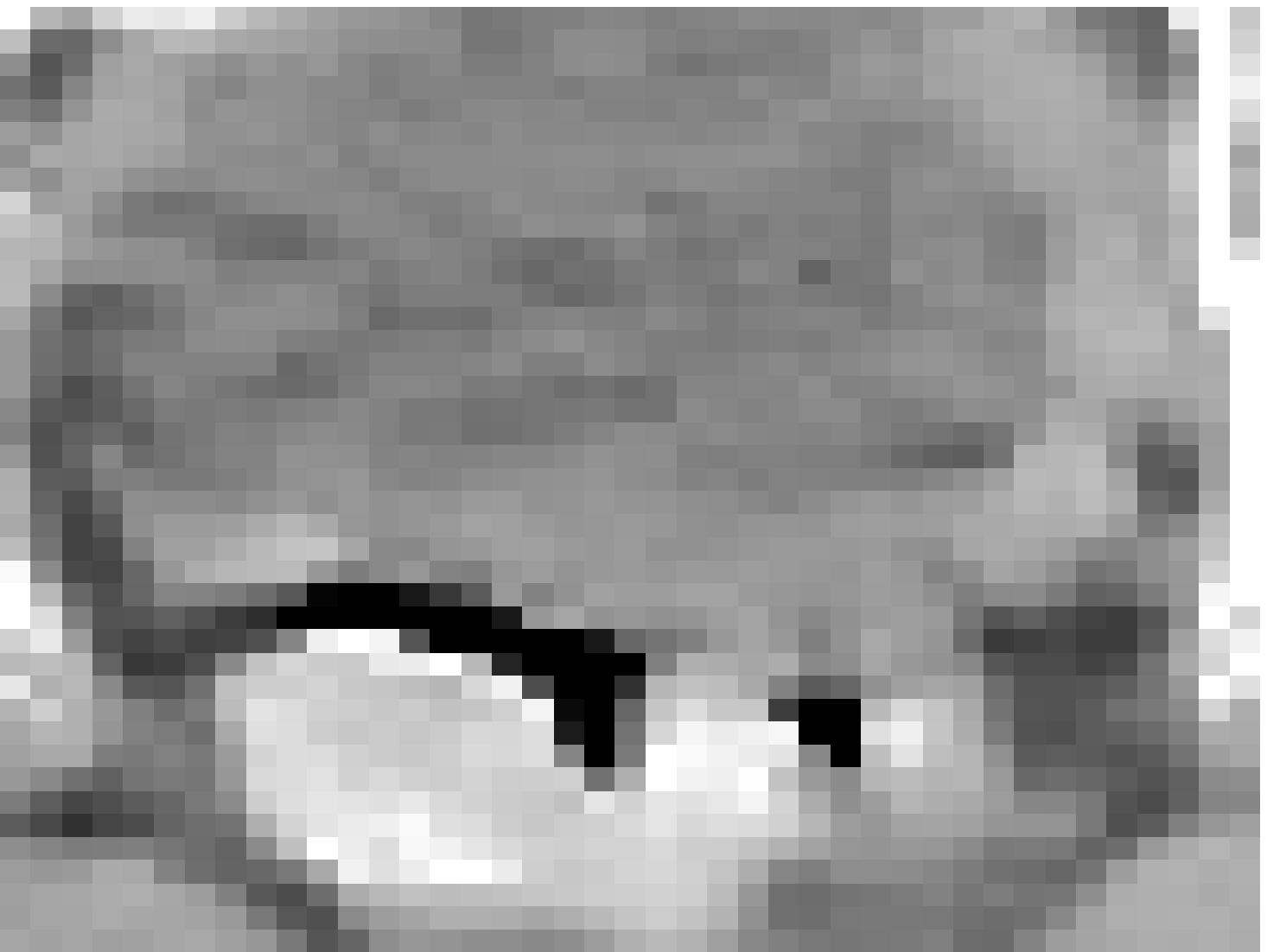}};
			\draw [-stealth, line width=2pt, cyan] (0.7,1.3) -- ++(-0.3,-0.3);
			\draw [-stealth, line width=2pt, cyan] (1.7,0.45) -- ++(-0.45,-0.0);
			\end{tikzpicture}\\		
			(a)&
			(b)&
			(c)&
			(d)&
			(e)
		\end{tabular}
		\caption{Zoom-in results of Fig.~\ref{ablationsliceU_10000} corresponding to the red boxes in Fig. \ref{Truth}.
			(a) No Filter; (b) Using $\nabla$;  (c) Learnable filters;  (d) Learnable HP; (e) AHP-net.
		}
		\label{ablationsliceZoomU_10000}
	\end{center}
\end{figure}

\begin{figure}
	\begin{center}
		\begin{tabular}{c@{\hspace{-1pt}}c@{\hspace{-1pt}}c@{\hspace{-1pt}}c@{\hspace{-1pt}}c@{\hspace{-1pt}}c@{\hspace{-1pt}}c}
			\begin{tikzpicture}
			\node[anchor=south west,inner sep=0] (image) at (0,0) {\includegraphics[width=.2\linewidth,height=.2\linewidth]{MoDLHUZoom_5000_5000_2_38.eps}};
			\draw [-stealth, line width=2pt, cyan] (0.7,1.3) -- ++(-0.3,-0.3);
			\draw [-stealth, line width=2pt, cyan] (1.7,0.45) -- ++(-0.45,-0.0);
			\end{tikzpicture}&
			\begin{tikzpicture}
			\node[anchor=south west,inner sep=0] (image) at (0,0) {\includegraphics[width=.2\linewidth,height=.2\linewidth]{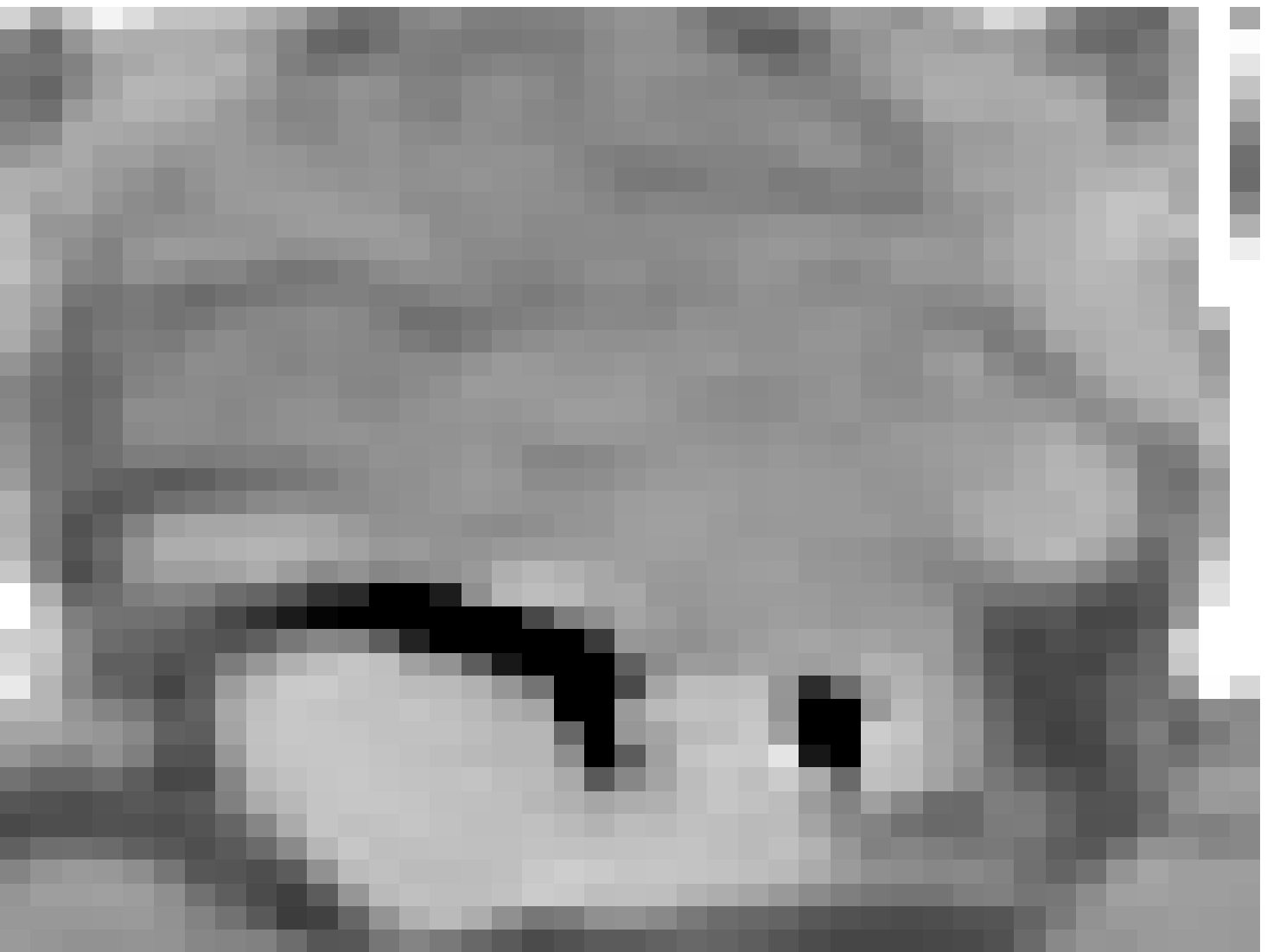}};
			\draw [-stealth, line width=2pt, cyan] (0.7,1.3) -- ++(-0.3,-0.3);
			\draw [-stealth, line width=2pt, cyan] (1.7,0.45) -- ++(-0.45,-0.0);
			\end{tikzpicture}
			&
			\begin{tikzpicture}
			\node[anchor=south west,inner sep=0] (image) at (0,0) {\includegraphics[width=.2\linewidth,height=.2\linewidth]{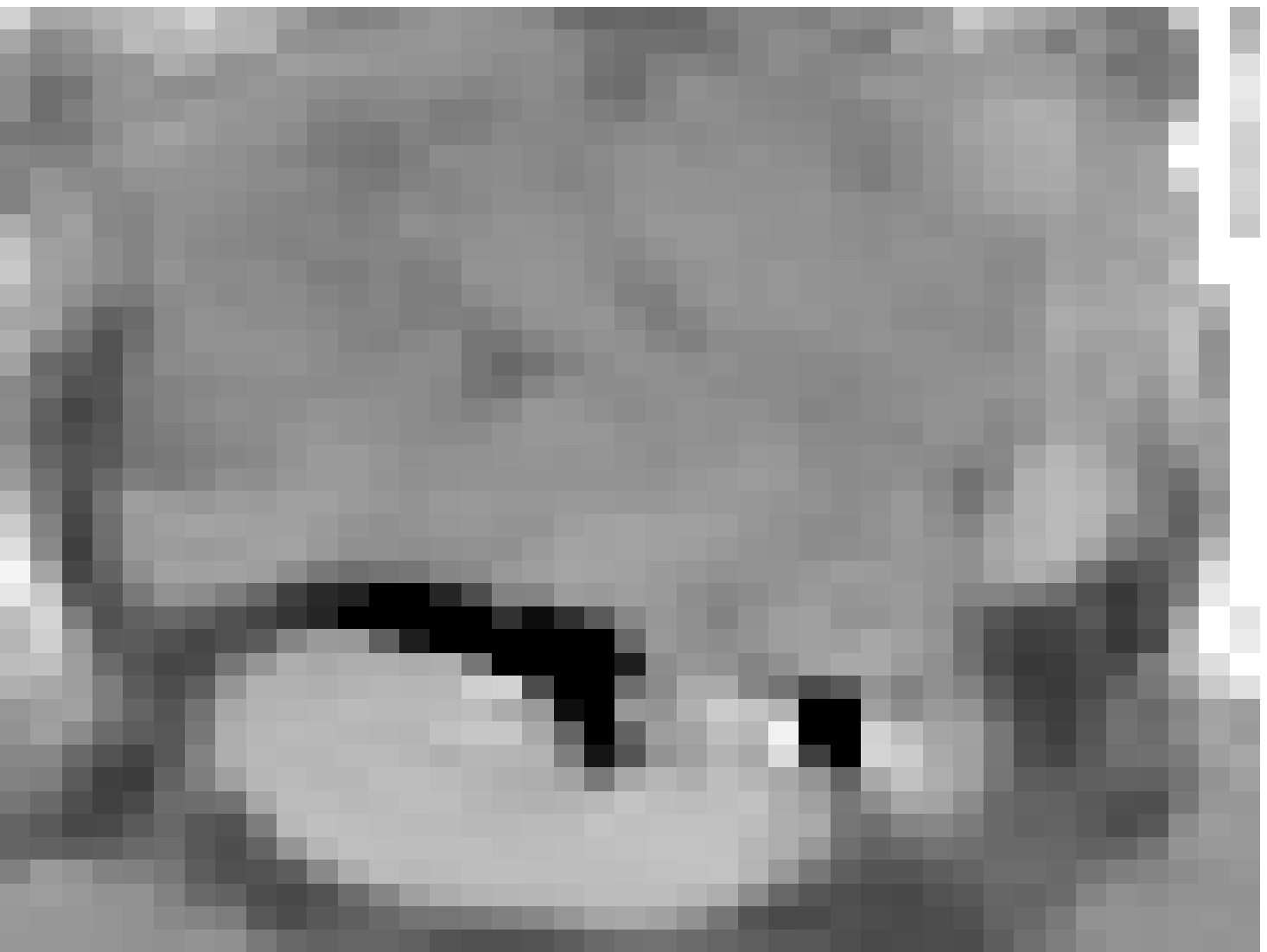}};
			\draw [-stealth, line width=2pt, cyan] (0.7,1.3) -- ++(-0.3,-0.3);
			\draw [-stealth, line width=2pt, cyan] (1.7,0.45) -- ++(-0.45,-0.0);
			\end{tikzpicture}&
			\begin{tikzpicture}
			\node[anchor=south west,inner sep=0] (image) at (0,0) {\includegraphics[width=.2\linewidth,height=.2\linewidth]{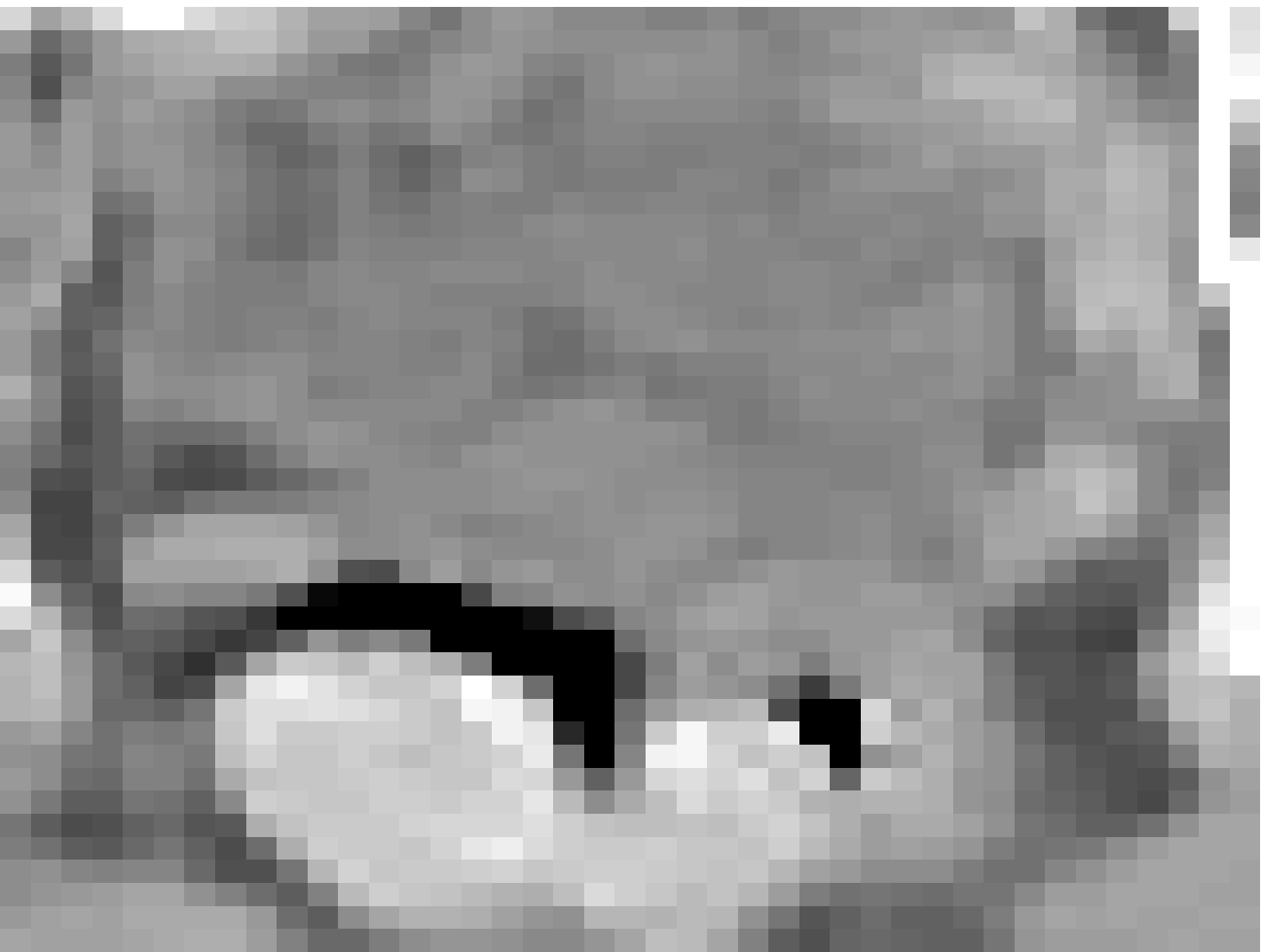}};
			\draw [-stealth, line width=2pt, cyan] (0.7,1.3) -- ++(-0.3,-0.3);
			\draw [-stealth, line width=2pt, cyan] (1.7,0.45) -- ++(-0.45,-0.0);
			\end{tikzpicture}&
			\begin{tikzpicture}
			\node[anchor=south west,inner sep=0] (image) at (0,0) {\includegraphics[width=.2\linewidth,height=.2\linewidth]{AHPHUZoom_5000_5000_2_38.eps}};
			\draw [-stealth, line width=2pt, cyan] (0.7,1.3) -- ++(-0.3,-0.3);
			\draw [-stealth, line width=2pt, cyan] (1.7,0.45) -- ++(-0.45,-0.0);
			\end{tikzpicture}\\		
			(a)&
			(b)&
			(c)&
			(d)&
			(e)
		\end{tabular}
		\caption{Zoom-in results of Fig.~\ref{ablationsliceU_5000} corresponding to the red boxes in Fig. \ref{Truth}.
			(a) No Filter; (b) Using $\nabla$;  (c) Learnable filters;  (d) Learnable HP; (e) AHP-net.
		}
		\label{ablationsliceZoomU_5000}
	\end{center}
\end{figure}

Fig. \ref{ablationsliceU_10000} and Fig. \ref{ablationsliceU_5000}    	show the  images reconstructed by different  versions of the AHP-Net  trained under same dose level
with the dose level of $I_i=1\times10^4$ and $I_i=5\times10^3$. 
And their zoomed-in images  are displayed in Fig. \ref{ablationsliceZoomU_10000} and Fig. \ref{ablationsliceZoomU_5000}.


\end{document}